\documentclass[twocolumn,tighten]{aastex62}

\begin{document}

\title{A Submillimeter Perspective on the GOODS Fields (SUPER GOODS).~III.
A Large Sample of ALMA Sources in the GOODS-S}


\author[0000-0002-6319-1575]{L.~L.~Cowie}
\affiliation{Institute for Astronomy, University of Hawaii,
2680 Woodlawn Drive, Honolulu, HI 96822, USA}

\author[0000-0003-3926-1411]{J.~Gonz{\'a}lez-L{\'o}pez}
\affiliation{N\'ucleo de Astronom\'ia de la Facultad de Ingenier\'ua y Ciencias, Universidad Diego Portales,
Av. Ej\'ercito Libertador 441, Santiago, Chile}
\affiliation{Instituto de Astrof\'isica and Centro de Astroingenier\'ia, Facultad de F\'isica, 
Pontificia Universidad Cat\'olica de Chile, Casilla 306, Santiago 22, Chile}

\author[0000-0002-3306-1606]{A.~J.~Barger}
\affiliation{Department of Astronomy, University of Wisconsin-Madison,
475 N. Charter Street, Madison, WI 53706, USA}
\affiliation{Department of Physics and Astronomy, University of Hawaii,
2505 Correa Road, Honolulu, HI 96822, USA}
\affiliation{Institute for Astronomy, University of Hawaii, 2680 Woodlawn Drive,
Honolulu, HI 96822, USA}

\author{F.~E.~Bauer}
\affiliation{Instituto de Astrof\'isica and Centro de Astroingenier\'ia, Facultad de F\'isica, 
Pontificia Universidad Cat\'olica de Chile, Casilla 306, Santiago 22, Chile}
\affiliation{Millennium Institute of Astrophysics (MAS), Nuncio Monse{\~{n}}or S{\'{o}}tero 
Sanz 100, Providencia, Santiago, Chile} 
\affiliation{Space Science Institute,
4750 Walnut Street, Suite 205, Boulder, Colorado 80301, USA} 

\author{L.-Y.~Hsu}
\affiliation{Institute for Astronomy, University of Hawaii,
2680 Woodlawn Drive, Honolulu, HI 96822, USA}

\author{W.-H.~Wang}
\affiliation{Academia Sinica Institute of Astronomy and Astrophysics,
P.O. Box 23-141, Taipei 10617, Taiwan}

\begin{abstract}
We analyze the $>4\sigma$ sources in the most sensitive 100~arcmin$^2$ 
area (rms $<0.56$~mJy) of a SCUBA-2 850~$\mu$m survey of the GOODS-S 
and present the 75 band~7 ALMA sources ($>4.5\sigma$) obtained from 
high-resolution interferometric follow-up observations. 
The SCUBA-2---and hence ALMA---samples should be complete to 2.25~mJy.
Of the 53 SCUBA-2 sources in this complete sample, only five have no ALMA 
detections, while 13\% (68\% confidence range 7--19\%)
have multiple ALMA counterparts. Color-based high-redshift dusty galaxy 
selection techniques find at most 55\% of the total ALMA sample. 
In addition to using literature spectroscopic and optical/NIR photometric redshifts, 
we estimate FIR photometric redshifts based on an Arp~220 template. 
We identify seven $z\gtrsim4$ candidates.
We see the expected decline with redshift of the 4.5~$\mu$m and 24~$\mu$m to 
850~$\mu$m flux ratios, confirming these as good diagnostics of $z\gtrsim4$
candidates. We visually classify 52 ALMA sources, finding 44\%
(68\% confidence range 35--53\%) to be apparent mergers.
We calculate rest-frame $2-8$~keV and $8-28$~keV luminosities using the 
7~Ms {\em Chandra\/} X-ray image.
Nearly all of the ALMA sources detected at $0.5-2$~keV
are consistent with a known X-ray luminosity to 850~$\mu$m 
flux relation for star-forming galaxies, while most of those detected at
$2-7$~keV are moderate luminosity AGNs that lie just above
 the $2-7$~keV detection 
threshold. The latter largely have substantial obscurations of $\log N_H=23-24$~cm$^{-2}$, 
but two of the high-redshift candidates may even be Compton thick.
\end{abstract}


\section{Introduction}
\label{secintro}
Deep imaging with the Submillimeter Common-User Bolometer Array (SCUBA; Holland et al.\ 1999) 
on the 15~m James Clerk Maxwell Telescope (JCMT) revolutionized studies of the distant universe 
by revealing the presence of distant, dusty,  extremely luminous galaxies 
(e.g., Smail et al.\ 1997; Barger et al.\ 1998; Hughes et al.\ 1998; Eales et al.\ 1999). 
These so-called submillimeter galaxies (SMGs; see Casey et al. 2014 for a review) 
are some of the most powerfully star-forming galaxies in the universe and
are significant contributors to the total star formation history from
$z\sim2$ to at least $z\sim5$ (Barger et al.\ 2000, 2012, 2014; Chapman et al.\ 2005; 
Wardlow et al.\ 2011; Casey et al.\ 2013; Swinbank et al.\ 2014; Cowie et al.\ 2017). 

SMGs have subsequently been detected by a wide range of single-dish submillimeter 
and millimeter cameras and telescopes. The second generation camera
\hbox{SCUBA-2} (Holland et al.\ 2013) is by far the most powerful of these, covering
50~arcmin$^2$ per pointing (16 times larger than SCUBA) 
with a sensitivity 5 times fainter than the Large APEX BOlometer CAmera (LABOCA;
Siringo et al.\ 2009) on the 12~m Atacama Pathfinder EXperiment (APEX). 
The angular resolution on the sky of \hbox{SCUBA-2} ($\sim14''$ FWHM PSF at 
850~$\mu$m) is also much better than that of LABOCA ($\sim19''$ at
870~$\mu$m), or than that of the space-based {\em Herschel\/} ($\sim35''$ at 
500~$\mu$m) and {\em Planck\/} ($\sim4.8'$ at 850~$\mu$m) satellites, yielding 
more accurate positions and suffering less source blending. The natural limit of 
single-dish submillimeter observations is the depth at which confusion---the blending 
of sources and/or where the noise is dominated by unresolved 
contributions from fainter sources---becomes important. 

The lack of positional accuracy is a major concern when trying to determine
the properties of SMGs. Historically, moderately deep radio interferometric 
images were used to identify counterparts 
(e.g., Barger et al.\ 2000; Ivison et al.\ 2002; Chapman et al.\ 2003).
Using the deepest Karl G. Jansky Very Large Array (VLA) 1.4~GHz image of the sky
(11.5~$\mu$Jy at $5\sigma$; Owen 2018), Cowie et al.\ (2017) showed that 
75\% of the 154 SCUBA-2 sources with 850~$\mu$m fluxes 
$>2$~mJy in the Great Observatories Origins Deep Survey-North field 
(GOODS-N; Giavalisco et al.\ 2004) had one or more radio counterparts
(66\% had single radio counterparts); however, 39
had no viable radio counterparts at all within the match radius.

Millimeter/submillimeter interferometry with the IRAM Plateau de Bure Interferometer,
the Submillimeter Array (SMA; Ho et al.\ 2004), 
and now the Atacama Large Millimeter/submillimeter Array (ALMA) have become essential tools
for obtaining precise positions and resolving single-dish submillimeter sources into multiple SMG
counterparts (e.g., Wang et al.\ 2011; Barger et al.\ 2012, 2014; Smol{\v c}i{\'c} et al.\ 2012; 
Chen et al.\ 2013; Karim et al.\ 2013; Hodge et al.\ 2013; Miettinen et al.\ 2015; Simpson et al.\ 2015).
Even SCUBA-2's angular resolution at 450~$\mu$m ($\sim7''$ FWHM PSF) is
insufficient for making accurate counterpart identifications at other wavelengths,
yet such identifications are critical for estimating photometric redshifts, modeling 
spectral energy distributions (SEDs), and determining basic properties.

Fully mapping the SMG population requires a ``wedding cake" approach. 
The bottom layer of the cake corresponds to the most wide-field (and shallowest) 
submillimeter surveys (e.g., South Pole Telescope, {\em Herschel\/}, {\em Planck}). 
Because these surveys primarily trace the most extreme (and rare) star-forming 
galaxies in the universe, they are intriguing for probing the upper bound of parameter 
space. However, they generally lie at a characteristic star formation rate (SFR) limit of
$\sim1000\,M_{\odot}$\,yr$^{-1}$ above which galaxies become
quite rare (Karim et al.\ 2013; Barger et al.\ 2014).

Thus, to assess how ``typical" massive galaxy formation occurs, it is critical to
characterize fainter SMGs (defined here as sources with 850~$\mu$m fluxes
$\sim2-5$~mJy), which comprise the knee of the luminosity distribution and 
produce 20$-30$\% of the extragalactic background light (EBL) at this 
wavelength. This middle layer of the wedding cake is best probed by SCUBA-2 
surveys, which can find sources down to the JCMT confusion limit of $\sim1.65$~mJy 
at 850~$\mu$m (Cowie et al.\ 2017). 
This corresponds to SFRs~$\ga220$\,$M_{\odot}$\,yr$^{-1}$ over
$z=1-8$ for a Kroupa (2001) initial mass function (IMF) and an Arp~220 template.
At these fluxes, the SFRs are higher than found 
in any extinction-corrected UV-selected sample (Reddy \& Steidel 2009; 
van der Burg et al. 2010). Thus, one can only discover these galaxies with 
submillimeter observations (Barger et al.\ 2014; Cowie et al.\ 2017). Moreover,
with a sky density of only $\sim1$ per 1~arcmin$^2$ at 2~mJy (e.g., Hsu et al.\ 2016),
the most efficient approach to studying these galaxies in detail is by targeting 
single-dish discovered SMGs with small field-of-view interferometers like ALMA, 
rather than by direct ALMA imaging searches.

Finally, the deepest but smallest area component---the top layer of the wedding 
cake---comes from direct submillimeter interferometric imaging surveys
(e.g., Aravena et al.\ 2016; Hatsukade et al.\ 2016; Dunlop et al.\ 2017; Ueda et al.\ 2017).
Such observations are best suited to probing the faintest SMGs, 
which begin to overlap with UV-selected samples.

We are writing a series of papers on our deep SCUBA-2 observations of the 
GOODS fields, which we call the SUbmillimeter PERspective on the 
GOODS fields (SUPER GOODS) series. In the first two papers in the series
(Cowie et al.\ 2017 or Paper~I, and Barger et al.\ 2017 or Paper~II), we presented
our SCUBA-2 observations of the GOODS-N/CANDELS/{\em Chandra\/} Deep
Field-North (CDF-N) complemented with targeted SMA submillimeter 
interferometry, Karl G. Jansky Very Large Array (VLA) radio interferometry, and 
{\em Herschel\/} imaging. 

In this third paper in the series, we present our 
SCUBA-2 observations of the GOODS-S/CANDELS/ {\em Chandra\/} Deep 
Field-South (CDF-S) complemented with targeted ALMA observations
and {\em Herschel\/} imaging. We made observations with SCUBA-2 to near 
the confusion limit for a $4\sigma$ detection.
This sample represents an order of magnitude increase 
in the surface density of sources 
compared to the wider and shallower LABOCA LESS survey
(Wei\ss\ et al.\ 2009) and provides the deepest SMG sample 
that can be generated with single-dish submillimeter telescope 
observations of blank fields. 
In ALMA Cycles~3 and 4, we targeted our 
SCUBA-2 850~$\mu$m sample with band~7 observations.
The SCUBA-2 sample contains an unbiased,
complete sample to 2.25~mJy, which means the 
ALMA sample above that flux should be as well, since the SCUBA-2
fluxes are upwardly Eddington biased and may also be composed
of blends; thus, the ALMA fluxes are expected to be lower relative to the
SCUBA-2 fluxes.
 
We chose the GOODS-S field due to the unparalleled breadth and
depth of the multiwavelength ancillary coverage (e.g., see Guo et al.\ 2013,
hereafter G13, for a detailed summary of the imaging data).
Our SCUBA-2 region boasts deep ground-based coverage in $U'UBVRIzYJHK_s\/$ 
plus many medium and narrowbands (e.g., Hildebrandt et al.\ 2006; Nonino et al.\ 2009;
Cardamone et al.\ 2010; Retzlaff et al.\ 2010; Hsieh et al.\ 2012; Fontana et al.\ 2014;
Straatman et al.\ 2016, hereafter S16); deep {\it HST} 
imaging from GOODS-S (Giavalisco et al.\ 2004), 
CANDELS (Grogin et al.\ 2011; Koekemoer et al.\ 2011), 
HUDF (e.g., Beckwith et al.\ 2006; Koekemoer et al.\ 2013; Teplitz et al.\ 2013),
HDUV (Oesch et al.\ 2015), and ERS (Windhorst et al.\ 2011);
ultradeep {\em Spitzer\/} (3.6--70~$\mu$m, e.g., Magnelli et al.\ 2011; Ashby et al.\ 2015) 
and {\em Herschel\/} (100--500~$\mu$m, 
Elbaz et al.\ 2011; Oliver et al.\ 2012; Magnelli et al.\ 2013) imaging; 
hundreds of hours of VLT and Keck multiobject spectroscopy 
(e.g., Szokoly et al.\ 2004; Le F\`evre et al.\ 2005; Popesso et al.\ 2009;
Treister et al.\ 2009; Balestra et al.\ 2010; Silverman et al.\ 2010; Cooper et al.\ 2012;
Suh et al.\ 2015) plus 3D-HST (Brammer et al.\ 2012; Momcheva et al.\ 2016) for 
redshifts of many 1000s of galaxies; and photometric redshift estimates from multiple
codes and template SED sets (e.g., Santini et al.\ 2009, 2015; Rafferty et al.\ 2011; 
Dahlen et al.\ 2013; Hsu et al.\ 2014; Skelton et al.\ 2014; S16).
Such extensive coverage allows immediate characterization of
ALMA-detected sources with sufficient sensitivity to constrain tightly
the dust SEDs (da Cunha et al.\ 2015; Scoville et al.\ 2016).

In addition, the SCUBA-2 data are well-matched to the deepest portion of 
the deepest X-ray image ever taken, the 7\,Ms {\em Chandra\/} 
exposure (Luo et al.\ 2017). These unique X-ray data make it possible for us 
to investigate the SFR versus X-ray luminosity relation for SMGs and to 
see which of the SMGs contain active galactic nuclei (AGNs).

In Section~\ref{secdata}, we describe the SCUBA-2 observations and reductions, 
construct the SCUBA-2 source catalog, and compare with previous single-dish 
observations. We also present the ALMA observations and reductions, 
construct the ALMA source catalog, and compare with the Franco et al.\ (2018,
hereafter, F18) ALMA 1.13~mm observations.
In Section~\ref{secALMAvsS2}, we compare the SCUBA-2 and ALMA
fluxes, astrometry, and relative positional offsets, and we determine 
the fraction of SCUBA-2 sources with multiple ALMA counterparts.
In Section~\ref{secancillary}, we describe the ancillary data and identify 
multiwavelength counterparts to the ALMA sources.  
We search the literature for optical/NIR spectroscopic and photometric redshifts, 
which we critically assess. We examine how well various color-based selections 
proposed in the literature do at finding high-redshift dusty galaxies. We estimate 
FIR photometric redshifts, which we compare with the spectroscopic and 
photometric redshifts. Finally, we show that the 4.5~$\mu$m and 24~$\mu$m to
850~$\mu$m flux ratios drop rapidly with increasing redshift. 
In Section~\ref{sechighz}, we tabulate candidate high-redshift ($z>4$) sources.
In Section~\ref{secsfh}, we study the flux dependence of the redshifts and 
construct the star formation history.
In Section~\ref{secmorph}, we visually inspect the morphologies and merger 
properties of all ALMA counterparts that are bright enough to do this.
In Section~\ref{secxray}, we calculate the X-ray luminosities,
examine whether the origin of the X-ray emission is predominantly due to
AGN activity or star formation, and use the X-ray photon indices to determine 
the column densities.
In Section~\ref{secsummary}, we summarize our results.

We assume the Wilkinson Microwave
Anisotropy Probe cosmology of $H_0=70.5$~km~s$^{-1}$~Mpc$^{-1}$,
$\Omega_{\rm M}=0.27$, and $\Omega_\Lambda=0.73$ 
(Larson et al.\ 2011) throughout.

\section{Data}
\label{secdata}

\subsection{SCUBA-2 Observations}
\label{secscuba2}
We made observations from late 2011 through early 2017
using SCUBA-2, which obtains simultaneous data at 450~$\mu$m 
and 850~$\mu$m. The mean aimpoint of our SCUBA-2 observations is 
R.A. 03:32:26.49, Decl.~$-27$:48:29.0 (J2000.0).
In the central region, we used the \textsc{CV Daisy} scan 
pattern---whose field size ($5\farcm5$ radius;
by this radius the noise is twice the central noise) 
is well matched to the deepest portion of the X-ray image---to obtain the 
maximum depth.
To cover the outer regions, we used the 
larger field \textsc{PONG-900} scan pattern ($10\farcm5$ radius; by this
radius the noise is twice the central noise), which finds brighter but rarer sources.
Detailed information on the SCUBA-2 scan patterns
can be found in Holland et al.\ (2013).

\begin{deluxetable}{ccc}
\tablecaption{CDF-S SCUBA-2 Observations}
\tablehead{Weather & Scan & Exposure\\ 
Band & Pattern & (Hr)}
\startdata
1  &    \textsc{CV Daisy}  &8.5   \\
2   &     \textsc{CV Daisy} & 45.0  \\
3  &    \textsc{CV Daisy}  &5.5        \\
1  &    \textsc{PONG-900}  & 6.0      \\
2  &    \textsc{PONG-900}  & 32.1       \\
3  &    \textsc{PONG-900}  & 12.0  
\enddata
\label{obs}
\end{deluxetable}

We summarize our SCUBA-2 GOODS-S observations in Table~1, 
where we give the weather conditions
(band~1, $\tau_{\rm 225~GHz}<0.05$; band~2, $0.05<\tau_{\rm 225~GHz}<0.08$;
band~3, $0.08<\tau_{\rm 225~GHz}<0.12$), scan patterns used, and exposure times. 
Including the poorer weather band~3 observations, we have
109~hr of exposure time in the final 850~$\mu$m reductions, roughly equally divided 
between the two scan patterns.  We did not use the small amount of band~3 data in
the 450~$\mu$m reductions, and with predominantly band~2 (rather than band~1) 
weather, the 450~$\mu$m observations are not very deep (central rms of
3.6~mJy at 450~$\mu$m versus 0.30~mJy at 850~$\mu$m).
Thus, we primarily focus here on the 850~$\mu$m results.

\begin{figure*}[ht]
\centerline{\includegraphics[width=14cm,angle=0]{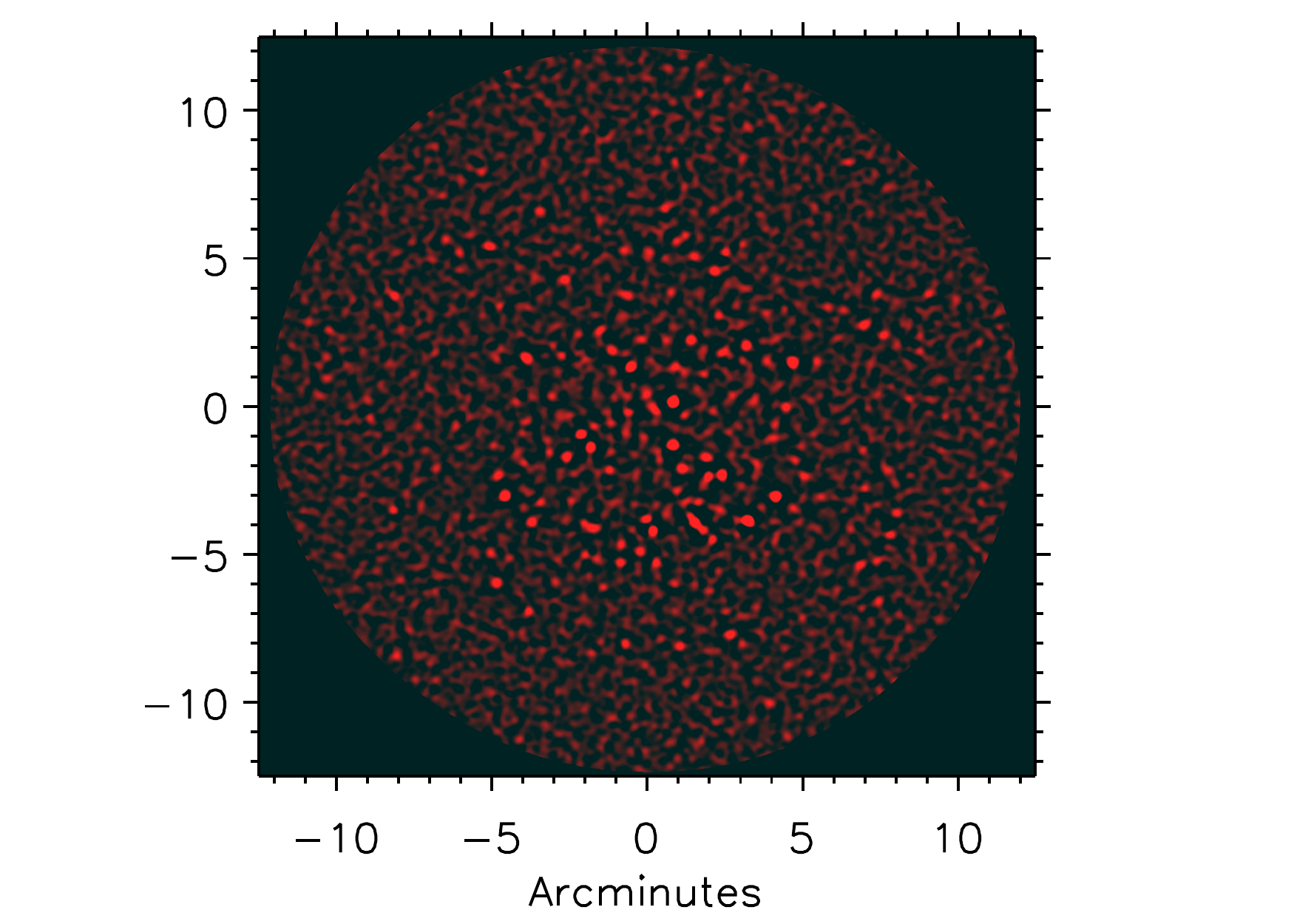}}
\caption{
850~$\mu$m matched filter S/N image of the CDF-S region.
The more sensitive central region (radius less than $\sim6'$)
is dominated by the \textsc{CV Daisy}
observations, while the outer region is covered by
the \textsc{PONG-900} observations.
\label{pretty_850}
}
\end{figure*}

\begin{figure*}
\includegraphics[width=8.5cm]{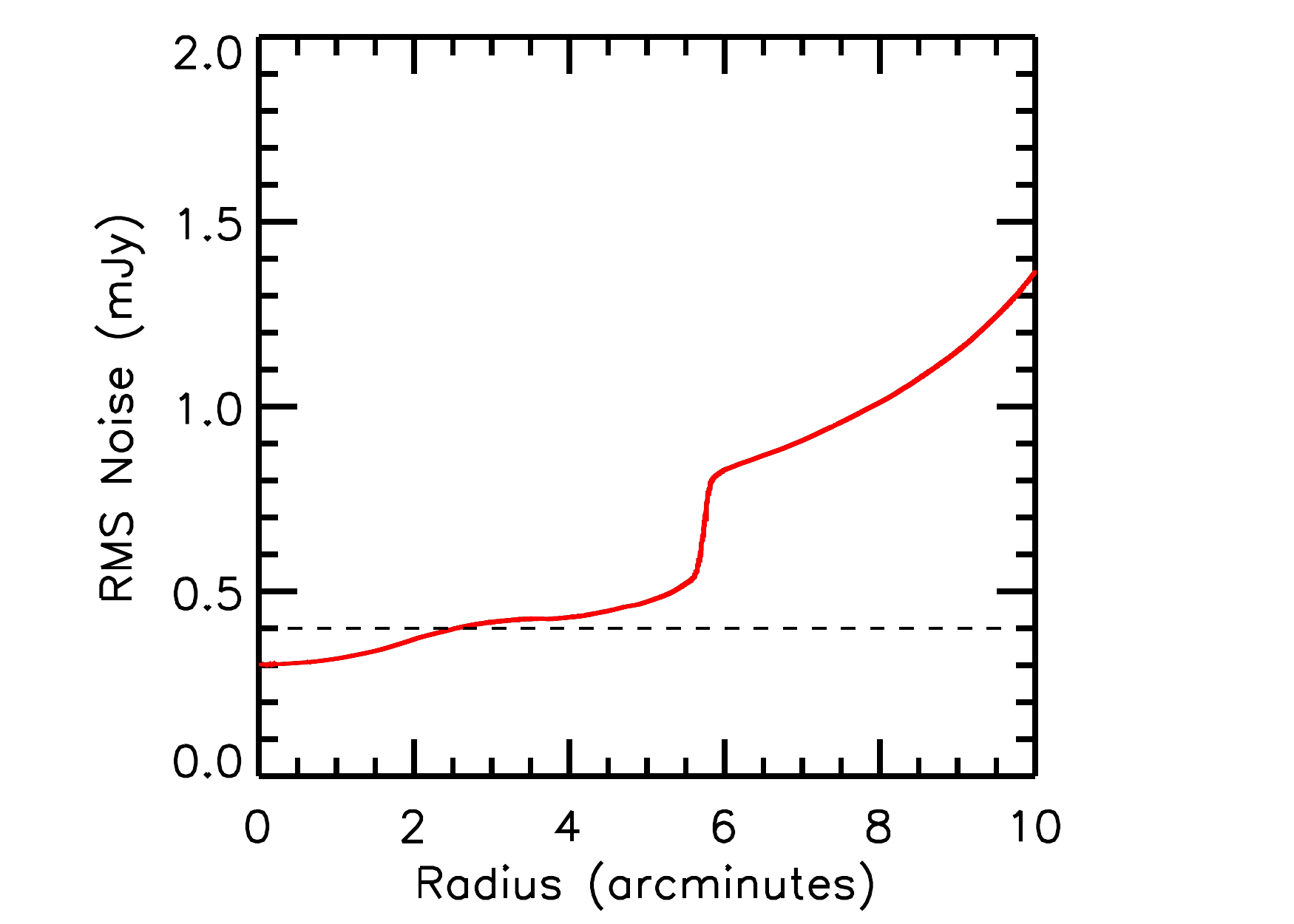}
\includegraphics[width=8.5cm]{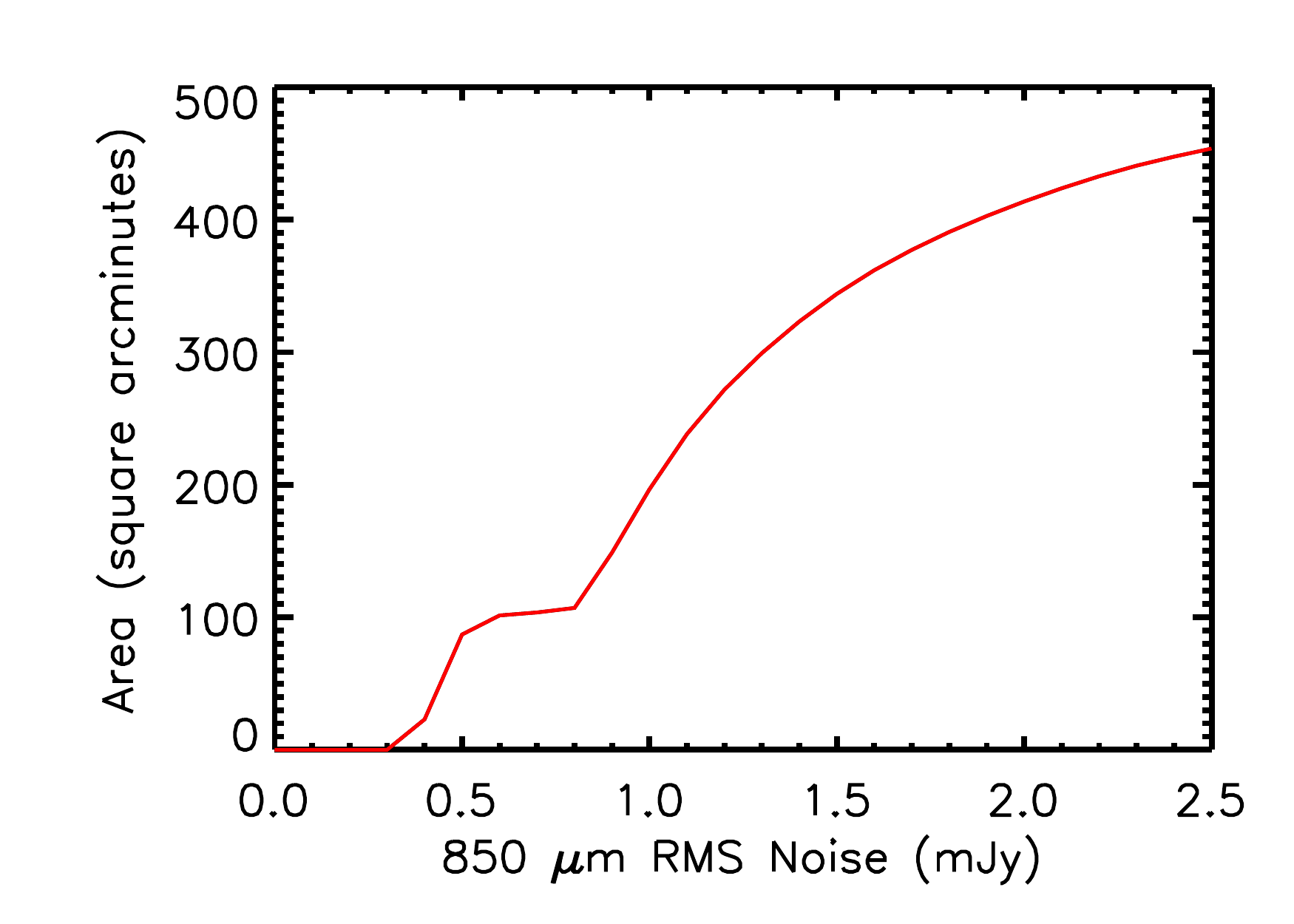}
\caption{
{\em (Left)\/} Azimuthally averaged 850~$\mu$m rms noise vs. radius.
The more sensitive central region (radius less than $\sim6'$)
is dominated by the \textsc{CV Daisy}
observations, while the outer region is covered by
the \textsc{PONG-900} observations. The black dashed horizontal
line shows the rms noise corresponding to a $4\sigma$ detection
threshold of 1.6~mJy, which is approximately the confusion limit
for the JCMT at 850~$\mu$m of $\sim$1.65~mJy. 
{\em (Right)\/} Cumulative area covered vs. 850~$\mu$m rms noise.
\label{noise_radius}
}
\end{figure*}

Our reduction procedures follow Chen et al.\ (2013)
and are described in detail in Paper~I. The 
galaxies are expected to appear as unresolved sources at the
$\sim 14''$ resolution of the JCMT at 850~$\mu$m.
We therefore applied a matched-filter to our maps, which
provides a maximum likelihood estimate of the source strength for
unresolved sources (e.g., Serjeant et al.\ 2003).

In Figure~\ref{pretty_850}, we show the 850~$\mu$m matched-filter 
S/N image of the CDF-S 
made from all of the observations, including the band~3 data.
In Figure~\ref{noise_radius}, we show the rms noise versus radius {\em (left)\/}, 
as well as the cumulative area observed below a given rms noise {\em (right)\/}.

\subsection{SCUBA-2 Source Catalog Construction}
\label{secscuba2cat}
As in Paper~I, we generated the source catalogs by
identifying the peak S/N pixel, subtracting this peak pixel
and its surrounding areas using the PSF scaled and centered
on the value and position of that pixel, and then searching
for the next S/N peak. We iterated this process until we
reached a S/N of 3.5. We then limited the sample
to the sources with a S/N above 4, giving a sample of 146
sources at 850~$\mu$m, 123 with fluxes above
2~mJy. 

We present and analyze this {\em full SCUBA-2 sample\/} in 
A. Barger et al.\ (2018, in preparation; Paper~IV).
Although we also occasionally make use of the full sample
in this paper (mainly when making comparisons with other samples), 
our primary goal here is to construct and analyze a complete SCUBA-2 sample.

We start with the 90 SCUBA-2 sources with fluxes $>1.65$~mJy that lie within 
the central region of the image where 
the SCUBA-2 noise is less than 0.56~mJy (a 100~arcmin$^2$ area, 
or roughly a radius of $5\farcm7$, though the area is not perfectly 
circular); we refer to this as our {\em full central SCUBA-2 sample}. 
Given the noise of 0.56~mJy at the outer edges of this region 
(see Figure~\ref{noise_radius}), the flux limit for our {\em complete central
SCUBA-2 sample\/} is then 2.25~mJy ($>4\sigma$). 

In Table~2, we list the 53 SCUBA-2 sources in this complete sample, along 
with their SCUBA-2 fluxes, their measured ALMA source positions 
(see Section~\ref{almaobs}), the offsets between the SCUBA-2 and 
ALMA source positions, and their measured ALMA fluxes. 
However, for 5 of the 53 sources, the ALMA observations yielded no ALMA 
detections; thus, for those 5 cases, we list the central 
$1\sigma$ noise in the ALMA image (enclosed in square brackets) instead of an ALMA 
flux detection. For three of these sources, the central noise in the ALMA image
is low enough ($\sim 0.1$~mJy) that we should easily detect the SCUBA-2 
source at the $4.5\sigma$ level in the ALMA image,
even if it were a double source where the two components lay near the half power
radius of the ALMA image. The remaining two sources have noisier ALMA
images, and the SCUBA-2 source might be missed if it had multiple components
or were positioned in less sensitive portions of the ALMA beam.

\subsection{Comparison with Previous Single-Dish Observations}

\begin{figure*}
\hskip-1.5cm
\includegraphics[width=12.5cm,angle=0]{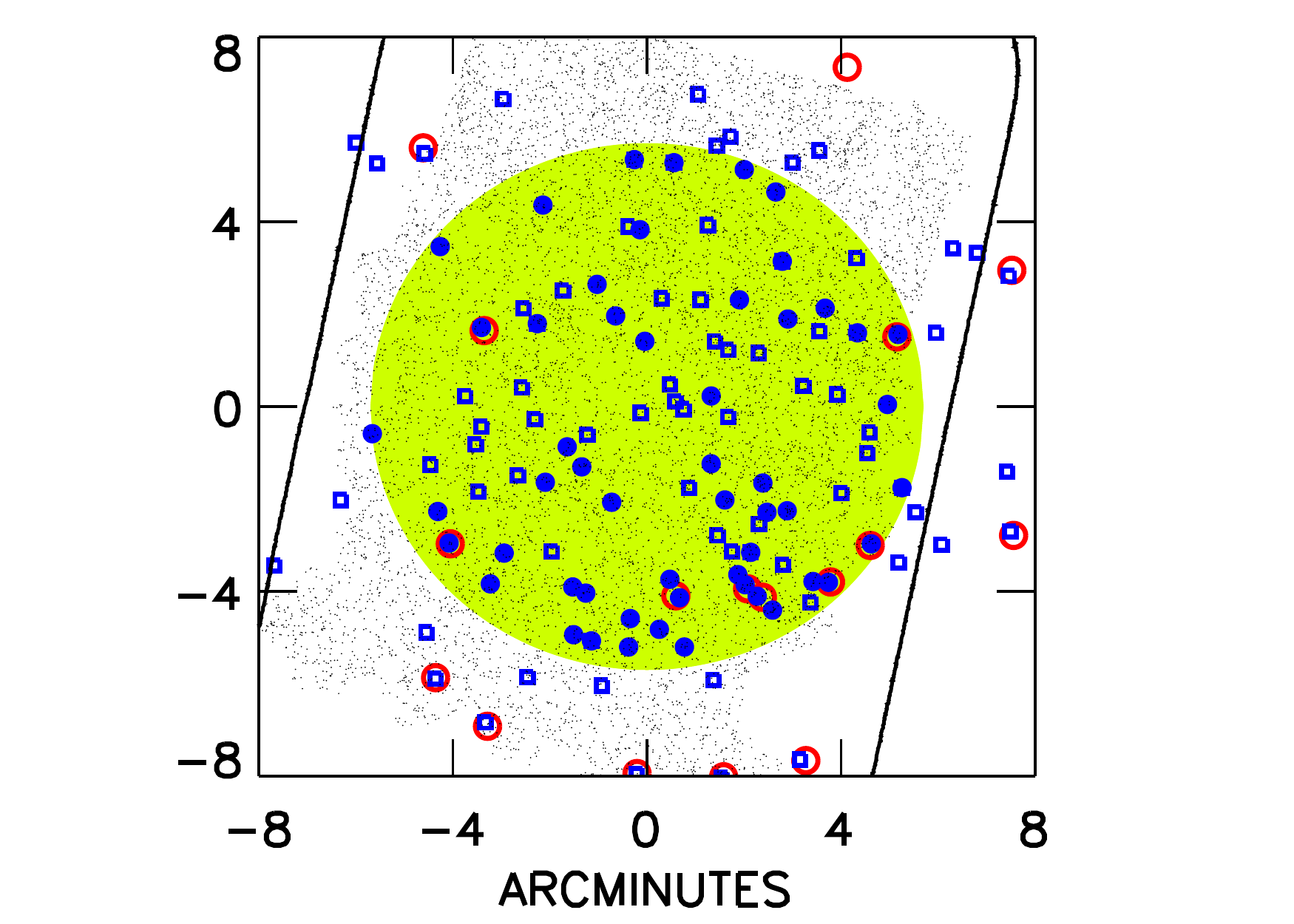}
\hskip-3.5cm
\includegraphics[width=12.5cm,angle=0]{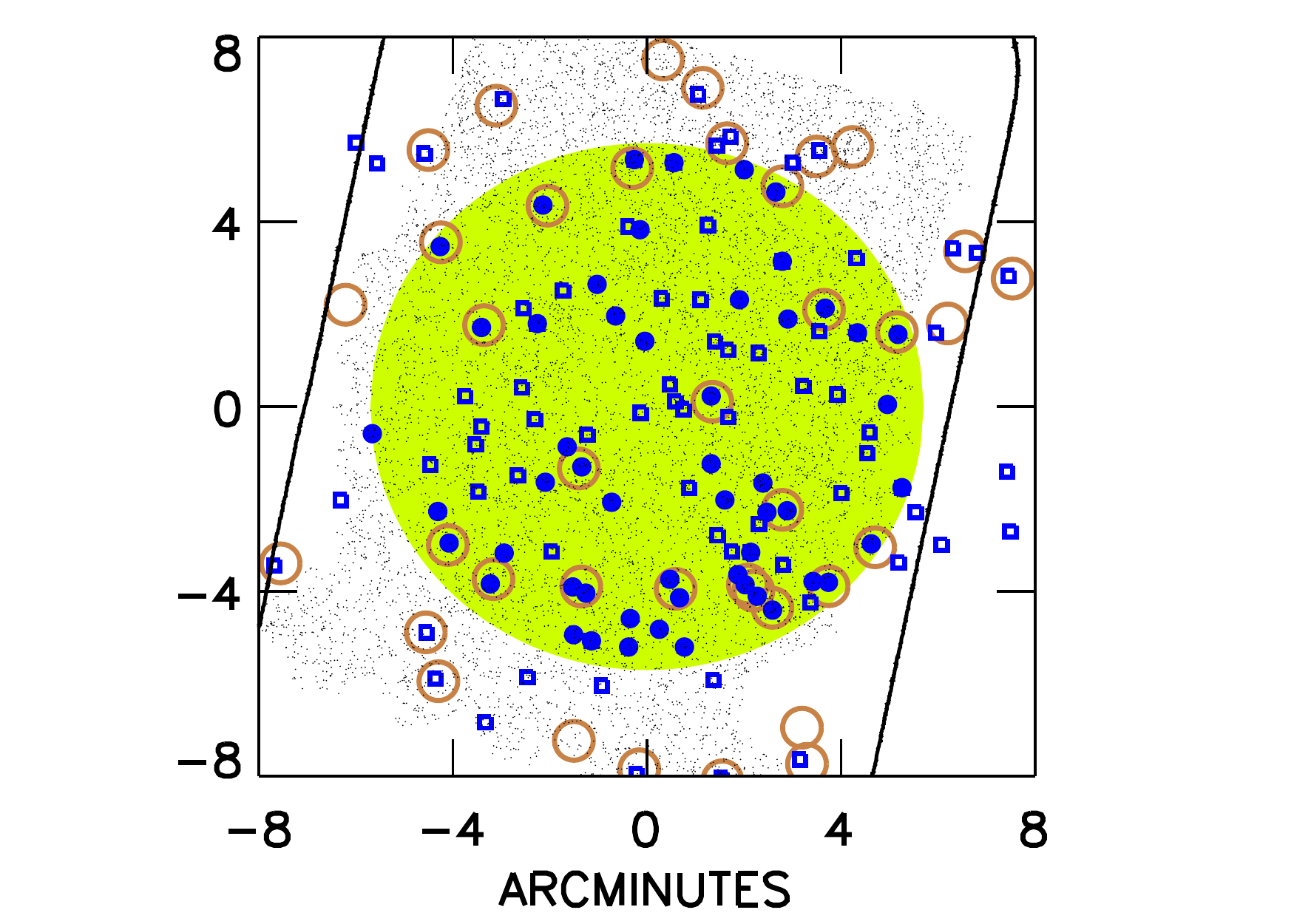}
\caption{
Comparison of our complete central SCUBA-2 sample (blue solid circles) 
with {\em (left)\/} the LABOCA 850~$\mu$m sample of Wei\ss\ et al.\ (2009; 
red circles whose size illustrates the $19\farcs2$ FWHM PSF) and with
{\em (right)} the AzTEC 1.1~mm sample of Scott et al.\ (2010; golden circles
whose size illustrates the $30\farcs0$ FWHM PSF).
In both panels, the green shading shows the central $5\farcm7$ radius region that
corresponds to the deepest SCUBA-2 data. Blue open squares show other secure
($>4\sigma$) SCUBA-2 sources from our full sample (these sources are either 
fainter than 2.25~mJy, or they lie outside the 
central region). The black speckled shading shows the {\em HST\/} ACS GOODS-S 
region (Giavalisco et al.\ 2004), and the black solid rectangle
shows the GOODS-{\em Herschel\/} region (Elbaz et al.\ 2011).
\label{laboca}
}
\end{figure*}

The CDF-S was previously observed at 850~$\mu$m with LABOCA
on the APEX telescope (Wei\ss\ et al.\ 2009) and at 1.1~mm with AzTEC 
on the 10~m Atacama Submillimeter Telescope Experiment (ASTE; Scott et al.\ 2010).
In Figure~\ref{laboca}, we compare these samples 
with our complete central SCUBA-2 sample (blue solid circles),
and with other secure ($>4\sigma$) SCUBA-2 sources from our full sample 
(blue open squares; these sources are either fainter than 2.25~mJy, or they lie 
outside the $5\farcm7$ central region shown in green). 

The LABOCA and AzTEC images are considerably shallower and lower 
resolution than the SCUBA-2 image, and all of their sources within the $5\farcm7$ 
radius region have counterparts in our full central SCUBA-2 sample. 
The LABOCA ($19\farcs2$ FWHM PSF) observations cover the much wider
Extended CDF-S (ECDF-S) region, but their limiting $4\sigma$ flux is around 
$5-6$~mJy, and they find only eight sources within the central region. 
The AzTEC ($30\farcs0$ FWHM PSF) observations are deeper and contain 
20 sources in the central region, but because of the lower resolution and
associated blending, the AzTEC sources positions
have substantial offsets relative to the SCUBA-2 source positions. 

\begin{figure}
\hskip-1.0cm
\includegraphics[width=11.5cm,angle=0]{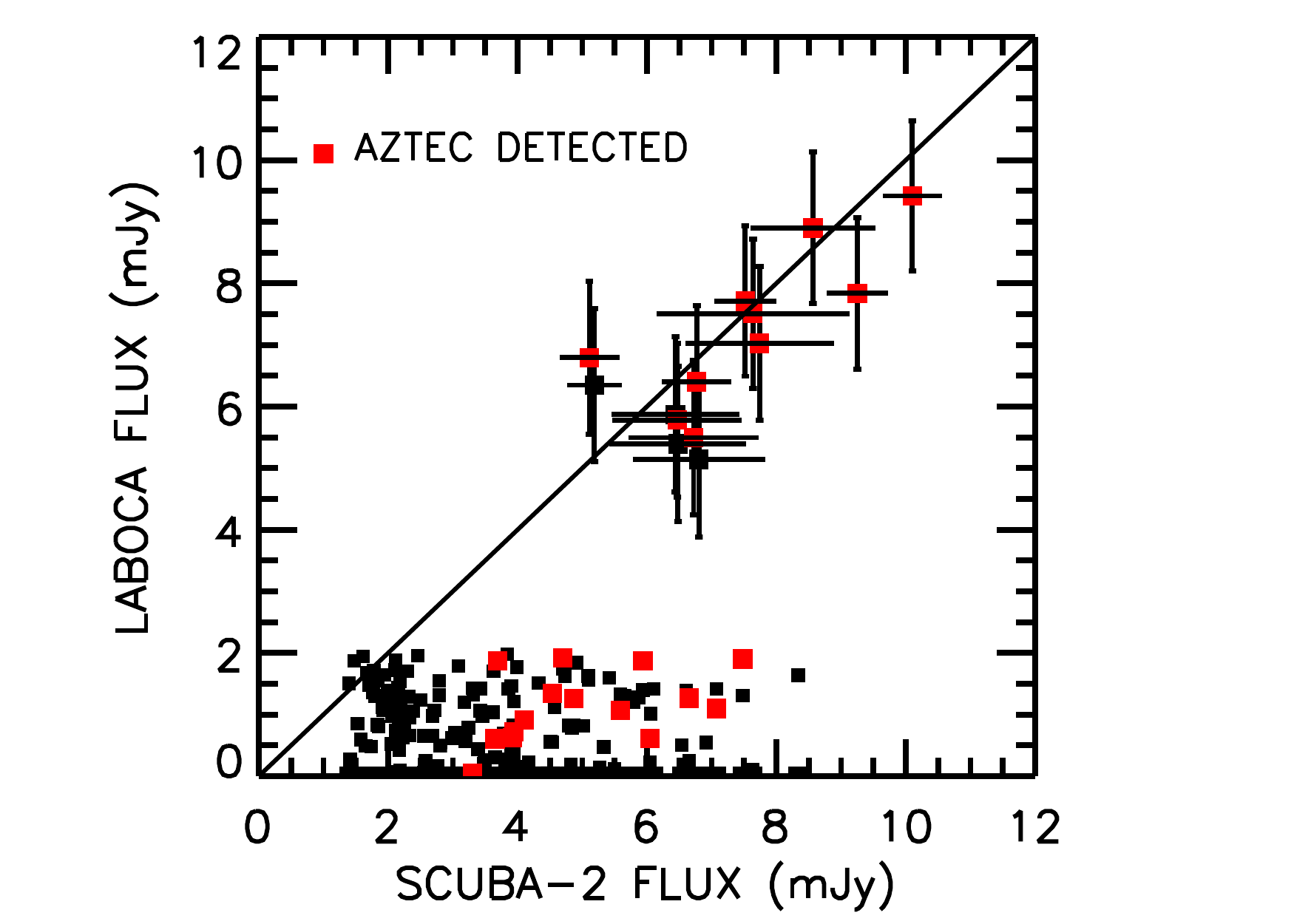}
\caption{
Comparison of the directly measured SCUBA-2 and LABOCA fluxes
for our full SCUBA-2 sample. We compare to the nearest LABOCA 
match (whenever present) within a $6''$ radius around each SCUBA-2 
source, and we do not account for any blending. 
SCUBA-2 sources without LABOCA detections
are shown randomly scattered at fluxes between 0 and 2~mJy at the 
bottom of the plot for clarity. Sources that are also detected in 
the longer wavelength AzTEC observations are shown in red. 
The mean ratio of the LABOCA fluxes to the SCUBA-2 fluxes is $0.96\pm0.04$.
\label{laboca_fluxes}}
\end{figure}

Using our full SCUBA-2 sample to give a substantial overlap,
in Figure~\ref{laboca_fluxes}, we compare the LABOCA fluxes with the 
SCUBA-2 fluxes. Each SCUBA-2 source was matched to the nearest LABOCA 
source within a $6''$ radius, whenever such a source is present. 
Note that the plotted fluxes are direct LABOCA and 
SCUBA-2 flux measurements without any accounting for blending.
Blending is only an issue for three sources in the LABOCA sample,
but it is much more severe for the lower resolution AzTEC data,
where 10 of the AzTEC sources correspond to multiple SCUBA-2 sources.
Sources detected with AzTEC are shown in red and include most 
of the brighter SCUBA-2 and LABOCA sources.

Where LABOCA detects the SCUBA-2 source, the mean ratio of the LABOCA 
fluxes to the SCUBA-2 fluxes is $0.96\pm0.04$, which is well within the 
$\sim10$\% calibration uncertainty on the SCUBA-2 data (Paper~I) and the
$\sim15$\% calibration uncertainty on the LABOCA data (Hodge et al.\ 2013). 
However, both LABOCA and AzTEC miss bright submillimeter sources that they 
should have been able to detect (see Figure~\ref{laboca_fluxes}). 
For example, LABOCA, with its rms of 1.2~mJy, should have 
detected sources above 6~mJy at the $5\sigma$ level, but of
the 16 SCUBA-2 sources with detected fluxes above 6~mJy to rms noise 
less than 1~mJy, only 11 are included in the LABOCA sample. This substantial 
incompleteness will carry through to the follow-up observations with ALMA (ALESS).

\subsection{ALMA Observations}
\label{almaobs}
As part of our ALMA program 
``BASIC: A \underline{B}right \underline{A}LMA \underline{S}urvey of SMGs
\underline{I}n the {\em \underline{C}handra\/} Deep Field-South" 
(ADS/JAO.ALMA \#2015.1.00242.S; PI: F.~Bauer),
we observed 61 fields in the CDF-S between July 13 and 
November 25, 2016 with the C40-4 array configuration in band~7 (hereafter, the 
cycle3 program). We also observed a substantially overlapped sample of 
60 fields (21 new and 39 repeats) in July and August, 2017
(ADS/JAO.ALMA \#2016.1.01079.S; PI: F.~Bauer) in the 
C40-3 array configuration in band~7 (hereafter, the cycle4 program). 
We refer to the combination of the cycle3 and cycle4 observations as 
``BASIC''. We analyzed an additional 12 fields observed as part of 
other ALMA band~7 projects (hereafter, ``archival'').
We processed all the ALMA data in a consistent manner.

We centered the BASIC observations on the targeted source positions and used a
spectral setup configured with four 1.875~GHz spectral windows (using time division mode) 
placed around a central frequency of 343.5~GHz in order to sample crudely the
same 850~$\mu$m continuum as SCUBA-2.
This returned a representative spectral resolution of $\sim28$~km~s$^{-1}$. 

The cycle3 program consisted of three complete and one partially 
complete executions (hereafter, executions~1 through 4), with combined 
integration times between 2.4~min and 3.2~min for each source. We calibrated the data
and made images using {\sc casa} version 4.7.0 and the 
included pipeline. We used source J0522-3627 as the bandpass calibrator, Ceres and J0334-4008 
as the flux calibrators, and J0329-2357 and J0348-2749 as the phase and gain calibrators. 
We performed manual flagging in addition to the pipeline processing to remove outlier baselines 
and/or antennas. We estimate the final error in the absolute flux calibration to be $\sim5$\%.

After data calibration, we discovered that executions 1 and 2 had worse phase calibration than 
executions~3 and 4, especially for the targets observed at the beginning of the executions. 
Execution~1 showed a problem with the phase calibration that could not be corrected by the 
available water vapor radiometer data, resulting in an observed wandering offset in the positions 
of the first several sources when compared to the other three executions. 
Execution~2 also showed indications of a phase calibration problem, but this problem was 
considerably less severe than for execution~1. 
Because of this behavior, for the first six targets observed,
we only used the data corresponding to executions $2-4$, while for the rest of the targets, 
we included all of the available data. 

The cycle4 program consisted of four complete executions, with a combined integration
time of 2.7~min for each source. We calibrated the data and made images using {\sc casa} 
version 4.7.2 and the corresponding included pipeline. 
We used sources J0522-3627 and J0006-0623 as the bandpass calibrators, 
J0334-4008 as the flux calibrator, and J0348-2749 as the phase and gain calibrator. 
The cycle4 observations did not show any of the problems seen in the cycle3 observations. 

A nominal natural-weighted beam of $0\farcs38\times 0\farcs3$ was achieved for the cycle3 
observations, while the cycle4 observations were taken with a slightly more extended 
antenna configuration, achieving a natural-weighted beam of $0\farcs15\times 0\farcs1$. 
We combined the visibilities for the fields that were observed in both cycles using the {\sc casa} 
task concatenate, thereby obtaining a combined beam similar to the one from cycle3
$(0\farcs38\times 0\farcs31)$. 
We made images of the targets using the task {\sc clean}. We made dirty 
images using natural weighting and a mild {\em uv}-taper to achieve a representative 
synthesized beam of $0\farcs53\times0\farcs48$ and a position angle of -87\fdg16.
Based on the size estimates from Simpson et al.\ (2015), we adopted a beam size of 
$\approx0\farcs5$, which offers the best trade-off between retaining a relatively low 
rms and increasing the sensitivity to extended sources in our data; tapering the data 
to larger beams means weighting toward shorter baselines and hence fewer antennas 
and lower sensitivity.

We included a further 14 sources from 12 band~7 archival images 
(there were 11 single sources and one triple source in these images). 
Since 11 of these images have extremely similar observational setups to BASIC, 
we used the same reduction procedures.
We acquired four archival sources from the ALESS sample (Hodge et al.\ 2013; 
ADS/JAO.ALMA \#2011.1.00294.S, ADS/JAO.ALMA \#2012.1.00307.S),
eight more from the X-ray AGN sample of 
Mullaney et al.\ (2015; ADS/JAO.ALMA \#2012.1.00869.S), and one from the red 
nugget sample of P.I. ~G.~Barro (ADS/JAO.ALMA \#2013.1.00205.S). 
The 14th source was observed in an extended configuration as part of the 
ALMA Redshift 4 Survey (AR4S) of 
Schreiber et al.\ (2017; ADS/JAO.ALMA \#2012.1.00983.S).
We reduced it to match the current work as best as possible. 

We calibrated the original ``raw'' datasets using {\sc casa} version 4.7.2 
and either the pipeline for newer datasets or the reduction script provided for 
older datasets (updated for the correct {\sc casa} version). As above, we performed 
extra manual flagging, as necessary. In cases where 
the natural beam of the archival data was smaller than $\approx0\farcs5$, 
we adopted a $uv$-taper to achieve a similar beam size. Otherwise, we used 
the natural-weighted images. 

We performed an initial source search on the dirty images of both the BASIC and 
archival datasets prior to primary beam correction, with the dual purpose of 
assessing the rms sensitivity and finding all secure detections. 
We produced cleaned images by placing $2''\times2''$ clean boxes around all secure 
sources detected with S/N$\geq5$ in the dirty images. We stopped the final cleaning 
process after 1000 iterations, such that most of the emission associated with the 
sources was recovered. 

The flux densities that we measured in the final images of the cycle3 program are in 
good agreement with the ones we obtained for the individual executions, suggesting 
that the phase calibration issue present in executions~1 and 2 does not affect our final 
flux density measurements.

\subsection{ALMA Source Catalog Construction}
We searched each of the $0\farcs5$ cleaned images
for significant sources. We restricted our search to the area contained 
within an $8\farcs75$ radius, which corresponds to the ALMA half
power. This also is well matched to the SCUBA-2
FWHM and should contain all sources contributing to
the corresponding SCUBA-2 flux. The total searched area is 
6.3~arcmin$^2$ for the 82 unique BASIC images and 12 archival images.

We selected all sources
with a peak flux S/N above 4.5 (see below). We determined the peak
flux noise using the dispersion in 100 independent
beam positions surrounding the source. 
We measured a median rms noise for the sources
in the cycle3 program of 0.15~mJy with
a minimum of 0.11~mJy and a maximum of 0.23~mJy.
Sources that were only observed in the cycle4 program have a higher
median rms noise of 0.29~mJy, with a range from 0.19 to 0.32~mJy.
Sources with observations in both cycle3
and cycle4 may have a noise as low as 0.095~mJy.

The total searched area corresponds to 50,000
independent beams. For a Gaussian distribution in the
noise, we expect 1.5 false sources above a S/N cut of 4, and 
0.26 false sources above a S/N cut of 4.5. Since the Gaussian 
assumption may be questionable, we also made an empirical test 
by applying our detection method to the negative of each image. 
This gives two sources at S/N$=4-4.5$ and none above 4.5, consistent 
with the Gaussian expectation. We therefore restricted our final
sample to the sources with S/N$>4.5$, which we expect 
to be highly robust.

\begin{figure}
\hskip -1.0cm
\includegraphics[width=11.5cm,angle=0]{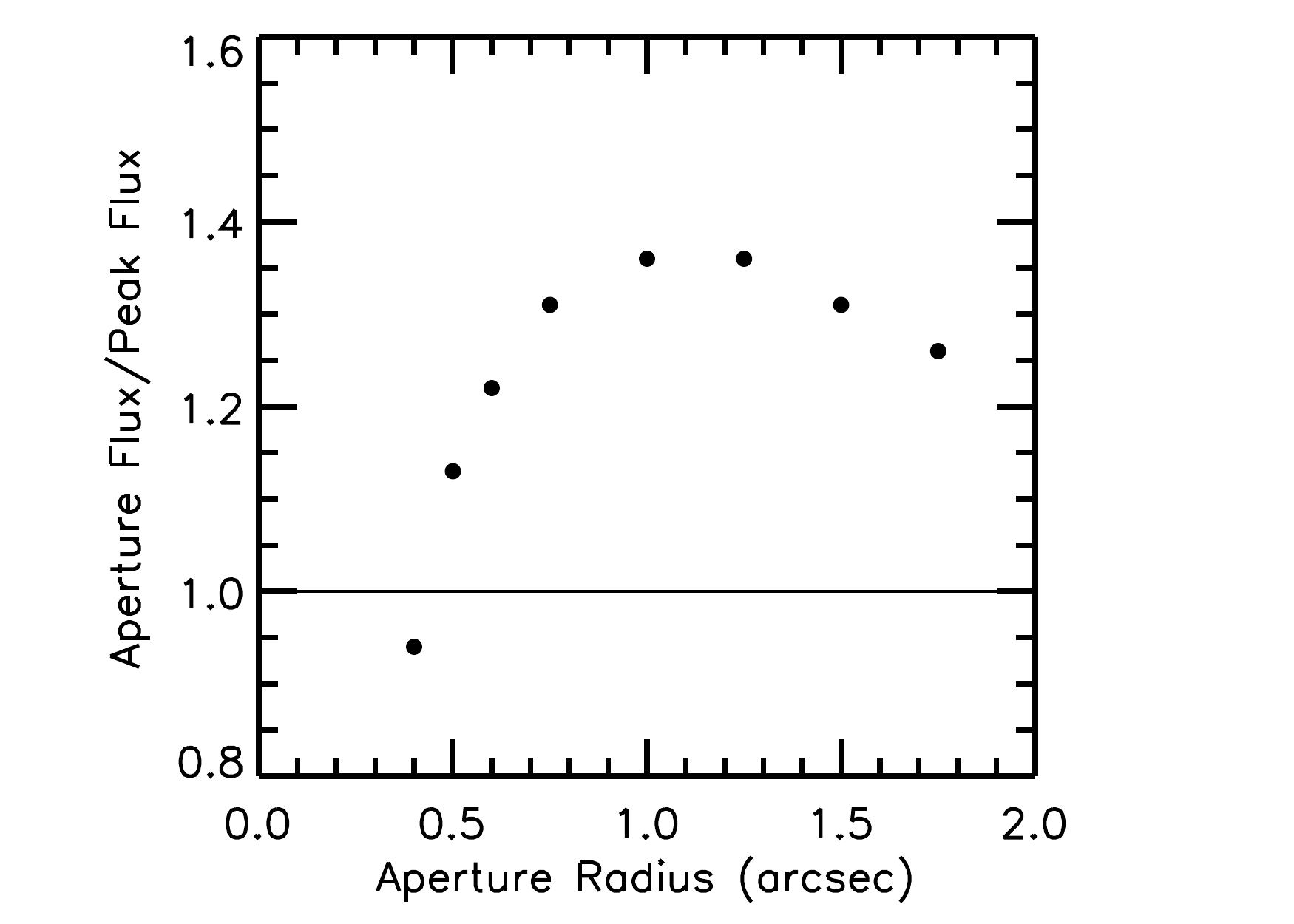}
\caption{
Median ratio of the aperture flux to the peak flux vs. the aperture radius
for the 75 ALMA sources.
\label{ap_corr}}
\end{figure}

We found that the peak fluxes, even in the $0\farcs5$ tapered images,
slightly underestimate the total fluxes. The reason for this is that
the sources are resolved. In order
to estimate the total fluxes, we used two methods.
In the first method, we took the ratio of aperture 
flux measurements made in a range of aperture radii to the peak flux 
measurements. As we show in 
Figure~\ref{ap_corr}, the median multiplicative correction asymptotes 
beyond a radius of $\approx0\farcs6$, and we used this as our preferred radius.
For the sources where the ratio of the aperture flux to the peak flux exceeds
1.4, we used the aperture flux. For the remaining sources, we used the
peak flux times the median correction of 1.22. We refer to
these as our {\em best flux estimates\/}. 

\begin{figure}
\hskip -1.25cm
\includegraphics[width=12cm,angle=0]{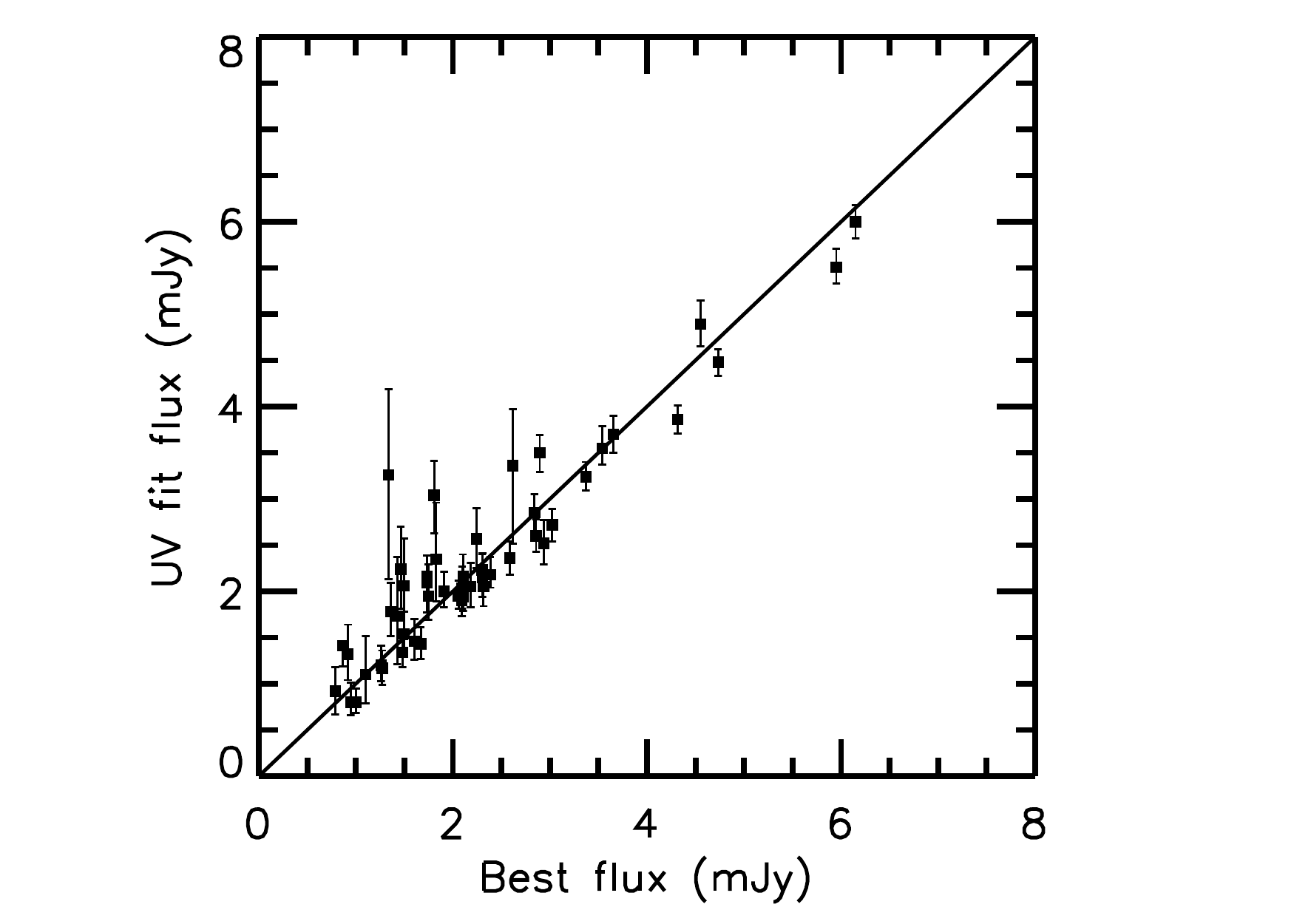}
\caption{
Comparison of the flux measured using fits to a Gaussian
spatial profile in the {\em uv\/} plane vs. the simpler best flux
estimate based on the peak and aperture fluxes, as described
in the text.
\label{jorge}}
\end{figure}

In the second method, we fitted the sources in the {\em uv\/} 
plane by adopting a Gaussian shape. 
J.~Gonz{\'a}lez-L{\'o}pez et al.\ (2018, in preparation) give
more details on these fits, which are used to estimate the source
shapes and sizes. As we show in Figure~\ref{jorge},
the fluxes that we estimate from the {\em uv\/}
fits agree extremely well with the more straightforward best flux 
estimates.

In Table~3, we give brief descriptions and numbers
for the various samples that we use in this paper.
The ALMA images cover all 53 $>2.25$~mJy 
sources in the complete central SCUBA-2 sample (Table~2), together 
with a number of fainter SCUBA-2 sources. 
We note that 
the ALMA band~7 primary beam of $\sim17''$ is larger 
than the SCUBA-2 beam of $\sim14''$, allowing ALMA 
to image all sources that contribute to the SCUBA-2 flux.

\begin{deluxetable*}{lcccc}
\setcounter{table}{2}
\tablecaption{CDF-S Central Region Details}
\tablehead{Description & Covered Area & \multicolumn{3}{c}{Number} \\ \cline{3-5}
\colhead{} & \colhead{(arcmin$^2$)} & \colhead{Total} & \colhead{Targeted} & \colhead{At least 1 ALMA} \\
\colhead{} & \colhead{} & \colhead{} & \colhead{by ALMA} & \colhead{$>4.5\sigma$ Detection}
}
\startdata
SCUBA-2 $>4\sigma$ sources with flux $>1.65$~mJy & 100 & 90 & 79 & 60 \\
$\dots$ with SCUBA-2 flux $>2.25$~mJy (see Table~2)  &   & 53 & 53 & 48 \\
$\dots$ with SCUBA-2 flux $>4$~mJy & & 16 & 16 & 16  \\
$\dots$ with SCUBA-2 flux $2.25-4$~mJy & & 37 & 37 & 32 \\
$\dots$ with SCUBA-2 flux $1.65-2.25$~mJy & & 37 & 26 & 10  \\
$\dots$ add't just outside central region & & 3 & 3 & 3 \\
Add't targets in central region but too faint for SCUBA-2 sample & & 3 & 3 & 3 \\
\hline
Total ALMA $>4.5\sigma$ detections (see Table~4) & 6.3 & 75 & & \\
$\dots$ with flux $>2.25$~mJy (see Table~4) & & 41 & & \\
$\dots$ from targeting SCUBA-2 $>4\sigma$ sources with flux $>1.65$~mJy & 5.3 & 69 & & \\
\enddata
\tablecomments{The 69 ALMA $>4.5\sigma$ sources from targeting 79
SCUBA-2 $>4\sigma$ sources with flux $>1.65$~mJy include one triple, 
7 doubles, 52 singles, and 19 fields with no detections.}
\label{tabsampledetails}
\end{deluxetable*}

In total, we have 75 $>4.5\sigma$ ALMA sources, which corresponds to a lower 
limit of 0.75 sources per arcmin$^2$ (since we
have not uniformly sampled the area with ALMA).
We summarize this sample, which we hereafter refer to as our {\em total ALMA sample},
in Table~4, where we provide the peak
and best ALMA fluxes and errors, the S/N, the F18
1.13~mm ALMA flux when available (see Section~\ref{franco}),
the SCUBA-2 source
number from Table~2 whenever there is such a source located within a
radius of $5''$ of the ALMA position, and the offset between 
the SCUBA-2 and ALMA source positions for these sources. 
Only one of the detected sources in Table~4
has a S/N between 4.5 and 5, while the remainder all have S/N above 5.

The effects of blending and Eddington bias will only reduce
the measured ALMA fluxes relative to the SCUBA-2 fluxes.
Thus, we hereafter refer to the ALMA sample of 41 sources above 2.25~mJy
as our {\em complete ALMA sample}.
The cumulative ALMA number density above 2.25~mJy is just over 
1500 sources per deg$^2$, compared with an 
expected value of just under 2200 based on incompleteness and 
Eddington bias corrected SCUBA-2 
850~$\mu$m number counts for a number of fields (e.g., Hsu et al.\ 2016). 
In a subsequent paper, we will provide incompleteness corrected ALMA counts 
to fainter fluxes and compare with the corrected SCUBA-2 counts in more detail.

\subsection{Comparison with 1.13~mm ALMA sample}
\label{franco}

F18 obatined a sample of 1.13~mm
galaxies based on a homogeneous scan of a 69~arcmin$^2$
region of the GOODS-S. This region is largely
contained within our total ALMA area, with all
20 sources in their main catalog
lying in our area, and only one of their
three supplemental sources lying outside our area.

In terms of detections, 17 of their 20 main catalog sources 
are in our total ALMA sample. The remaining
three are not detected in our SCUBA-2 image,
and, indeed, correspond to sources that F18 consider likely 
to be false (these are their sources AGS14, AGS16, and AGS19). 
Their two supplemental sources that lie in our area
are also detected in our total ALMA sample.

In Table~4, we give the F18 fluxes based on
their GALFIT values and without applying any deboosting.
However, in the region where there are also
longer wavelength ALMA observations from 
Dunlop et al.\ (2017), we find that the F18 fluxes
lie below the interpolation between our 850~$\mu$m values
and the Dunlop et al.\ 1.3~mm values, suggesting that the 
1.13~mm data of F18 need to be corrected for extended flux.
In our analysis (but not in Table~4), we apply a multiplicative 
correction of 1.35 to allow for this.

\section{ALMA versus SCUBA-2}
\label{secALMAvsS2}

\begin{figure}
\hskip -1.0cm
\includegraphics[width=11.5cm,angle=0]{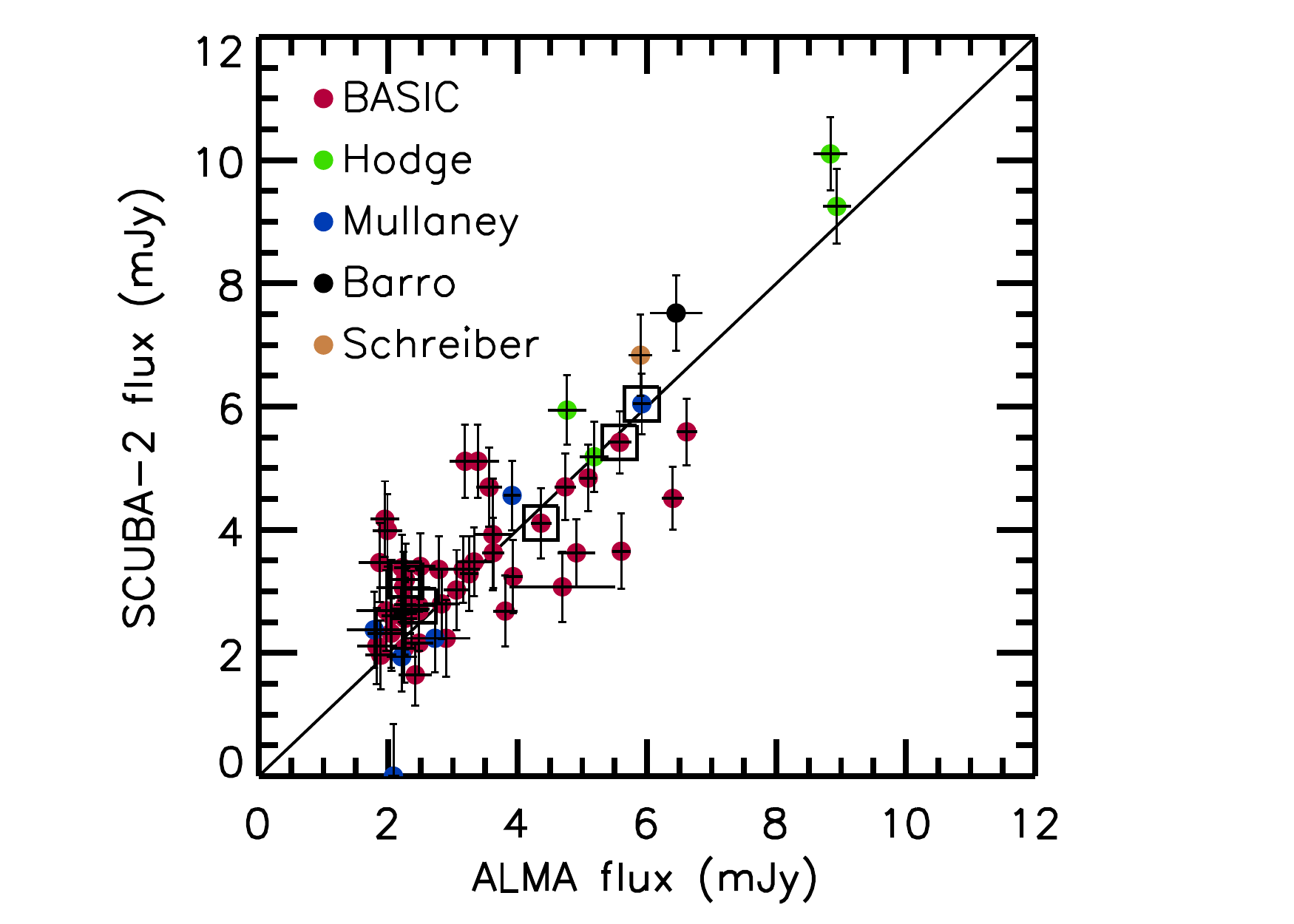}
\caption{
Comparison of the ALMA fluxes with the SCUBA-2 fluxes measured at the ALMA positions.
Where there are multiple ALMA sources, we have added the individual ALMA fluxes
and marked them with a large open square. The BASIC sources are shown
in red, while the archival sources are color coded according to the legend.
\label{alma_flux_compare}
}
\end{figure}

\vskip 0.4cm
\begin{figure*}
\hskip -0.5cm
\hspace*{0.2in}\centerline{\includegraphics[angle=180,width=9in]{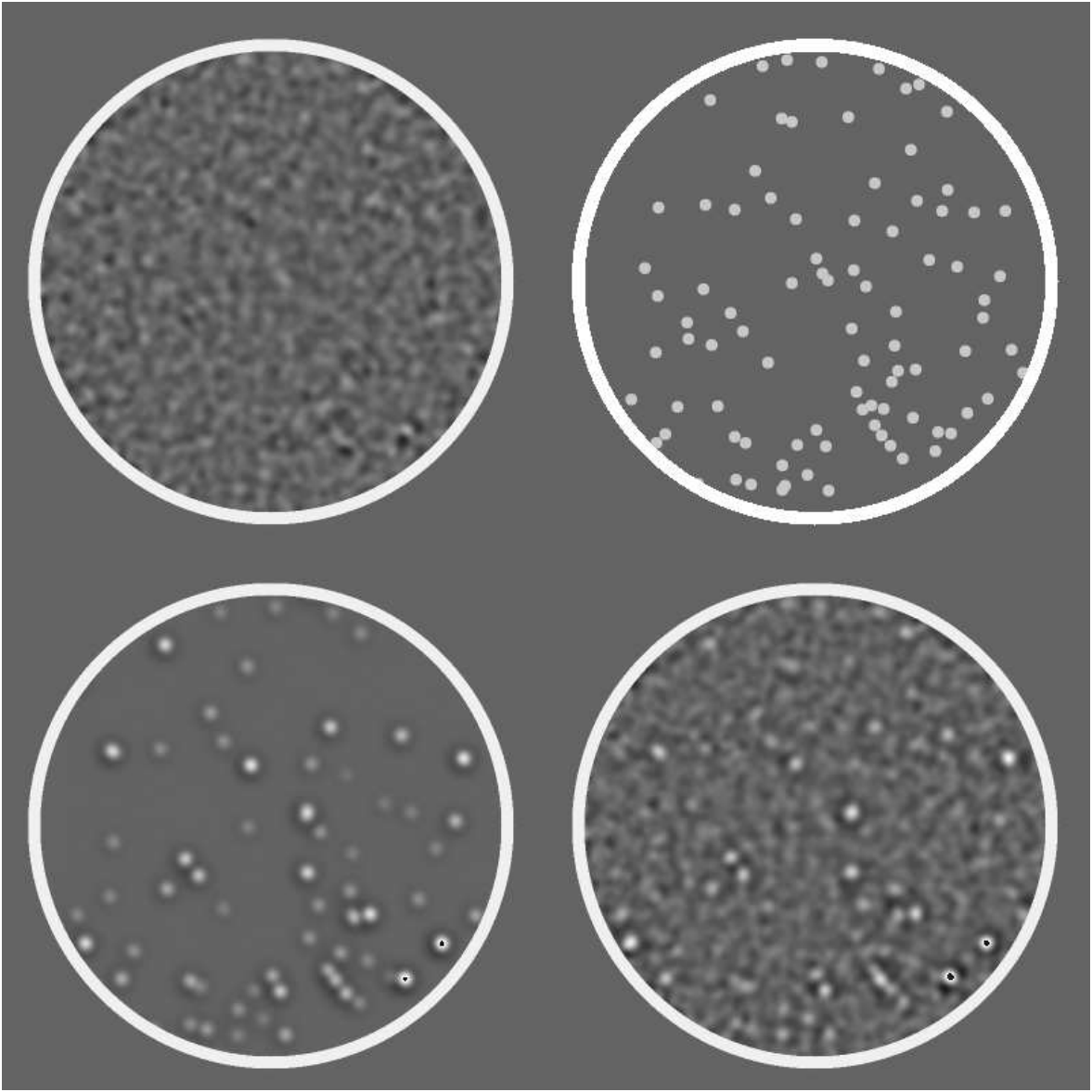}}
\vskip -1.0cm
\caption{
{\em Top left:\/} ALMA-constructed SCUBA-2 image subtracted from the actual 
SCUBA-2 image.
{\em Top right:\/} ALMA pointings in the field.
{\em Bottom left:\/} ALMA-constructed SCUBA-2 image. 
{\em Bottom right:\/} Actual SCUBA-2 image. 
\label{alma_positions}
}
\end{figure*}

In Figure~\ref{alma_flux_compare}, we show a simple flux comparison 
between the ALMA sources and the SCUBA-2 sources by plotting
the nearest peak SCUBA-2 flux against the
ALMA flux. Where more than one ALMA source corresponds
to a SCUBA-2 source, we have added the ALMA fluxes together and 
surrounded the symbol with an open box.
All the ALMA fluxes are consistent with the SCUBA-2 fluxes within our 10\% 
estimated error in the SCUBA-2 fluxes and the similar estimated error
in the ALMA fluxes.

We make a more precise comparison that properly allows
for blending by constructing a simulated SCUBA-2 image that
would be produced by the observed ALMA sources. Each ALMA
source is convolved with the SCUBA-2 matched-filter PSF
and added to form a final image. We show the simulated
SCUBA-2 image in the bottom left panel of Figure~\ref{alma_positions}
compared with the true SCUBA-2 image in the bottom right panel. 

We subtracted the ALMA-constructed SCUBA-2 image from the real 
SCUBA-2 image to form a residual image, which we show in the upper
left panel of Figure~\ref{alma_positions}. We show the ALMA pointings
in the upper right panel. We searched the differenced
image the same way that we did when we generated the catalog
of SCUBA-2 sources from the observed SCUBA-2 image. 
This procedure detects seven sources with SCUBA-2 fluxes
above 2.25~mJy in the differenced image. 
The only source in the differenced image
with a flux substantially larger than 2.5~mJy has a
flux of 3.4~mJy. This source is a close
neighbor of the brightest source in the field and may
be a residual from imprecise subtraction of the bright source.
This is covered by an archival ALMA pointing, and while
there may be a weak (0.7~mJy) source near this position,
there is nothing (i.e., individual or a small grouping of sources)
that would produce this much flux.

The remaining six sources have fluxes between 2.25 and
just over 2.5~mJy. All of these correspond to sources
in Table~2, and all were observed as part of BASIC. For source~15 in Table~2,
the ALMA detection is faint compared to the SCUBA-2
flux, and for the remaining sources
(44, 46, 49, 50, and 53), no ALMA sources were detected.
This suggests that there may be several ALMA counterparts 
to these SCUBA-2 sources, which we have not been able
to detect.

Based on the mean offset of 39 isolated ALMA sources with fluxes greater
than 2.25~mJy from the nearest
SCUBA-2 peak, we find an absolute astrometric offset of $0\farcs9$ in R.A. and 
$-0\farcs3$ in Decl., which we applied to the SCUBA-2 positions.
This is well within the expected uncertainty in the absolute 
SCUBA-2 astrometry. 

In Figure~\ref{alma_offsets}, we show the offsets between the ALMA and SCUBA-2 
positions for the complete central SCUBA-2 sample (Table~2), excluding only the
5 sources with no ALMA detections. Where there is a single ALMA counterpart,
we show the offset with a red square, and where there is more than one ALMA
counterpart, we show the offsets with connected blue diamonds. The mean
offset of the individual sources (i.e., the scatter) is $2\farcs2$ for all SCUBA-2 
sources with fluxes above 2.25~mJy and a single ALMA counterpart.

The overall percentage of SCUBA-2 sources above 2.25~mJy
with multiple ALMA counterparts is 13\% (68\% confidence range
7--19\%). Here each of the counterparts has to be above the
ALMA $4.5\sigma$ detection threshold. The median $4.5\sigma$ detection limit 
is 1.04~mJy, and 52 of the 75 sources have limits less than 1.25~mJy. Thus,
at the 2.25~mJy limit, we could detect two roughly equal sources, while at 3~mJy, 
we could detect a multiple with two sources with a 2:1 flux ratio.
If the two sources in Table~2 with SCUBA-2 fluxes
above 2.25~mJy, no ALMA counterpart, and a high noise ALMA
image were also multiples, then this would rise to 16\%.

This is a considerably smaller multiplicity than the $\sim30$\% found
in the ALESS followup of the LABOCA sample (Hodge et al.\ 2013; Simpson
et al.\ 2015). In substantial part, this is due to the smaller SCUBA-2 beam, which has
64\% of the area of the LABOCA beam (LABOCA was on the 12~m APEX telescope).
The smaller beam will pick up a smaller fraction of multiples.
There could also be a dependence on flux, and the lower multiplicity
could be partly a consequence of the depth of the 
SCUBA-2 sample. If we restricted to sources with SCUBA-2 fluxes
greater than 4~mJy, then the fraction would be 27\%, with a range
of 16 to 42\%. This is highly uncertain, because of the
small sample size, and, indeed, there could be no variation
in the multiplicity as a function of flux. Note that Hill et al.\ (2018) found that
$\lesssim15$\% of 105 sources from the SCUBA-2 Cosmology Legacy Survey
observed with the SMA at fluxes $\gtrsim10$~mJy 
were multiples (defined as bright SMGs that resolve into two or more
galaxies with flux ratios close to 1).

\begin{figure}
\hskip -1.0cm
\includegraphics[width=11.5cm]{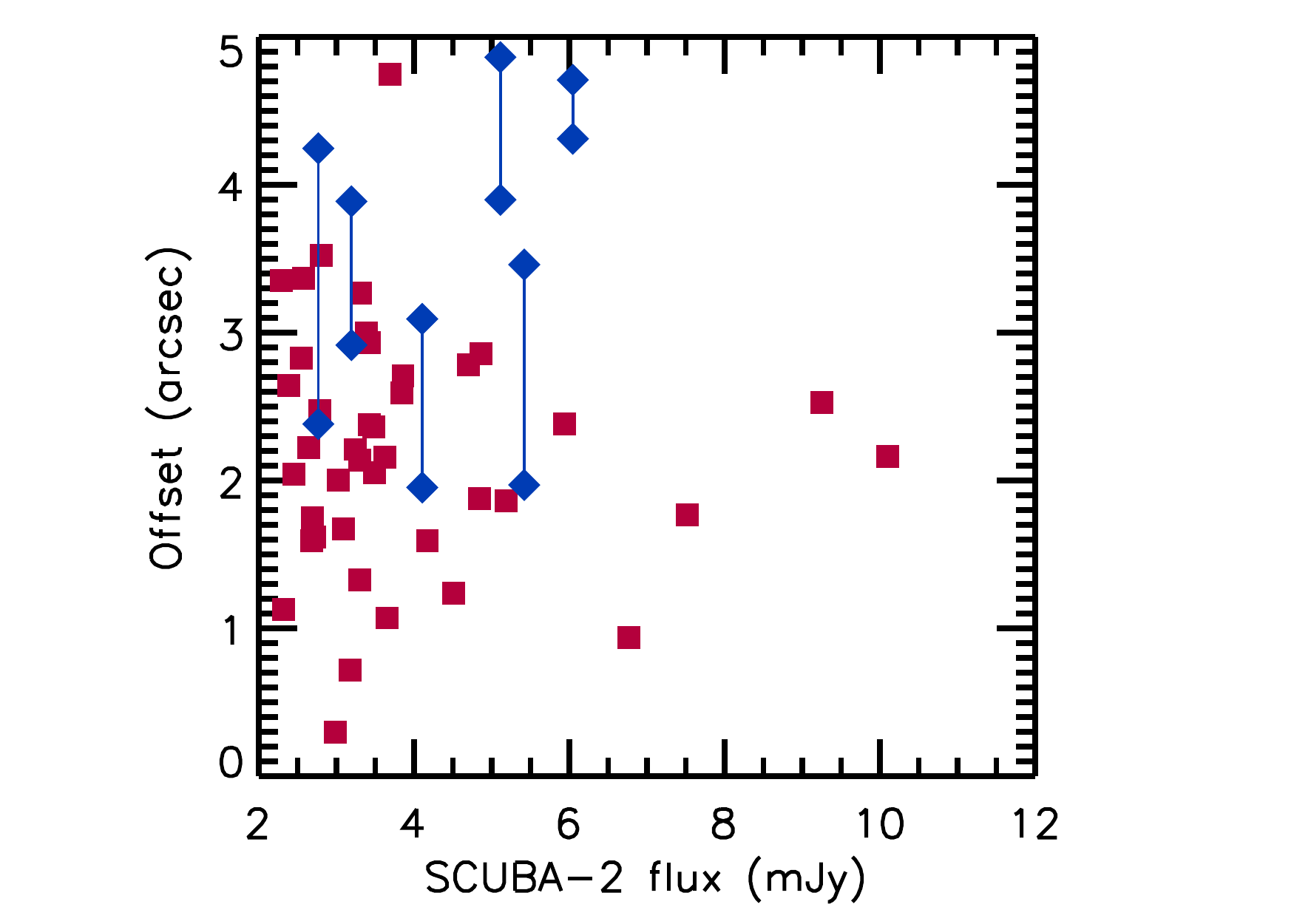}
\caption{
Offsets between the ALMA and SCUBA-2 positions vs. SCUBA-2 flux
for the complete central SCUBA-2 sample, excluding only the 5 sources with
no ALMA detections. Red squares show the SCUBA-2 
sources with unique ALMA counterparts. Blue diamonds show the SCUBA-2 sources 
with multiple ALMA counterparts; here the offsets are joined by blue lines.
The median offset is $2\farcs2$ for the SCUBA-2 sources with unique 
ALMA counterparts.
The fraction of sources with multiple counterparts is 13\%, with a 68\% confidence 
range of 7 to 19\%.
\label{alma_offsets}
}
\end{figure}

\section{Ancillary information}
\label{secancillary}
In this section, we consider the 75 ALMA sources in Table~4
(i.e., our total ALMA sample).
While we acknowledge that this is not a complete sample below 2.25~mJy,
completeness is not needed when trying to understand the
properties of these securely detected ($>4.5\sigma$) ALMA
sources. Indeed, the precise positions allow us to exploit the
wealth of additional information available
in the CDF-S/GOODS-S region, which is one of 
the most intensively studied regions on the sky. 
In particular, we can find the
optical, near-infrared (NIR), and mid-infrared (MIR) counterparts 
to the SMGs, determine whether there are previous spectroscopic 
identifications, and calculate photometric redshifts
for the sources.

\subsection{Optical and F160W Counterparts}
\label{secopt}
All but one source in our total ALMA sample lie within
the deep GOODS {\em HST\/} ACS optical
imaging of Giavalisco et al.\ (2004) and the deep 
CANDELS {\em HST\/} WFC3 ultraviolet/optical channel and WFC3 infrared
channel (WFC3/IR) imaging of Grogin et al.\ (2011) and Koekemoer et al.\ (2011).

In Figure~\ref{basic_alma_images}, we show color 
thumbnail images made from the version~1.5 data release of the 
Hubble Legacy Fields by Illingworth et al.\ (2016), which combines 
2442 orbits of the GOODS-S/CDF-S region ACS and WFC3/IR images.
We overlay the ALMA data as white contours. 
The one ALMA source that lies outside the GOODS-S and CANDELS 
regions (source~27 or ALMA033203-275039 in Table~4) is omitted 
from the figure.


\begin{figure}
\setcounter{figure}{10}
\hskip -0.5cm
\includegraphics[width=11cm]{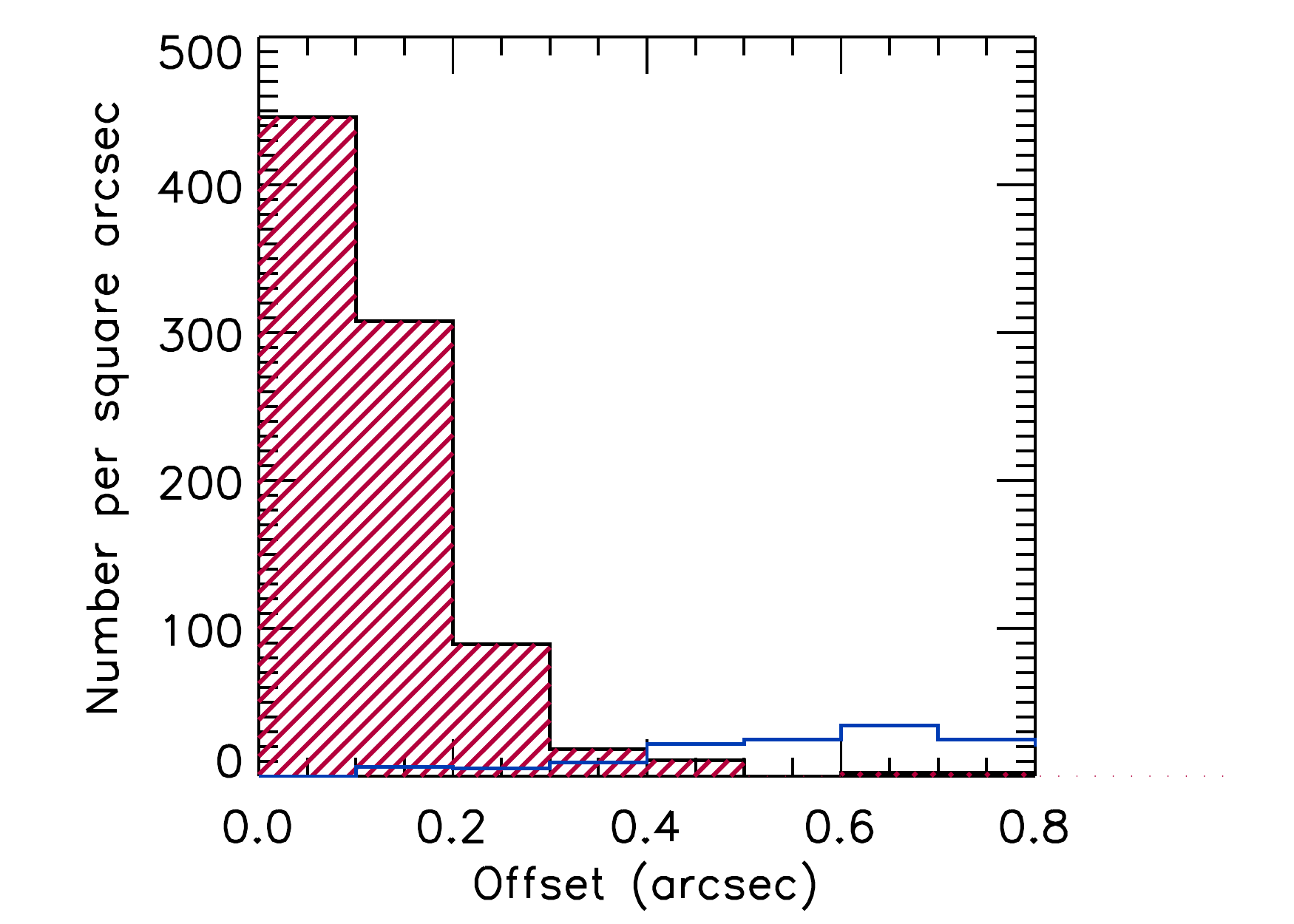}
\caption{Distribution of offsets between the ALMA and
G13 positions (red hatched histogram).
The numbers are per unit area in $0\farcs1$ 
stepped annuli. The blue histogram shows the expected
distribution for a randomized sample with the same number
of sources.
\label{false_match}
}
\end{figure}

\begin{figure}
\hskip -1.0cm
\includegraphics[width=11.5cm]{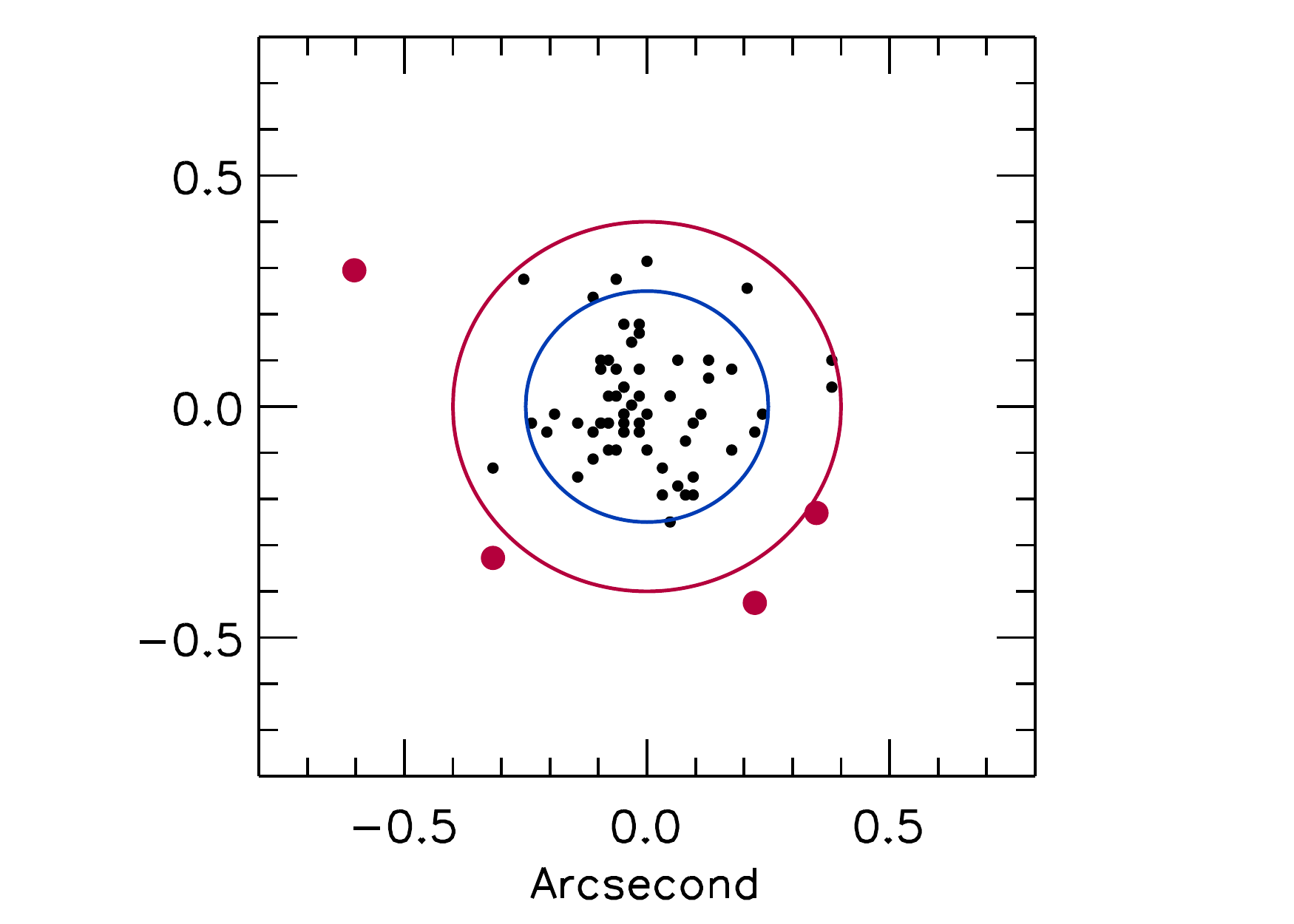}
\caption{Offset in R.A. ($x$ axis) and Decl. ($y$ axis)
between the ALMA position and the nearest
NIR (F160W) peak. The blue large circle corresponds to $0\farcs25$ 
and the red large circle to $0\farcs4$.
Sources with separations larger than $0\farcs4$ are shown with red 
circles.
The positional uncertainties on both the ALMA sources and the
F160W counterparts are much smaller than the measured offsets.
The offsets may reflect whether the peak star formation is separated
from the center of the NIR image (rest-frame optical at the redshifts 
of the sources), or, in some cases, whether the nearest NIR galaxy 
is actually the correct counterpart.
\label{displace}
}
\end{figure}

Comparing the ALMA positions with the CANDELS coordinates measured from the 
Hubble Legacy Fields F160W image gives an absolute astrometric offset of $0\farcs06$ 
in R.A. and an offset of $-0\farcs24$ in Decl. These are very similar
to the offsets found by Chen et al.\ (2015) from their analysis
of the ALESS sample in the ECDF-S ($0\farcs11$ and $-0\farcs25$).
We apply these offsets throughout in comparing the various samples,
including for the overlays in Figure~\ref{basic_alma_images}.

After excluding source~27, 66 of the remaining 74 ALMA sources have 
counterparts within a $1''$ matching radius in the CANDELS catalog of G13,
after applying the small astrometric correction. In Figure~\ref{false_match},
we show the distribution of the offsets between the ALMA and G13 
positions (red hatched histogram). (We include these offsets in Table~5.)
We also show the expected distribution
of a random sample of the same size (blue histogram).
The mean of the offsets between the ALMA and G13 positions 
is $0\farcs19$, with 49 of the ALMA sources lying within $0\farcs25$. 
Beyond $0\farcs4$, there is a high probability that the source
is a false match, and 5 of the 66 sources fall in this category.

The sources without counterparts (8 of them) or those that are probable false 
matches (5 of them) are sources~1, 2, 6, 13, 19, 29, 44, 51, 58, 61, 64, 68, and 75
in Table~4 (see Figure~\ref{basic_alma_images}). For these sources, we 
measured the magnitudes directly at the ALMA positions using the imaging 
data, matching the zero points and the apertures to the G13 sample.
However, for the 61 sources with counterparts closer than $0\farcs4$, we 
adopted the optical and NIR magnitudes from G13. 

In some cases, the absence of a near counterpart in G13 to the ALMA source
is due to the blending of a red galaxy with a bright near neighbor 
(e.g., source~6 or ALMA033246-275120 and source~29 or ALMA033232-274540; 
see Figure~\ref{basic_alma_images}), which results 
in an overly large offset. Thus, as a second approach, we decided to measure 
directly from the image the offsets of the nearest F160W peaks 
from the ALMA positions. We only measured 
these offsets for sources that were clearly detected in the NIR, and we 
again excluded source~27 or ALMA033203-275039. We found that
67 of the remaining 74 ALMA sources have counterparts within a 
$1''$ matching radius. We show these offsets in Figure~\ref{displace}. 
In this case, we find
54 ALMA sources with a NIR peak within $0\farcs25$ (blue circle in 
Figure~\ref{displace}), as compared to 49 when we matched  
to the G13 catalog. Note that the error in the determination of the ALMA centroids 
is extremely small by comparison, with the measured scatter typically
around $0\farcs015$ (see J.~Gonz{\'a}lez-L{\'o}pez et al.\ 2018, in preparation, 
for a more detailed discussion). Thus, the measured offsets relate primarily to a 
physical offset of the dust-emitting regions from the NIR peaks in the galaxies.

We also find that 4 of the 67 ALMA sources have a NIR peak separation
larger than $0\farcs4$ (red circles in Figure~\ref{displace}), as compared 
to 5 of the 66 when we matched to the G13 catalog. 
For these sources (1 or ALMA033207-275120\footnote{This source is 
part of the ALESS survey. Hodge et al.\ (2013) assumed that the NIR peak
was the correct counterpart and used its corresponding spectroscopic
redshift from Danielson et al.\ (2017). 
See Section~\ref{secFIRz} for another argument against this identification.}, 
13 or ALMA033217-275233, 
14 or ALMA033222-274936, and 32 or ALMA033211-274615;
see Figure~\ref{basic_alma_images}), either the strongest star 
formation region is more strongly obscured and not coincident
with the NIR peak (sources~14 and 32 might fall
into this category), or we may be misidentifying a
projected neighbor object as the counterpart. It is also possible
that the ALMA source could be a background object lensed by a 
neighbor object, as, for example, might be the case for source~1. 

In summary, the ALMA sources have extremely diverse optical/NIR counterparts
that range from bright, low-redshift galaxies
to sources invisible in the optical/NIR.
These latter sources are the most plausible candidates for very high-redshift 
SMGs, and we shall return to them later in the paper.

\subsection{$K_s$ Counterparts}
\label{seckband}
In order to compare uniformly the total ALMA sample to the main
ALESS sample of Hodge et al.\ (2013) and the CDF-N sample of
Paper I,
we measured $K_s$ magnitudes directly from the Canada-France-Hawaii Telescope
$K_s$ image of Hsieh et al.\ (2012). This $K_s$ image covers the
total ALMA sample (including source 27), as well as
the ALESS main sample. We used $3''$ diameter apertures, 
which provide a better approximation to the total
magnitudes than smaller apertures.
We chose the zeropoint to match the Very Large Telescope based
$K_s$ magnitudes given in the G13 CANDELS catalog. 

\begin{figure}
\hskip -1.0cm
\includegraphics[width=10cm]{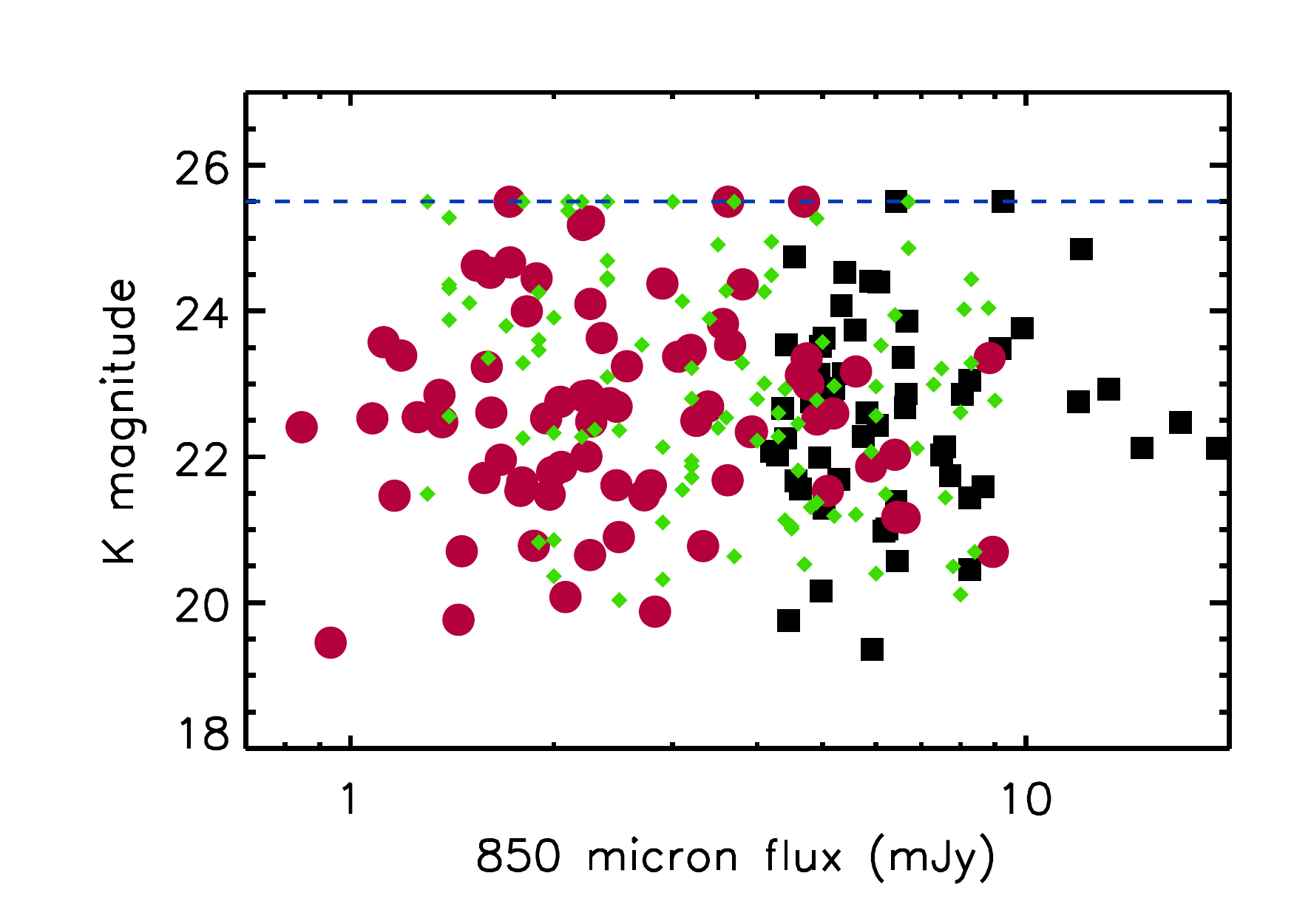}
\caption{$K_s$ magnitude vs. 850~$\mu$m flux for
the total ALMA sample (red circles), the ALESS sample (green
diamonds), and the GOODS-N sample down to 5~mJy (black squares). 
Sources fainter than $K_s = 25.5$ (blue dashed line) are shown at that value.
\label{smm_kmg}
}
\end{figure}

In Figure~\ref{smm_kmg},
we compare the $K_s$ magnitudes for the total ALMA sample (red circles)
and the ALESS sample (green diamonds). 
We also compare with the GOODS-N SCUBA-2 850~$\mu$m sample 
from Paper~I down to the 5~mJy level (black squares), which
has nearly complete, high-precision submillimeter positional 
identifications from SMA imaging. There is broad consistency among
the samples, with all showing a wide range of $K_s$ magnitudes
from brighter than 20 to fainter than 25 at all submillimeter fluxes.

\subsection{Redshifts}
\label{secredshifts}
We searched the literature for spectroscopic redshifts (hereafter, speczs) 
for the sources in the total ALMA sample.
We summarize these in Table~5, noting the reference for each specz
in the table notes. We excluded speczs where the measured galaxy
was close to but appeared not to be the ALMA counterpart,
or where we considered the spectral identification to be insecure.
We rejected redshifts for sources 1, 6, and 29.

There have been numerous photometric redshifts (hereafter, photzs)
compiled for the field (see Section~\ref{secintro}). We take our photzs from S16, 
who used the EAZY code (Brammer et al.\ 2008) to fit the extensive and 
current ZFOURGE catalog from 0.3 to 8~$\mu$m. 
Of the 75 sources in the total ALMA sample, 
73 have counterparts within $1\farcs0$ in the S16 catalog,
after applying the astrometric corrections.
However, a number of these are tagged with a very poor quality 
flag $(Q)$ from the EAZY code, reflecting the unusual SEDs of these
high-redshift dusty galaxies and the limited number of band detections.
If we restrict to the high-quality ($Q<3$) photzs, then this gives 
photzs for 59 of the ALMA sources, which we give in Table~5, together 
with the 95\% confidence range.

\begin{figure*}
\hspace{-0.75cm}\includegraphics[width=10cm]{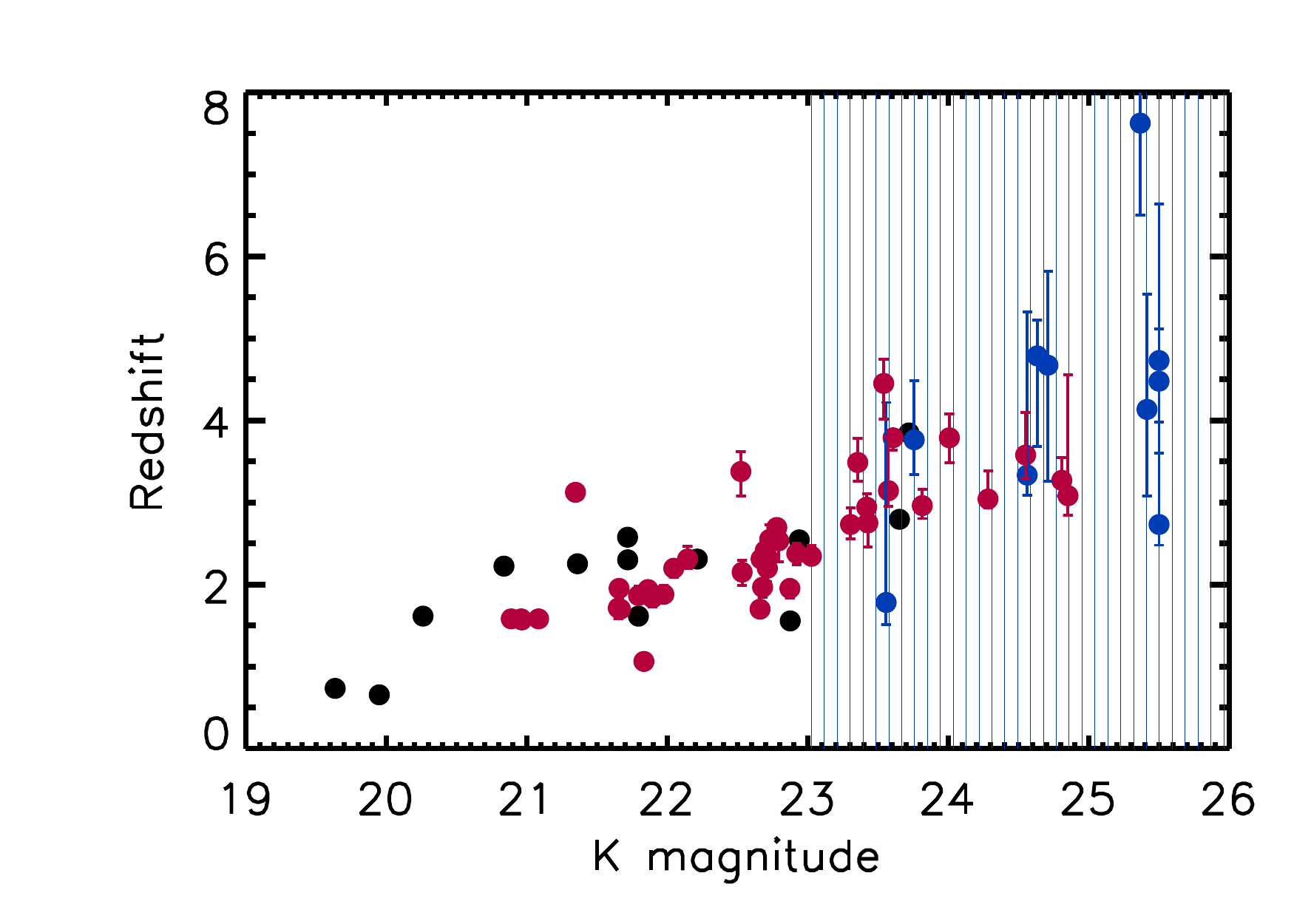}
\hspace{-0.85cm}\includegraphics[width=10cm]{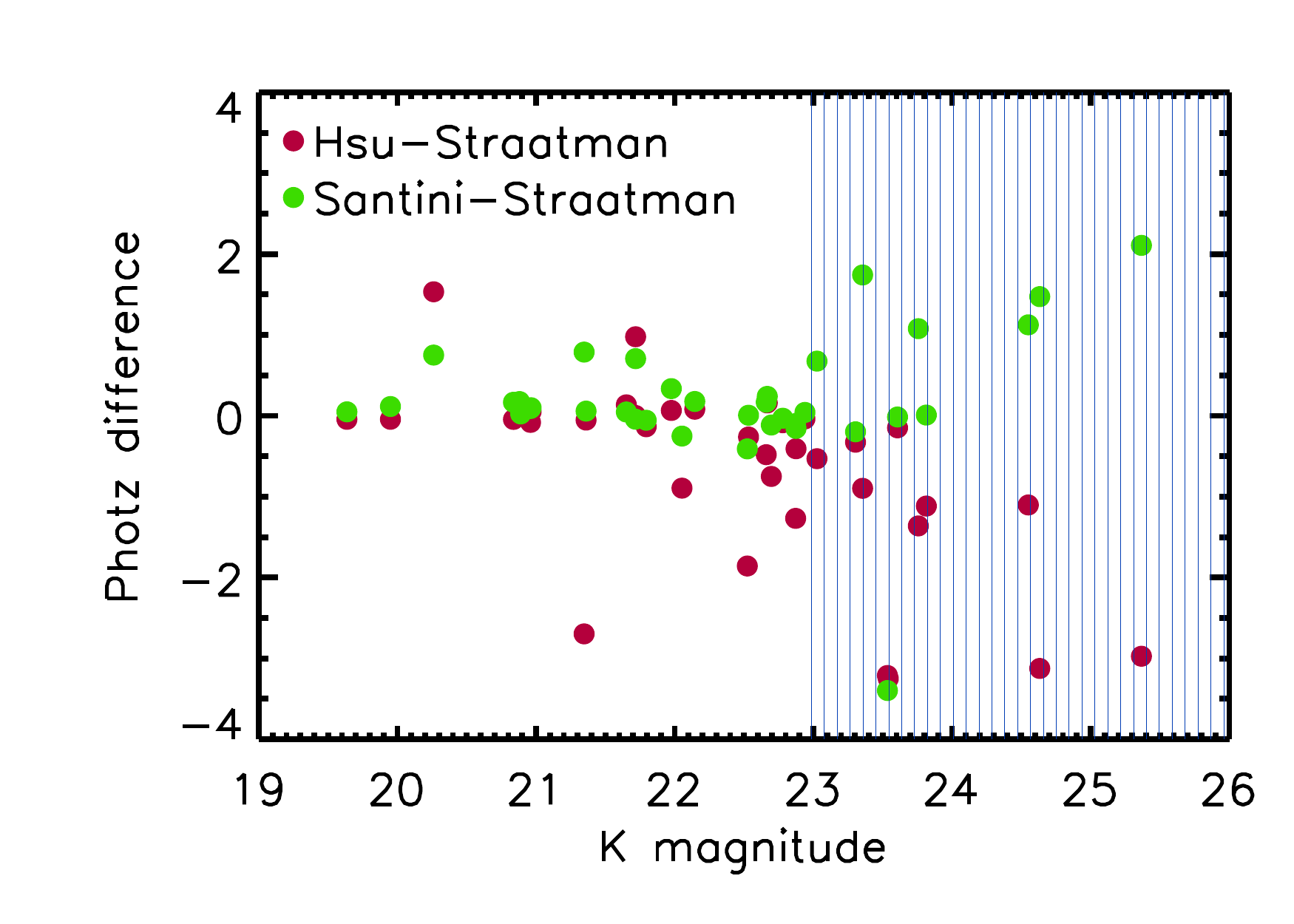}
\vskip -0.5cm
\hspace{-0.85cm}\includegraphics[width=10cm]{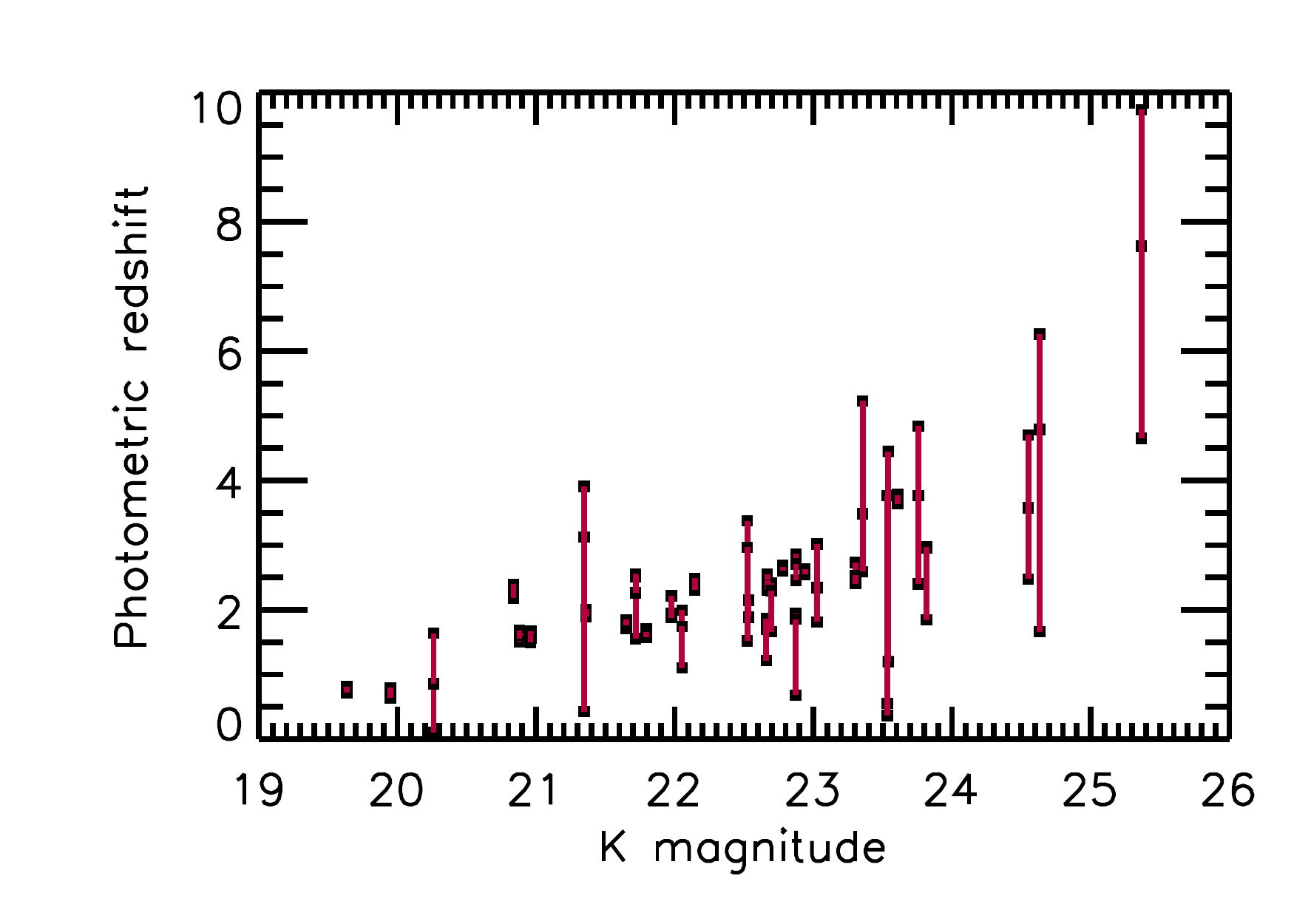}
\hspace{-0.75cm}\includegraphics[width=10cm]{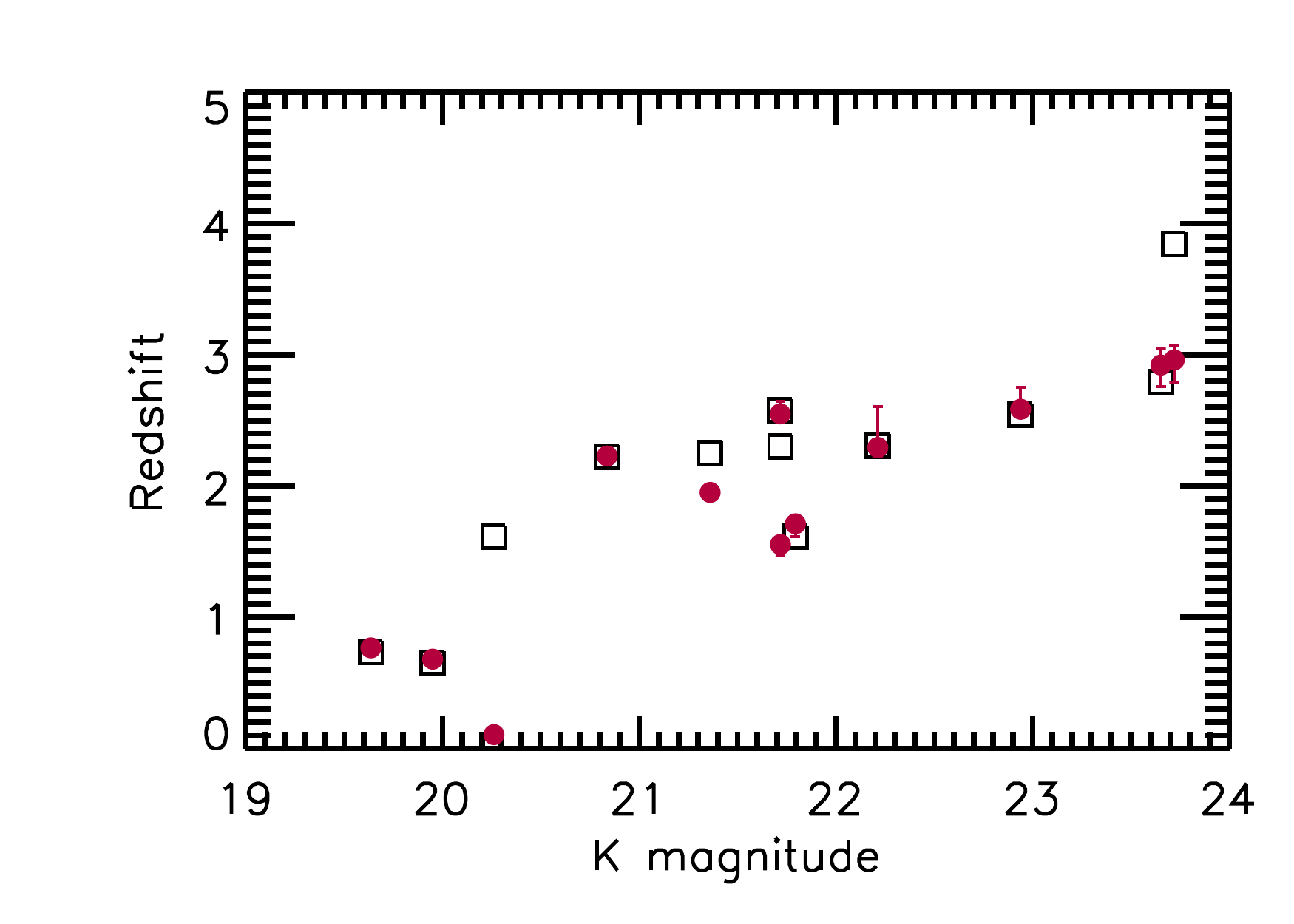}
\caption{Redshift vs. $K_s$-band magnitude. 
{\em Top left:\/} Black circles show speczs. Where there are no
speczs, red circles show photzs with $Q<3$ (high-quality) and blue circles 
show photzs with $Q>3$ (poor) from S16. The error
bars show the 95\% confidence range for the photzs.
In the remaining panels, we use only the high-quality photzs from S16.
{\em Top right:\/} 
Red circles show the Hsu et al.\ (2014) photzs minus the S16 photzs,
and green circles show the Santini et al.\ (2015) photzs minus the S16 photzs. 
Blue shaded region (this panel and top left panel) shows the $K_s>23$ 
magnitude range above which the photzs become very uncertain.
{\em Bottom left:\/}
For each source, the photzs from all three papers are
shown simultaneously (black squares), connected by red lines.
{\em Bottom right:\/} 
Even at bright $K_s<22$ magnitudes, the photzs
do not always reproduce the speczs, as demonstrated
over the $K_s$ range where there are speczs (black open squares); these
are compared with the photzs from S16 (red circles).
\label{kmg_z}
}
\end{figure*}

In the top left panel of Figure~\ref{kmg_z}, we plot redshift versus $K_s$ 
magnitude for sources with speczs (black circles). For sources where there 
are no speczs, we plot photzs with $Q<3$ (red circles) and photzs with 
$Q>3$ (poor fits; blue circles) from S16. We hereafter use only the high-quality
photzs from S16, unless specifically stated otherwise.
The redshifts show a correlation with the $K_s$ magnitudes, which has been
demonstrated to be a crude redshift estimator (Barger et al.\ 2014;
Simpson et al.\ 2014). This can also be seen when we use other photometric
redshifts estimates, such as those of Hsu et al.\ (2014) or
Santini et al.\ (2015). However, there are large differences between
the various photz estimates at the fainter magnitudes, as 
we illustrate in the top right panel of Figure~\ref{kmg_z} by plotting
the difference between the various photzs as a function of $K_s$ 
magnitude. Particularly at $K_s > 23$ (blue shading), there 
are considerable uncertainties in the photzs and substantial
variations between the groups' estimates. Despite this, the overall trend to
higher redshifts with increasing $K_s$ magnitude can still be seen in
the bottom left panel, where we show simultaneously the three photz 
estimates for each source (black squares), connected with red lines.
Finally, in the bottom right panel, we illustrate how even at bright $K_s<22$
magnitudes, there are some large discrepancies between the speczs 
and the photzs.

\begin{figure}
\hspace{-0.75cm}\includegraphics[width=9.5cm]{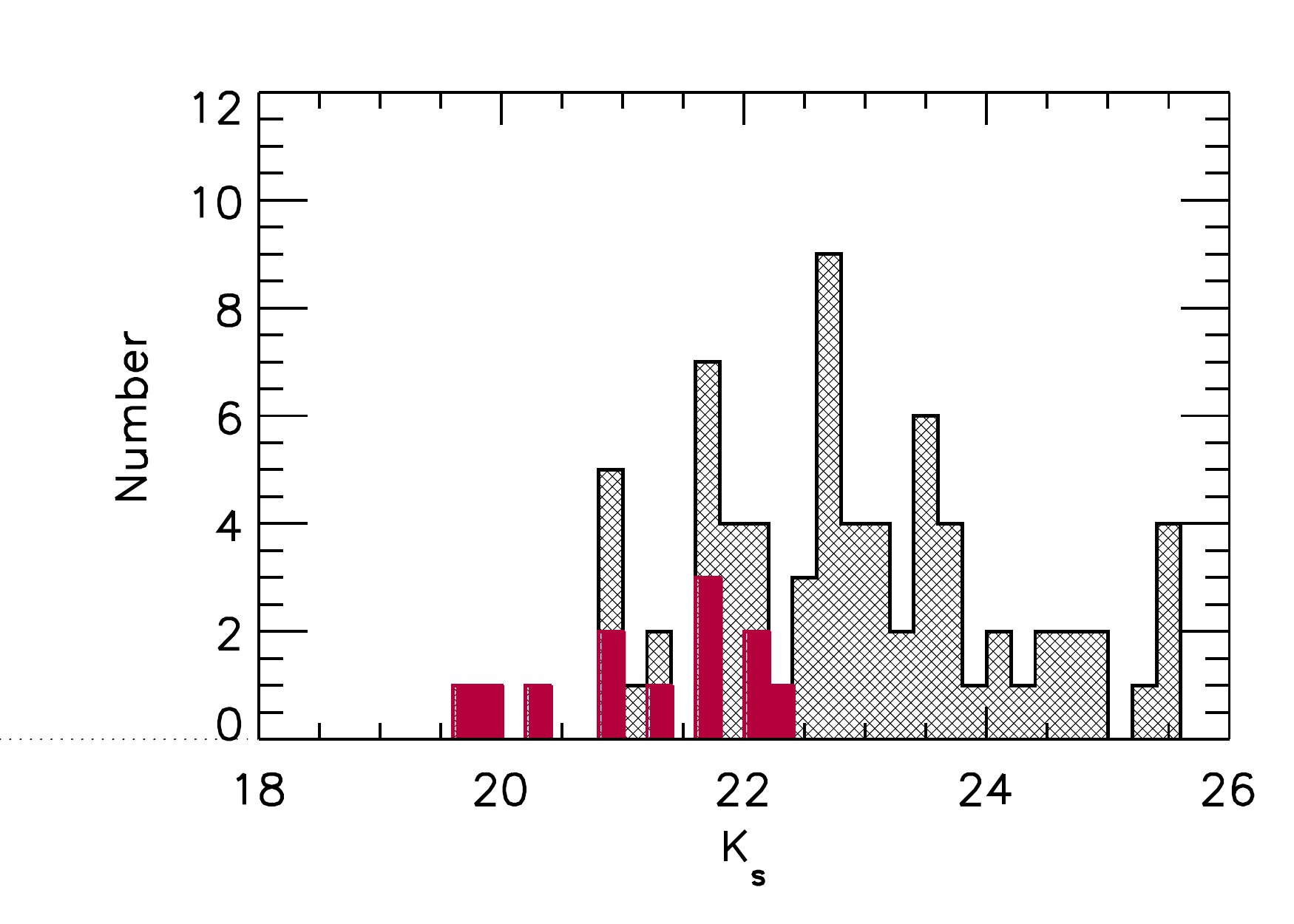}
\vskip -0.9cm
\hskip -0.75cm
\includegraphics[width=9.5cm]{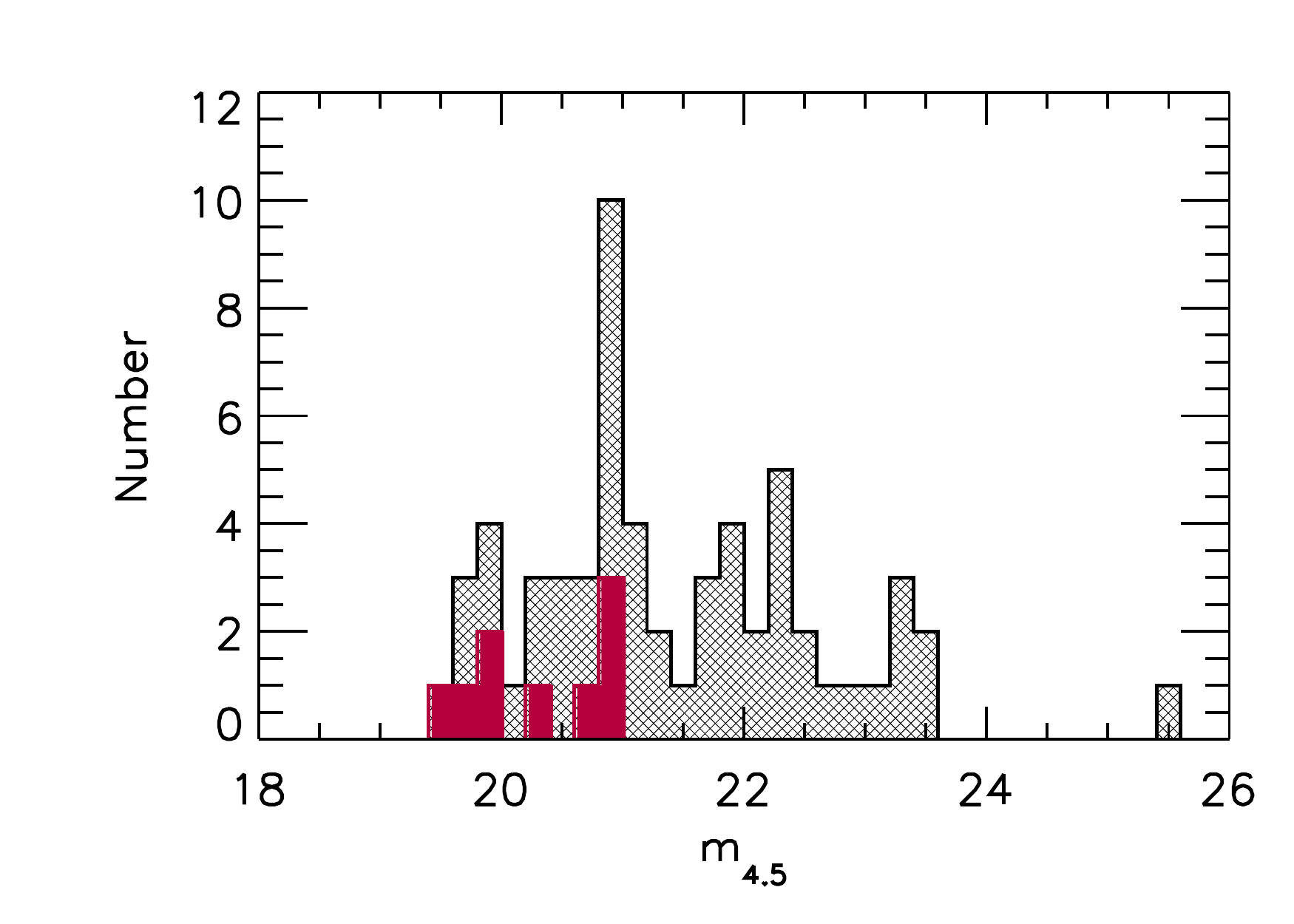}
\caption{Histograms of the (a) $K_s$ and (b) $m_{4.5}$ magnitudes
for the total ALMA sample. In (b), only the 60 ALMA
sources that are isolated in the S-CANDELS data are shown
in order to avoid photometric contamination.
Undetected sources are shown at 25.5 in both panels.
Red shading shows sources with speczs. 
\label{hists}
}
\end{figure}

In Figure~\ref{hists}(a), we show the distribution of $K_s$ magnitudes 
for the total ALMA sample (gray shading) and for those sources with 
speczs (red shading). We show undetected sources at a nominal 
magnitude of $K_s=25.5$. Three of the sample are undetected in $K_s$ 
(sources~13 or ALMA033217-275233, 19 or ALMA033226-275208, and 
58 or ALMA033244-275011). 
These are a subset of the sources that are very faint in the F160W band.

\begin{figure}
\hspace{-0.75cm}\includegraphics[width=11.0cm]{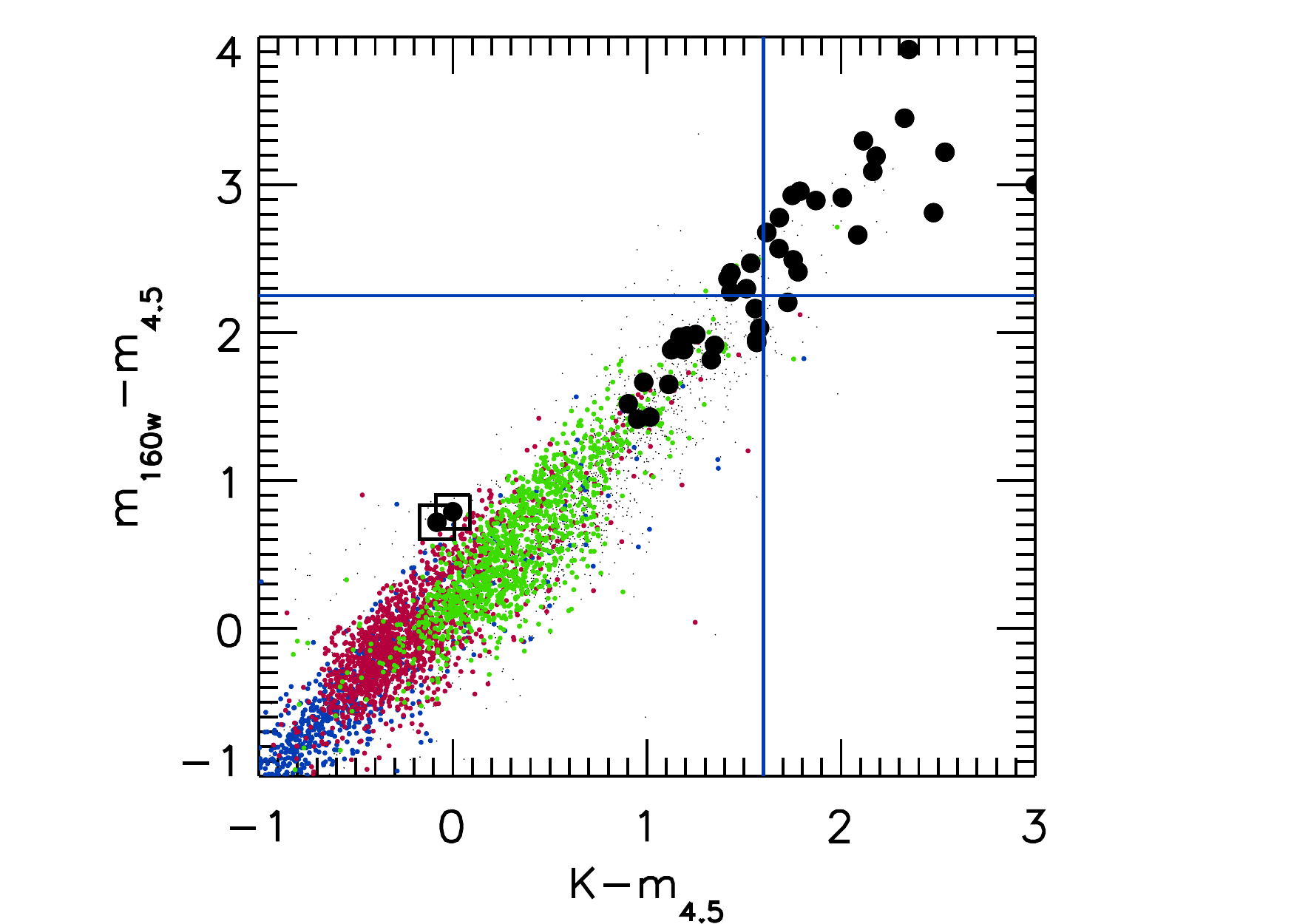}
\caption{Comparison of the HIERO ($m_{\rm 160W} - m_{4.5}>2.25$)
and KIERO ($K-m_{4.5}>1.6$) selections (blue lines). The black large circles
show the 42 ALMA sources that have measured $K_s$ magnitudes in the
G13 catalog and are isolated in the IRAC bands.
The different colored dots show the full GOODS-S data with $K=18-24$ divided 
into photometric redshift intervals
(blue is $z=0-0.5$, red is $z=0.5-1$, green is $z=1-1.5$,
and black for the remainder). 
The two ALMA sources with specz $0<z\le1$ are marked with open squares.
\label{hiero_kiero}
}
\end{figure}

\begin{figure}
\includegraphics[width=10.0cm]{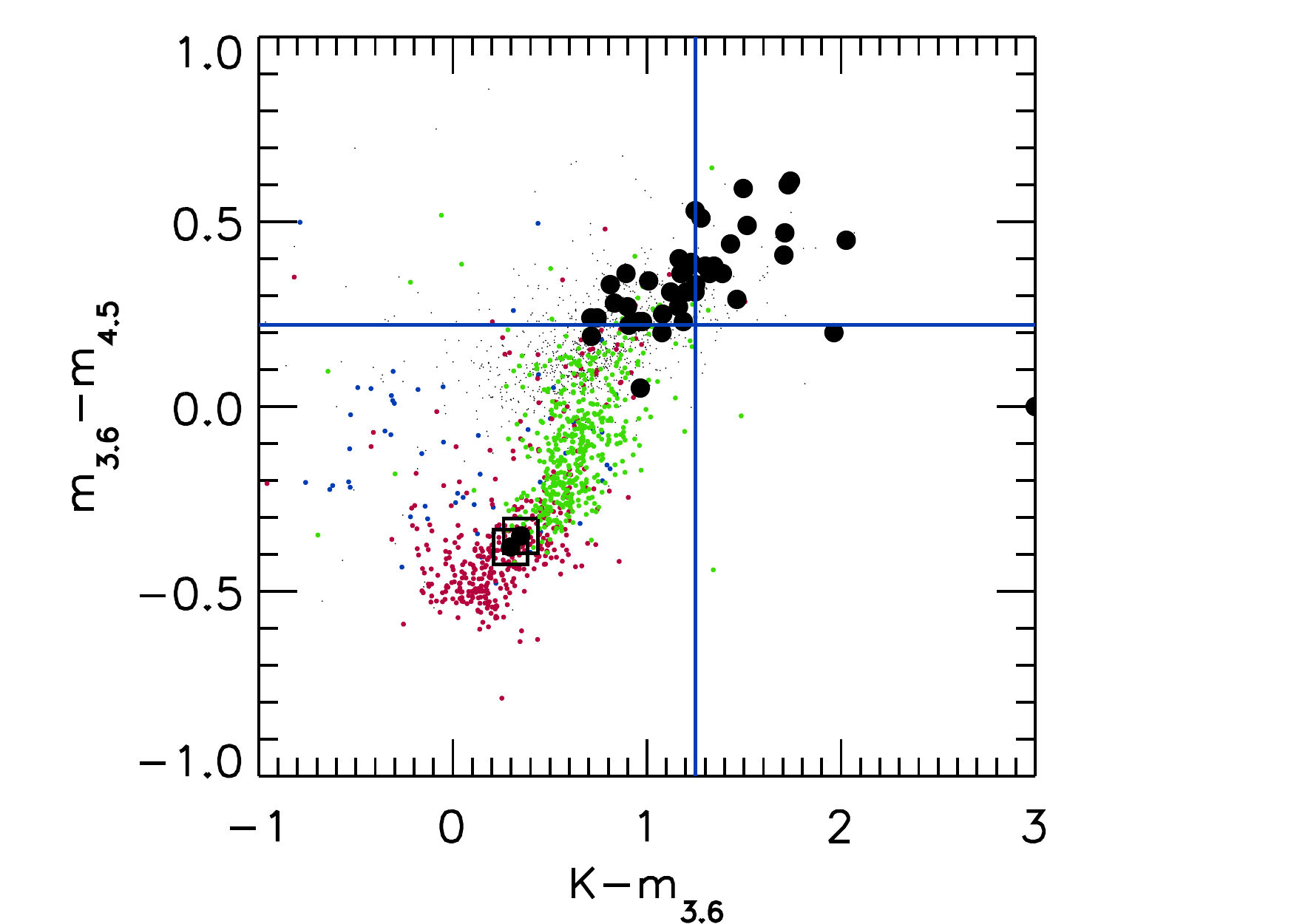}
\caption{Chen et al.\ (2016) three color selection. Only
sources with $m_{\rm 850LP}-K >1.1$ are shown. 
The plotted data are 42 ALMA sources with counterparts in 
the G13 catalog and that are isolated in the IRAC bands 
(black large circles), and GOODS-S sources with $K=18-24$ (dots);
the latter are divided into photometric
redshift intervals (blue is $z=0-0.5$, red is $z=0.5-1$, green is $z=1-1.5$, 
and black for the remainder). The remaining two selections 
($K_s-m_{3.6}>1.25$ and $m_{3.6}-m_{4.5} > 0.22$)
are shown by the blue lines.
The two ALMA sources with specz $0<z\le1$ are marked with
enclosing large open squares.
\label{selection_plot}
}
\end{figure}

\subsection{MIR Counterparts and SMG Color Selections}
\label{mir}
We obtained 3.6~$\mu$m and 4.5~$\mu$m fluxes for the total ALMA sample
from the S-CANDELS catalog of Ashby et al.\ (2015), supplemented with 
S-CANDELS fluxes from NED IPAC for the small number of sources not 
included in the Ashby et al.\ catalog. 
(The latter fluxes do not appear to have been published elsewhere.)
We used the $2\farcs4$ diameter aperture fluxes, 
which closely match the 3.6~$\mu$m and 4.5~$\mu$m fluxes of G13. 
We restricted to the 60 ALMA sources that are isolated in 
the S-CANDELS data to avoid photometric contamination.
All but one of these sources are detected in both the 3.6~$\mu$m and 
4.5~$\mu$m bands. The remaining source (source~75 or ALMA033217-274713) 
is not detected in either, but this is the faintest of the ALMA sources and may 
be spurious. In Figure~\ref{hists}(b), we show the distribution of the 4.5~$\mu$m 
magnitudes for the 59 isolated and detected sources, plus the undetected
source~75 plotted at a nominal 4.5~$\mu$m magnitude of $m_{4.5}=25.5$.

Infrared colors have been used to select high-redshift dusty
galaxies (Wang et al.\ 2012; Chen et al.\ 2016; Wang et al.\ 2016).
These methods rely on the fact that the colors asymptote
at high redshifts (see Figure~4 of Wang et al.\ 2012), and high
values can only be obtained with very large reddening. Wang et al.\ (2012)
used a condition $K_s-m_{4.5} > 1.6$ (KIEROs), while Wang et al.\ (2016)
used $m_{\rm 160W}-m_{4.5} > 2.25$ (HIEROs). These selections are extremely
similar, as can be seen from Figure~\ref{hiero_kiero}, where we plot 
$m_{\rm 160W}-m_{4.5}$ versus $K_s-m_{4.5}$. For these analyses,
in order to obtain optimally matched fluxes,
we use only sources with counterparts in the G13
catalog, and we take all of the magnitudes from this reference.
We show the KIEROS and
HIEROs selections as blue lines, and we use different colored dots to show 
the GOODS-S data between $K_s=18-24$ divided into photometric redshift intervals. 
We only plot the 42 ALMA sources that have measured $K_s$ magnitudes in the 
G13 catalog and are isolated in the IRAC bands (black large circles).
We find that 18 of these are selected by the KIEROs criterion, 22 by the 
HIEROs criterion, and 17 by both. 

As an aside, we note that two of the fainter SMGs in Table~4 
and Figure~\ref{hists} (source~66 or ALMA033210-274807 and source~74 
or ALMA033222-274935) fall at much bluer
colors in Figure~\ref{hiero_kiero} and Figure~\ref{selection_plot}
than might be expected. We mark them with open squares in both figures.
Both have spectroscopic redshifts
($z=0.654$ and $z=0.732$, respectively, with source 66 also being 
serendipitously detected in CO in BASIC) and bright 
$K_s$ magnitudes ($K_s<20$).
These sources lie above the main track of the galaxies in 
Figure~\ref{hiero_kiero} with red $m_{160w}-K_s$ colors.

In Figure~\ref{selection_plot}, we illustrate the three color selection of Chen et al.\ (2016) 
($m_{\rm 850LP}-K_s >1.1$, $K_s-m_{3.6}>1.25$, and $m_{3.6}-m_{4.5} > 0.22$)
by plotting $m_{3.6}-m_{4.5}$ versus $K_s-m_{3.6}$ and only showing sources 
that satisfy $m_{\rm 850LP}-K_s >1.1$. The plotted data are 42 ALMA sources 
with counterparts in the G13 catalog and that are isolated in the IRAC bands
(black large circles), and GOODS-S sources with $K=18-24$, divided into 
photometric redshift intervals (colored dots). We find that 15 of
the ALMA sources are selected. These are a subset of the HIEROS selection.

However, the Chen et al.\ (2016) selection is sensitive to the $K_s-m_{3.6}$
color and hence to the precise details of the ground-based and {\em Spitzer\/} 
photometry. For instance, Laporte et al.\ (2017) found a better selection with
$K_s-m_{3.6}>0.8$ in their ALMA studies of the Frontier Fields. Using 
$K_s-m_{3.6}>1$ would result in 26 of the ALMA sources being selected.

In conclusion, infrared color selection methods are useful for narrowing 
down the fraction of the population that could be SMGs. However, photometric uncertainties 
remain a major drawback toward obtaining complete or even representative samples.
There is also the issue of how many of the selected sources are not SMGs.

To examine the latter for each of the infrared color selection methods, 
we consider the GOODS-S region that is covered by the ALMA observations.
Here the G13 catalog
gives 31 KIEROs and 44 HIEROs with $K_s$ magnitudes between 18 and
24. Of the 31 KIEROs, we have ALMA detections for 13, and a further
four are detected above a $2\sigma$ threshold in the SCUBA-2 image cleaned 
of ALMA sources. In combination, this gives a 55\% submillimeter
detection rate. Of the 44 HIEROs, we have ALMA detections for 12, and a
further seven have $>2\sigma$ SCUBA-2 detections.
In combination, this gives a 43\% submillimeter detection rate.

Likewise, the G13 catalog gives 39 sources for the Chen et al.\ (2016) three color 
selection, of which we have ALMA detections for 15, and a further three have 
$>2\sigma$ SCUBA-2 detections. 
In combination, this gives a 46\% submillimeter detection rate. 
If we relax the three color selection
criterion to $K_s-m_{3.6}>1$, as discussed above, then the G13 catalog gives 
93 sources, of which we have ALMA detections for 26,
and a further 12 have $>2\sigma$ 
SCUBA-2 detections. This reduces the combined detection rate to 41\%.

We conclude that all three of the infrared color selection methods are comparable
in their SMG selections and contamination rates. We note that the KIEROs method
may be more versatile, since it uses only ground-based and {\em Spitzer\/}
data and hence avoids the need for deep {\em HST\/} data.

\begin{figure*}
\hspace{-2cm}\includegraphics[width=5.5in,angle=0]{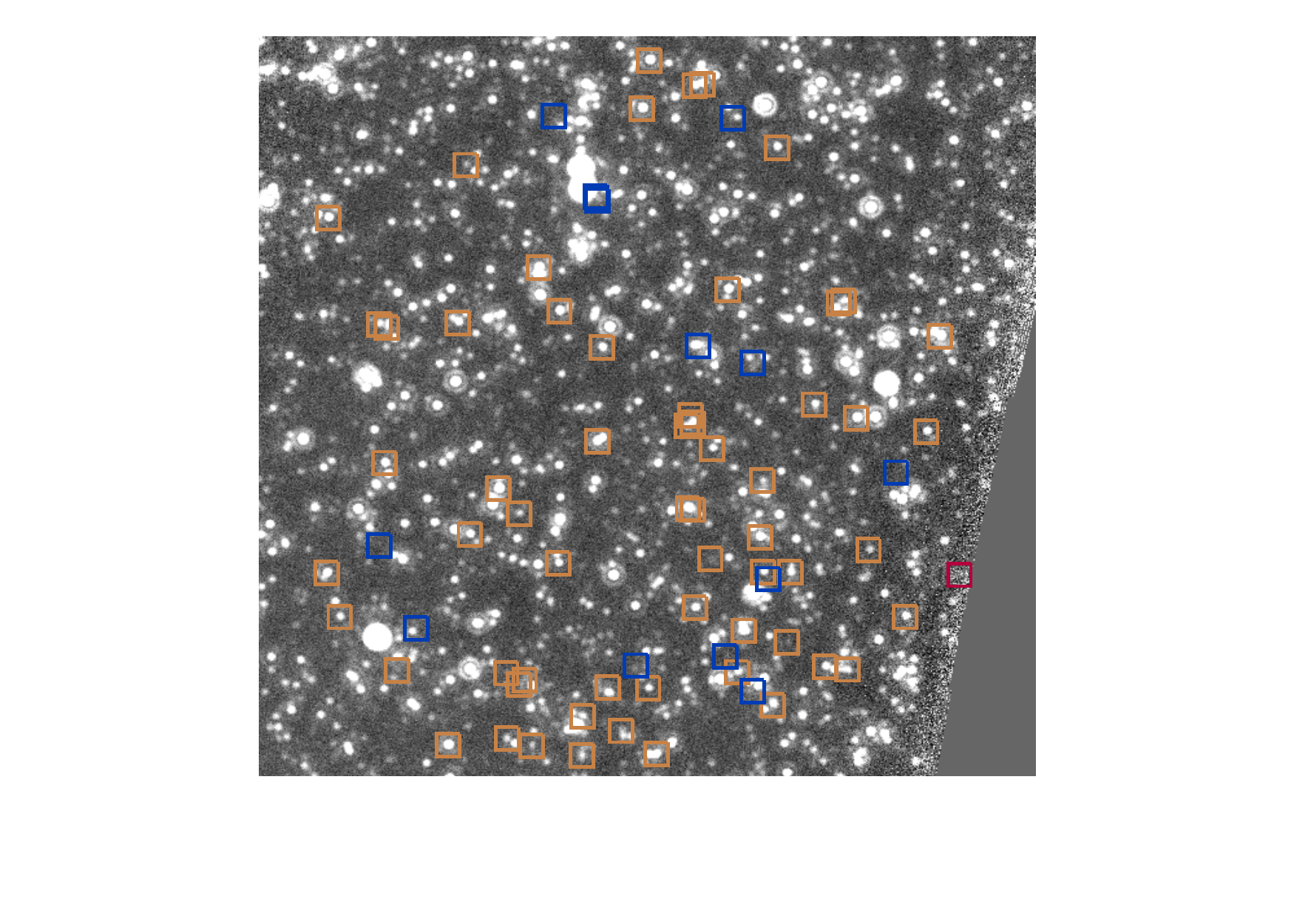}
\hspace{-5.5cm}\includegraphics[width=5.5in,angle=0]{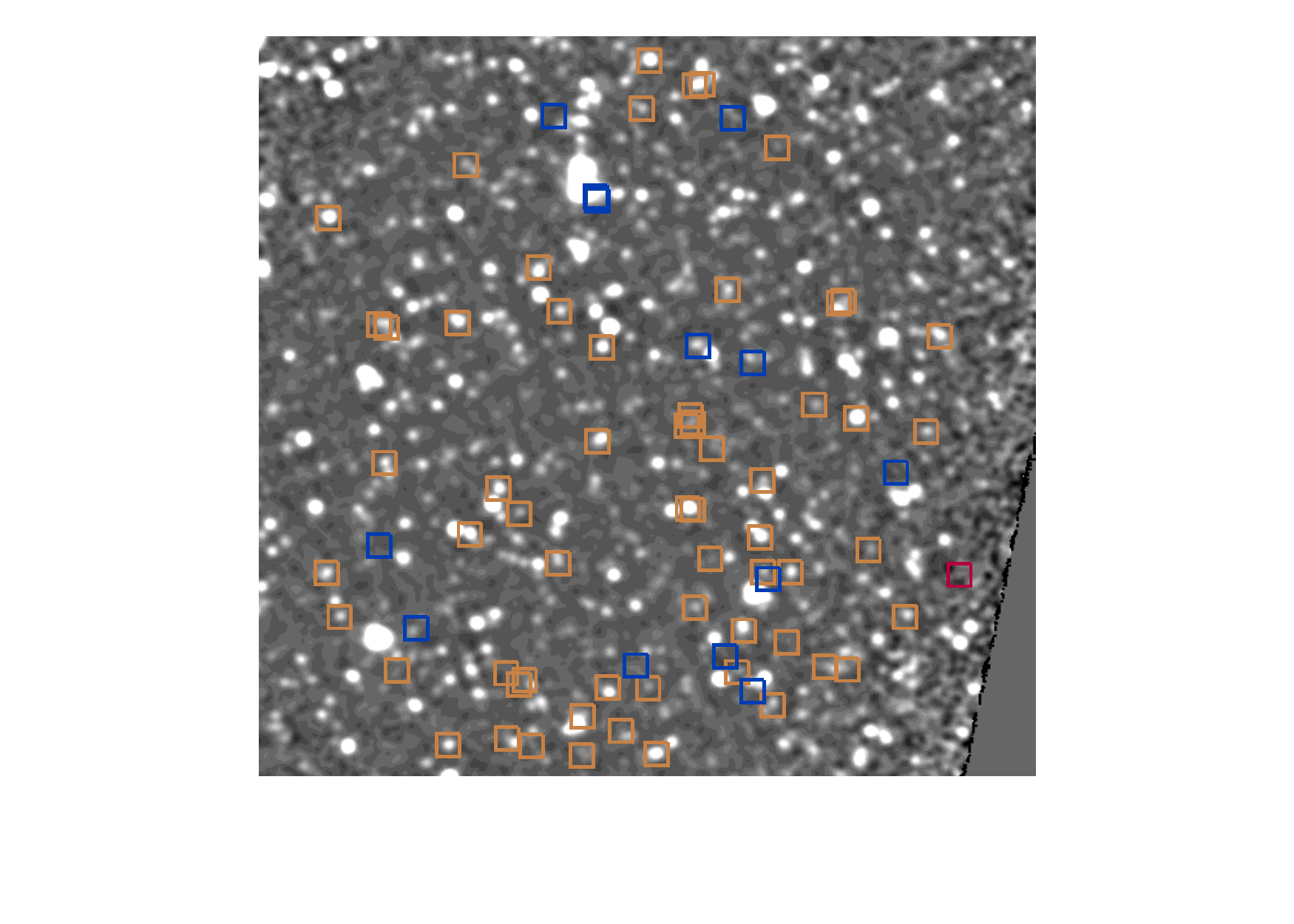}
\vskip -1.25cm
\caption{
Overlay of the ALMA sources on the {\em (left)\/} 24~$\mu$m and
{\em (right)\/} 100~$\mu$m images of the field. Sources with 24~$\mu$m 
counterparts in the Elbaz et al. (2011) catalog within a matching
radius of $1\farcs5$ and with fluxes $>20~\mu$Jy are shown in gold, while other 
ALMA sources are shown in blue. As discussed in the text, the differentiation
between the blue and gold squares is not sensitive to the choice of matching radius. 
We exclude one source (27) that lies at the edge of the field; it is shown 
with a red square.
\label{24_160}
}
\end{figure*}

\begin{figure*}
\includegraphics[width=2.5in,angle=0]{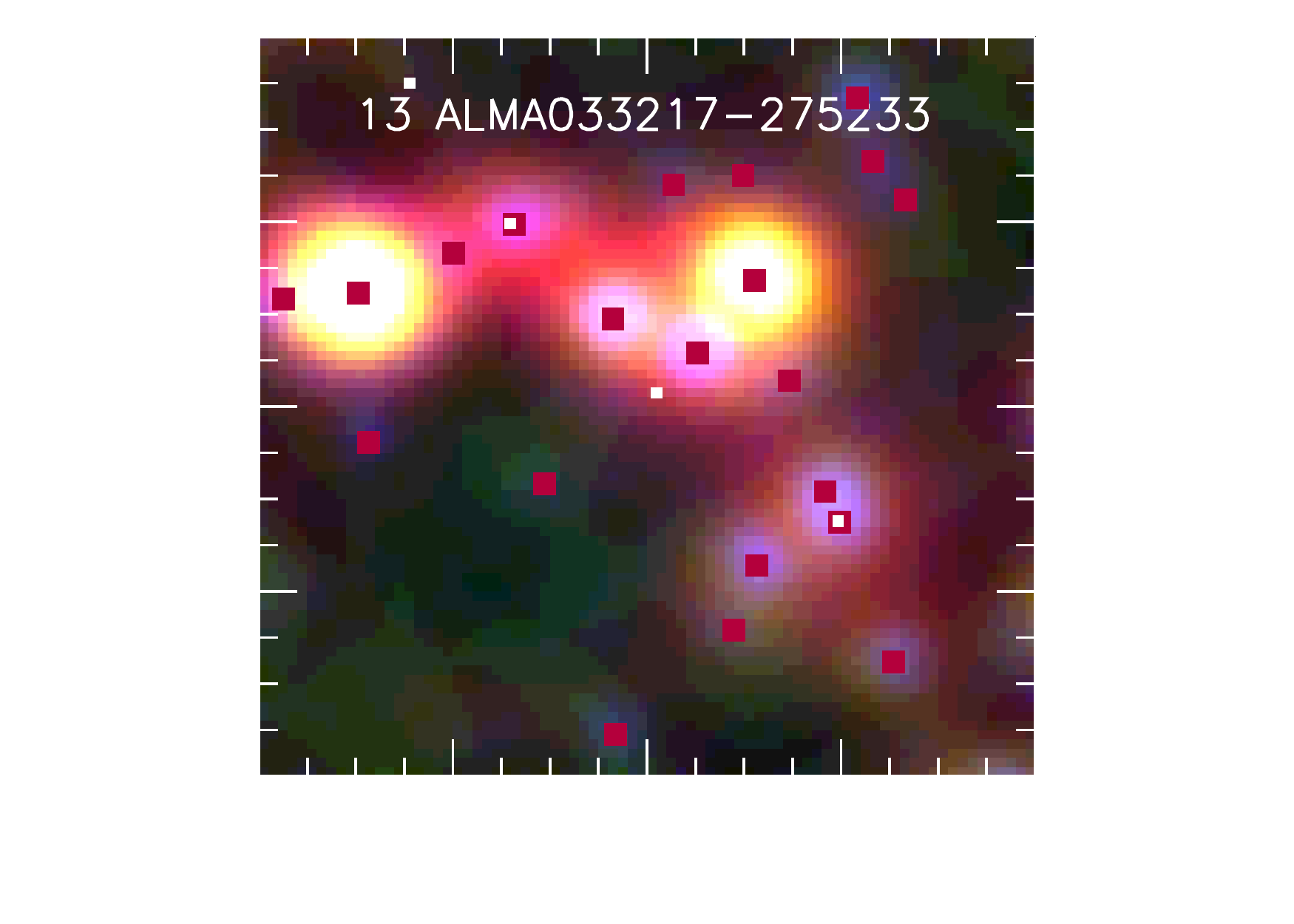}
\hspace{-3.3cm}\includegraphics[width=2.5in,angle=0]{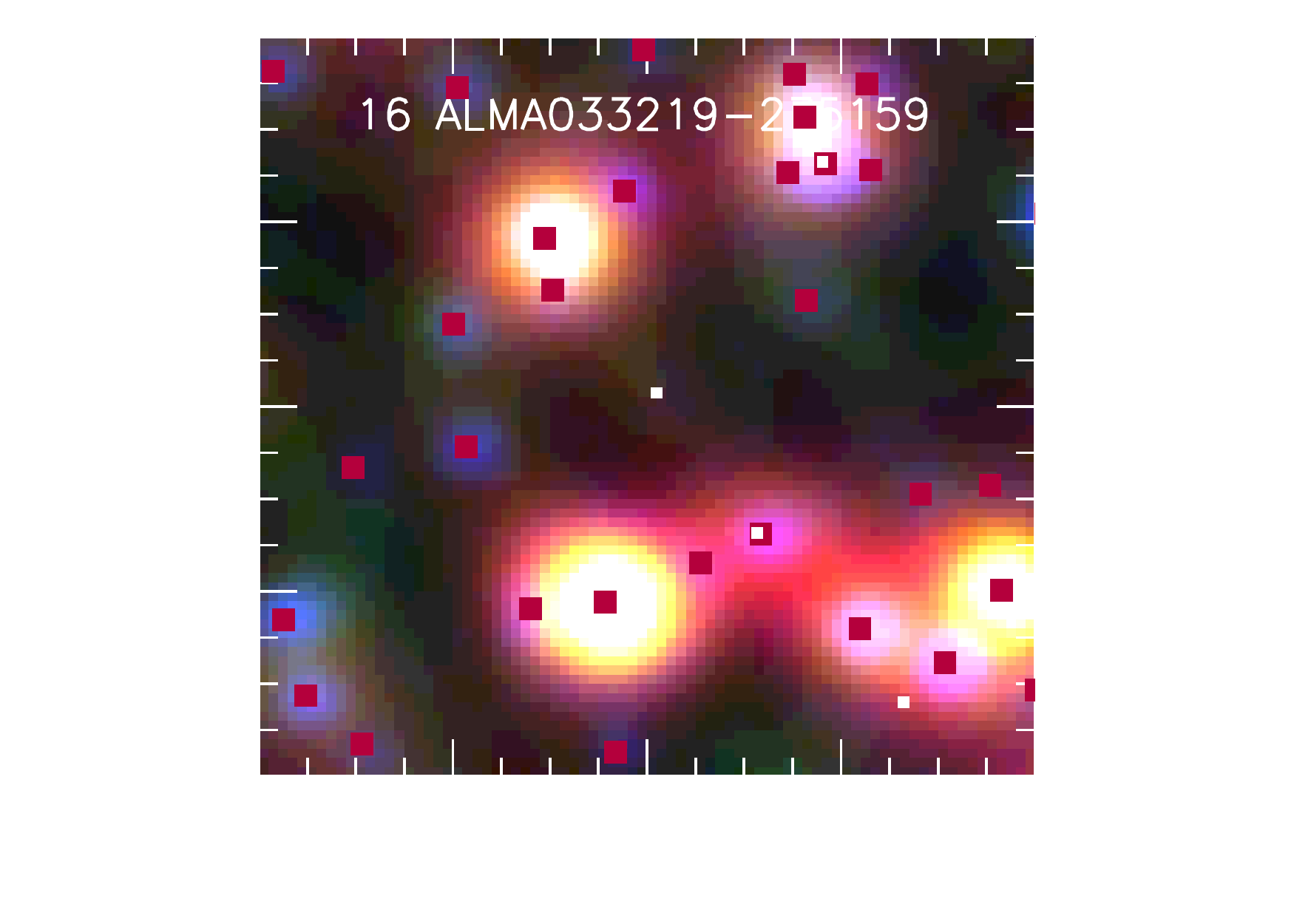}
\hspace{-3.3cm}\includegraphics[width=2.5in,angle=0]{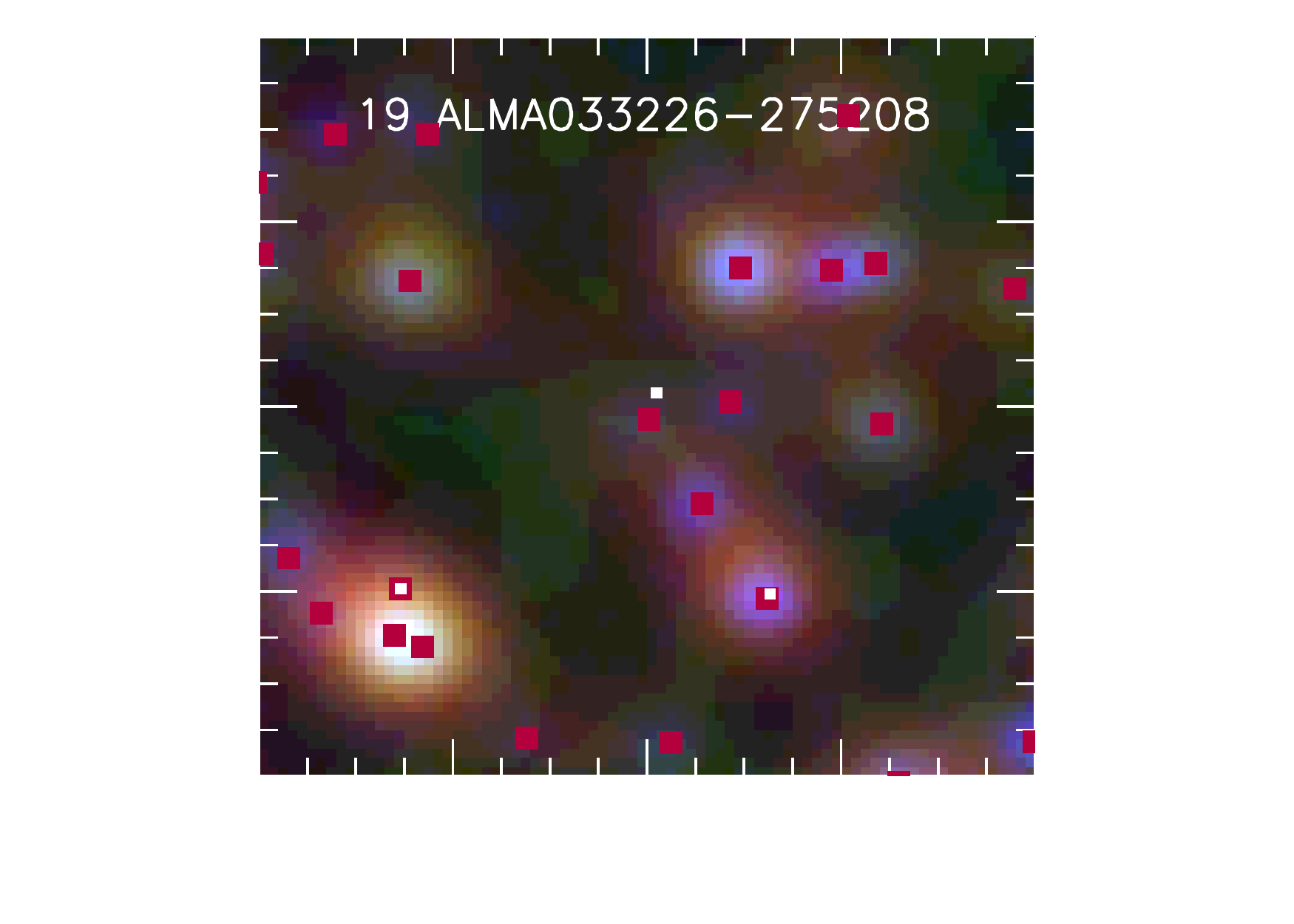}
\hspace{-3.3cm}\includegraphics[width=2.5in,angle=0]{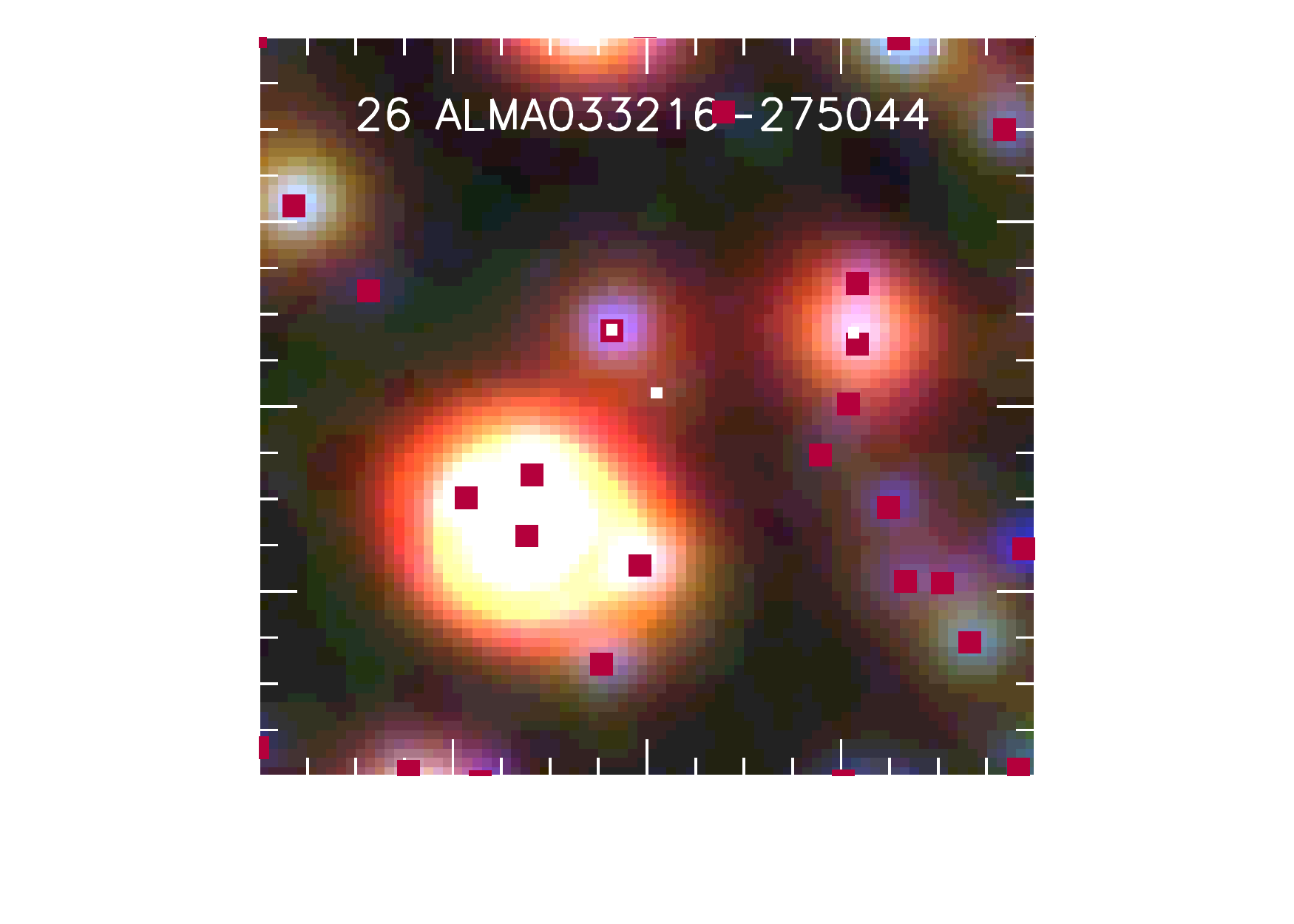}
\includegraphics[width=2.5in,angle=0]{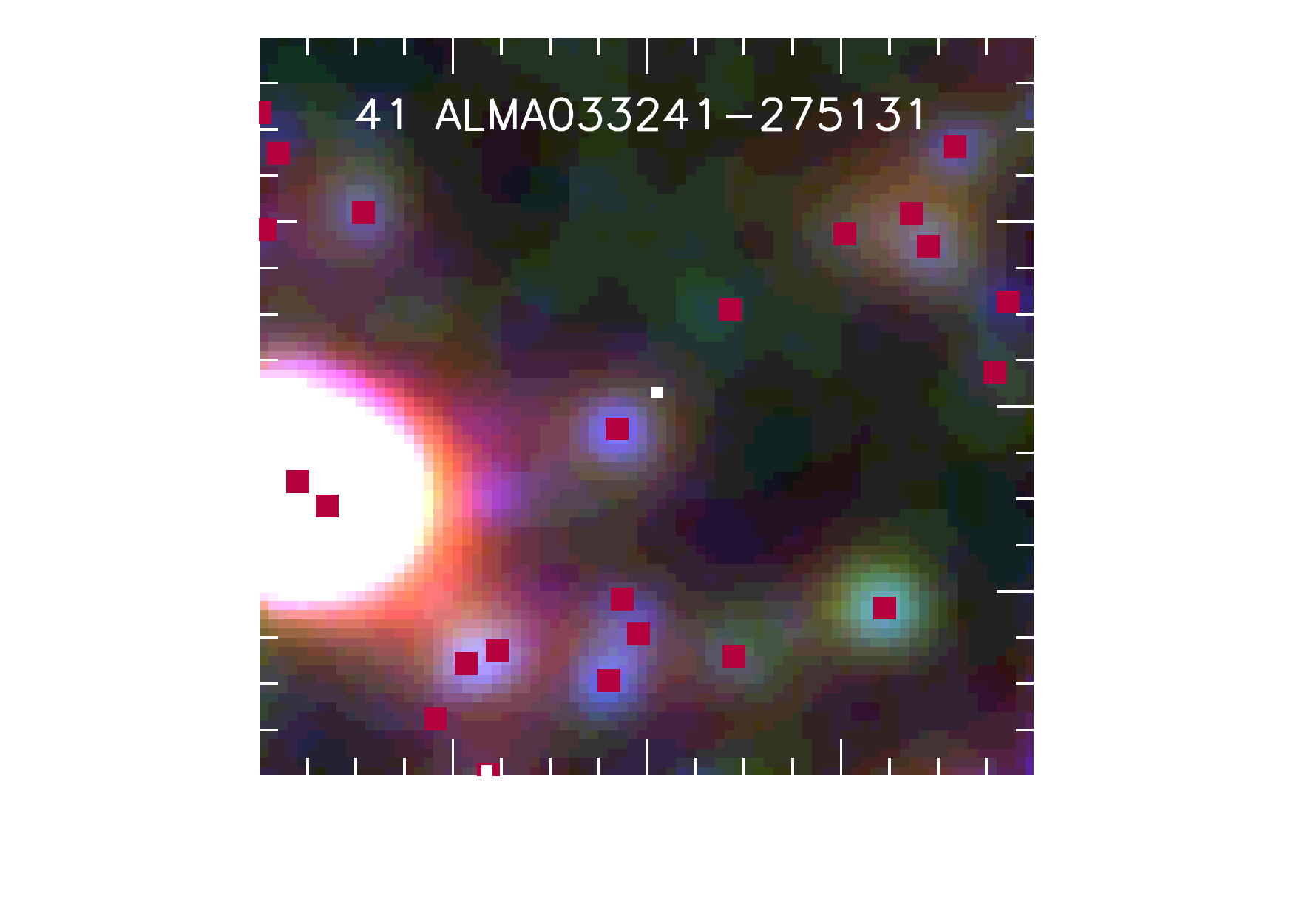}
\hspace{-3.3cm}\includegraphics[width=2.5in,angle=0]{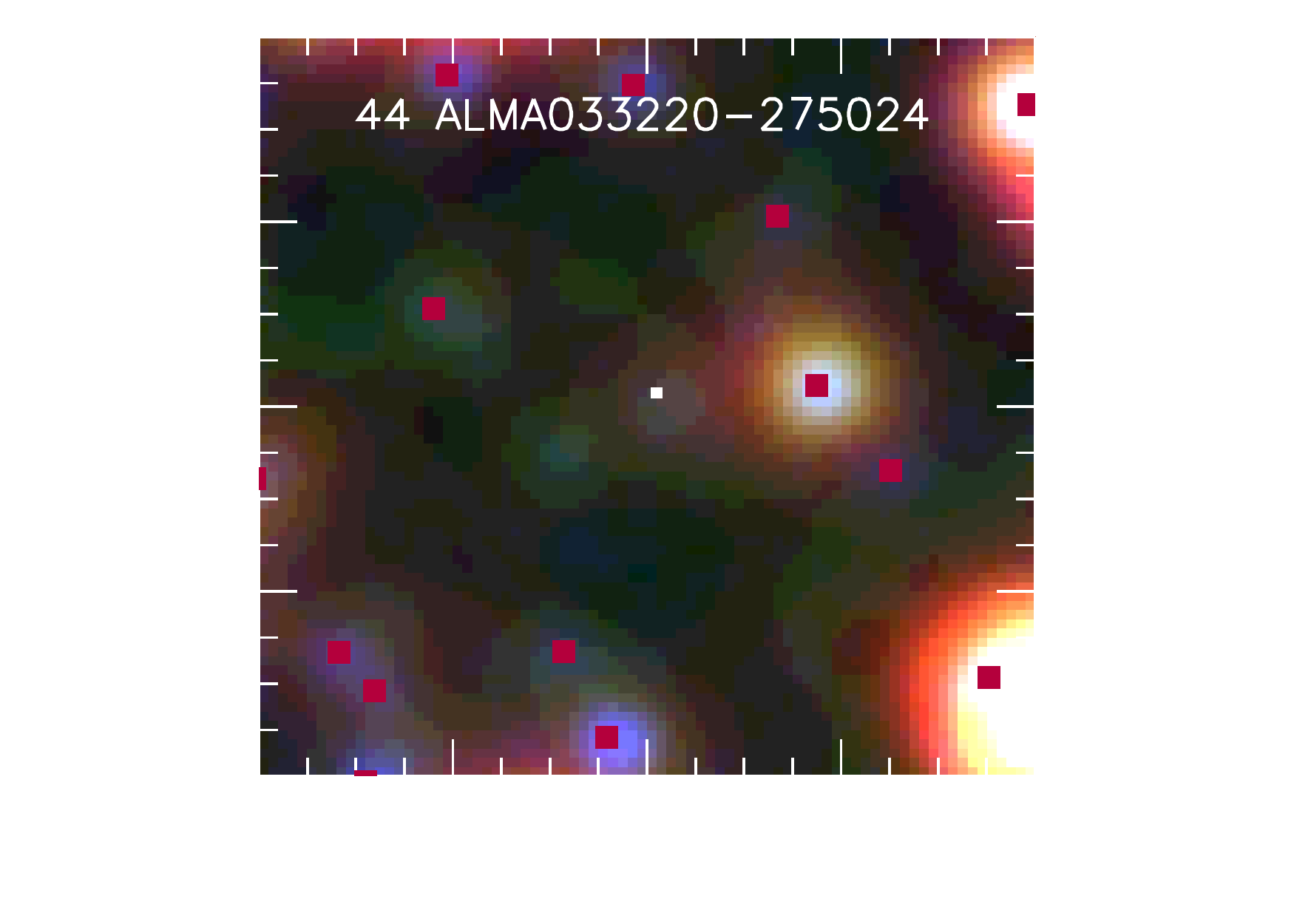}
\hspace{-3.3cm}\includegraphics[width=2.5in,angle=0]{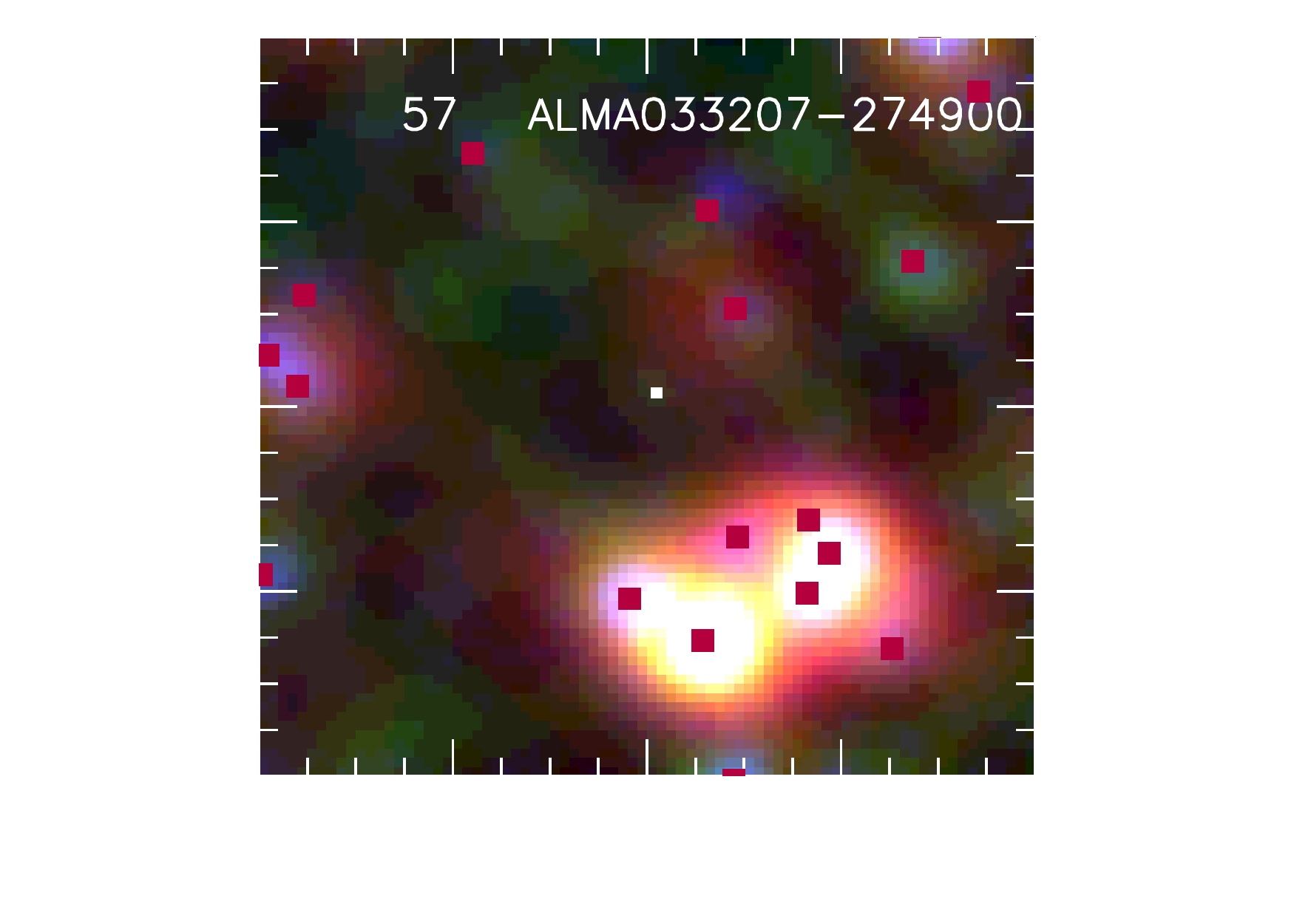}
\hspace{-3.3cm}\includegraphics[width=2.5in,angle=0]{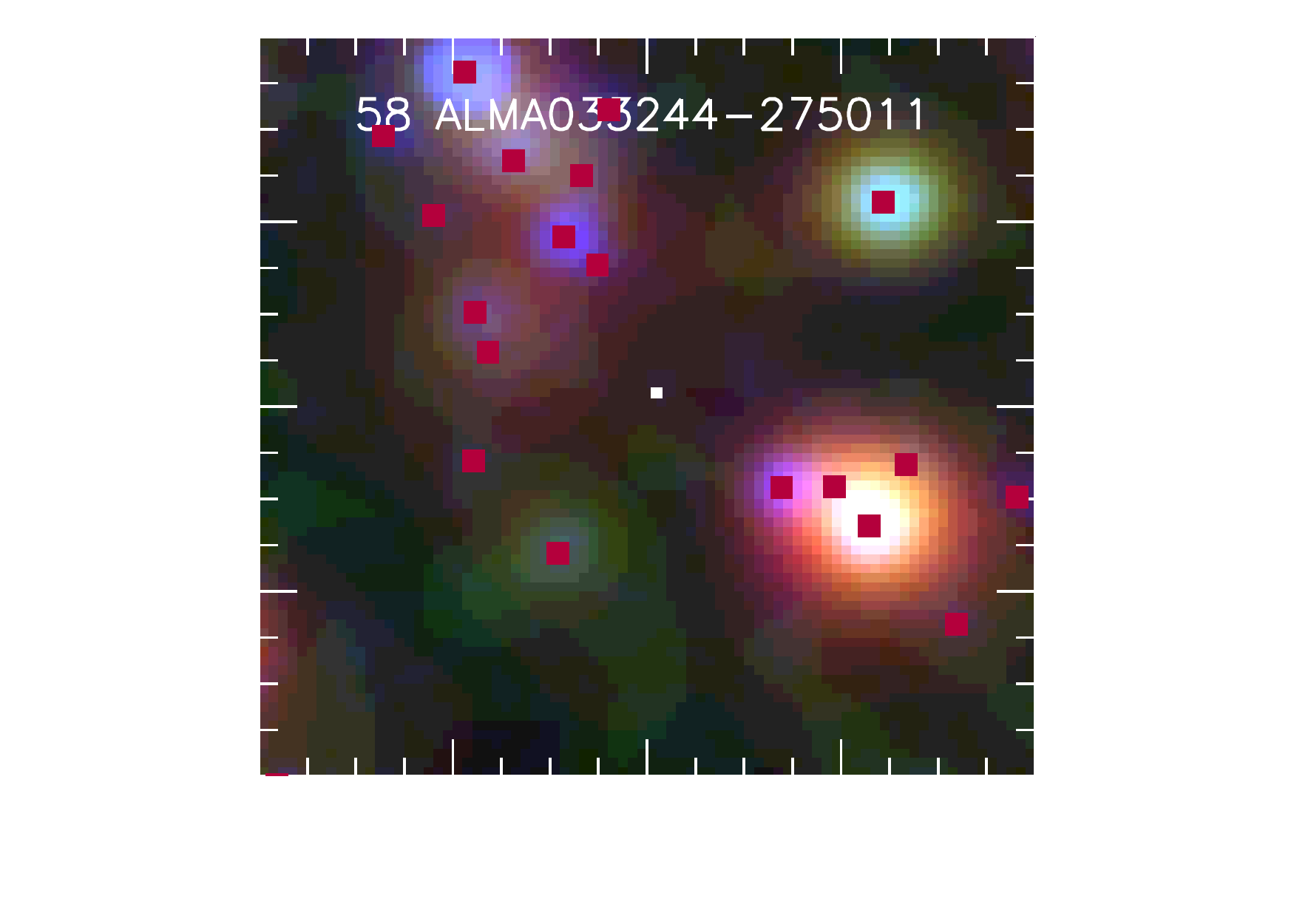}
\includegraphics[width=2.5in,angle=0]{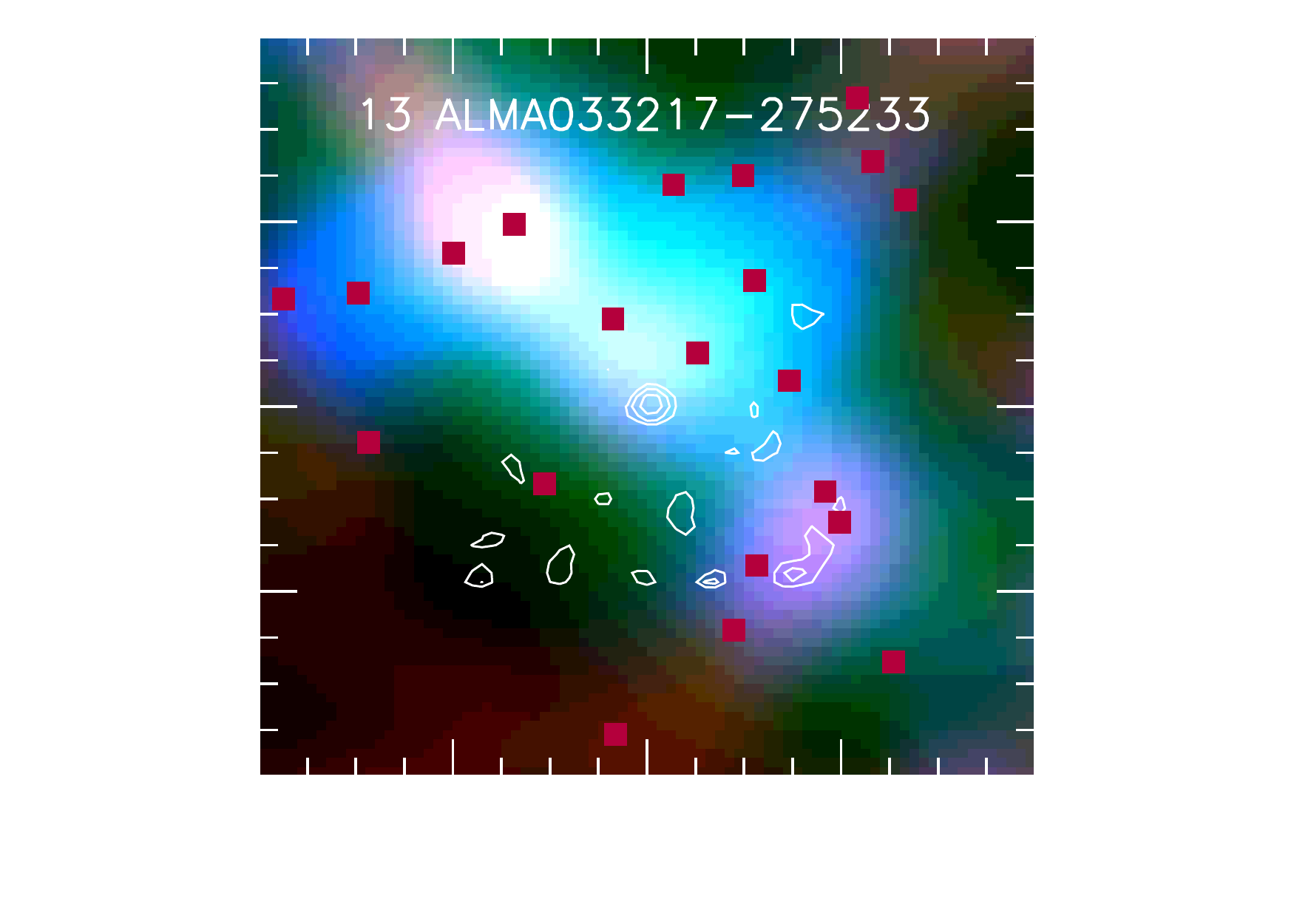}
\hspace{-3.3cm}\includegraphics[width=2.5in,angle=0]{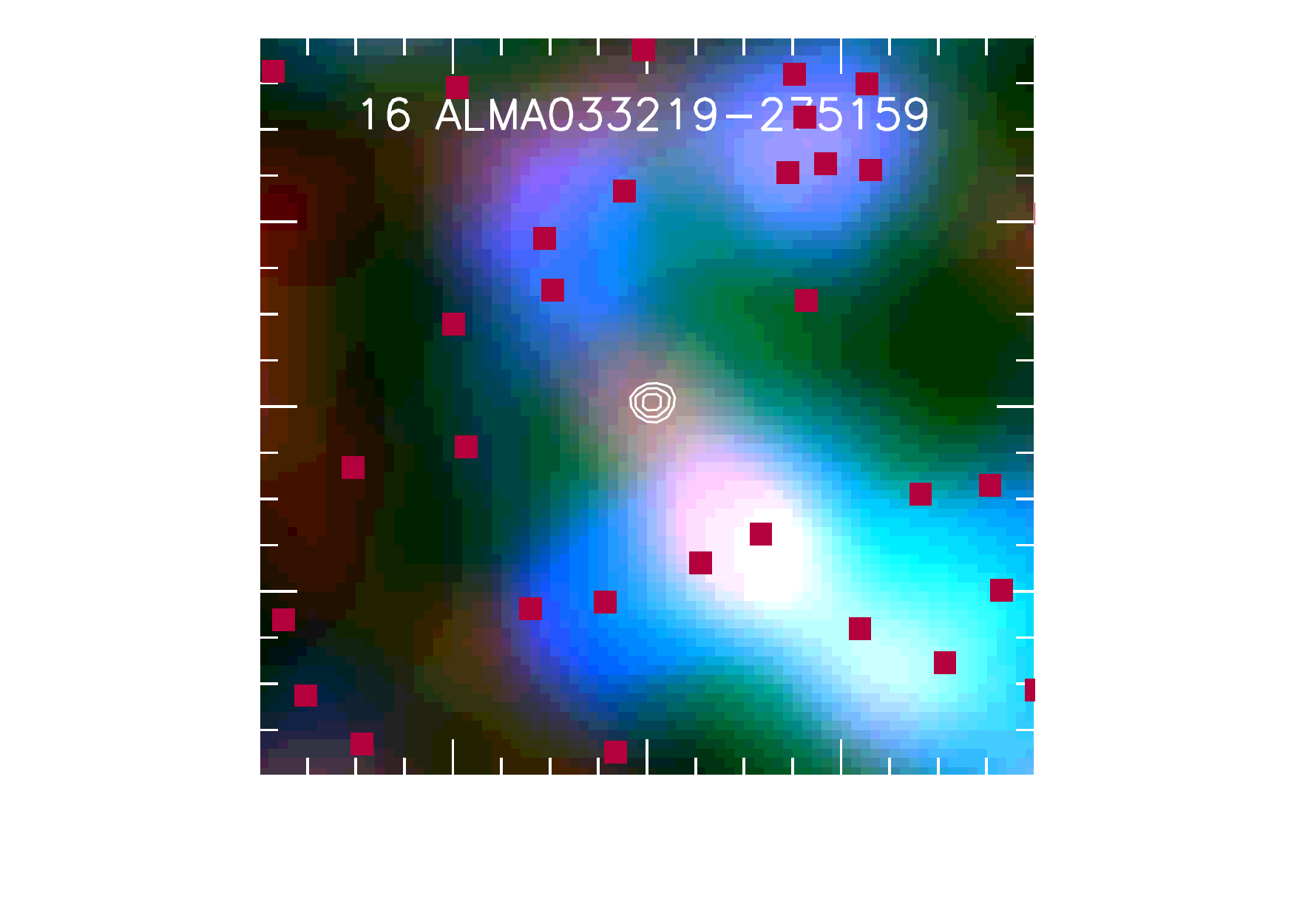}
\hspace{-3.3cm}\includegraphics[width=2.5in,angle=0]{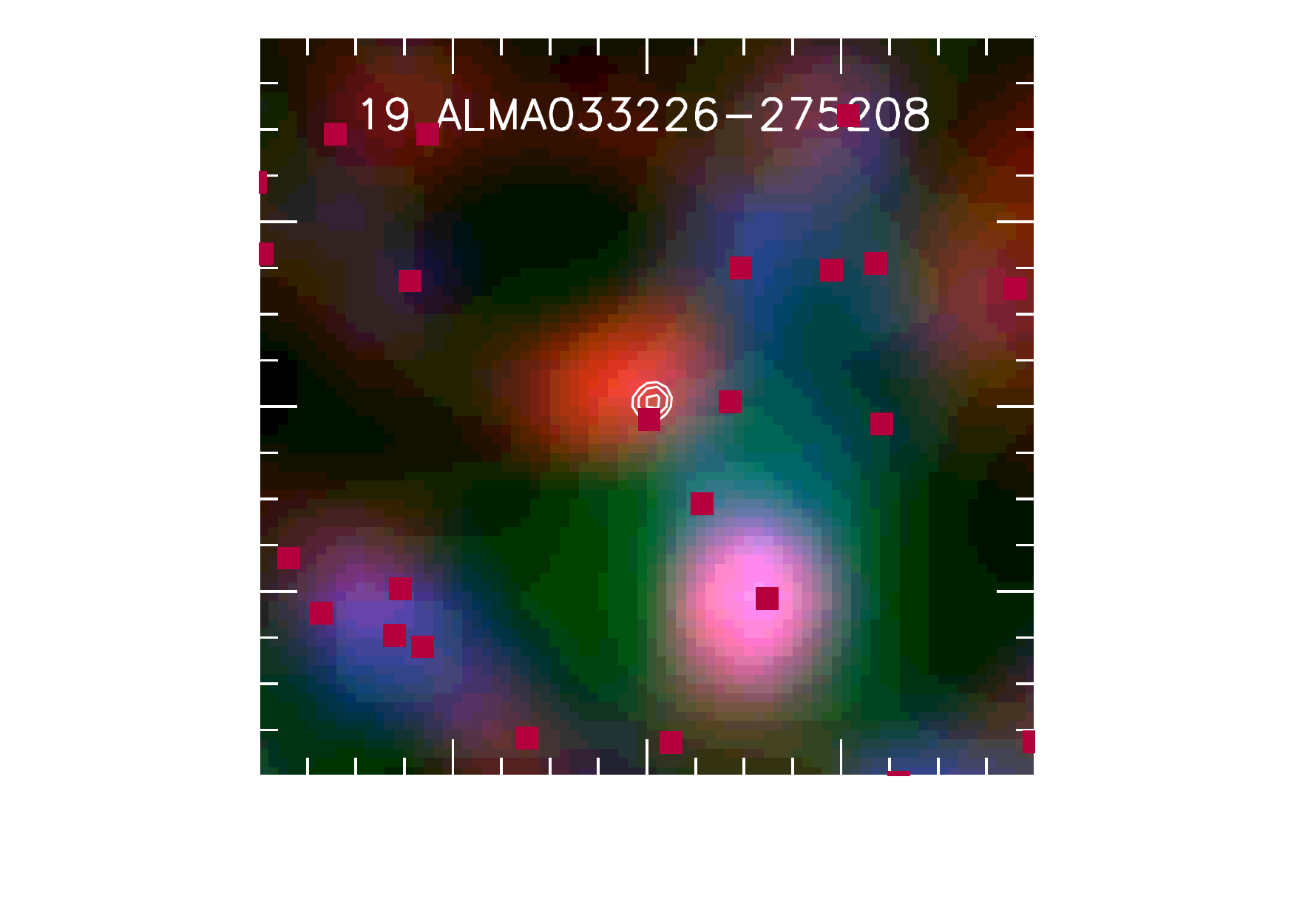}
\hspace{-3.3cm}\includegraphics[width=2.5in,angle=0]{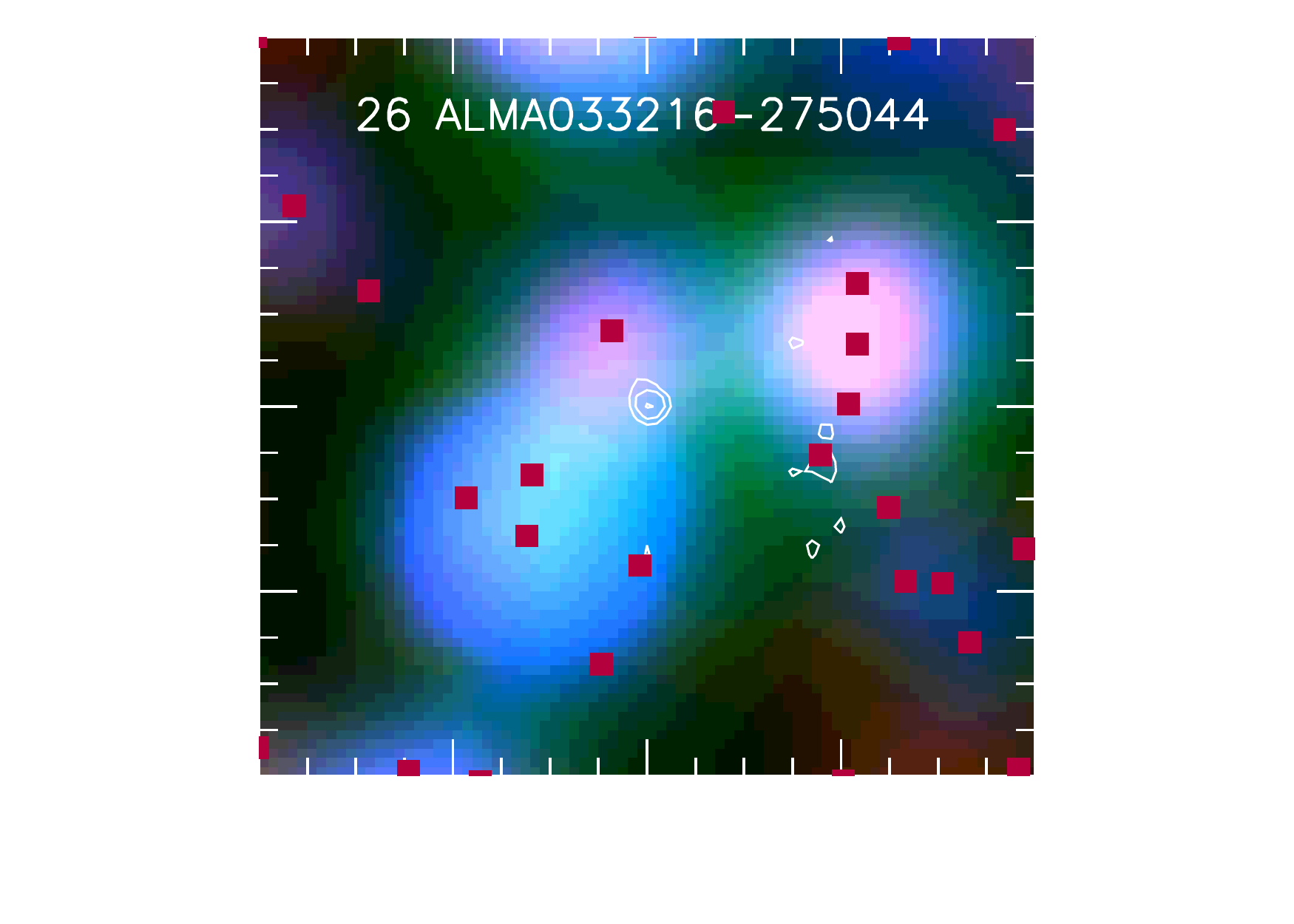}
\includegraphics[width=2.5in,angle=0]{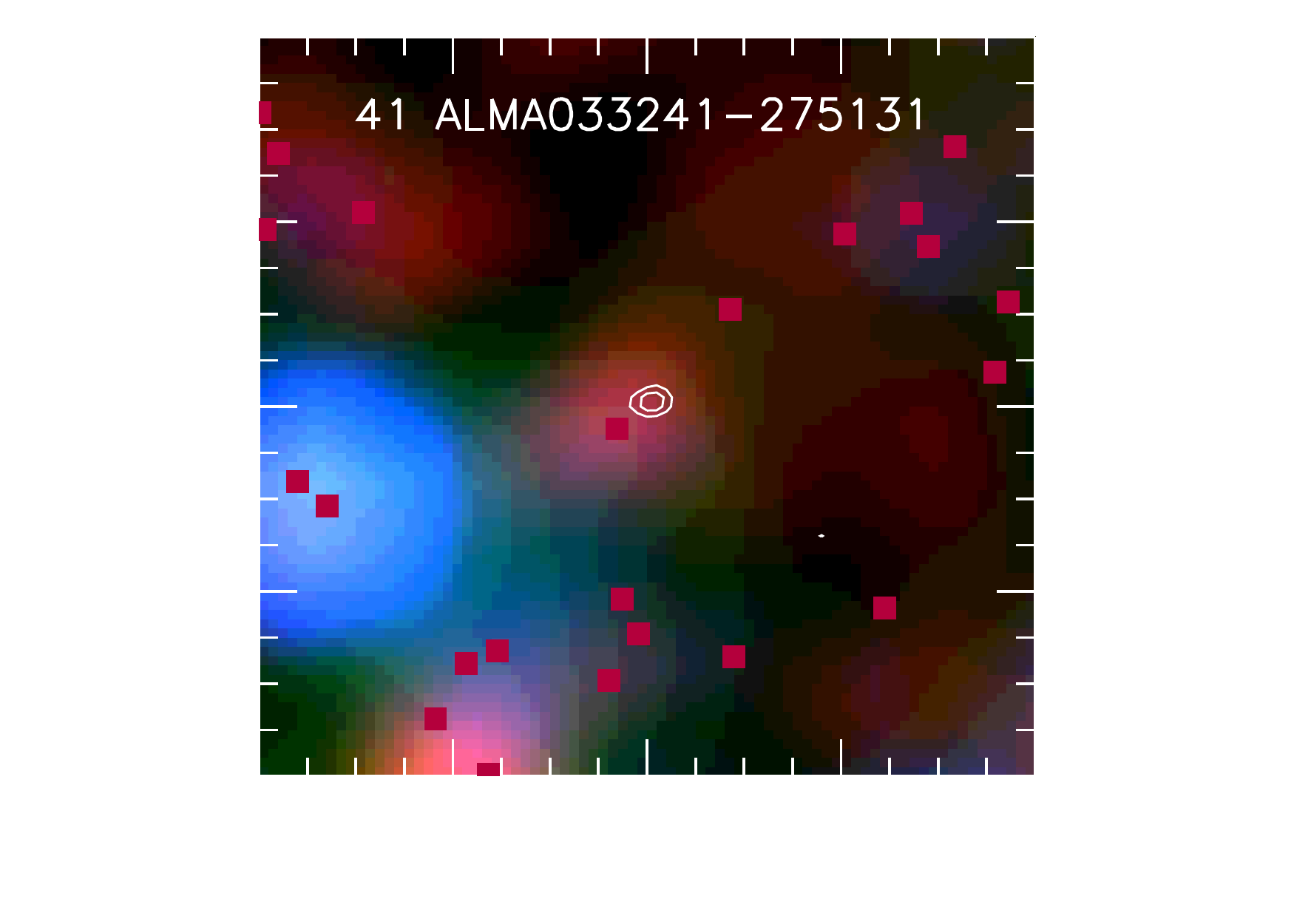}
\hspace{-2.55cm}\includegraphics[width=2.5in,angle=0]{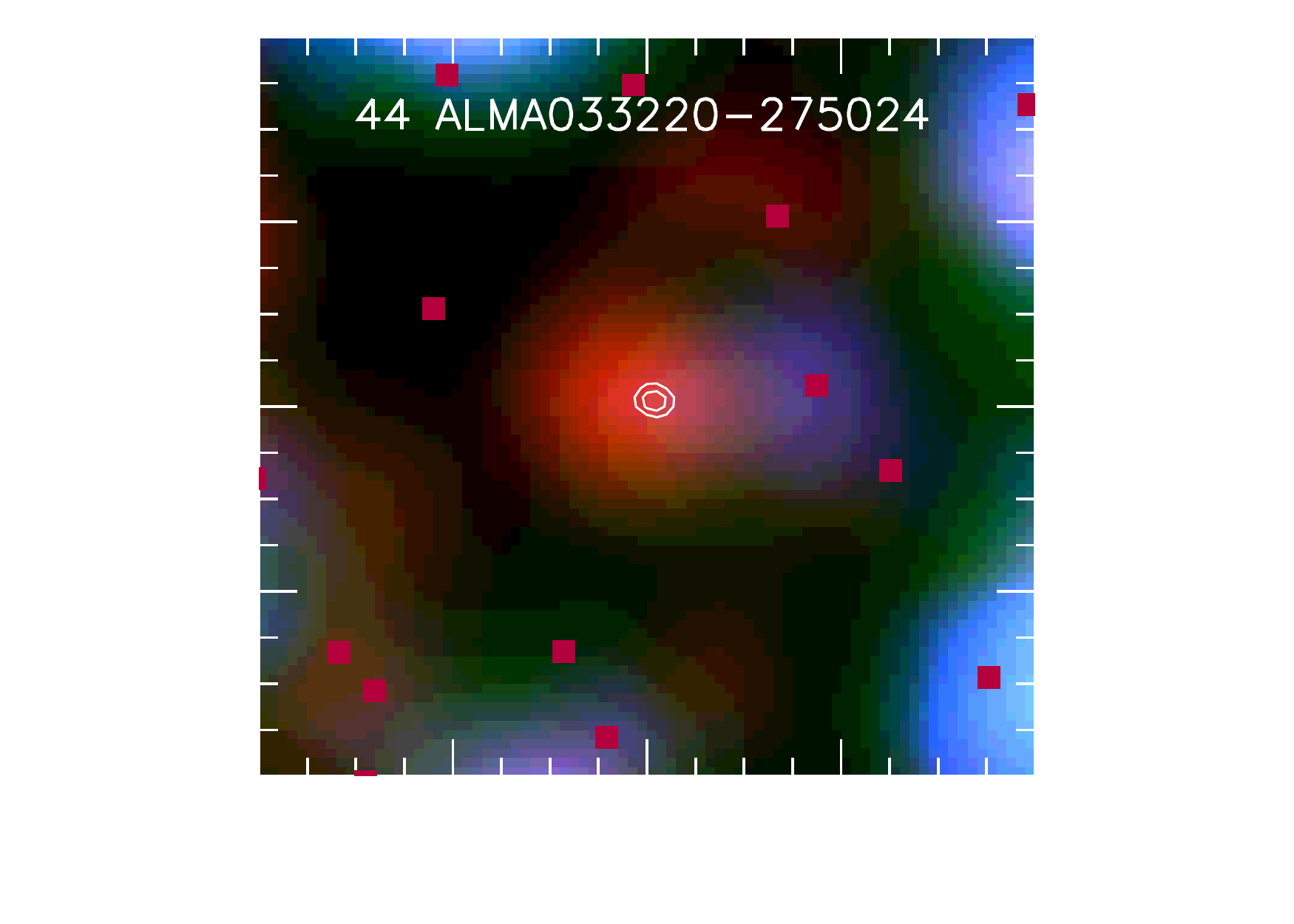}
\hspace{-2.55cm}\includegraphics[width=2.5in,angle=0]{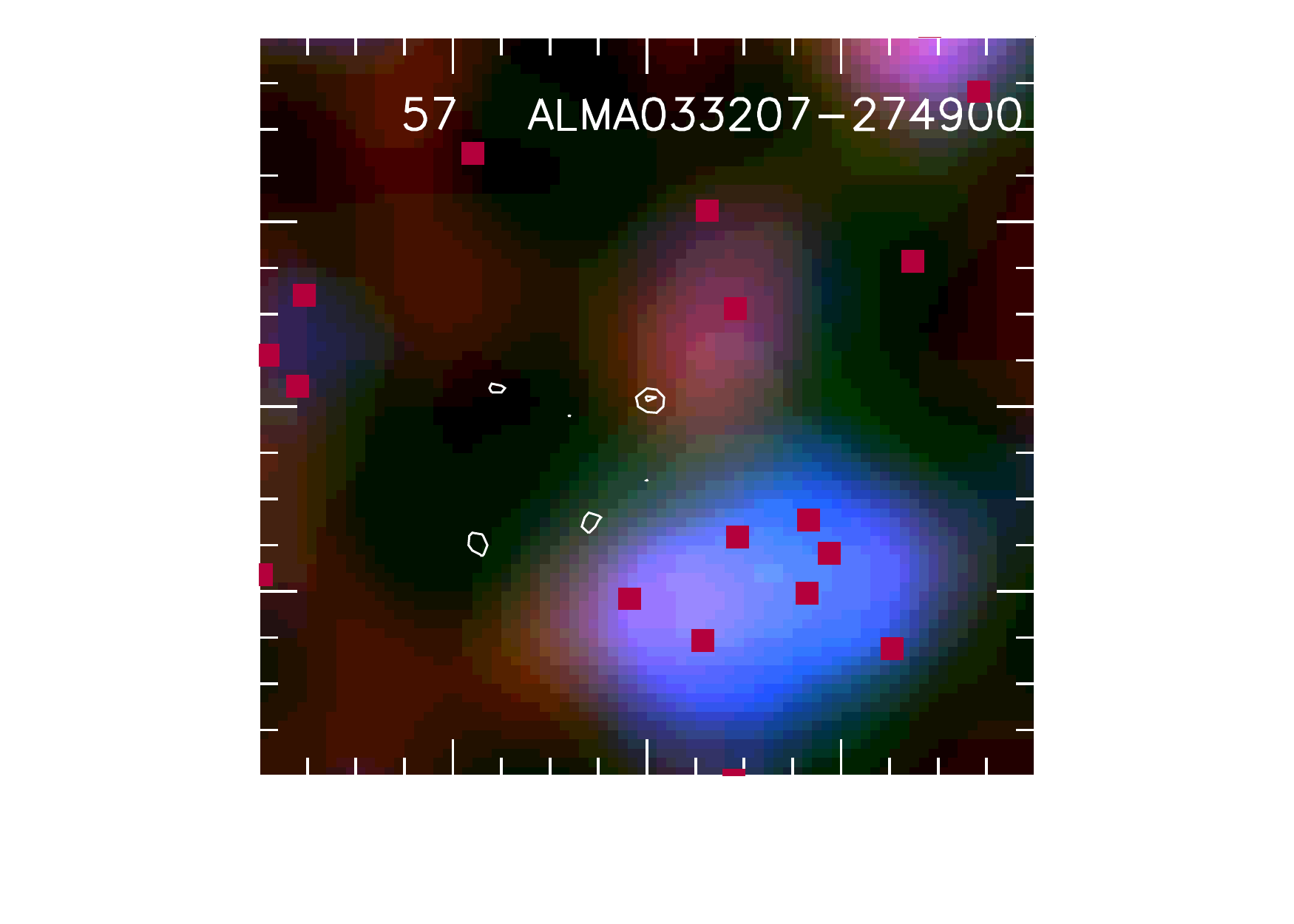}
\hspace{-2.55cm}\includegraphics[width=2.5in,angle=0]{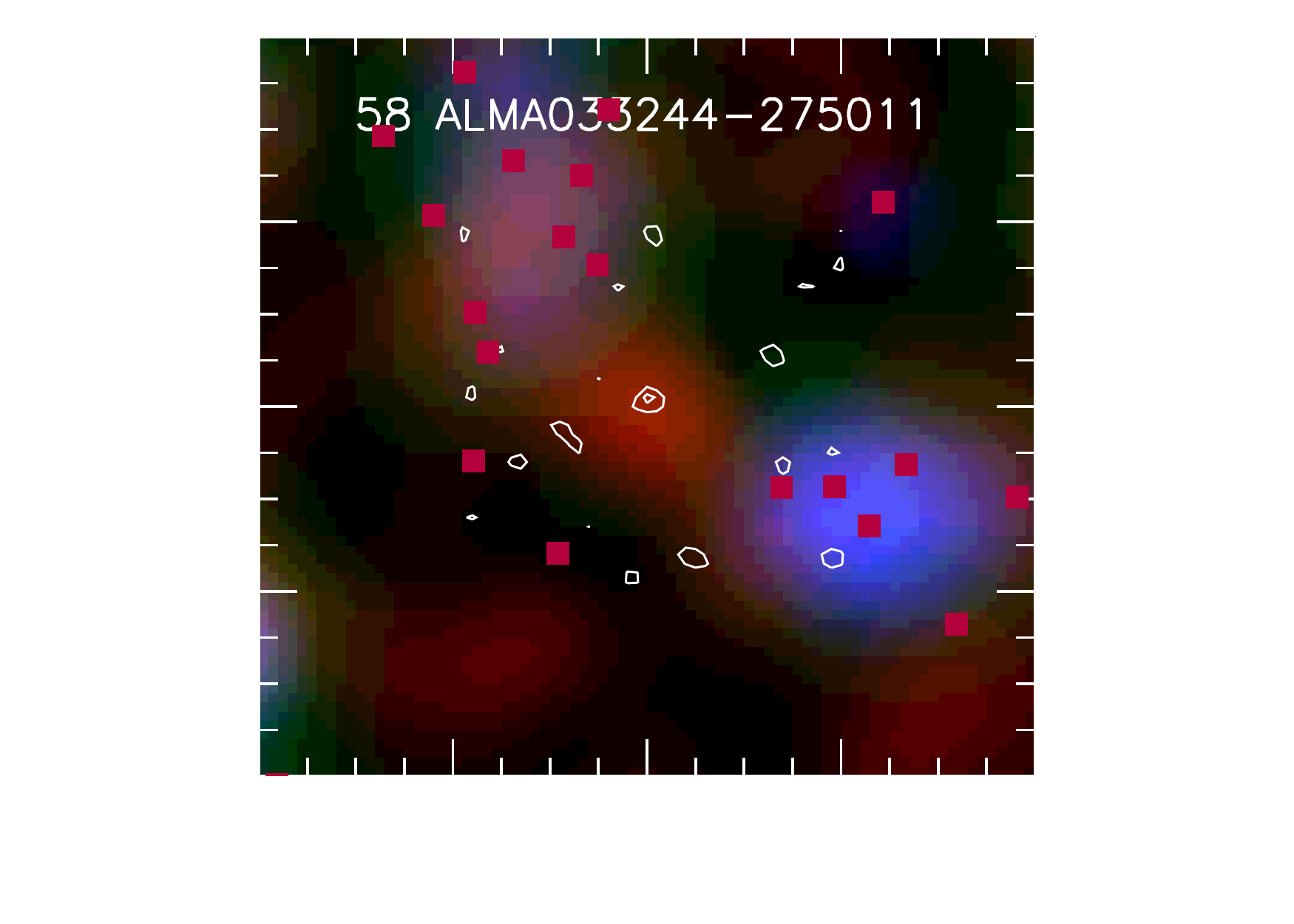}
\caption{{\em Upper 8 panels:\/} Three-color {\em Spitzer\/} and {\em Herschel\/} 
images (24, 100, and 160~$\mu$m) for the brightest ALMA sources  with
24~$\mu$m fluxes less than 20~$\mu$Jy. The thumbnails are $80''$ on a side.
The ALMA sources are shown with white squares, and the sources in the field
with 24~$\mu$m fluxes $>20~\mu$Jy (Elbaz et al.\ 2011 used these as priors 
for the {\em Herschel\/} PACS 100~$\mu$m data) are shown with red squares.
We mark the ALMA positions with squares rather than overlaying contours
to more clearly show the relative positions of the 24~$\mu$m sources. 
The numbers refer to Table~4.
{\em Lower 8 panels:\/} Three-color {\em Herschel\/} and SCUBA-2 images
(250, 350, and 850~$\mu$m) for the same sources as above.
The $0\farcs5$ FWHM tapered ALMA images are overlaid as white contours
at 0.6, 1.2, 2.4, and 4.8 mJy/beam.
\label{basic_alma_fir}}
\end{figure*}

\subsection{FIR Counterparts}
\label{fir}
All but one source in the total ALMA sample lie within both the 
deep 24~$\mu$m {\em Spitzer\/} MIPS image 
(Giavalisco et al.\ 2004) and the deep 100~$\mu$m and 160~$\mu$m 
{\em Herschel\/} PACS images of the GOODS-S (Elbaz et al.\ 2011). 
All lie within the deep longer wavelength {\em Herschel\/} SPIRE 
images of the field (Elbaz et al.\ 2011; Oliver et al.\ 2012). 
In Figure~\ref{24_160}, we mark the ALMA sources by overlaying open 
squares on the 24~$\mu$m (left) and 100~$\mu$m (right)
images. We exclude the one source that lies at the edge of these images 
(source~27; red square in Figure~\ref{24_160}) 
from our MIR analysis. This source also lies at the edge of the ACS and 
CANDELS images (see Section~\ref{secopt}).

Where possible, we used the catalog from Elbaz et al.\ (2011)
to obtain 24~$\mu$m and $100-500~\mu$m fluxes.
We first cross-identified the total ALMA sample against the 24~$\mu$m
sources with fluxes $>20~\mu$Jy, which are what Elbaz et al.\ used as 
priors in measuring the {\em Herschel\/} PACS 100~$\mu$m
fluxes. Most of the {\em Herschel\/} sources have
24~$\mu$m counterparts (Magnelli et al.\ 2011). We find that 61 of the 
remaining 74 ALMA sources (after excluding source~27) 
have a 24~$\mu$m prior in the Elbaz et al.\ (2011) catalog within a
$1\farcs5$ radius. This result is not sensitive
to the choice of matching radius; for example, we recover the same
result with a $2\farcs5$ matching radius. 

In Figure~\ref{24_160},
ALMA sources with 24~$\mu$m fluxes from the Elbaz et al.\ (2011) catalog
of $>20~\mu$Jy sources are shown as gold squares, and ALMA sources with 
measured 24~$\mu$m fluxes fainter than this are shown as blue squares. 
Many of the latter are also faint at 100~$\mu$m, as seen in the
right panel of Figure~\ref{24_160}.
All very high redshift ($z>4$) sources within the ALMA sample
are likely to be contained in this subsample. We shall return
to consider this in detail  in the discussion.

For the remaining unmatched sources, we measured fluxes in the 
24~$\mu$m images at the ALMA positions using a $3''$ diameter aperture, 
and in the longer wavelength data in matched-filter images after first
removing the sources in the Elbaz et al.\ (2011) catalog from
the images. We aperture corrected these 24~$\mu$m fluxes to match the 
measured 24~$\mu$m fluxes in the Elbaz et al.\ catalog. We excluded 
four sources (14, 34, 68, and 72) that are contaminated by a neighboring 
bright source in the 24~$\mu$m sample.

In Figure~\ref{basic_alma_fir}, we show individual images for the eight
ALMA sources with an 850~$\mu$m flux $>1.65$~mJy, which is roughly 
the confusion limit of the SCUBA-2 image, and a measured
24~$\mu$m flux $<0.02$~mJy.
We show {\em Spitzer\/} and {\em Herschel\/} three color images in the 
top panels (24, 100, and 160~$\mu$m) 
and {\em Herschel\/}  and SCUBA-2 three color images in the bottom 
panels (200, 350, and 850~$\mu$m). 
In the top panels, we mark the ALMA sources with white squares and 
the 24~$\mu$m sources with fluxes $>20~\mu$Jy
with red squares.

\begin{figure}
\hspace{-0.5cm}\includegraphics[width=3.6in]{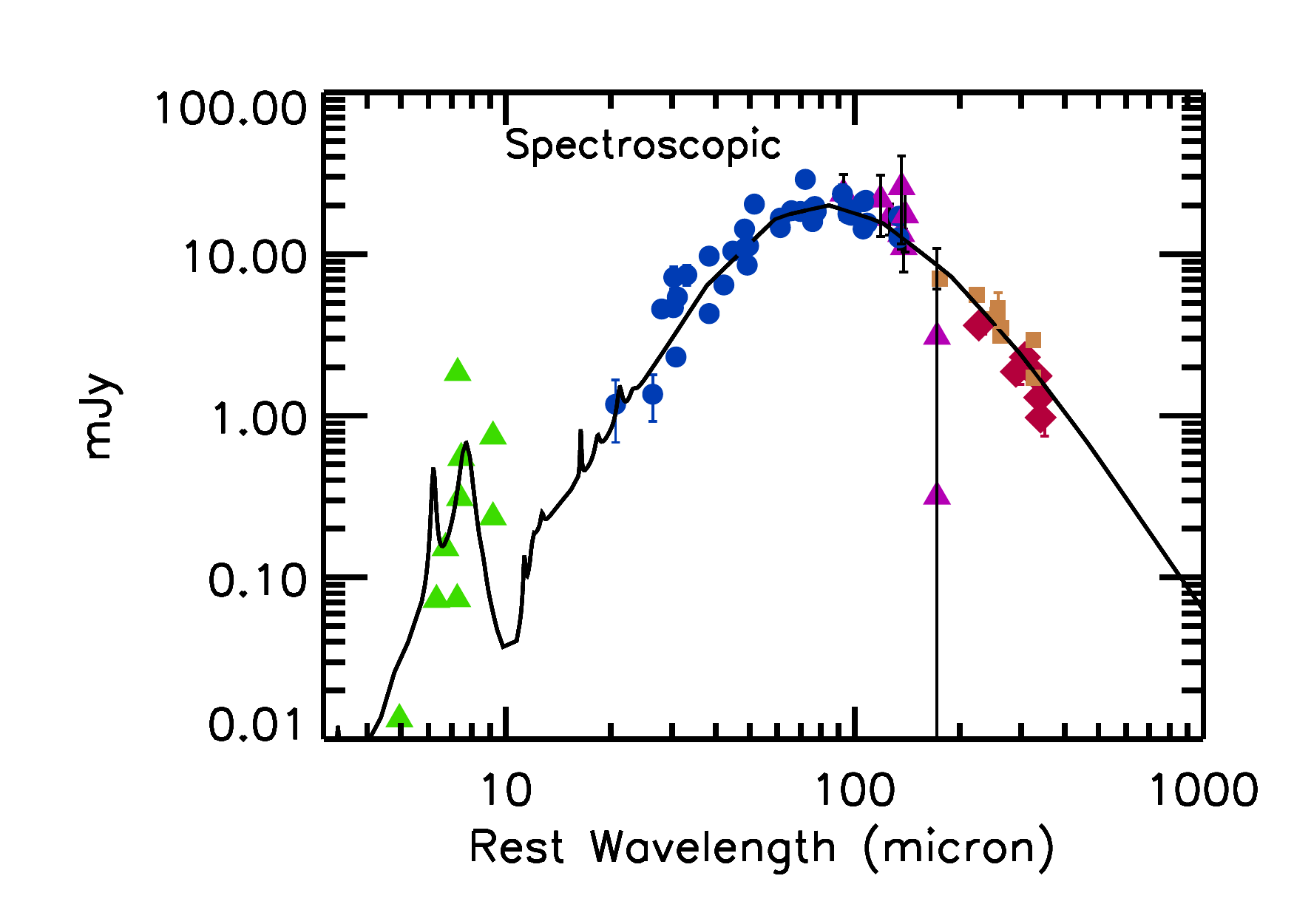}
\caption{Flux vs. rest-frame wavelength for the 9 uncontaminated sources 
with secure speczs $z>1$ and 24~$\mu$m priors. 
Gold squares show our ALMA 850~$\mu$m fluxes, red diamonds the ALMA 1.13~mm
fluxes from F18, purple triangles our SCUBA-2 450~$\mu$m fluxes,
blue circles the {\em Herschel\/} fluxes, and green triangles the {\em Spitzer\/}
24~$\mu$m fluxes. All of the SEDs
are normalized to a peak flux of 20~mJy. The black overlay
shows the Arp~220 SED from Silva et al.\ (1998).
\label{spectral_sed}
}
\end{figure}

\subsection{FIR Photometric Redshifts}
\label{secFIRz}
In Barger et al.\ (2012, 2014) and Paper~I, we found that almost all 
of the spectroscopically
identified SMGs with fluxes greater than 2~mJy in the GOODS-N 
can be reasonably fit by an Arp~220-like FIR SED. This also appears to
be true in the GOODS-S. In Figure~\ref{spectral_sed}, we show
the rest-frame SEDs for the 9 uncontaminated sources with secure 
speczs $z>1$ and 24~$\mu$m priors (4, 5, 9, 18, 25, 35, 40, 46, and 56).
(We excluded from this list source~59 or ALMA033222-274815, which 
is confused with source~38 or ALMA033222-274811.
We also excluded source~1 or ALMA033207-275120, which we discuss below.)
Since we are only interested in the relative shapes, we
normalized the SEDs to all have a peak flux of 20~mJy. 
We overlay the Arp~220 SED of Silva et al.\ (1998) in black.

\begin{figure}
\hspace{-0.7cm}\includegraphics[angle=0,width=11.0cm]{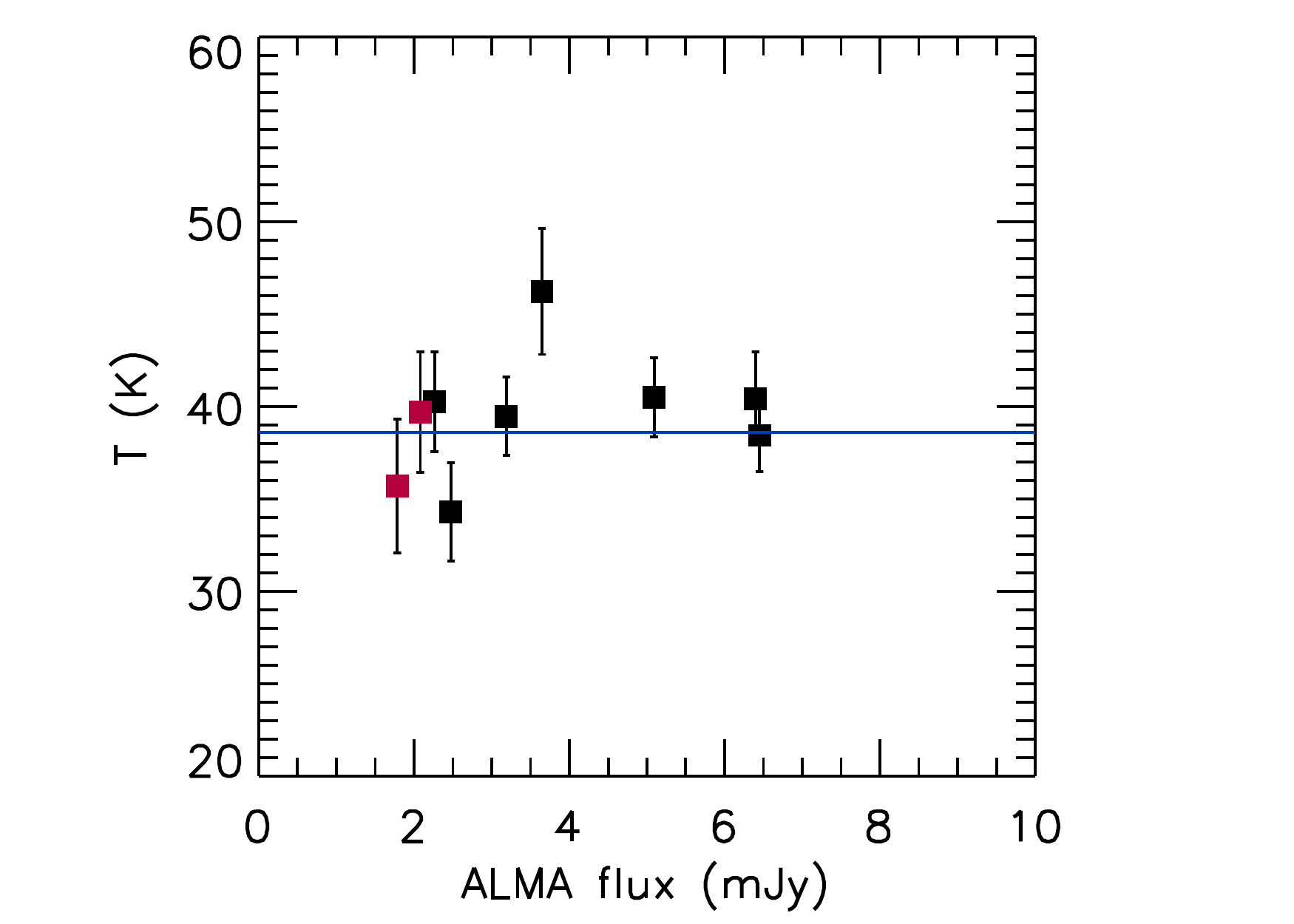}
\caption{Dust temperatures for the 9 uncontaminated sources with 
secure speczs $z>1$ and 24~$\mu$m priors.
The values are derived from a graybody fit with an emissivity $\beta=1.5$,
and the error bars are $1\sigma$ around the best fit value.
Only data at rest-frame wavelengths above 40~$\mu$m were fitted.
The blue line shows the value derived by fitting
the Arp~220 SED from Silva et al.\ (1998) in the same way.
Sources with rest-frame $2-8$~keV luminosities above 
$10^{43}$~erg~s$^{-1}$ are shown in red.
\label{fitit}
}
\end{figure}

In order to quantify the above, we measured the dust temperatures
for each of the nine sources assuming a graybody with an emissivity
$\beta=1.5$. We fitted only above a rest-frame wavelength of 40~$\mu$m
to minimize the effects of short wavelength power law contributions.
In Figure~\ref{fitit}, we show these dust temperatures. The mean (median) 
is 39.4 (40.3)~K, as compared with 38.6~K for Arp~220 fitted in the same way. 
Only one source (source~18 or ALMA033222-274804) 
has a temperature ($T=46.2\pm3.4$~K) 
that is more than $2\sigma$ different than that of Arp~220.
This is the only $z>3$ source among the nine. 
It may be that higher redshift sources are slightly hotter. Otherwise, the sources 
do not show any clear temperature dependence
on the submillimeter flux or on the presence of an X-ray
AGN (see Figure~\ref{fitit}).

\begin{figure}
\hspace{-1.0cm}
\includegraphics[width=10.0cm]{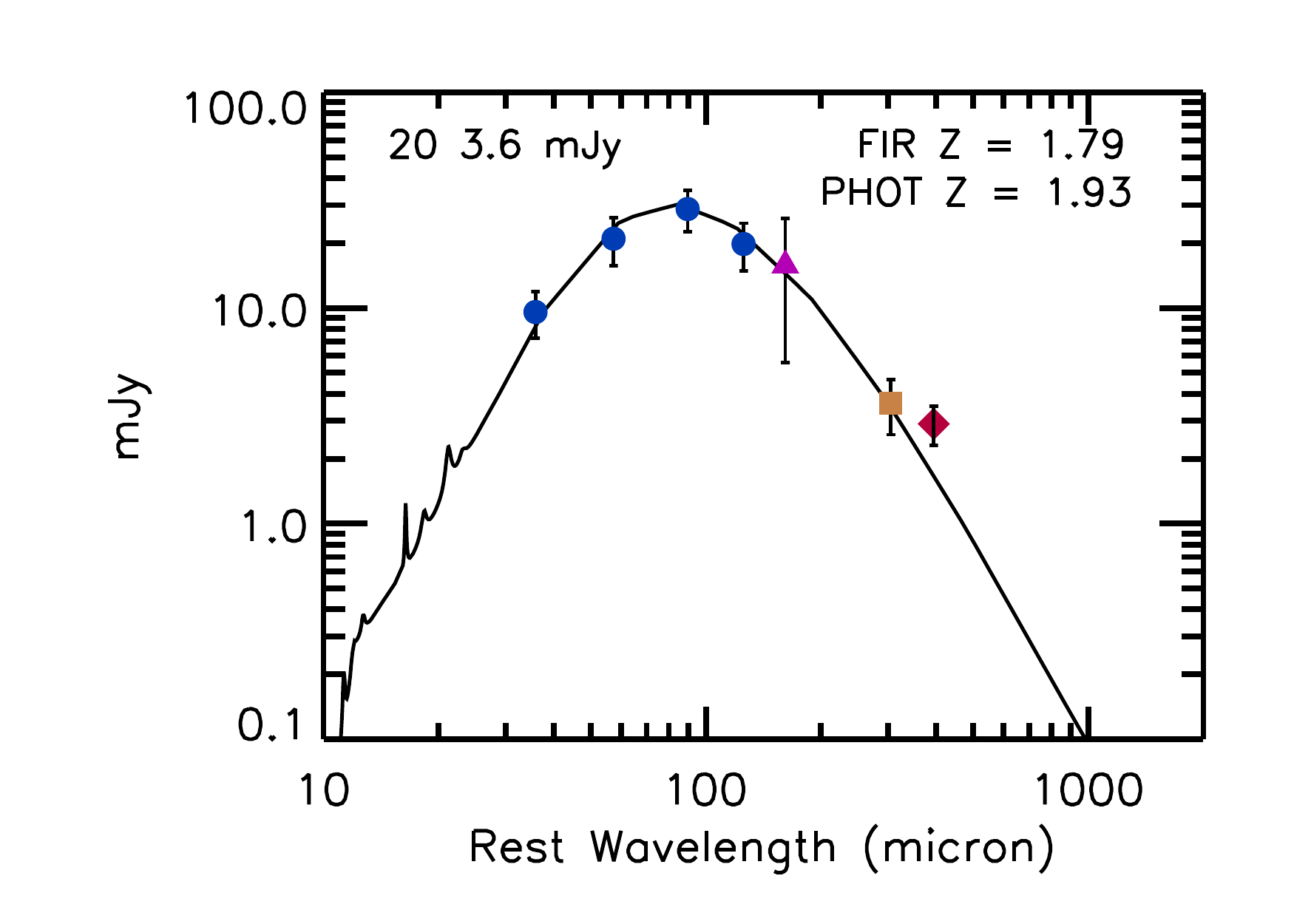}
\vskip -0.9cm
\hspace{-1.0cm}
\includegraphics[width=10.0cm]{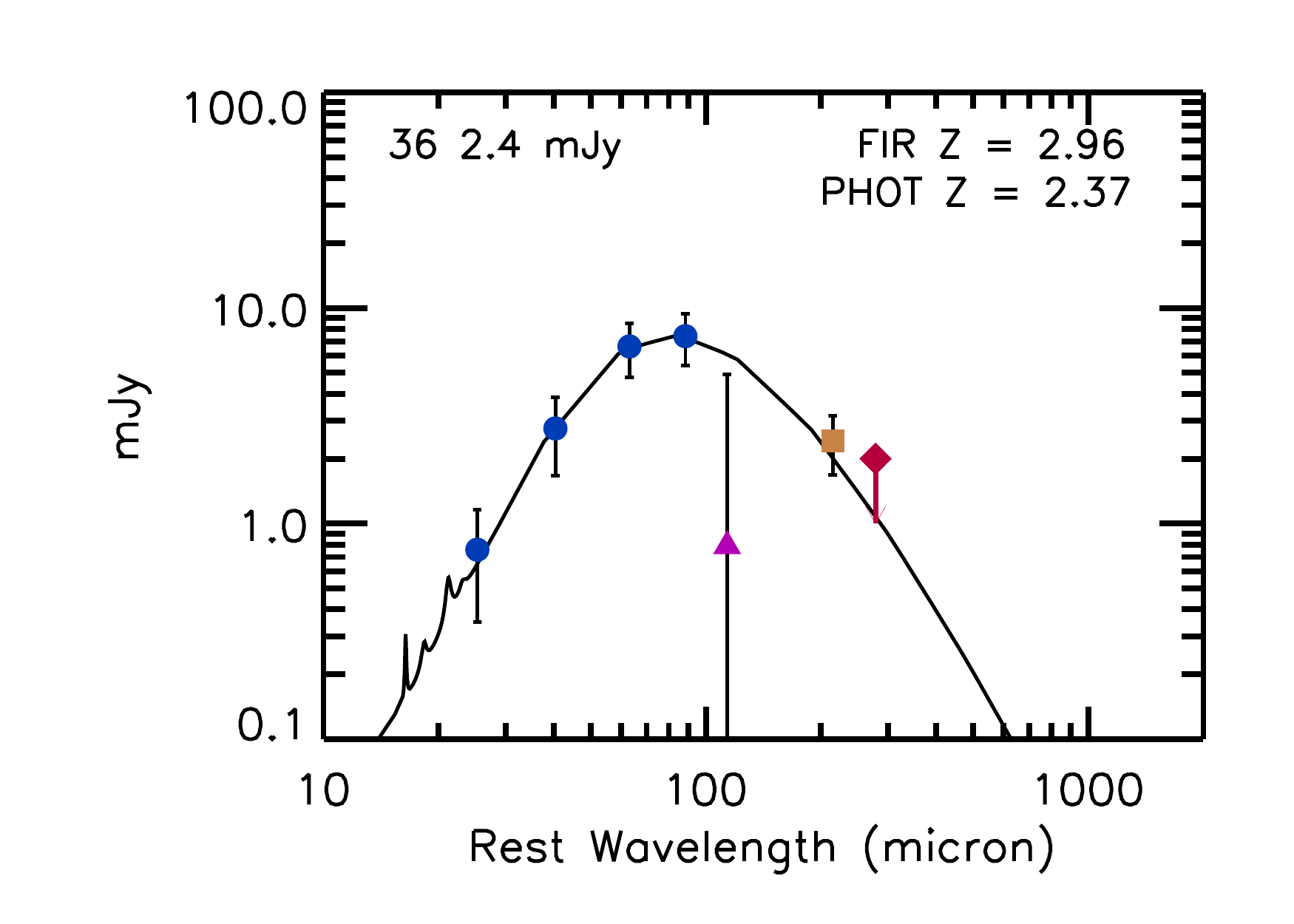}
\caption{Sample fits used to determine the FIRzs, which are listed
in the upper right corner of each plot, along with the photzs. 
Gold squares show our ALMA 850~$\mu$m fluxes, red diamonds the 
F18 or AzTEC 1.1~mm fluxes, purple triangles our SCUBA-2 450~$\mu$m
fluxes, and blue circles the {\em Herschel\/} fluxes. The errors 
are $\pm1\sigma$. Source~36 or ALMA033220274836
is not in the F18 or AzTEC catalogs, so 
we show a $2\sigma$ upper limit (downward pointing arrow) at 1.1~mm 
based on the AzTEC data.
The black overlay shows the Arp~220 template fit.
\label{firz_sample_plot}
}
\end{figure}

Given this, for the 59 sources with ALMA fluxes
above 1.65~mJy (this flux limit was chosen to provide high S/N
throughout the SED), we used the Arp~220 template to estimate 
FIR photometric redshifts (hereafter, FIRzs)
using a $\chi^2$ fit to the combined
{\em Herschel\/}, SCUBA-2 450~$\mu$m, ALMA 850~$\mu$m, and 
F18 ALMA 1.13~mm (if available) data for 
each source. In addition to the measured noise, we also included
a systematic error of 20\% of the flux to allow for relative
calibration errors between the bands.
We exclude seven sources (26, 29, 37, 41, 57, 58, and 59)
where the $\chi^2$ fit is poor (probability less than 5\%)
and where there is likely contamination of the photometry from
neighbors. We give the estimated FIRzs in
Table~5, together with the 95\% confidence range. The range of
temperatures for the nine spectroscopic sources discussed above
relative to that of Arp~220, which was used in the fitting,
would add a further roughly $\pm10$\% error to the FIRzs.
In Figure~\ref{firz_sample_plot}, we show two sample fits, one at 
$z\sim 2$ and one at $z\sim 3$.

\begin{figure}
\hspace{-1.0cm}\includegraphics[width=11.0cm]{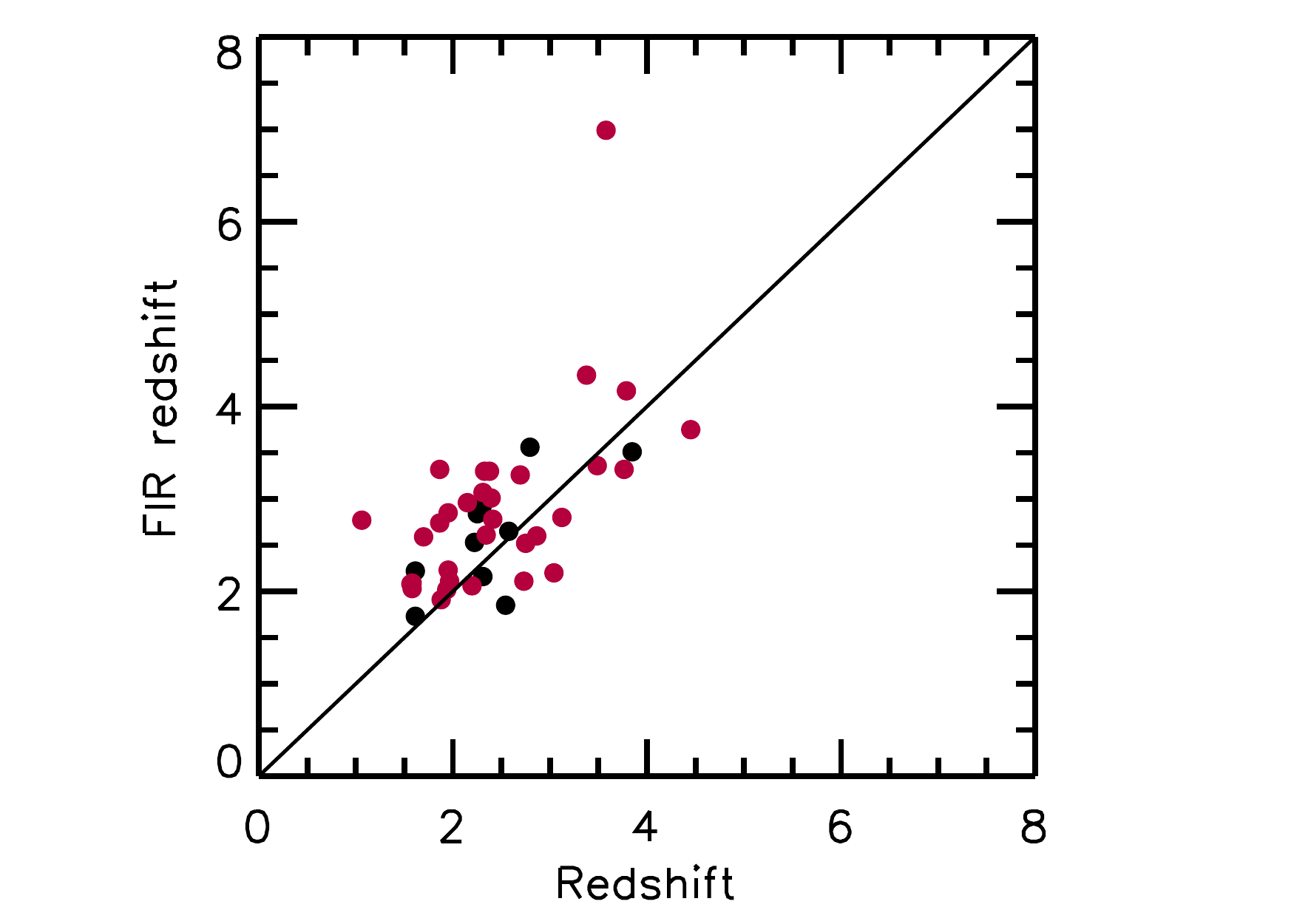}
\caption{Comparison of FIRz with specz (black circles) or photz (red circles). 
Only sources from the total ALMA sample with fluxes above 1.65~mJy and 
where there is a FIRz and either a specz or a photz
are shown. The one source (source~17 or ALMA033235-275215) 
with a very high FIRz estimate is shown at a nominal FIRz of 6.9 for clarity.
Sources with FIRzs above 4 that do not have reliable photzs 
are not shown here.
\label{firz_test}
}
\end{figure}

For source~1 or ALMA033207-275120, the FIRz of $z=3.26$ is significantly
higher than the specz of the nearby bright galaxy,
which Danielson et al.\ (2017) place at $z=1.539$.
Removing photometric contamination by the bright galaxy in the shorter wavelength
{\em Herschel\/} bands would only increase our estimate of the FIRz and hence the tension.
Given that there is also a significant spatial offset of the bright galaxy
and the ALMA source (see Section~\ref{secopt} and Figure~\ref{basic_alma_images}), 
it is possible that the ALMA detection is of a background source. However,
there appears to be an extension of the ALMA emission towards the bright
galaxy, which may suggest that we are seeing emission from the bright galaxy,
too. We therefore exclude this source from our subsequent analysis.

In Figure~\ref{firz_test}, we show FIRz versus specz (black circles) or photz
(red circles) for sources with ALMA 850~$\mu$m fluxes above 1.65~mJy and
where there is a FIRz and a specz or a photz with quality flag $Q<3$
in S16. While there is some scatter, the overall correlation is reasonable
for most of the sources. There is one seriously
discrepant source (source~17 or ALMA033235-275215) 
where the FIRz places the galaxy at a very high redshift;
we discuss this source in Section~\ref{sechighz}, together with
other sources where the FIRz gives a high redshift
but for which there is no reliable photz.

\begin{figure*}
\includegraphics[width=7.25cm]{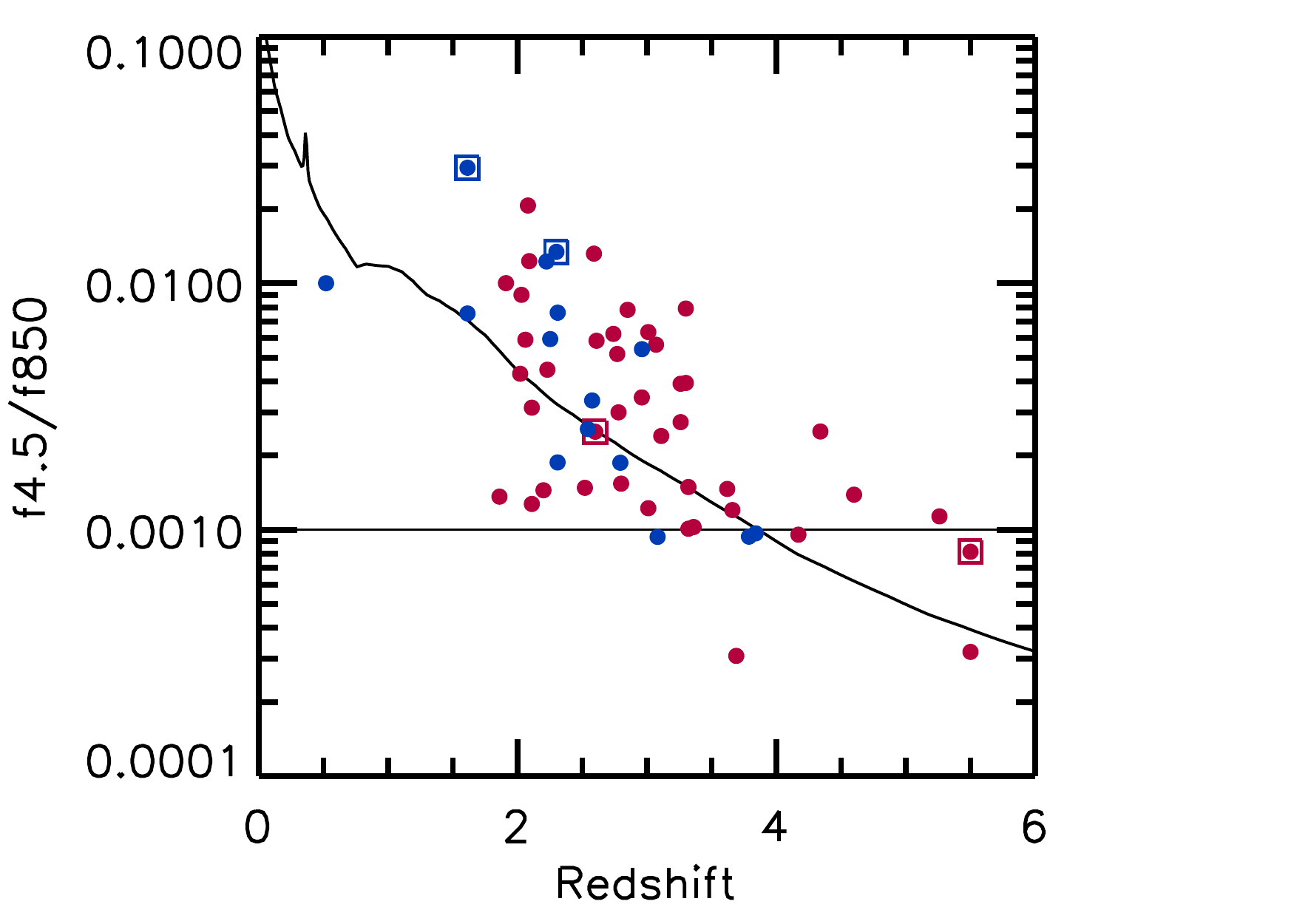}
\hspace{-1.5cm}\includegraphics[width=7.25cm]{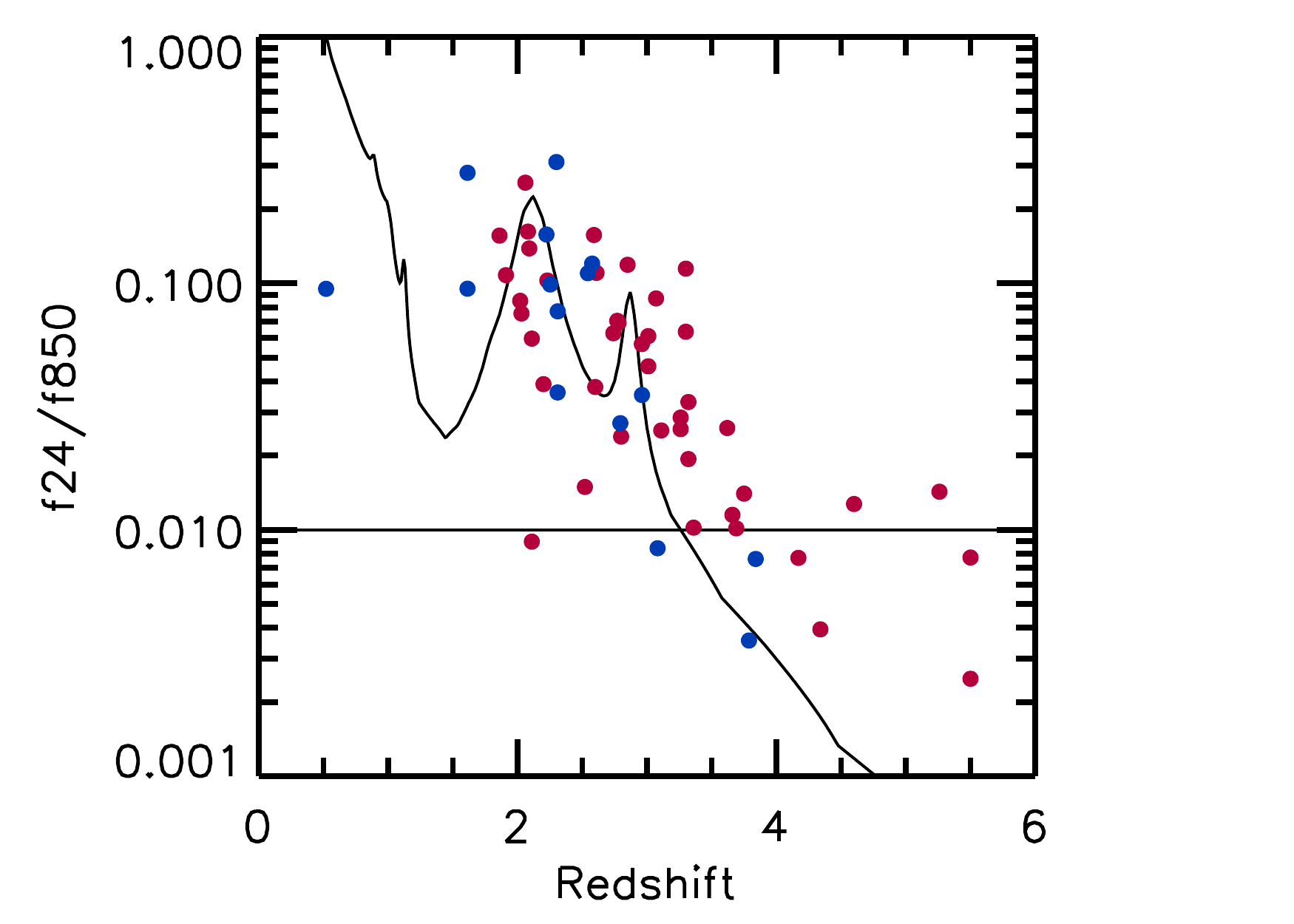}
\hspace{-1.5cm}\includegraphics[width=7.25cm]{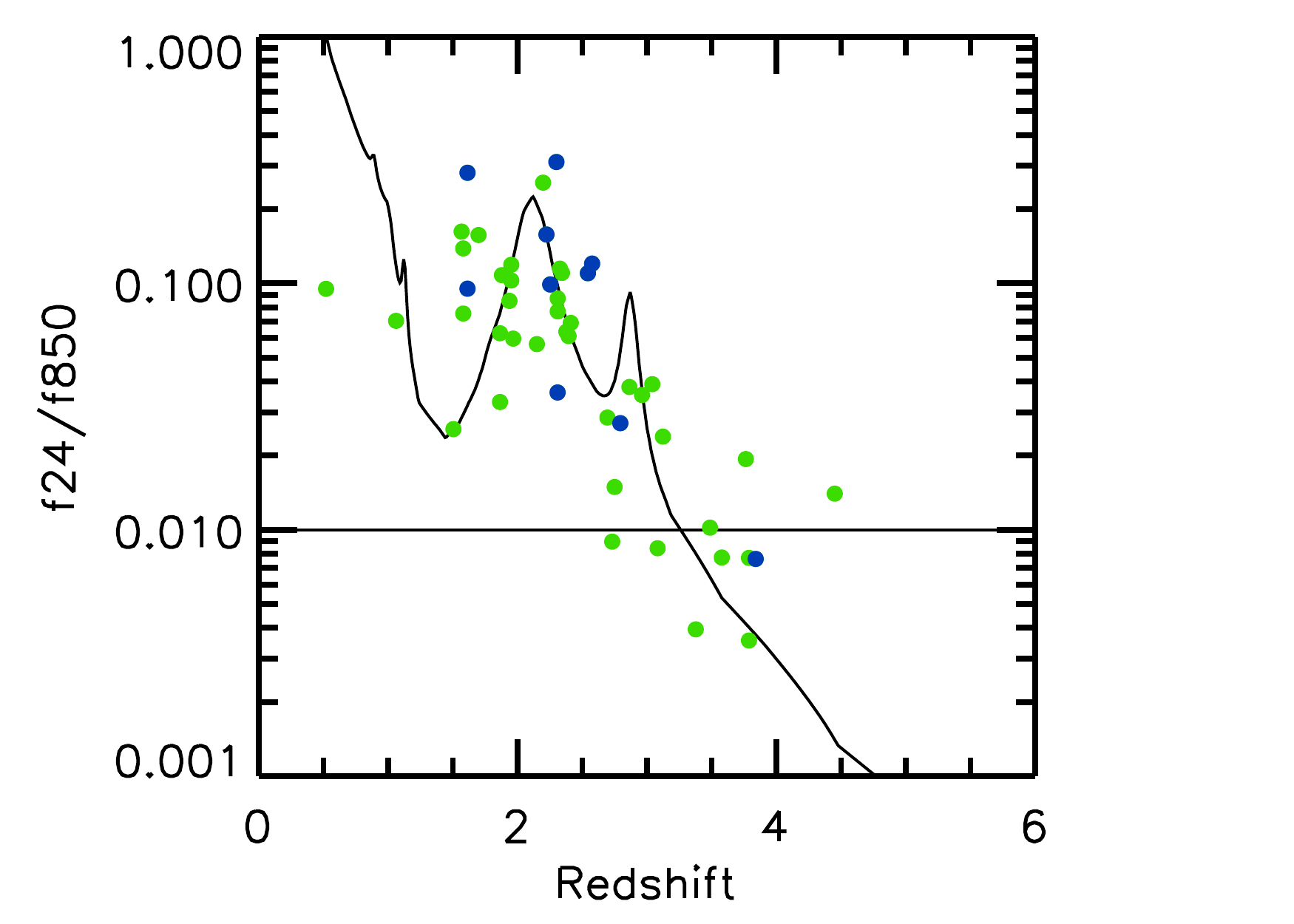}
\caption{(a) 4.5~$\mu$m to 850~$\mu$m and
(b) 24~$\mu$m to 850~$\mu$m flux ratios vs. redshift.
We use speczs (blue circles) where possible, then FIRzs (red circles), 
and finally photzs (blue circles). 
Sources with $z>5.5$ are shown at a nominal $z=5.5$. 
In (a), we enclose sources with rest-frame 
$2-8$~keV luminosities above $10^{43}$~erg~s$^{-1}$ in large squares.
(c) 24~$\mu$m to 850~$\mu$m flux ratio vs. redshift, where
this time we show only the speczs (blue circles) or the photzs (green circles).
In each panel, the black curve shows the path of the
Arp~220 template with redshift, and the horizontal line shows the rough
value of the ratio that can be used to select high-redshift sources.
\label{smm_z}
}
\end{figure*}

The FIRzs are only based on the SEDs above an observed wavelength
of 100~$\mu$m (the shortest wavelength {\em Herschel\/} data
available), and we may compare the results with the shorter 
wavelength bands from {\em Spitzer\/}. Because of the negative $K$-correction at 
850~$\mu$m, we expect both the 4.5~$\mu$m to 850~$\mu$m flux ratio and the 
24~$\mu$m to 850~$\mu$m flux ratio to drop rapidly with increasing redshift 
(see also Shu et al.\ 2016, who considered the 24~$\mu$m to 500~$\mu$m flux ratio). 

In Figure~\ref{smm_z}, we plot the two ratios versus redshift, along with the 
expectation from the Arp~220 template (black curve). In the first two panels, we 
use the speczs (blue circles), then the FIRzs for sources with no specz (red circles), 
and finally the photzs (blue circles again) for sources with neither a specz nor a FIRz. 
In the last panel, we show the 24~$\mu$m to 850~$\mu$m flux ratio vs. redshift for a
second time, but here we use only the speczs (blue circles) followed by the photzs (green circles). 
As an aside, we note that for a given redshift, most of the luminous X-ray AGNs 
(large squares in first panel) have high 24~$\mu$m to 850~$\mu$m flux ratios; 
however, there are only four such sources, and the value of one is low.

While the dispersion in the plots is too large to
estimate accurate redshifts for individual sources,
we see the expected decline with redshift. The highest redshift sources 
($z\gtrsim4$) are faint at both 4.5~$\mu$m and 24~$\mu$m, as we 
illustrate with the horizontal lines plotted at the values
of 0.001 (4.5~$\mu$m to 850~$\mu$m flux ratio plot) and 0.01 
(24~$\mu$m to 850~$\mu$m flux ratio plots).
The 24~$\mu$m plot is more complex than the 4.5~$\mu$m plot,
because of the presence of the PAH 8~$\mu$m features. 
At $z\gtrsim4$, the 24~$\mu$m band samples rest-frame wavelengths 
below the 8~$\mu$m PAH peak, making the sources very faint. 

For the last panel, the potential high-redshift sources
seen in the other two panels (based on the FIRzs) are lost, because 
they do not have high-quality photz estimates. This emphasizes
the difficulties with trying to find high-redshift sources using photzs.
We conclude that when one has high-resolution 850~$\mu$m observations 
that can be accurately matched to the 4.5~$\mu$m or 24~$\mu$m images, 
then one has a strong diagnostic of possible high-redshift 
sources without the need for a full long-wavelength SED.

\begin{figure}
\hspace{-1.0cm}\includegraphics[width=10.0cm]{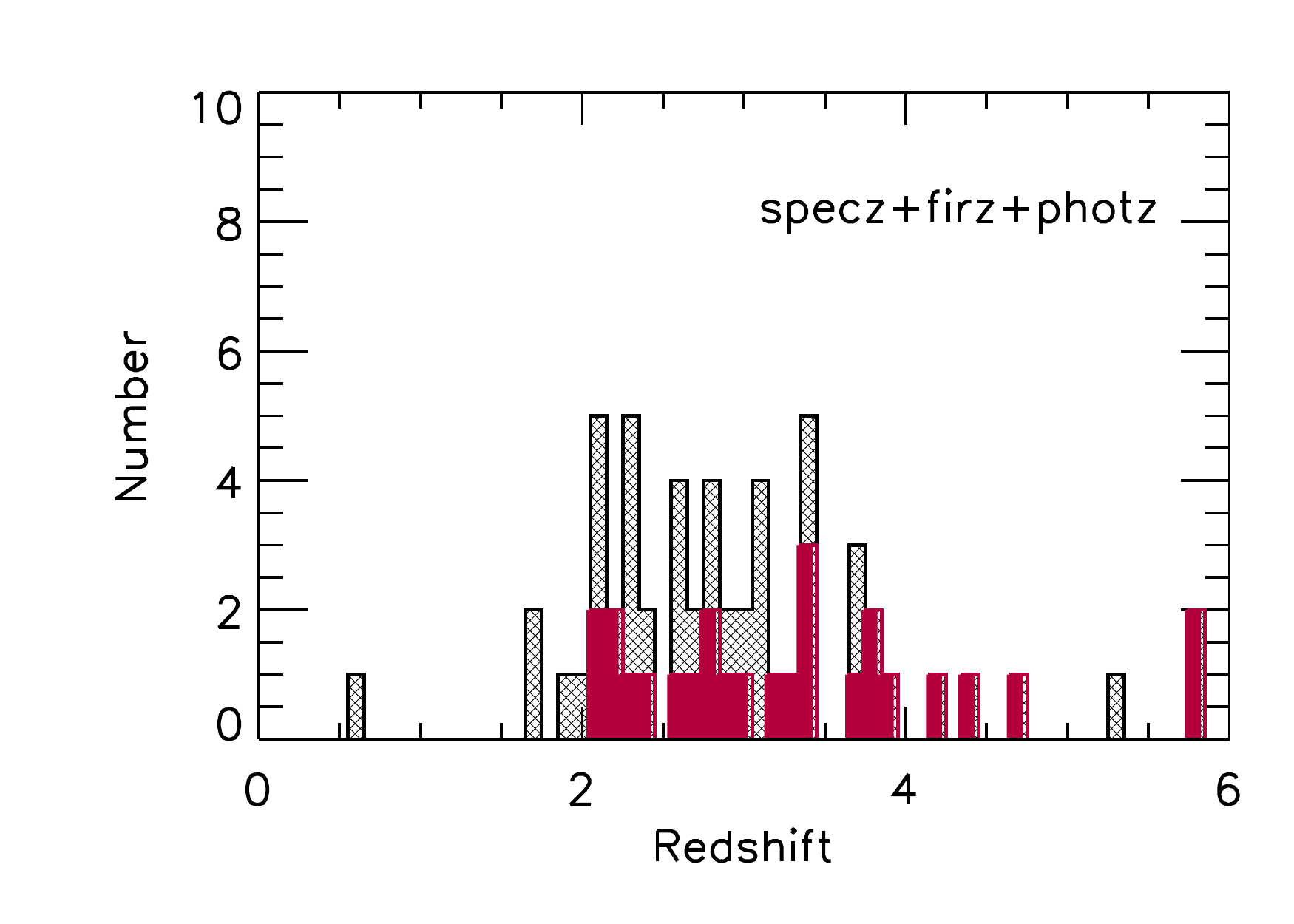}
\vskip -0.9cm
\hspace{-1.0cm}\includegraphics[width=10.0cm]{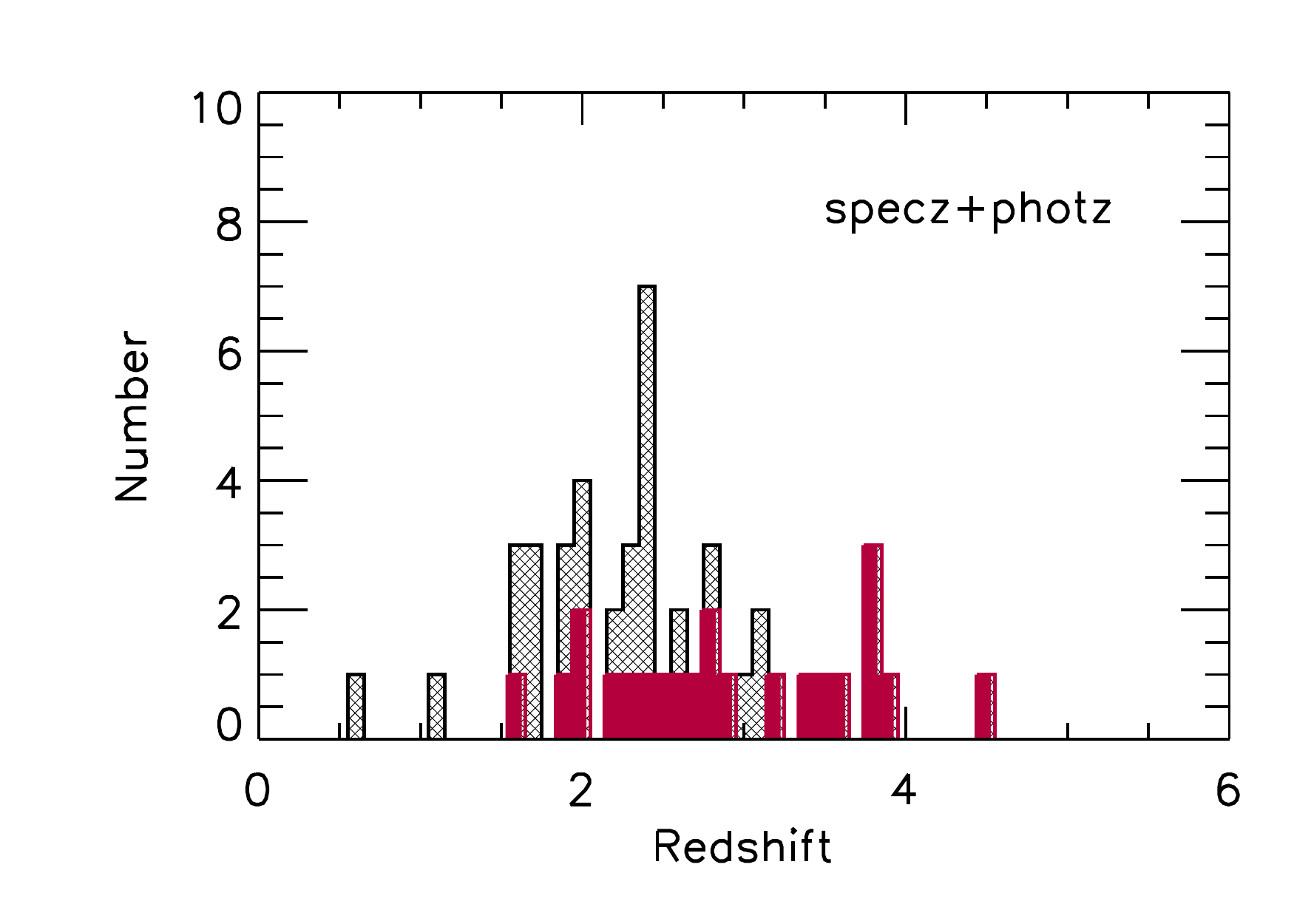}
\caption{Redshift distributions for the sources in
the total ALMA sample with fluxes above 1.65~mJy (gray shading).
The red shading shows the distributions for sources brighter than 3~mJy. 
{\em Top:\/} We use speczs where possible, then FIRzs,
and finally photzs, providing redshifts
for 55 sources above 1.65~mJy and 25 above
3~mJy.  Very high-redshift sources
(FIRzs $>5.5$) are shown at a nominal value of $z=5.7$.
{\em Bottom:\/} We use only the speczs and photzs, providing
redshifts for 47 sources above 1.65~mJy and 22 above 3~mJy.
\label{z_hist}
}
\end{figure}

\section{Candidate High-Redshift Sources}
\label{sechighz}
In Figure~\ref{z_hist}, we show redshift distributions for the sources in the 
total ALMA sample above 1.65~mJy (gray shading) and above 3~mJy (red shading).
In the top panel, where there is a specz (except for source~1 or ALMA033207-275120;
see Section~\ref{secFIRz}), we use this, then the FIRz,
and finally the photz. The median redshift is $z=2.74$, while
above 3~mJy, the median redshift is $z=3.26$.
In the bottom panel, we use only the speczs and photzs. 
The median redshift is $z=2.32$, while above 3~mJy, the median redshift is $z=2.86$. 

The FIRzs show a number of high redshifts for sources where the
photzs are of poor quality ($Q>3$ from S16).
The FIRzs place six of the sources with 850~$\mu$m fluxes
above 1.65~mJy at $z>4$ and two at $z>5.5$ (the latter are shown at 
a nominal redshift of $z=5.7$ in the figure). The photzs place seven sources 
at $z>4$, though six of these are of poor quality. Only one source
(source~19 or ALMA033226-275208) is found by both to be at $z>4$. 

\begin{figure}
\hspace{-1.0cm}\includegraphics[width=9.5cm]{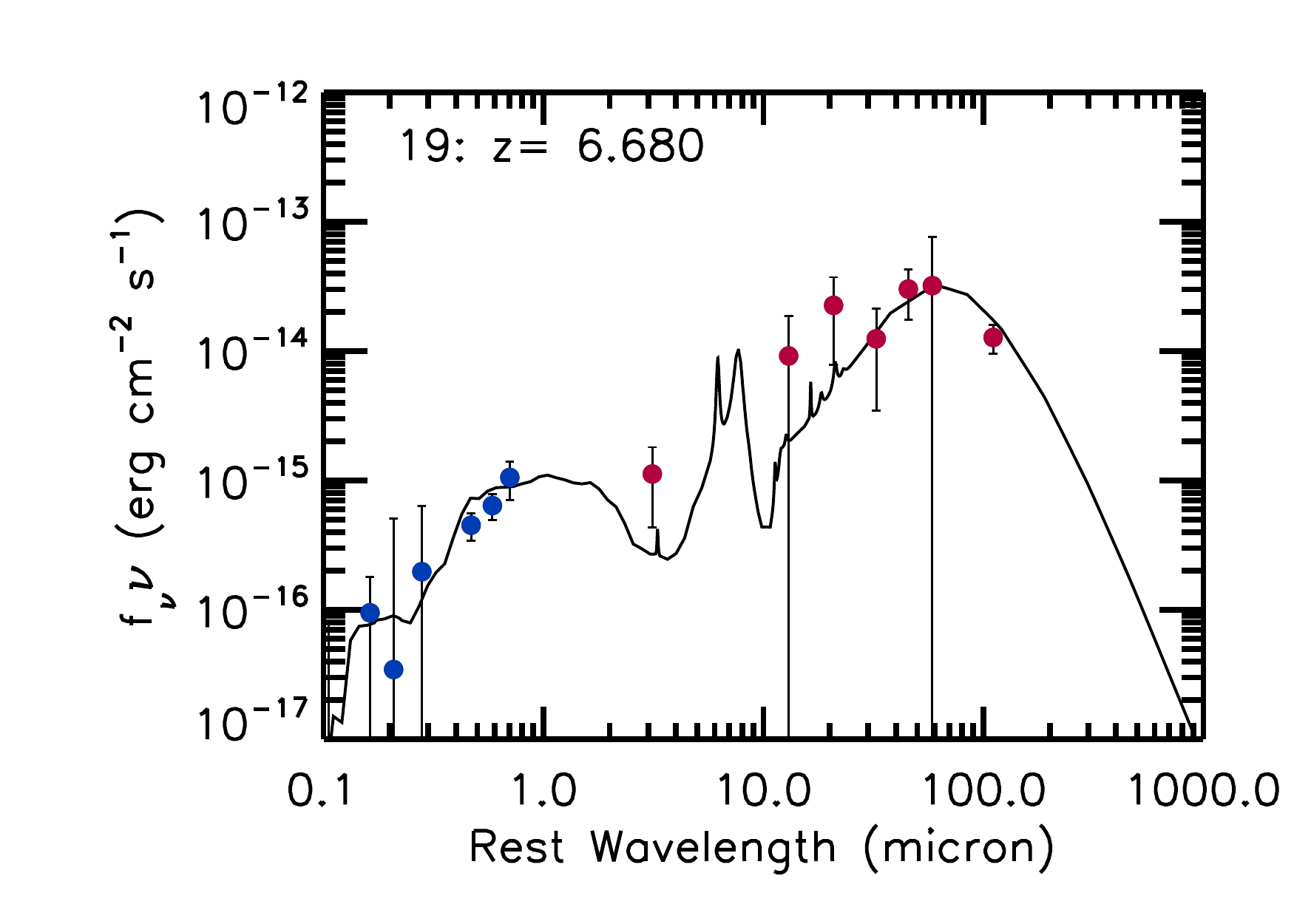}
\caption{Fit to the SED of source~19 or ALMA033226-275208 shown
in the frame of the FIRz fit. The blue circles show the data 
on which the photz is based, and the red circles the data on which the FIRz
is based. We show the Arp~220 SED for comparison (black curve).
\label{all_sed_19}
}
\end{figure}

We consider the six FIRz selected sources and the one high-quality
photz (source~2 or ALMA033211275212)
to be the most likely high-redshift candidates in the sample.
We summarize the properties of these seven sources in Table~6, where
we give the ratios of the 4.5, 24, 100, and 250~$\mu$m fluxes
to the ALMA 850~$\mu$m flux, the $2-7$~keV flux, and the 95\% confidence 
ranges for the photzs and the FIRzs. Source~17 or ALMA033235-275215 is an 
AGN, which we discuss further in Section~\ref{secxray}. The most compelling 
very high-redshift candidate is source~19, where both the poor quality photz 
and the FIRz estimate are consistent within the uncertainties with $z>6$. 
We show the full SED of this source in Figure~\ref{all_sed_19}. 
While they were not chosen in this way, all of the candidates
have low flux ratios in 24/850 (at or near 0.01) and 4.5/850 (at or near 0.001),
consistent with the high-redshift interpretation (see Section~\ref{secFIRz}).

\section{Flux Dependence of Redshifts and Star Formation History}
\label{secsfh}
\begin{figure}
\hspace{-1.0cm}\includegraphics[width=10.0cm]{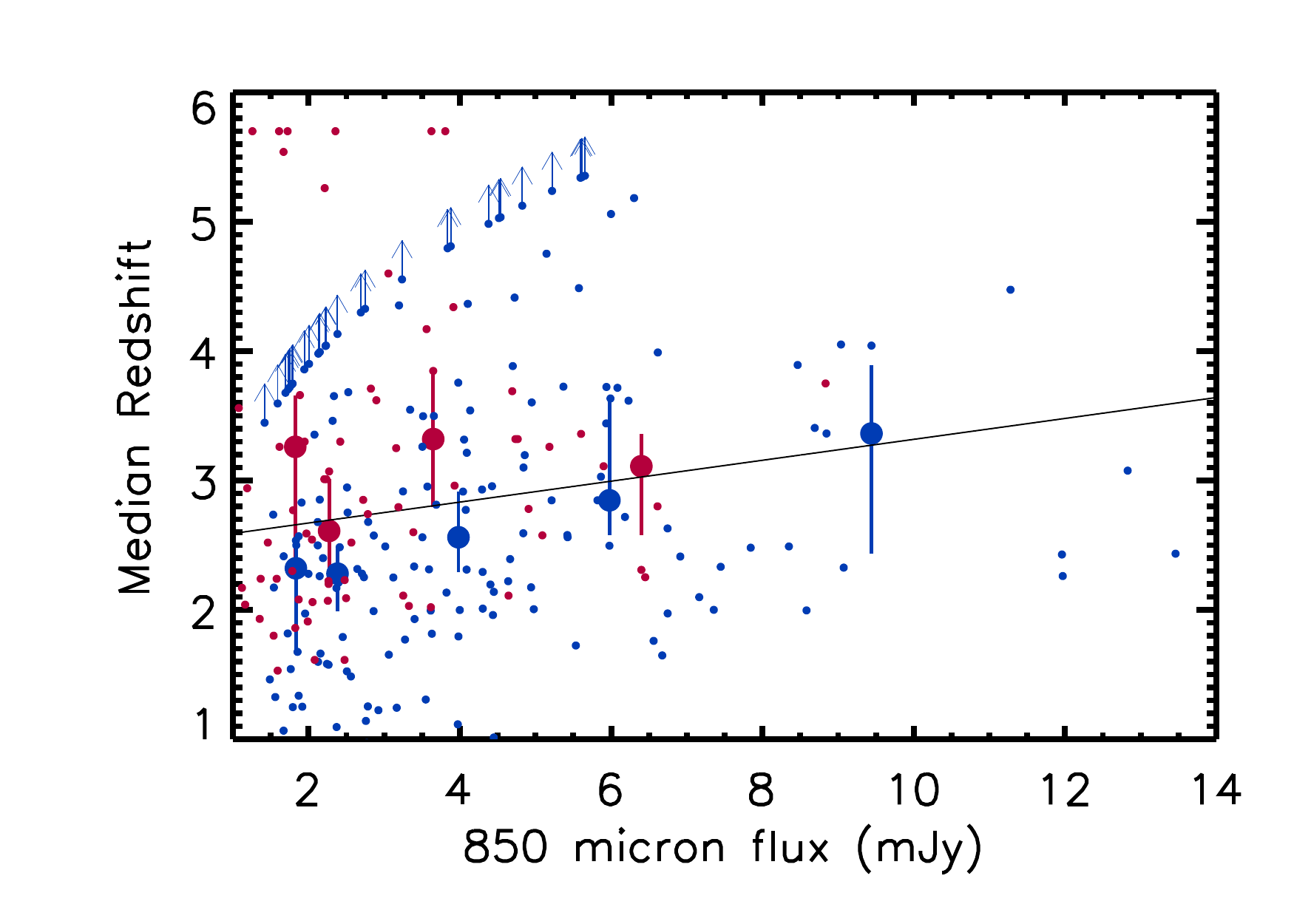}
\caption{Median redshift vs. 850~$\mu$m flux for the total ALMA 
sample (red large circles) and the wider GOODS-N sample (blue large
circles; Paper~I). The error bars show the 68\% confidence 
intervals on the medians. The smaller circles indicate individual sources. 
The blue points with upward pointing arrows correspond to $2\sigma$ 
lower limits on the redshifts in the GOODS-N sample.
The black line shows a linear least squares fit (Equation~\ref{red_fit}) 
to the combined samples. \label{smm_redsh}
}
\end{figure}

In Figure~\ref{smm_redsh}, we show the median redshift with
68\% confidence as a function of 850~$\mu$m flux for the 
total ALMA sample (red large circles) and for the wider sample of 
186 sources in the GOODS-N (blue large circles; Paper~I).
We also show the individual data points for each sample
(small circles). For the total ALMA sample, we plot sources that 
the FIRzs place at $z>5.5$ at a nominal redshift of $z=5.7$.
The GOODS-N redshifts contain a much higher fraction
of speczs, including a number at $z>4$, while the other 
redshifts for this field are based solely on the ratio of the 250~$\mu$m to 
850~$\mu$m flux. The upper envelope seen in the figure
consists of galaxies that have only a lower ($2\sigma$)
limit on the redshift. We show these with upward pointing arrows. 

A least-squares fit to the individual data points
in the combined samples gives the relation
(black line in Figure~\ref{smm_redsh})
\begin{equation}
z_{median}=2.40 + 0.09 (\pm0.03) \times S_{850}\ {\rm (mJy)} \,.
\label{red_fit}
\end{equation}
This is consistent with the Equation~7 relation given in Paper~I.
The redshift distribution is clearly increasing with increasing flux.
A Mann-Whitney test gives only a 0.05\% probability
that the combined $4-16$~mJy sample is drawn from the same redshift 
distribution as the combined $2-4$~mJy sample. 
These results are broadly consistent with recent modeling by 
B{\'e}thermin et al.\ (2015), who find lower redshift distributions
for lower fluxes.

However, there are significant issues of cosmic variance with
GOODS field sizes, and it is possible there may be
evolution in the dust temperature with redshift (see, e.g., 
Schreiber et al.\ 2018 and also Section~\ref{secFIRz}), which could 
change the redshift estimates. We postpone a more detailed discussion 
of this issue to a later paper in the series.

\begin{figure}
\hspace{-0.25cm}\includegraphics[width=10.5cm]{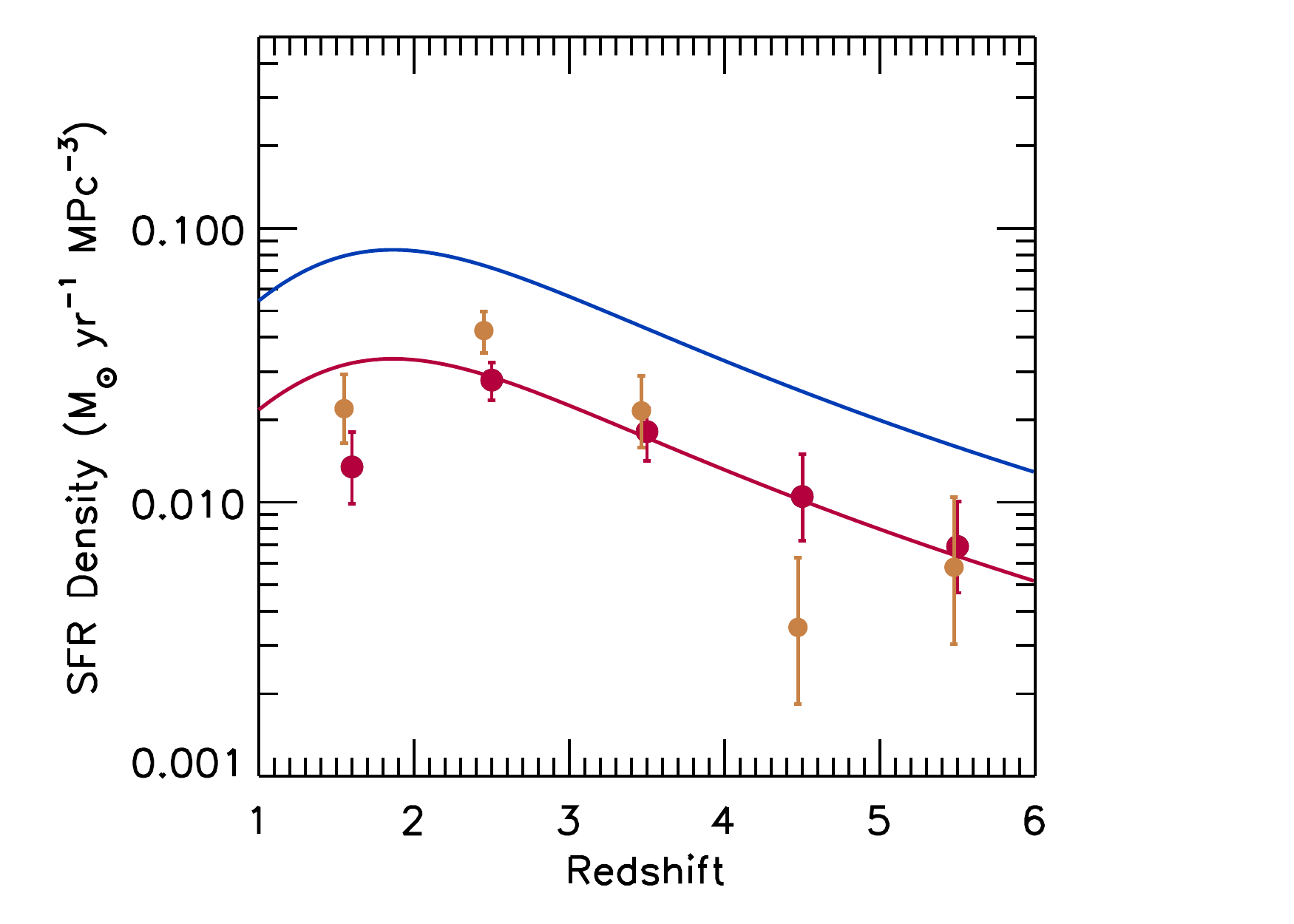}
\caption{SFR density per unit comoving volume vs. redshift for the 
total ALMA sample for a Kroupa (2001) IMF (gold circles; the
$1\sigma$ error bars are Poissonian based on the number of sources
in each bin). The computations were made at $z=1-2$, $2-3$, $3-4$,
$4-5$, and $5-6$ and are plotted at the mean redshift of each bin.
The blue curve shows the SFR density history computed by 
Madau \& Dickinson (2014), after conversion to a Kroupa IMF.
We also show the SFR density history for sources above 2~mJy in the GOODS-N 
(Paper~I; red circles), which we matched by multiplying the Madau \& Dickinson
results by 0.4 (red curve). However, once we take into account the different
FIR calibrations used in both works, the GOODS-N data are only 0.29 times
the Madau \& Dickinson curve.
\label{sfr_den_z}
}
\end{figure}

The total ALMA sample is still relatively small, and we postpone a more detailed
discussion of the SFR density history to A. Barger et al.\ (2018, in preparation), 
where we combine the full SCUBA-2 samples from both fields. 
Here we simply show that the star formation history computed from
the ALMA sources is fully consistent with that derived from
the GOODS-N observations in Paper~I. 

In Figure~\ref{sfr_den_z}, we plot the SFR density history for sources 
above 2~mJy in the GOODS-N (red circles). 
The computations were made at $z=1-2$, $2-3$, $3-4$,
$4-5$, and $5-6$ and are plotted at the mean redshift of each bin.
After adding the total ALMA sample using the same redshift intervals to 
the figure (gold circles), we can see that, within the uncertainties, 
the ALMA values are consistent with the GOODS-N values.

As noted in Paper~I, above $z\sim2$, the SFR density history corresponds
to 40\% (red curve) of the total SFR density history compiled by 
Madau \& Dickinson (2014), after converting their adopted Salpeter (1955) IMF
to a Kroupa (2001) IMF (blue curve). 
However, as described in Paper~I, we also need to 
take into account Paper~I's conversion from $L_{8-1000~\mu{\rm m}}$ to SFR, namely,
\begin{equation}
\log {\rm SFR}(M_\odot \ {\rm yr}^{-1}) = \log L_{8-1000\,\mu{\rm m}} \ {\rm (erg\ s}^{-1}) - 43.41 \,,
\label{eqconversion}
\end{equation}
when comparing to Madau \& Dickinson, who used a slightly lower conversion of $-43.55$
(after the IMF conversion). Once we do so, then we find that
the GOODS-N and GOODS-S data are only 0.29 times the Madau \& Dickinson curve.

\section{Morphologies and Mergers}
\label{secmorph}
The total ALMA sample is the largest existing sample with deep optical
and NIR data from {\em HST}, with roughly double the number of sources in
the ALESS based sample of Chen et al.\ (2015; though five of the brightest 
sources overlap with the Chen et al.\ sample) and much more extensive
wavelength coverage. The bulk of the ALMA sources are detected in the 
F160W band, which is the longest wavelength with high spatial resolution imaging. 

\begin{figure}
\hspace{-0.9cm}\includegraphics[width=9.5cm]{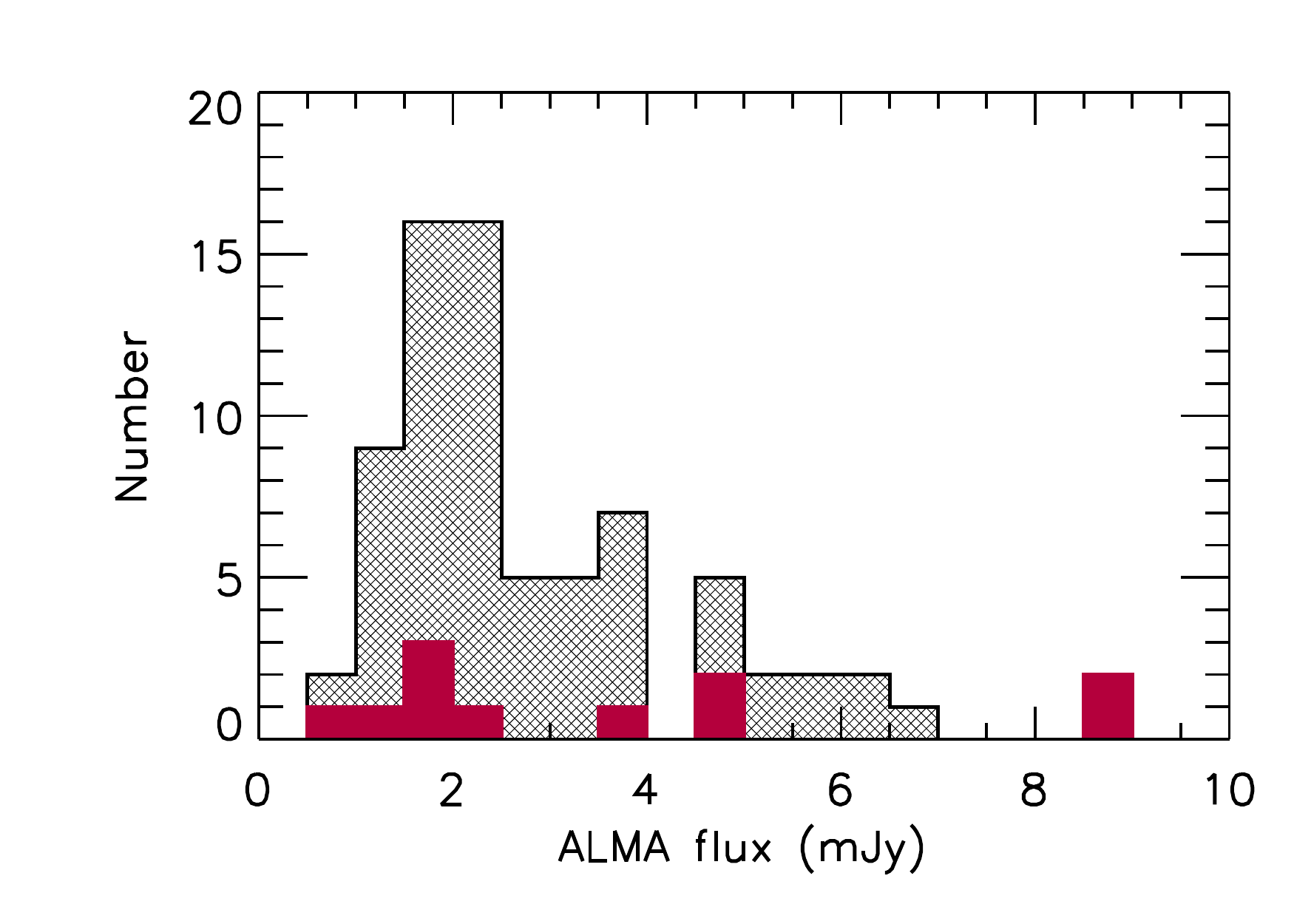}
\caption{
Histogram of the 850~$\mu$m flux distribution of the total ALMA sample (gray shading),
excluding source~27 or ALMA033203-275039. 
The red shading shows the distribution of the 11 sources without 
F160W counterparts.
\label{sm_hist}
}
\end{figure}

In Figure~\ref{sm_hist}, we show the 850~$\mu$m flux
distribution for the 74 ALMA sources (again excluding source~27 or ALMA033203-275039,
which lies off the {\em HST\/} images). We use red shading to indicate
the sources that are either undetected in F160W  
(sources~2, 19, 44, 61, 68, 75) or whose F160W counterparts may be
misidentifications (sources~1, 11, 13, 58, 64). 
Although the latter are substantially
overlapped with the sources whose separations between the F160W peak flux
and the ALMA position are $>0\farcs4$, as discussed in 
Section~\ref{secopt}, not all of those sources are included here, since, in some
cases, the galaxy still extends across the ALMA emission
(e.g., source~14 or ALMA033222-274936
and source~32 or ALMA033211-274615; see Figure~\ref{basic_alma_images}).
We note that source~11 or ALMA033219-275214 has two low-redshift sources
in its vicinity, neither of which is likely the correct counterpart.
We hereafter refer to these 11 sources as our undetected sample.

We do not detect at F160W 29\% (68\% confidence range 15--51\%) of the sources 
above 4~mJy, and 12\% (68\% confidence range 7--18\%) of the sources below 4~mJy.
If we take the 5 sources 
where we suspect we do not have the correct counterpart out of the undetected category, 
then we do not detect at F160W 7\% (68\% confidence range 1--23\%)
of the sources above 4~mJy and 8\% (68\% confidence range 5--14\%)
of the sources below 4~mJy.
In either case, we do not see any significant increase
in the fraction of  undetected sources as we move to lower submillimeter 
fluxes.  This differs from Hodge et al.\ (2013) and Chen et al.\ (2015),
who found a substantial increase in the fraction of F160W undetected sources
in the ALESS sample at fainter fluxes relative to higher fluxes.
Quantitatively, Chen et al.\ (2015) found that 36\% 
(68\% confidence range 25--52\%) of the 850~$\mu$m sources below 4~mJy 
were undetected, which is significantly higher than measured here.
However, all of the samples are small, and the depth of the 
F160W observations may also affect the results. Larger samples
are clearly required to resolve the issue of whether
the fraction of F160W faint sources increases with decreasing
submillimeter flux.

Even with the very deep F160W data, morphological classification
of the galaxies is extremely difficult. This is true whether
we use visual or automated classification schemes
(Chen et al.\ 2015). In the present paper, we 
only visually classify the galaxies. We postpone a quantitative 
morphological analysis to L.~Jones et al.\ (2018, in preparation).

\begin{figure}
\hspace{-0.9cm}\includegraphics[width=11.0cm]{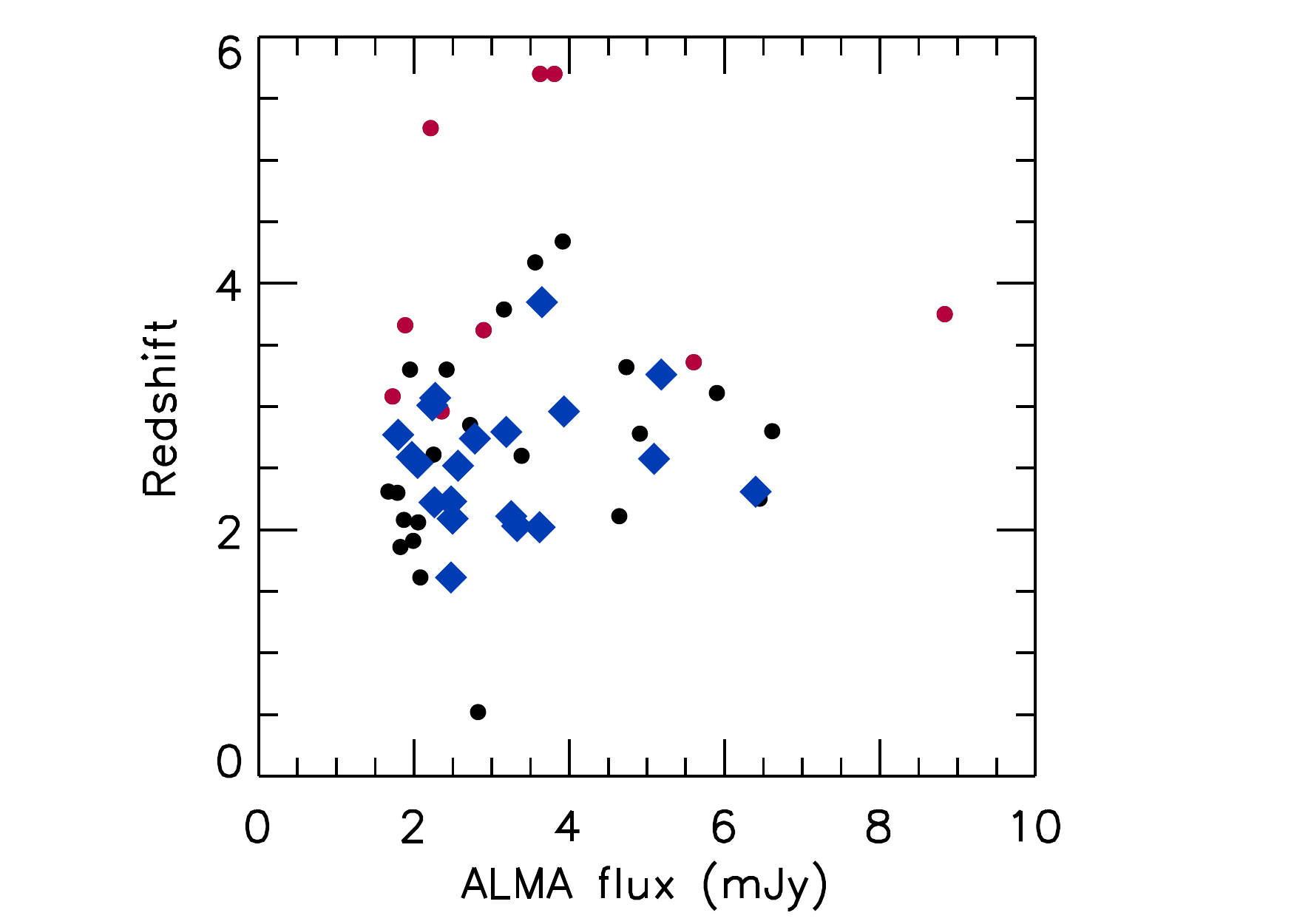}
\caption{
Redshift vs. 850~$\mu$m flux for the sources in the total ALMA sample  
with fluxes above 1.65~mJy, excluding only the 5 sources
whose F160W counterparts we consider to be
misidentifications (sources~1, 11, 13, 58, 64).
We use speczs when available, followed by FIRzs, then photzs.
We used the 1.65~mJy flux limit, because only those sources have
FIRzs.
Red circles show sources whose F160W fluxes are too faint to classify,
blue diamonds sources with a merger signature, and black
circles the remaining sources. 
\label{sm_firz}
}
\end{figure}

We find that,
in addition to the undetected sample, the F160W images of a 
further 11 galaxies are too faint to classify ($m_{160}>25$).
This leaves 52 galaxies that we can morphologically classify, of which
24 show what appears to be clear 
evidence of merging in the F160W images, either as extended tidal
tails, multiple nuclei with bridges, or highly asymmetric
images. Some sources, such as the ring galaxy
(source~23 or ALMA033237-275000) or the extended tidal tail
galaxy (source~43 or ALMA033220-275247), are impressive examples
of merging, while other sources are more subtle, and the subjective
nature of the classification should be kept in mind. In particular,
multiple nuclei may be confused with clumpy disks. 

To illustrate which sources can be classified, 
in Figure~\ref{sm_firz}, we show redshift versus flux for the
sources in the total ALMA sample with fluxes above 1.65~mJy,
excluding only the 5 sources whose F160W counterparts we consider 
to be misidentifications (sources~1, 11, 13, 58, 64).
We restricted to fluxes $>1.65$~mJy, because sources fainter than
that do not have FIRzs. We show the galaxies that are too faint to
classify with red circles. These galaxies lie 
at $z \gtrsim 3$, where the observed 1.6~$\mu$m corresponds to a 
rest-frame wavelength that lies below the 4000~\AA\ break. 
Chen et al.\ (2015) found a similar result for their ALESS sample and hence
truncated their sample to consider only $z<3$ galaxies. We did not make
such a redshift cut, but we restricted to the sources that can be classified. 
In practice, as the figure shows, the two selections are very similar.
The blue diamonds show the sources with clear evidence of merging.
The remaining galaxies (black circles) we have not attempted to classify, noting 
only that they are primarily extended and contain a mixture of smooth and 
irregular structures. Only one source (source~26 or ALMA033216-275044) 
appears compact based on the SExtractor star-galaxy classifier. However, 
we note the curious chain galaxy (Cowie et al.\ 1995) structure of 
source~22 or ALM033244-274635. While this may be a chance projection, 
the morphologies of the components are curiously similar, and the sequence 
terminates in the ALMA source. Barger et al.\ (2012) noted a similar source in
the GOODS-N (GOODS-7). It is possible that these sources are sequential 
star formers, with the SMG being the most recently formed stage. 

Overall, we obtain a 
merger fraction of 44\% (68\% confidence range 35--53\%).
The errors are Poissonian based on the number of sources and do
not include uncertainties in the classifications. We should also bear in 
mind that at least some of the unclassified sources may also
be mergers. 

The merger fraction is smaller if we consider only the sources 
with 850~$\mu$m fluxes $>4$~mJy. Here we have 3 mergers out of 10 classified sources,
or a  merger fraction 25\% (68\% confidence range 12--49\%).

Our high merger fraction is at odds with recent quantitative analyses, 
which have generally found the SMGs to be massive disks (e.g., Targett et al.\ 2013) 
and not preferentially major mergers (Swinbank et al.\ 2010). 
As Chen et al.\ (2015) argue, the quantitative methods can
miss many merging or disturbed sources, which are easily
distinguished in a visual inspection. It is not straightforward
to directly compare our results with Chen et al.,
who only give the fraction of disturbed (irregular
or merging systems) rather than attempting to distinguish
mergers. However, their high fraction of such systems
(82\%) also suggests that mergers are common. However,
once again, we emphasize the uncertainties in the classifications.
It is likely that this issue will only be resolved with JWST.

A detailed comparison of the ALMA sizes and morphologies 
with the optical and NIR images of the galaxies
is given in J. Gonz{\'a}lez-L{\'o}pez et al.\ (2018, in preparation).

\section{X-ray Luminosities and AGN Activity vs. Star Formation}
\label{secxray}
The total ALMA sample lies in the deepest region of 
the 7~Ms {\em Chandra\/} X-ray exposure (Luo et al.\ 2017). At
the largest off-axis angle for the ALMA sources, the detection threshold
in the Luo et al.\ catalog is $10^{-16}$~erg~cm$^{-2}$~s$^{-1}$ 
in the $2-7$~keV band, and $2\times10^{-17}$~erg~cm$^{-2}$~s$^{-1}$ 
in the deeper $0.5-2$~keV band. In the centermost regions,
the detection threshold is roughly three times lower than
these values. Just over half (41) of the ALMA sources
are detected in the Luo et al.\ catalog using a $1\farcs5$
matching radius: 39 are detected in the $0.5-2$~keV band, and 24 are 
detected in the shallower $2-7$~keV band. 
Only two sources are detected in the $2-7$~keV band and not
in the $0.5-2$~keV band. The matching is relatively insensitive to the 
choice of radius. If we instead used a $1\farcs0$ radius
(based on the 2$\sigma$ positional uncertainties of
the fainter {\em Chandra\/} sources in the Luo et al.\ region),
then we would only reduce the number of matches by one source.

\begin{figure}
\hspace{-1.0cm}\includegraphics[width=10.0cm]{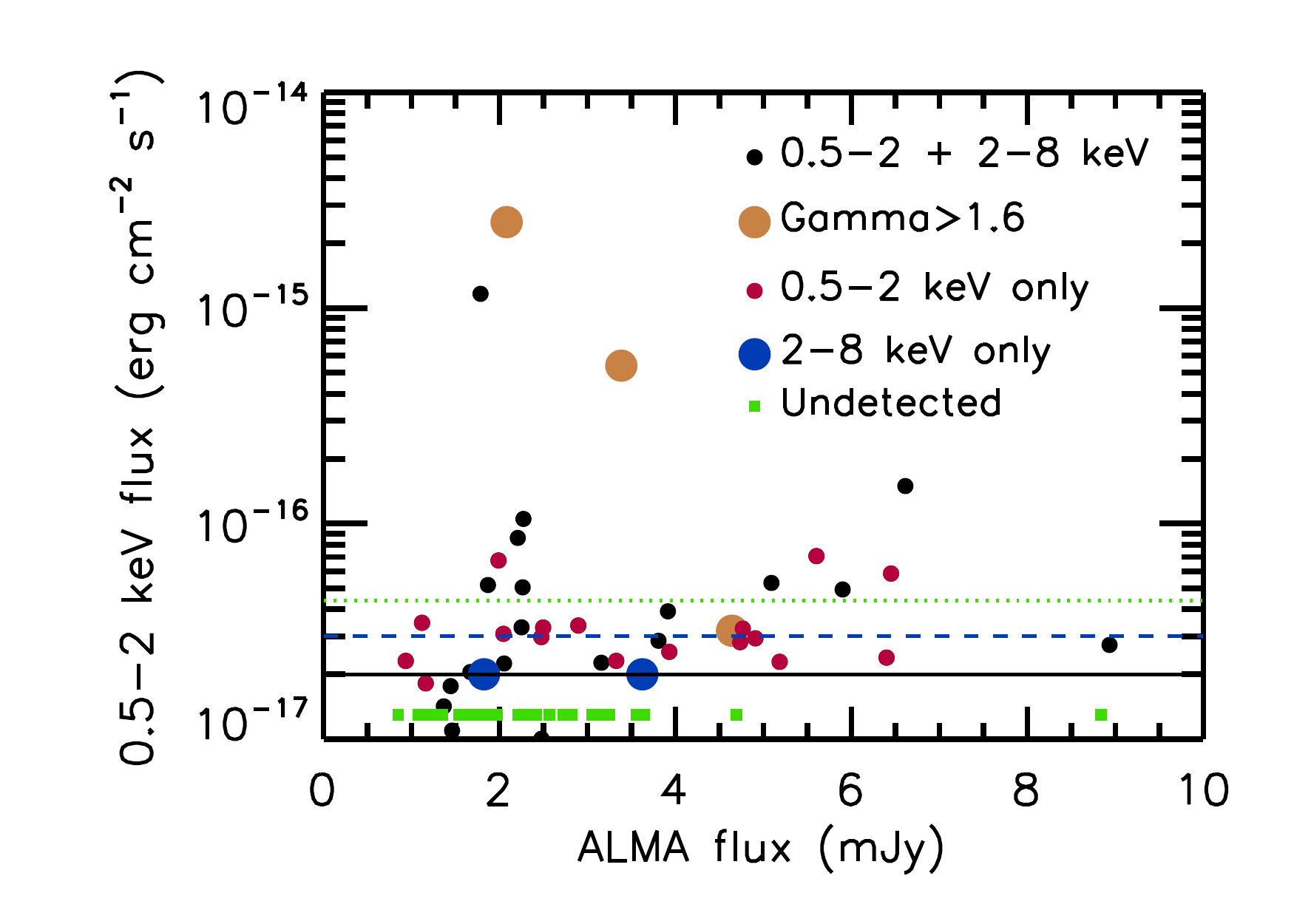}
\caption{$0.5-2$~keV flux vs. 850~$\mu$m flux for the total ALMA sample.
Black circles show sources detected at both $0.5-2$~keV and $2-7$~keV,
with those having $\Gamma>1.6$ denoted by 
gold large circles. Red circles show sources detected only at $0.5-2$~keV, 
and blue large circles show sources detected only at $2-7$~keV and 
placed at a nominal soft flux limit of $2\times10^{-17}$~erg~cm$^{-2}$~s$^{-1}$.
This is the 7~Ms CDF-S detection threshold from Luo et al.\ (2017; black line). 
The 4~Ms CDF-S flux detection threshold from Xue et al.\ (2011;
blue dashed line) and the 2~Ms CDF-N flux detection threshold from
Xue et al.\ (2016; green dotted line) are shown for comparison.
Sources not detected in either band are shown as green squares at 
a nominal flux of $1.3\times10^{-17}$~erg~cm$^{-2}$~s$^{-1}$. 
}
\label{xray_select}
\end{figure}

In Figure~\ref{xray_select}, we show the observed $0.5-2$~keV flux versus 
the 850~$\mu$m flux for the total ALMA sample. The black line shows the
soft X-ray flux limit above which sources would have been included
in the Luo et al.\ (2017) catalog. The green dotted line shows the same for the
2~Ms CDF-N (Xue et al.\ 2016), and the blue dashed line for the 
4~Ms CDF-S (Xue et al.\ 2011).
We see immediately that the bulk of the X-ray counterparts to the ALMA
sources could only be detected in a field as deep as the 7~Ms image. 
At 2~Ms, where the flux limits are 3.5 times higher, most of the sources would be missed. 

We note that the fraction of detected sources
is much higher at higher submillimeter fluxes.
Above 4~mJy, 12 of the 14 ALMA sources
have X-ray counterparts (86\%, with a 68\% confidence range 67--95\%),
while below this, it drops to 29 out of 61 (48\%, with a 68\% confidence range
38--57\%). (See also Ueda et al.\ 2017.)
The dependence on the submillimeter flux suggests that we
are seeing either star formation contributions to the
X-ray emission or enhanced AGN activity in the stronger SMGs.
 
Given that the median redshift of the ALMA sources is just
below $z=3$ (we have used the specz, if available, followed
by the FIRz, and then the photz), we compute the rest-frame $2-8$~keV
luminosities, $L_{2-8~{\rm keV}}$, from the observed-frame $0.5-2$~keV fluxes 
with no absorption correction and $\Gamma=1.8$ using
\begin{equation}
L_{2-8~{\rm keV}} = 4\pi d_L^2 f_{0.5-2\,{\rm keV}} ((1+z)/4)^{\Gamma-2}~{\rm erg~s^{-1}} \,.
\end{equation}
Using the $0.5-2$~keV flux to calculate the $2-8$~keV luminosities 
minimizes the $K$-corrections for sources at these redshifts.
This is important, given the variation in $\Gamma$. 
We take $L_{2-8~{\rm keV}}>10^{44}$~erg~s$^{-1}$ 
as the threshold for a source to be classified as an X-ray quasar.

The poorer sensitivity $2-7$~keV observations probe higher luminosities. 
They correspond to higher rest-frame energies,
\begin{equation}
L_{8-28~{\rm keV}} = 4\pi d_L^2 f_{2-7\,{\rm keV}} ((1+z)/4)^{\Gamma-2}~{\rm erg~s^{-1}} \,.
\end{equation}

\begin{figure}
\hspace{-1.0cm}\includegraphics[width=10.0cm]{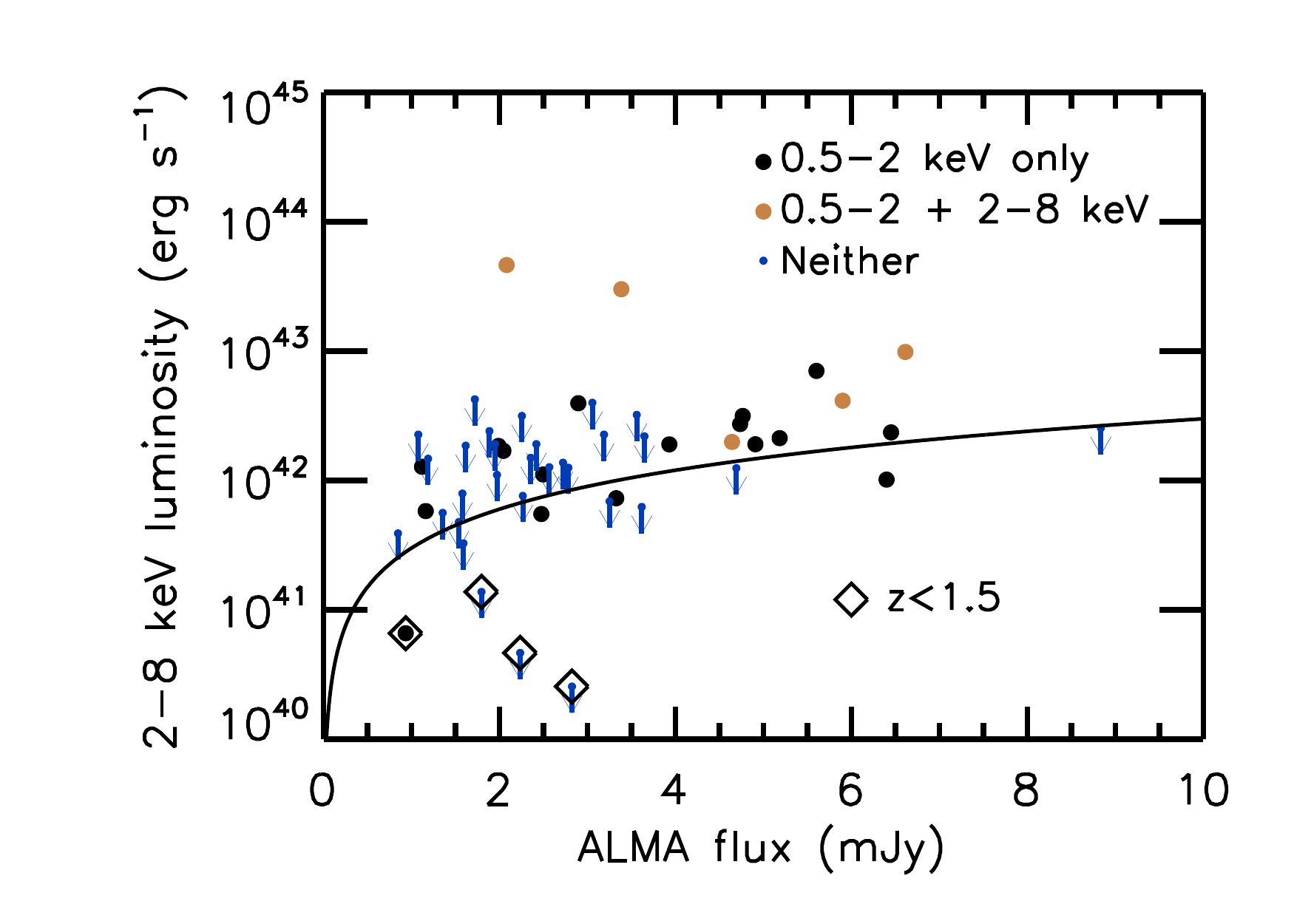}
\caption{Rest-frame $2-8$~keV luminosity vs. 850~$\mu$m flux for sources 
in the total ALMA sample that do not have hard X-ray spectra.
Black circles show sources detected at $0.5-2$~keV but not at $2-7$~keV.
Gold circles show sources with detections in both bands but having
$\Gamma > 1.2$. 
Blue downward pointing arrows show upper limits for sources 
not detected in either X-ray band, assuming the
Luo et al.\ (2017) catalog $0.5-2$~keV flux detection threshold of
$2\times 10^{-17}$~erg~cm$^{-2}$~s$^{-1}$.
Sources with $z<1.5$ are marked with open diamonds.
Black curve shows the relation between $L_{\rm 2-8~keV}$ 
and 850~$\mu$m flux for star-forming galaxies discussed in the text
(Equation~\ref{eqxlum850reln}).
\label{xray_smm}
}
\end{figure}

We expect the X-ray binaries produced during star formation 
to have soft X-ray photon indices. (Sazonov \& Khabibullin 2017
give an X-ray photon index of $2.1\pm0.1$ for the collective X-ray spectrum 
of luminous high-mass X-ray binaries.) Thus, in Figure~\ref{xray_smm},
we plot $L_{\rm 2-8~keV}$ versus ALMA flux for the sources in the total 
ALMA sample that are detected only in the $0.5-2$~keV band, or, if detected 
in both bands, have $\Gamma > 1.2$, or are not detected in either band
(shown as upper limits).

Combining the SFR versus X-ray luminosity relation of Mineo et al.\ (2014) 
with the SFR versus 850~$\mu$m flux relation of Barger et al.\ (2014) gives 
the relation (plotted as the black curve in Figure~\ref{xray_smm})
\begin{equation}
L_{\rm 2-8~keV} = 3 \times 10^{41} f_{850} {\rm(mJy)}~{\rm erg~s^{-1}} \,.
\label{eqxlum850reln}
\end{equation}
Since the submillimeter flux is not
linearly related to the SFR at $z\lesssim1.5$, we mark the
low-redshift sources in Figure~\ref{xray_smm} with open diamonds;
nearly all the other sources are consistent with the star formation relation. 
This suggests that 
the 7~Ms {\em Chandra\/} X-ray exposure is finally deep enough to start to 
probe the star formation taking place in the most intensely star-forming 
galaxies in the universe. However, two of the $\Gamma > 1.2$ sources 
detected in both bands (two of the gold circles; source~22 or ALMA033244274635 
and source~46 or ALMA033225274219) are too luminous to be interpreted 
in this way and appear to be relatively unobscured AGNs.

\begin{figure}
\hspace{-1.0cm}\includegraphics[width=10.0cm]{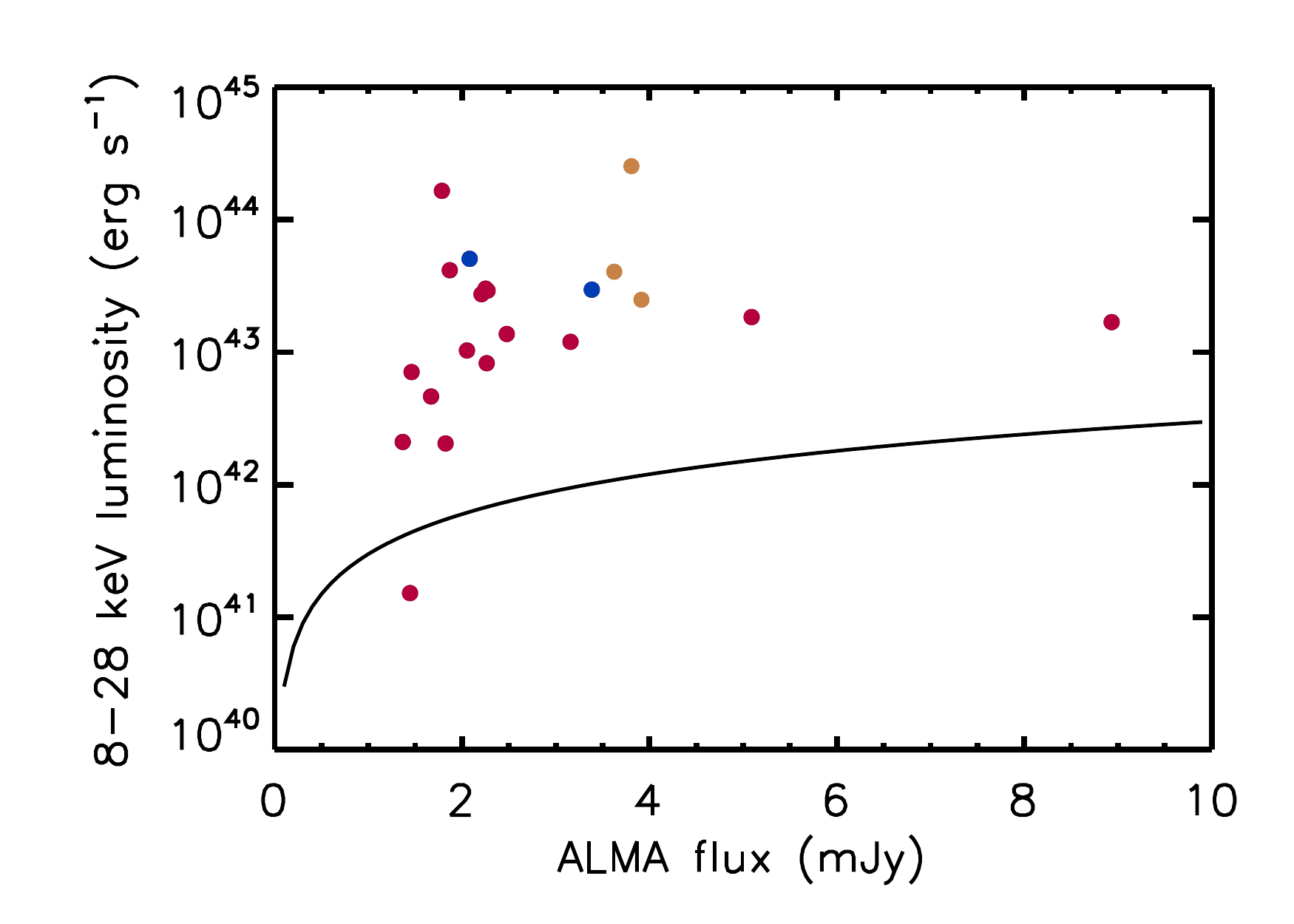}
\caption{Rest-frame $8-28$~keV luminosity for 21 of the 24 ALMA sources
detected in the observed-frame $2-7$~keV band, excluding the three sources 
found to be consistent with being star-forming galaxies from 
Figure~\ref{xray_smm}. Blue circles show relatively
unobscured sources ($\log N_H<22$~cm$^{-2}$), and gold circles 
show potentially Compton thick sources (also high redshift; see text). 
The remaining sources are shown in red. Black curve shows the relation 
between $L_{\rm 2-8~keV}$ and 850~$\mu$m flux for star-forming galaxies 
discussed in the text (Equation~\ref{eqxlum850reln}).
\label{xray_hard_lum}
}
\end{figure}

Of the 24 ALMA sources with hard band detections, we show 21 in 
Figure~\ref{xray_hard_lum}, excluding the three sources found to be 
consistent with being star-forming galaxies from Figure~\ref{xray_smm} 
(i.e., the gold circles that lie relatively close to the black curve). 
We plot rest-frame $8-28$~keV luminosity versus 850~$\mu$m flux
for the 21 sources. We summarize the properties of these 21 sources in Table~7.

The flux detection threshold in the $2-7$~keV band corresponds 
to a luminosity threshold of near $10^{43}$~erg~s$^{-1}$ for all but the 
lowest redshift sources.
This is well above the luminosity that could be accounted
for by star formation based on the Mineo et al.\ (2014) relation (black curve),
which does not change appreciably for higher X-ray luminosities.
The lowest luminosity source is the low-redshift source~66
or ALMA033210-274807, which has a spectroscopic redshift of $z=0.654$. 
Its optical spectrum is characteristic of a star-forming galaxy and does not show
any AGN signatures. 
We will consider it to be a star-forming galaxy subsequently.

Of the remaining 20 sources, most are moderate luminosity AGNs that 
lie just above the detection threshold. 
The optical spectra for two (source~9 or ALMA033235-274916
and source~46 or ALMA033225-274219) classify them as type~2 AGNs, and
for two others (source~40 or ALMA033231-274623 and source~59
or ALMA033222-274815) as star-forming galaxies. 
For these last two galaxies we do not see AGN signatures
in the optical spectra.
Only two have quasar luminosities (source~17 or ALMA033235-275215 and 
source~56 or ALMA033225-274305). For source~17,
such a high luminosity is a consequence of the high-redshift FIRz 
(see Section~\ref{sechighz}).

\begin{figure}
\hspace{-1.0cm}\includegraphics[width=10.0cm]{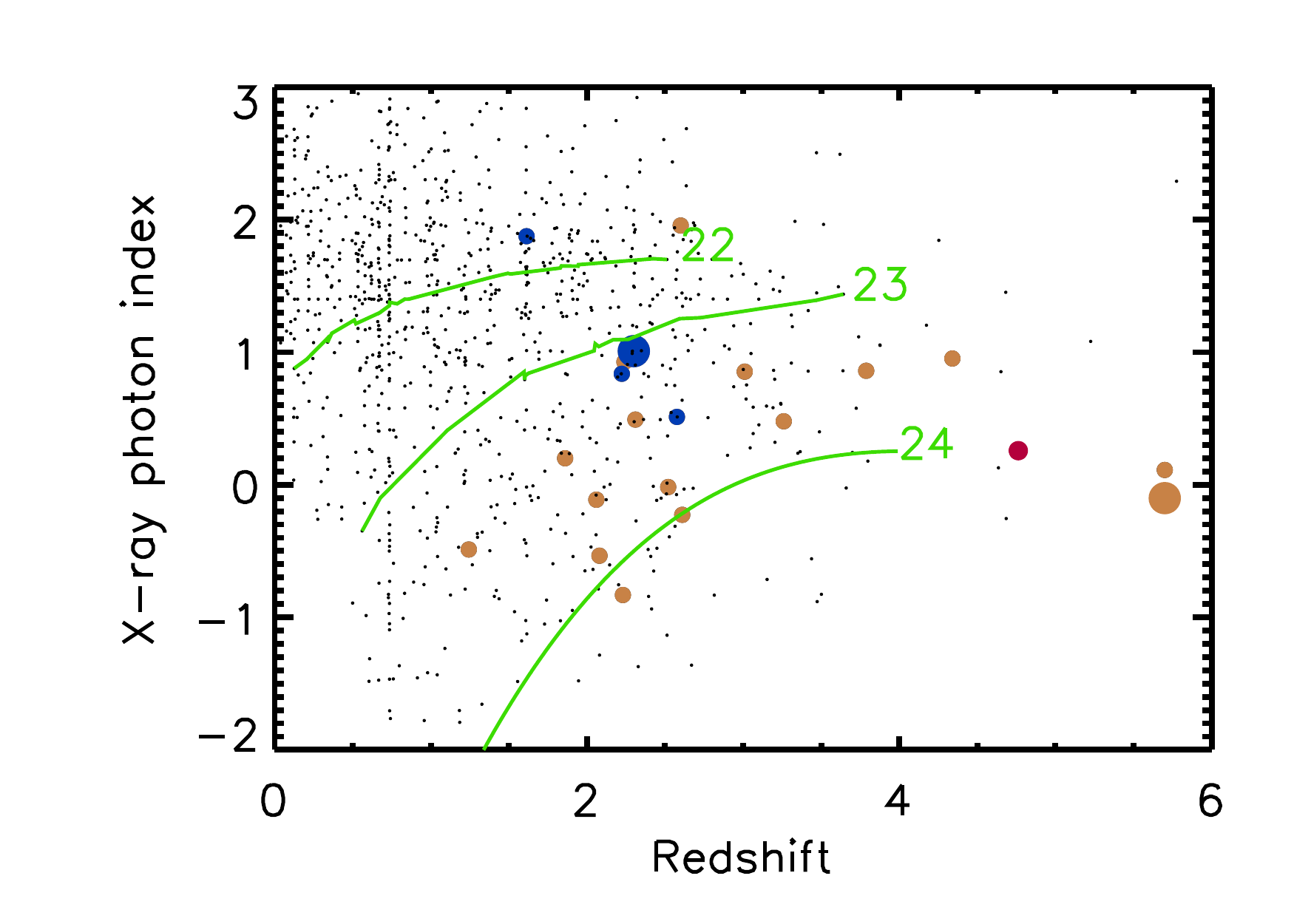}
\caption{Distribution of X-ray photon index vs. redshift for 20 of the 24 ALMA
sources detected in the $2-7$~keV band, excluding the three sources 
found to be consistent with being star-forming galaxies from 
Figure~\ref{xray_smm} and the one low-redshift source below the star formation
relation in Figure~\ref{xray_hard_lum} 
with a star-forming galaxy optical spectrum. Gold circles denote sources
with a FIRz or a photz, and blue circles denote those with a
specz. Sources with a very high redshift FIRz estimate
are shown at a nominal redshift of 5.7. Larger circles show sources with quasar X-ray
luminosities. The black dots show the full Luo et al.\ (2017)
sample. The green curves from top to bottom show absorption of 
$\log N_H=$22, 23, and 24~cm$^{-2}$.
\label{xray_gamma}
}
\end{figure}

Most are also substantially obscured 
($\log N_H\gg22$~cm$^{-2}$)
but not Compton thick. Only two are unobscured (blue circles in
Figure~\ref{xray_hard_lum}), and only the two potentially high-redshift ones
(source~17 and source~19 or ALMA033226-275208; see
Section~\ref{sechighz}) could be Compton thick (gold circles), though this
depends on the FIRzs being correct for these sources.

We show this in more detail in Figure~\ref{xray_gamma}, where we
plot X-ray photon index versus redshift for the 20 sources.
Larger symbols denote the two sources with quasar luminosities. 
Sources with a specz are denoted by blue circles, and sources with a FIRz
or a photz are denoted by gold circles. We also show
all of the sources from the Luo et al.\ (2017) catalog using the
redshifts from their catalog (black dots), distinguishing
the Gilli et al.\ (2011) high-redshift,
Compton thick source with a red circle. The green curves
correspond to obscuring column densities of $\log N_H=$ (top
curve), 23 (middle), and 24~cm$^{-2}$ (bottom).

Nearly all of the submillimeter sources lie at positions consistent with 
$\log N_H=23-24$~cm$^{-2}$, and the median $\log N_H$ of the sample 
is 23.5~cm$^{-2}$. Only the two potentially high FIRz sources (17 and 19) 
would be identified as Compton thick. 
Indeed, a substantial fraction (17\%) of the 106 hard X-ray selected AGNs with
$\log N_H$ $>23$~cm$^{-2}$ in the central SCUBA-2 region are detected in 
the submillimeter, while none of the 90 hard X-ray sources in the region with 
$\log N_H = 22-23$~cm$^{-2}$ are. This suggests that for these sources, the 
submillimeter emission is related to the AGN activity,
or, perhaps more simply, to the obscuration level.

\begin{figure}
\hspace{-1.0cm}\includegraphics[width=9.5cm]{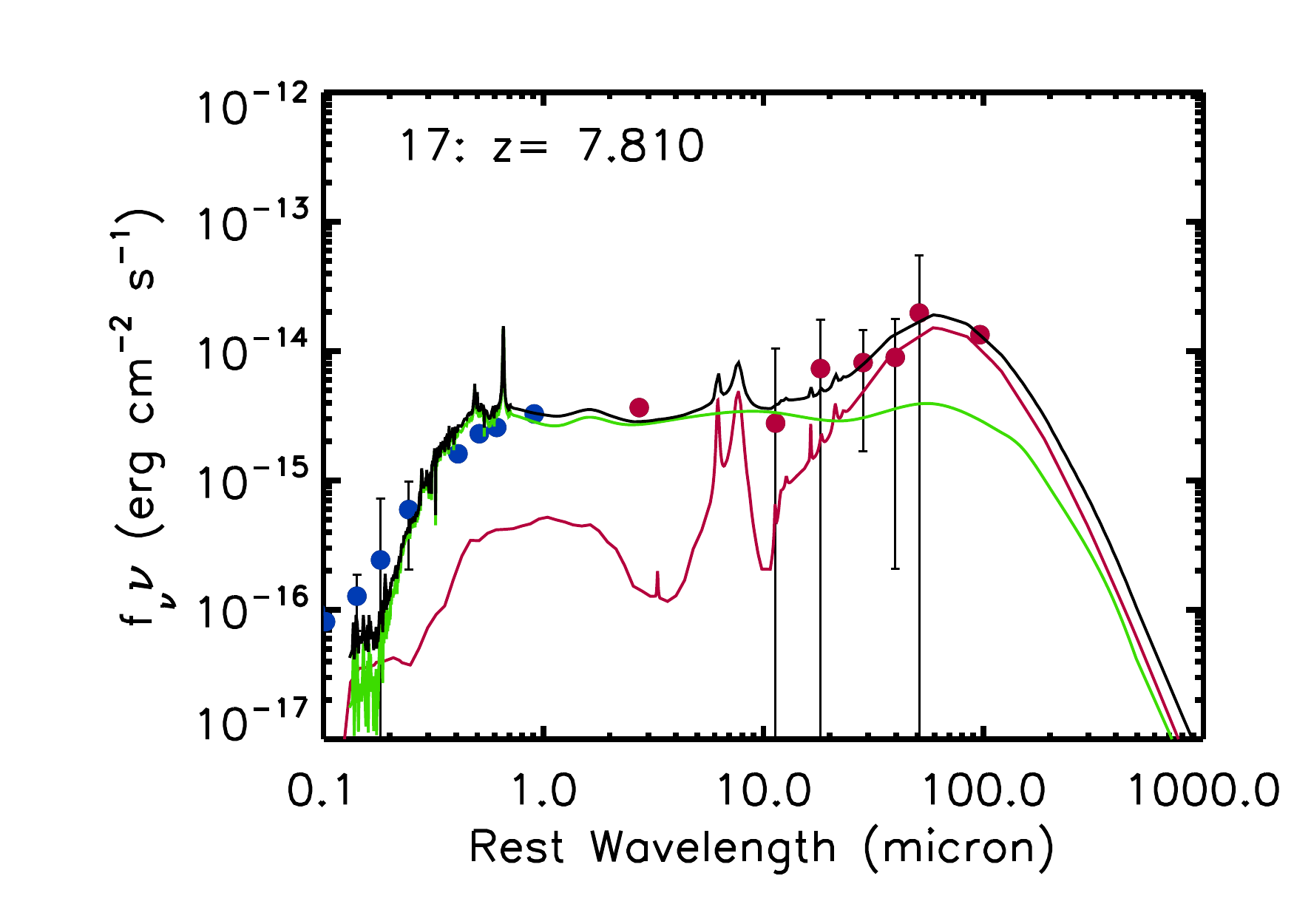}
\caption{Fit to the SED of source~17 or ALMA033235-275215. 
The blue circles show the data 
on which the photz is based, and the red circles the data on which the FIRz
is based. The fit is for the FIRz and shows that a combination of a
type~2 quasar (green curve) and a star former (red curve) can provide
a good fit to all the observations (black curve) at this redshift.
\label{alma_17}
}
\end{figure}

In order for the sources (sources~17 and 19)  not to be Compton thick, we would have
to reduce their redshifts to $z<2.5$. This is extremely
difficult to do, given that, in each case, the ALMA flux is brighter
than the shorter wavelength {\em Herschel\/} fluxes. 
Including an AGN contribution, which contributes preferentially to 
rest-frame wavelengths shortward of 100~$\mu$m, 
will only make this worse. 

In Figure~\ref{alma_17}, we show
a fit to source~17 with contributions from 
a type~2 quasar and a star former at longer wavelengths. 
This provides a good description of the SED for the high FIRz.
We conclude that both sources~17 and 19 may be high-redshift
Compton thick sources, similar to the high-redshift Compton
thick source of Gilli et al.\ (2011). However, the fits
depend on the assumed temperatures and the template,
making them substantially uncertain. Millimeter spectroscopic
measurements of the redshifts will ultimately determine
whether these sources are indeed Compton thick AGN.

\section{Summary}
\label{secsummary}

In this third paper in the SUPER GOODS series on powerfully star-forming
galaxies in the two GOODS fields, we analyzed the $>4\sigma$ sources in
the most sensitive 100~arcmin$^2$ area (rms $<0.56$~mJy)
of a deep SCUBA-2 survey of the GOODS-S and presented the
75 band~7 ALMA sources ($>4.5\sigma$) obtained from 
high-resolution interferometric follow-up observations. 
Based on the noise level of the SCUBA-2 image in the 100~arcmin$^2$ area, we
expect the SCUBA-2 sample to be complete to the $4\sigma$ level of 2.25~mJy. 
Since the SCUBA-2---but not ALMA---flux may be biased upwards due to Eddington bias 
and source blending, we also expect the ALMA sample to be complete to this level. 
The main results of our analysis are as follows:

\begin{itemize}

\item[$\bullet$]
Only five of the 53 SCUBA-2 sources in the complete sample above 2.25~mJy
have no ALMA counterparts, and only 13\% (68\% confidence range 7--19\%)
have multiple counterparts.
Given the larger primary beam size of $\sim17''$ for ALMA in band~7 
versus $\sim14''$ for SCUBA-2 at 850~$\mu$m, ALMA should be able 
to image all the sources that contribute to the SCUBA-2 flux. Indeed, when
we summed the individual ALMA fluxes in cases where there were multiple ALMA 
counterparts, we found that the ALMA and SCUBA-2 fluxes were consistent with one another.
This also emphasizes that our best ALMA fluxes are not missing flux due to resolution issues
(that is, we have not resolved out the flux), so they can be used to generate
accurate ALMA number counts.

\item[$\bullet$]
The ALMA sources have diverse optical/NIR counterparts ranging from bright, 
low-redshift galaxies to sources which remain undetected even in extremely deep
optical/NIR images. The latter sources are the most plausible candidates for very 
high-redshift SMGs.

\item[$\bullet$]
Using known spectroscopic and photometric redshifts from the literature,
we confirmed the known correlation of redshift with $K_s$ magnitude for the ALMA sources. 
Thus, as previously noted (e.g., Barger et al.\ 2014; Simpson et al.\ 2014), 
$K_s$ magnitude can be used as a crude redshift estimator.

\item[$\bullet$]
We compared and tested the KIEROS (Wang et al.\ 2012), HIEROS (Wang et al.\ 2016), 
and three color (Chen et al.\ 2016) techniques for selecting high-redshift dusty galaxies.
Although all of the methods are very comparable in their ability to select SMGs, their
detection rates are only at the $40-55$\% level and depend strongly on the photometry
methods employed.

\item[$\bullet$]
Consistent with Barger et al.\ (2012, 2014) and Paper~I,
we found that an Arp~220 template provides a good representation of the 
multiwavelength SEDs of the ALMA sources, so we used it to estimate FIR 
photometric redshifts. 
While there is some scatter, the overall correlation of these redshifts with the 
spectroscopic and photometric redshifts is reasonable for most of the sources. 

\item[$\bullet$]
We also compared the
4.5~$\mu$m and 24~$\mu$m to 850~$\mu$m flux ratios to the redshifts. 
We found a tight correlation, with both ratios showing the expected decline with redshift,
and the highest redshift sources being faint at both 4.5~$\mu$m and 24~$\mu$m.
These flux ratios can therefore be used to find high-redshift ($z\gtrsim4$) candidates,
even in the absence of full long-wavelength SEDs.

\item[$\bullet$]
For the ALMA sources above 1.65~mJy (the flux limit above which we had
high S/N throughout the SEDs and hence could
estimate FIR photometric redshifts), the median redshift is $z=2.74$, and for
those above 3~mJy, the median redshift is $z=3.26$. However, these rely on
mostly estimated redshifts, which may be problematic if there is AGN activity or
if sources lie at high redshift. Based on both the FIR and standard 
photometric redshifts, we found a total of seven candidate $z\gtrsim4$ sources.

\item[$\bullet$]
Consistent with Paper~I, we found that the redshift distribution of the ALMA sources 
increases with increasing 850~$\mu$m flux, and the contribution of dusty,
powerfully star-forming galaxies observed here by SCUBA-2 and ALMA
to the overall star formation history is impressively large (of order 30\%).

\item[$\bullet$]
All but 11 of the 74 ALMA sources (excluding the one source that is not on the 
{\em HST\/} GOODS-S area) are detected in the F160W band, the longest
wavelength at which we can make high spatial resolution imaging. Unlike previous 
work (Hodge et al.\ 2013; Chen et al.\ (2016), we see little dependence of the F160W
undetected fraction on submillimeter flux.  We made visual morphological
classifications using the F160W data of the 52 sources that were bright enough 
to do so. We found that 24 showed clear evidence of merging for a merger 
rate of at least (since some of the unclassified sources may also be mergers)
44\% (68\% confidence range 35--53\%).

\item[$\bullet$]
We found that just over half (41) of the total ALMA sample are detected in the 
Luo et al.\ (2017) X-ray catalog of the 7~Ms CDF-S. The fraction of X-ray
detected sources is much higher at higher submillimeter fluxes:
86\% (68\% confidence range 67--95\%) above 4~mJy versus
48\% (68\% confidence range 38--57\%) below 4~mJy. 
Without such deep data, many, if not most, X-ray detections of ALMA
sources would be missed (e.g., the 2~Ms data in the CDF-N).

\item[$\bullet$]
We computed the rest-frame $2-8$~keV luminosities, $L_{\rm 2-8~keV}$, from the
observed $0.5-2$~keV fluxes and the rest-frame $8-28$~keV luminosities, 
$L_{\rm 8-28~keV}$, from the shallower observed $2-7$~keV fluxes for the ALMA
sources. Since X-ray
binaries produced during star formation should have soft X-ray photon indices,
we first considered only the sources detected in the soft band (39), or, if detected
in both bands, then having a photon index $\Gamma>1.2$ (5). Combining the 
Mineo et al.\ (2014) SFR versus X-ray luminosity relation with the 
Barger et al.\ (2014) SFR versus 850~$\mu$m flux relation, we obtained 
$L_{\rm 2-8~keV}=3\times 10^{41}~f_{850}$(mJy) erg~s$^{-1}$. We found that
nearly all of the $z>1.5$ sources are consistent with this star formation relation. 
This suggests that the 7~Ms image is deep enough to start to probe the star formation 
taking place in the most intensely star-forming galaxies in the universe.

\item[$\bullet$]
The $L_{\rm 8-28~keV}$ for 20 of the 24 hard band detections (the other four were consistent 
with the star formation relation discussed in the previous bullet) are all well above the 
luminosity that could be accounted for by star formation. Most are moderate luminosity
AGNs that lie just above the detection threshold. They have substantial obscuration 
($\log N_H=23-24$~cm$^{-2}$) but are not Compton thick.

\item[$\bullet$]
Two of our high-redshift candidates could be Compton thick, similar to the
high-redshift Compton thick source found by Gilli et al.\ (2011). The only
way for these two sources not to be Compton thick would be if their redshifts were
$z<2.5$. However, their ALMA fluxes are considerably brighter
than the shorter wavelength {\em Herschel\/} fluxes, making such redshifts
difficult to achieve.

\end{itemize}

\vskip 0.4cm
\acknowledgements
We gratefully acknowledge support from NASA grant NNX17AF45G (L.~L.~C.),
NSF grants AST-1313309 (L.~L.~C.) and AST-1313150 (A.~J.~B.),
CONICYT grants Basal-CATA PFB-06/2007 (F.~E.~B, J.~G.-L.), 
FONDECYT Regular 1141218 (F.~E.~B, J.~G.-L.)
and Programa de Astronomia FONDO ALMA 2016 31160033 (JG-L),
and the Ministry of Economy, Development, and Tourism's Millennium Science 
Initiative through grant IC120009, awarded to The Millennium Institute of Astrophysics, 
MAS (F.~E.~B.).
A.~J.~B. acknowledges additional support from the John Simon Memorial 
Guggenheim Foundation, the Trustees
of the William F. Vilas Estate, and the University of Wisconsin-Madison
Office of the Vice Chancellor for Research and Graduate Education
with funding from the Wisconsin Alumni Research Foundation.
ALMA is a partnership of ESO (representing its member states), 
NSF (USA) and NINS (Japan), together with NRC (Canada), 
MOST and ASIAA (Taiwan), and KASI (Republic of Korea), in cooperation 
with the Republic of Chile. The Joint ALMA Observatory is operated by 
ESO, AUI/NRAO and NAOJ.
The James Clerk Maxwell Telescope is operated by the East Asian Observatory 
on behalf of The National Astronomical Observatory of Japan, Academia Sinica 
Institute of Astronomy and Astrophysics, the Korea Astronomy and Space 
Science Institute, the National Astronomical Observatories of China and the 
Chinese Academy of Sciences (Grant No. XDB09000000), with additional funding 
support from the Science and Technology Facilities Council of the United Kingdom 
and participating universities in the United Kingdom and Canada.
The W.~M.~Keck Observatory is operated as a scientific
partnership among the California Institute of Technology, the University
of California, and NASA, and was made possible by the generous financial
support of the W.~M.~Keck Foundation.
The authors wish to recognize and acknowledge the very significant 
cultural role and reverence that the summit of Mauna Kea has always 
had within the indigenous Hawaiian community. We are most fortunate 
to have the opportunity to conduct observations from this mountain.


\begin{figure*}
\setcounter{figure}{9}
\includegraphics[width=2.5in,angle=0]{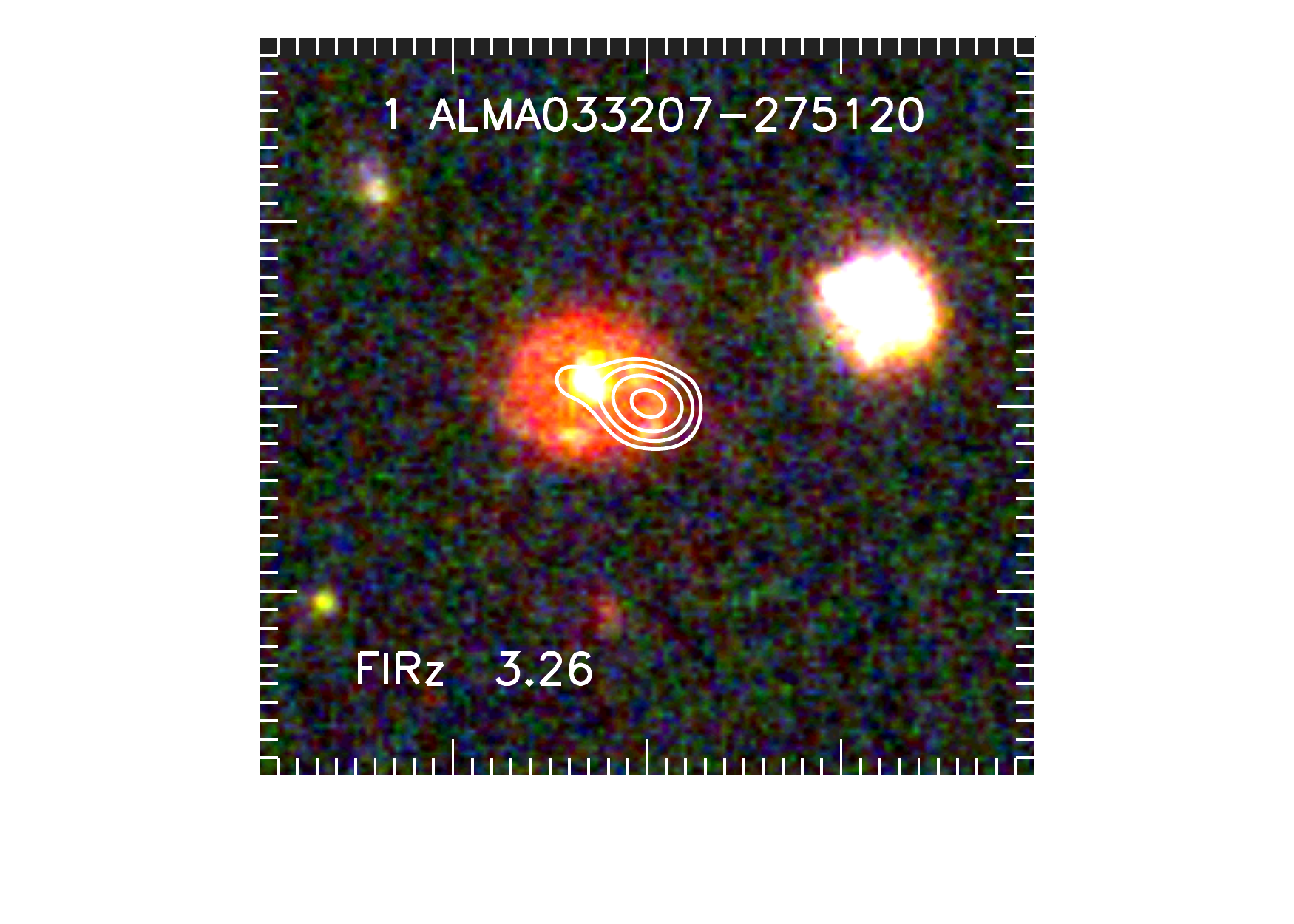}
\hspace{-3.3cm}\includegraphics[width=2.5in,angle=0]{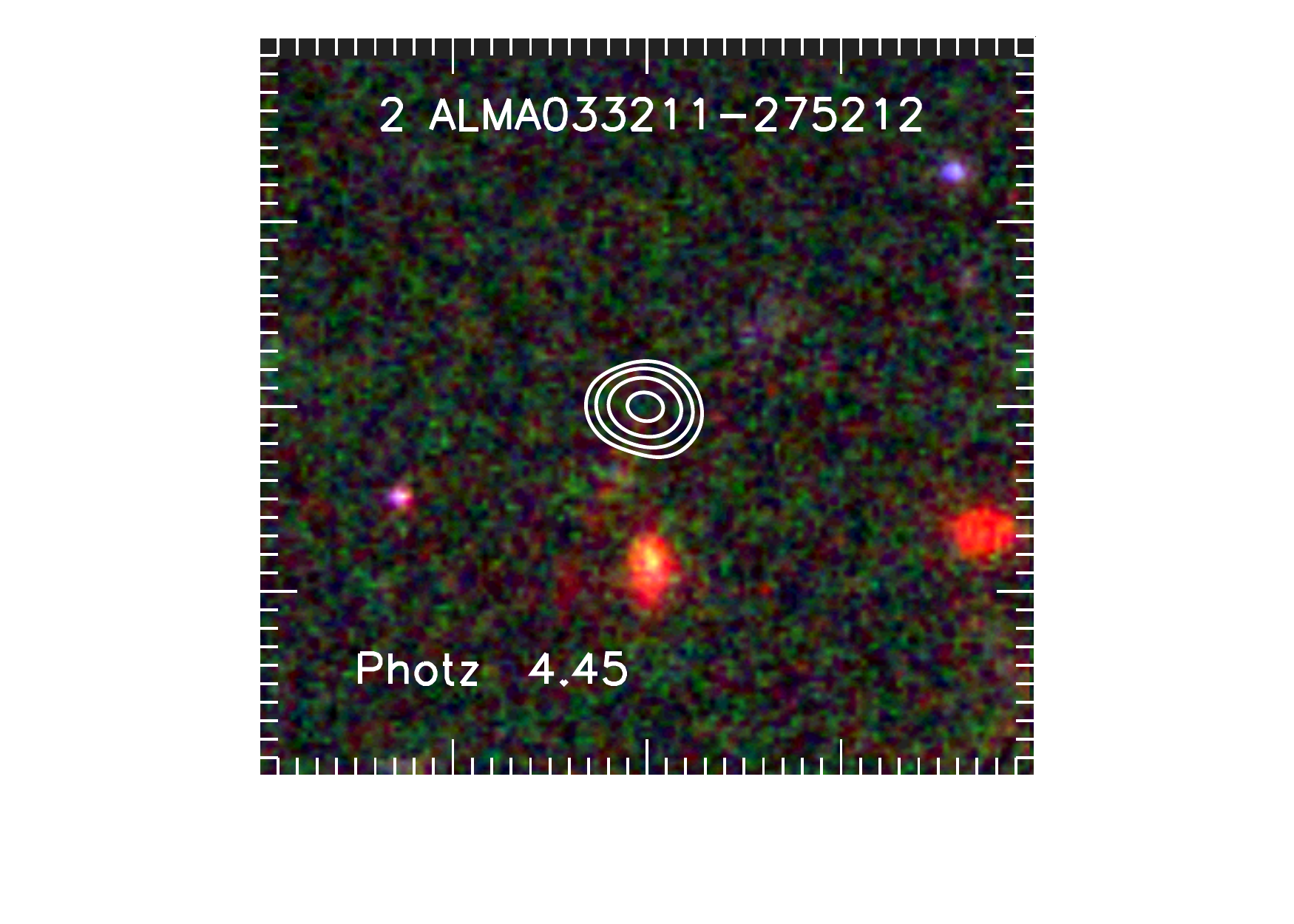}
\hspace{-3.3cm}\includegraphics[width=2.5in,angle=0]{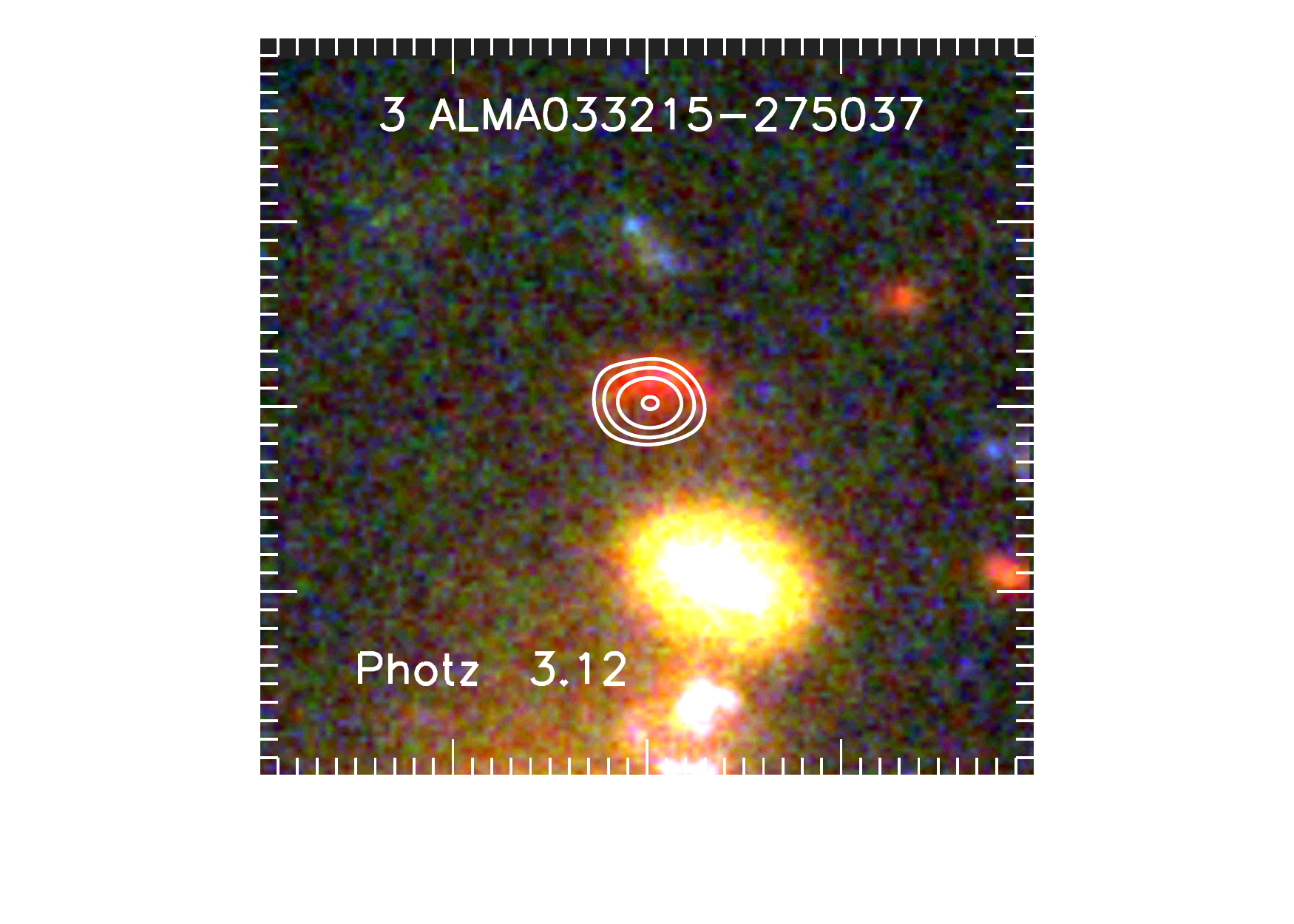}
\hspace{-3.3cm}\includegraphics[width=2.5in,angle=0]{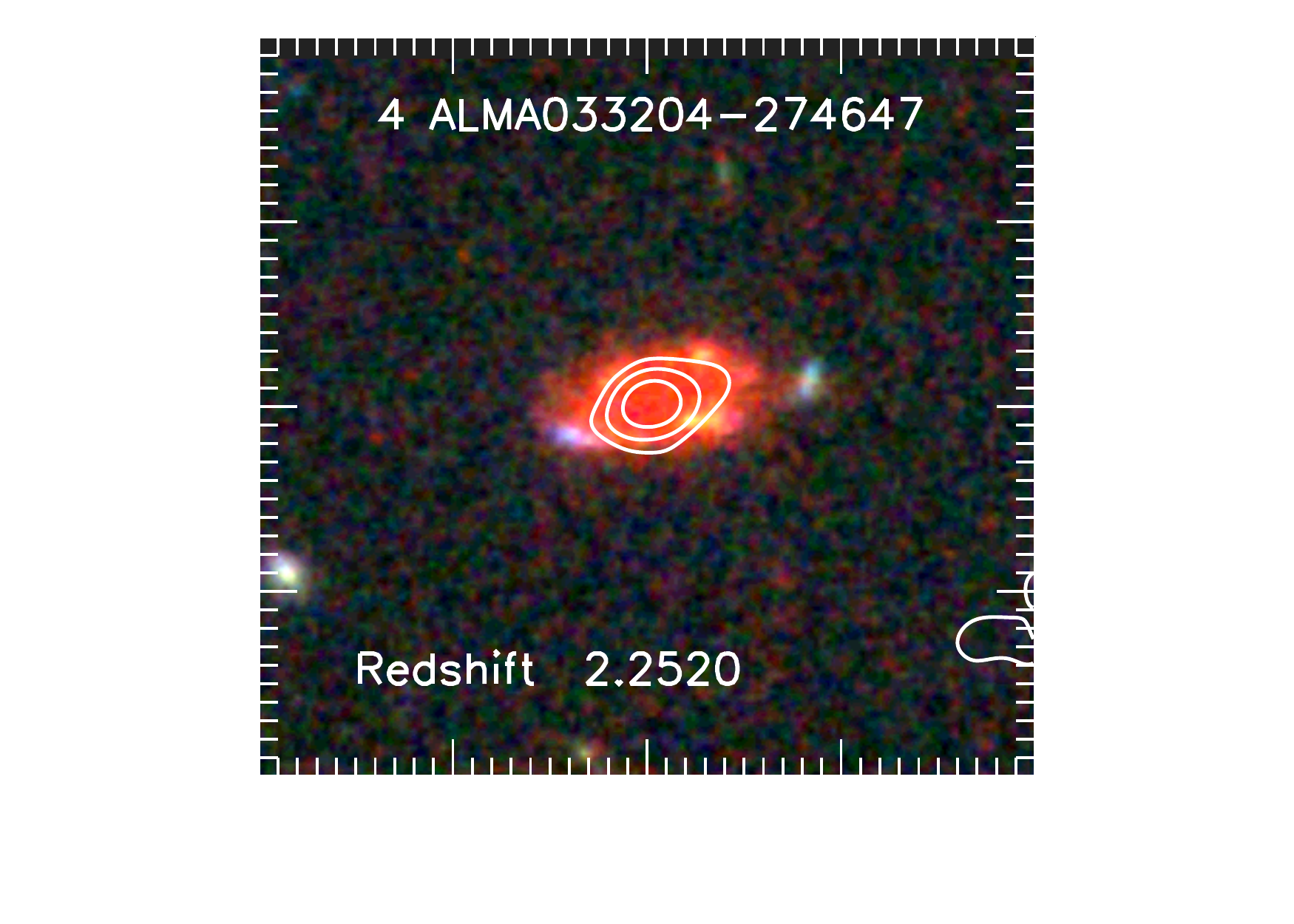}
\includegraphics[width=2.5in,angle=0]{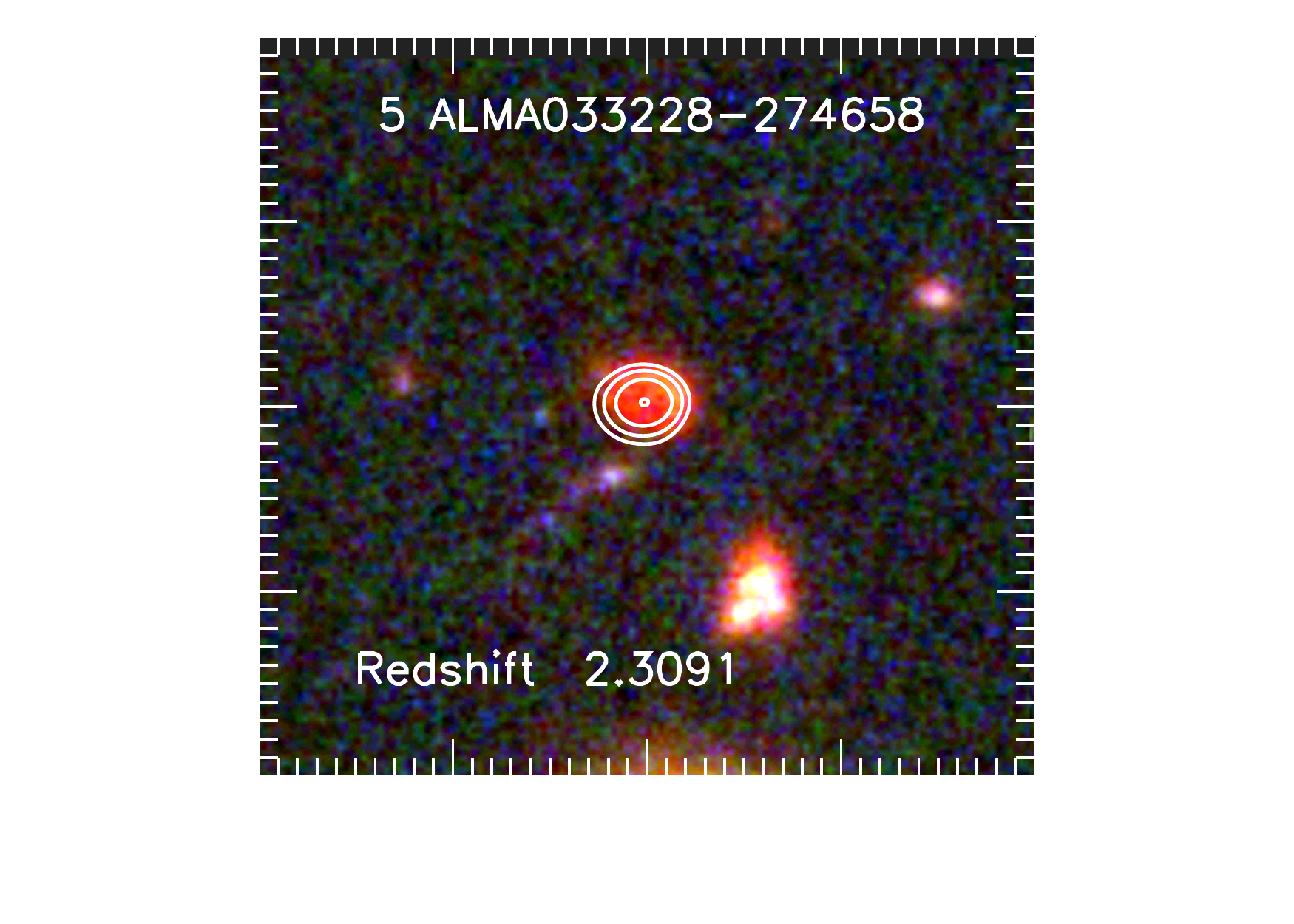}
\hspace{-3.3cm}\includegraphics[width=2.5in,angle=0]{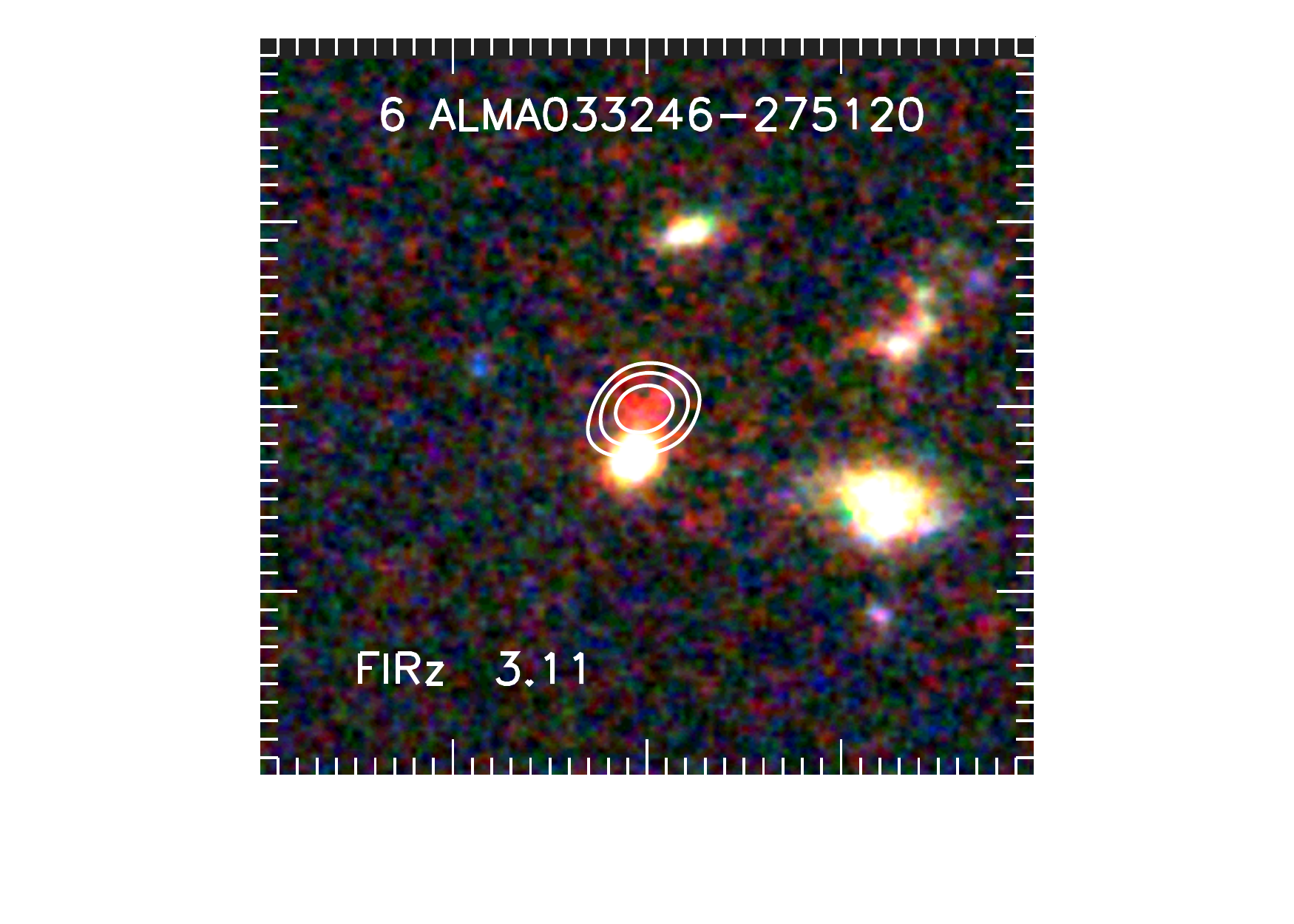}
\hspace{-3.3cm}\includegraphics[width=2.5in,angle=0]{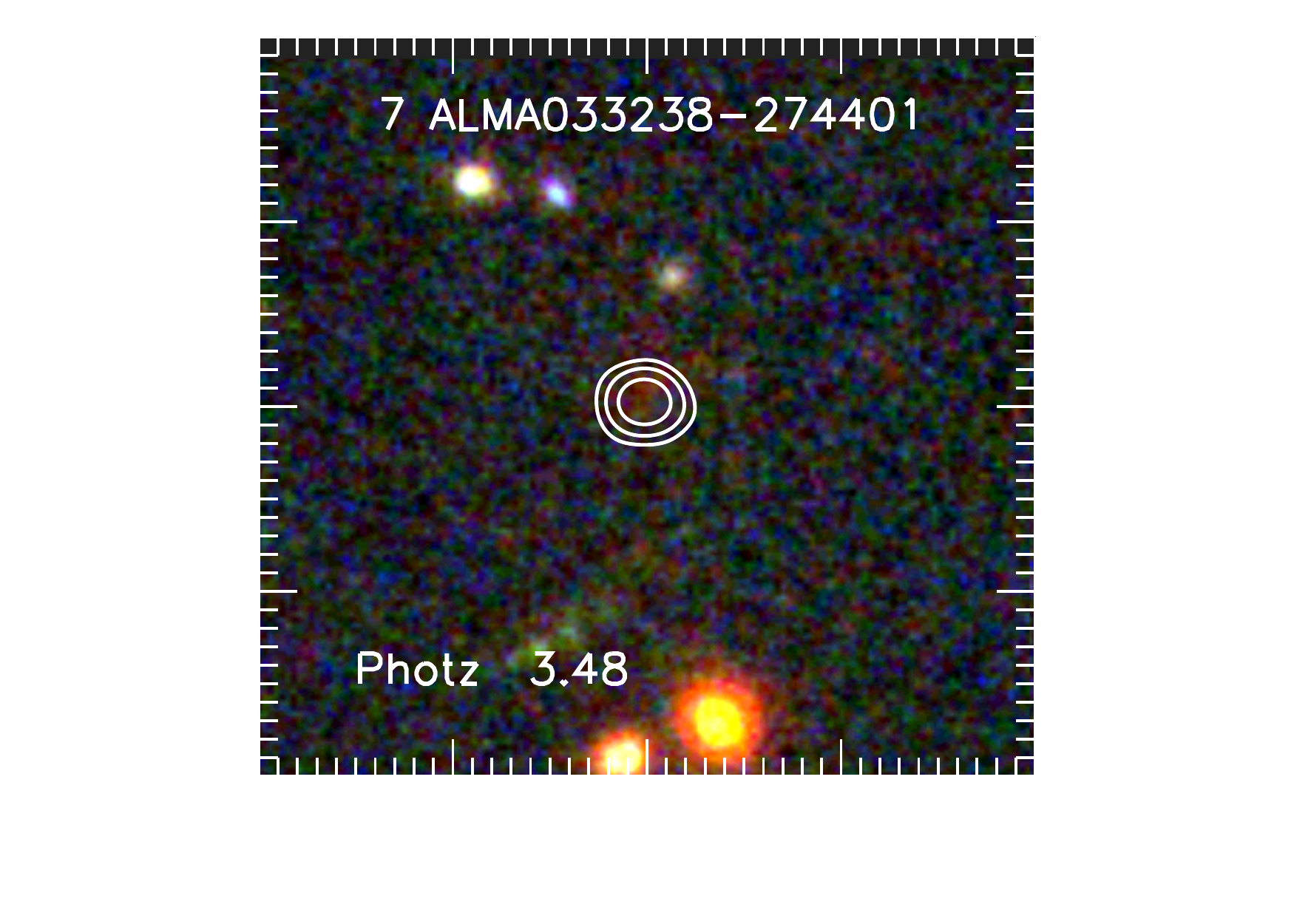}
\hspace{-3.3cm}\includegraphics[width=2.5in,angle=0]{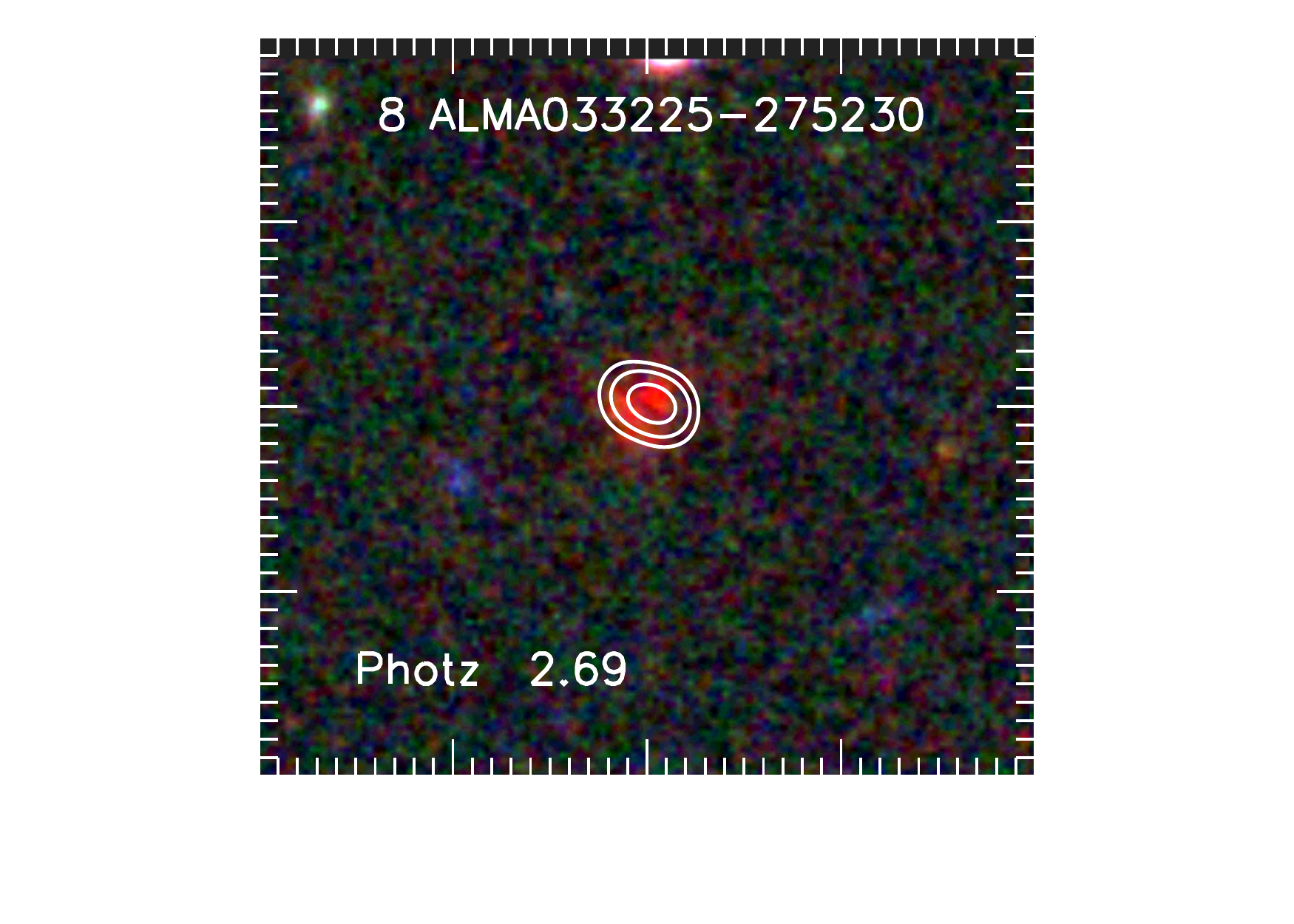}
\includegraphics[width=2.5in,angle=0]{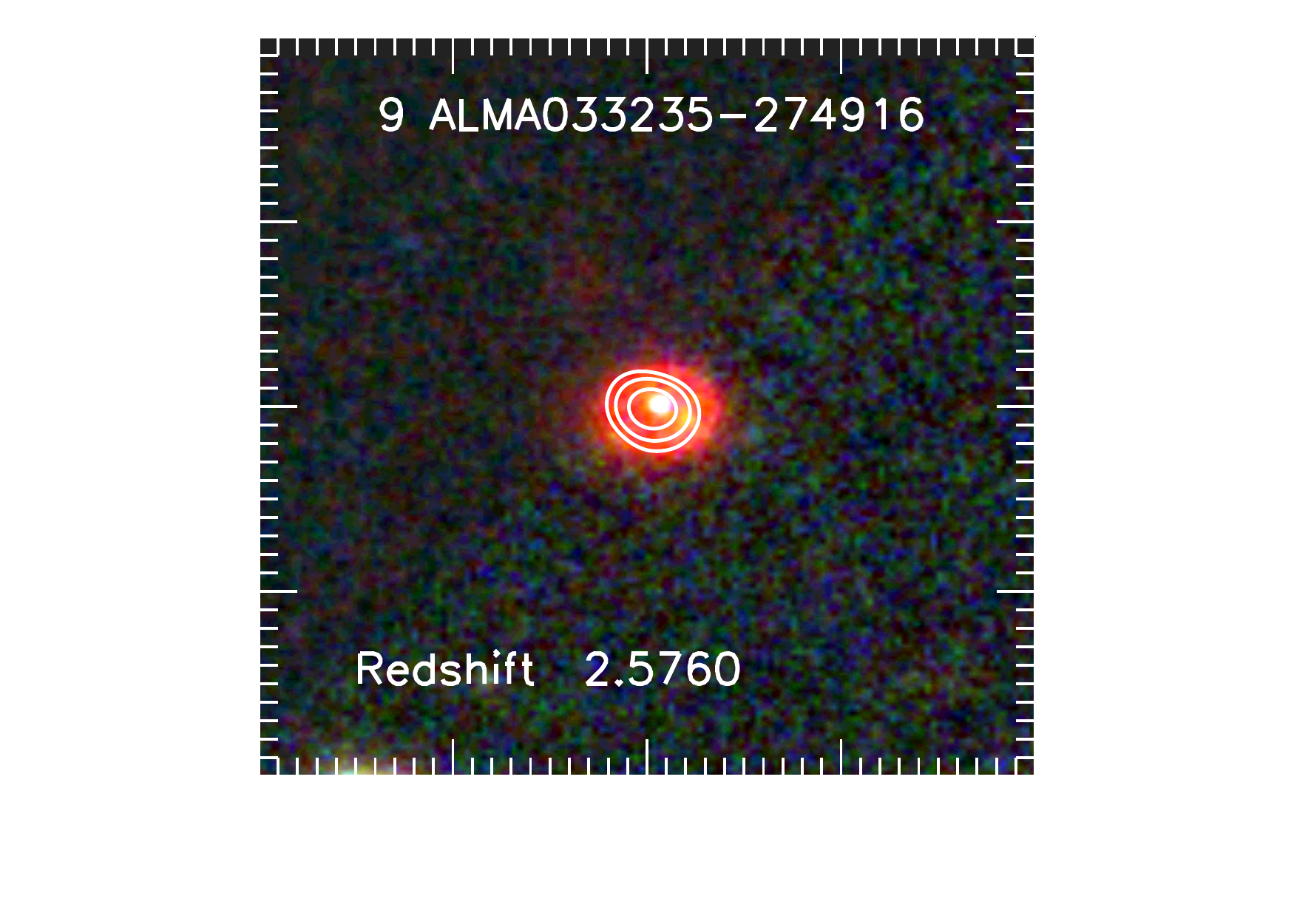}
\hspace{-3.3cm}\includegraphics[width=2.5in,angle=0]{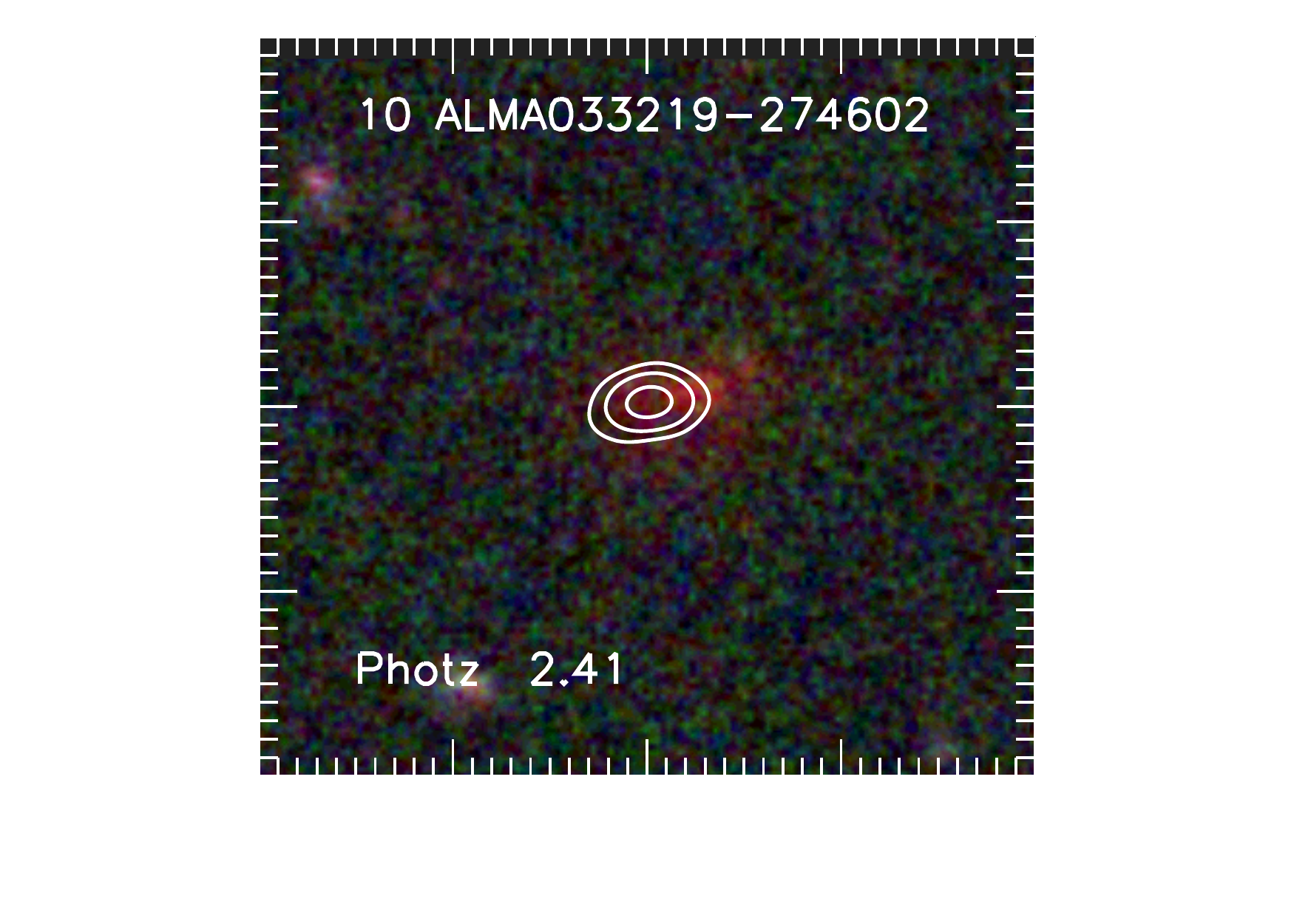}
\hspace{-3.3cm}\includegraphics[width=2.5in,angle=0]{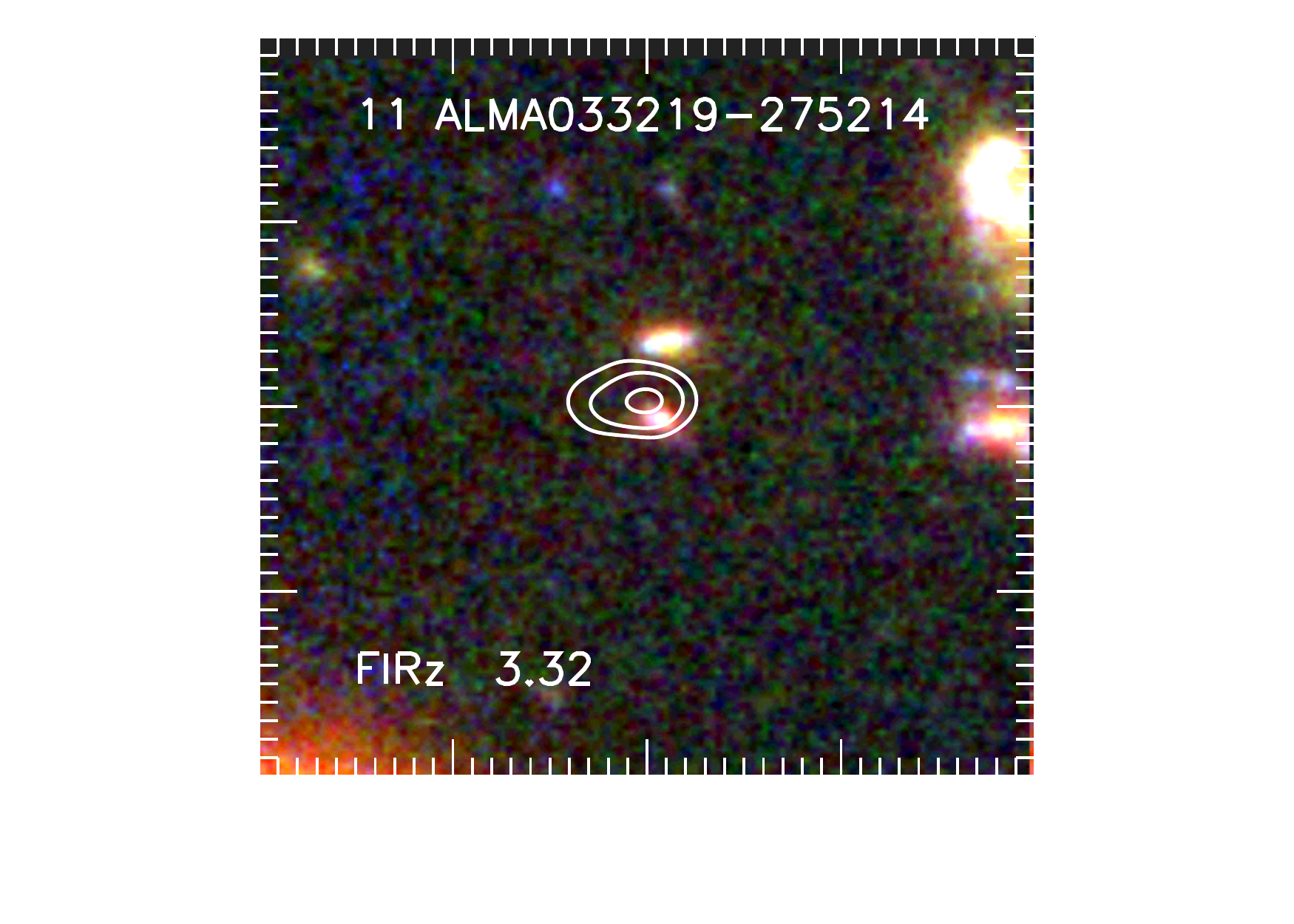}
\hspace{-3.3cm}\includegraphics[width=2.5in,angle=0]{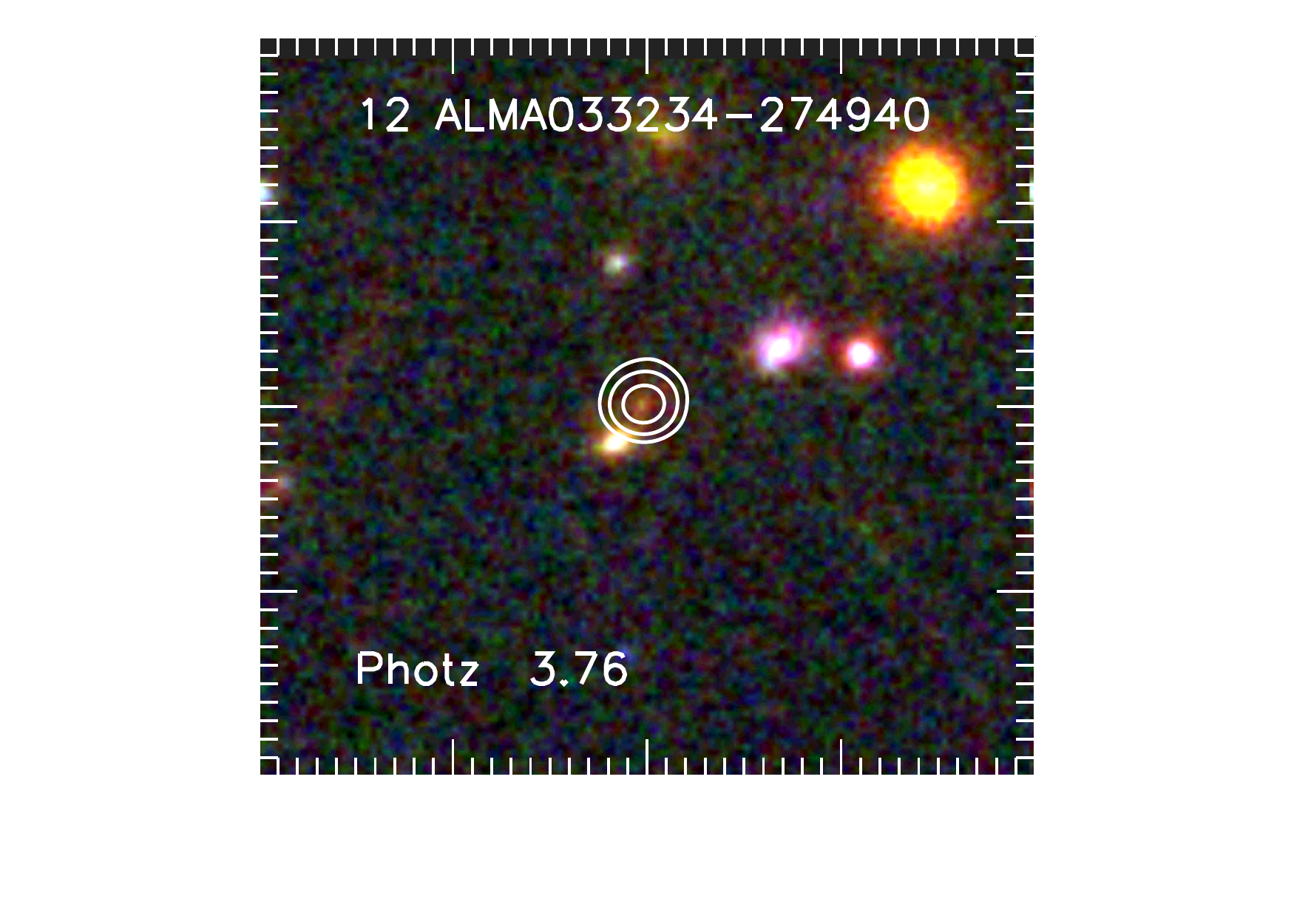}
\includegraphics[width=2.5in,angle=0]{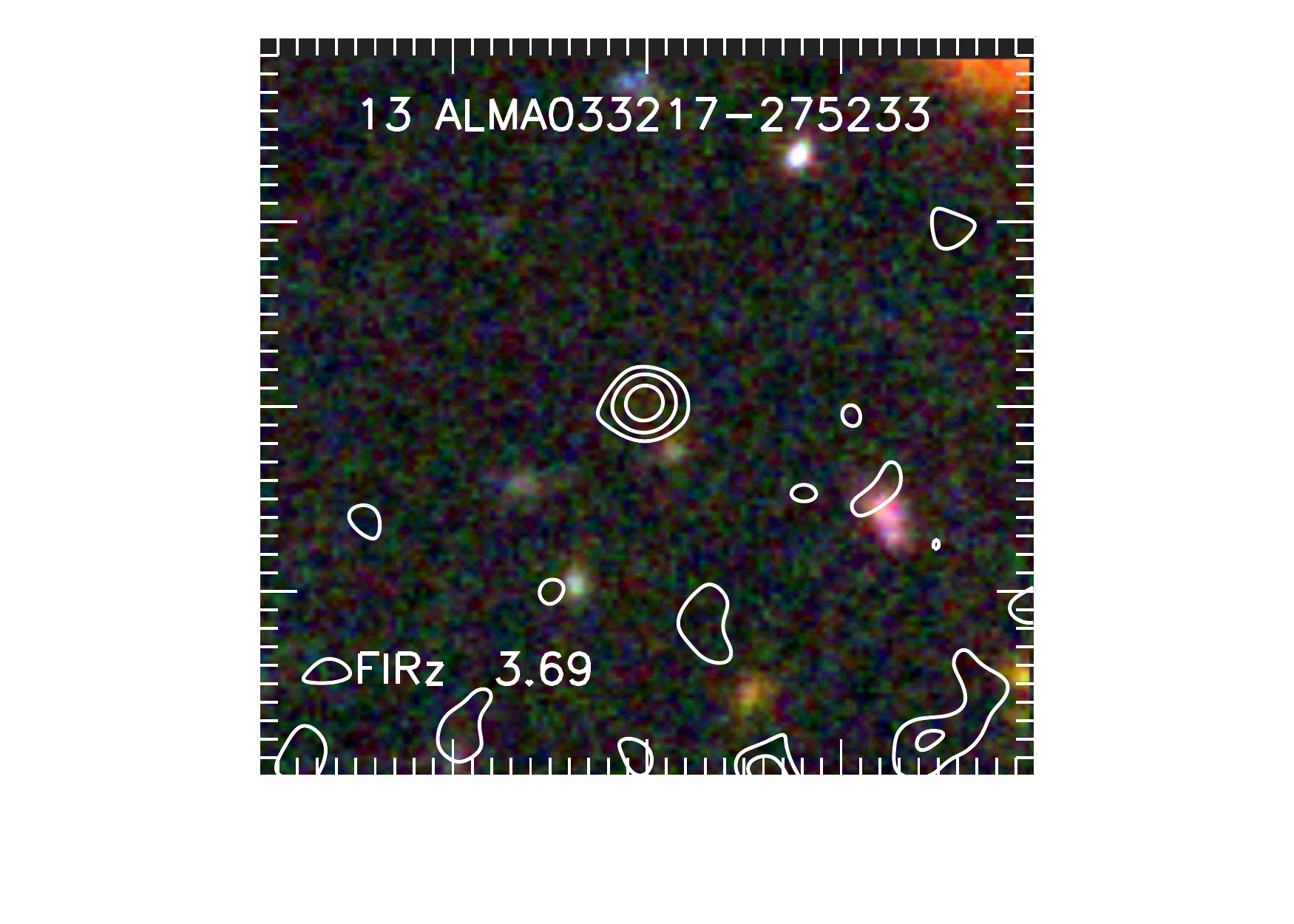}
\hspace{-3.3cm}\includegraphics[width=2.5in,angle=0]{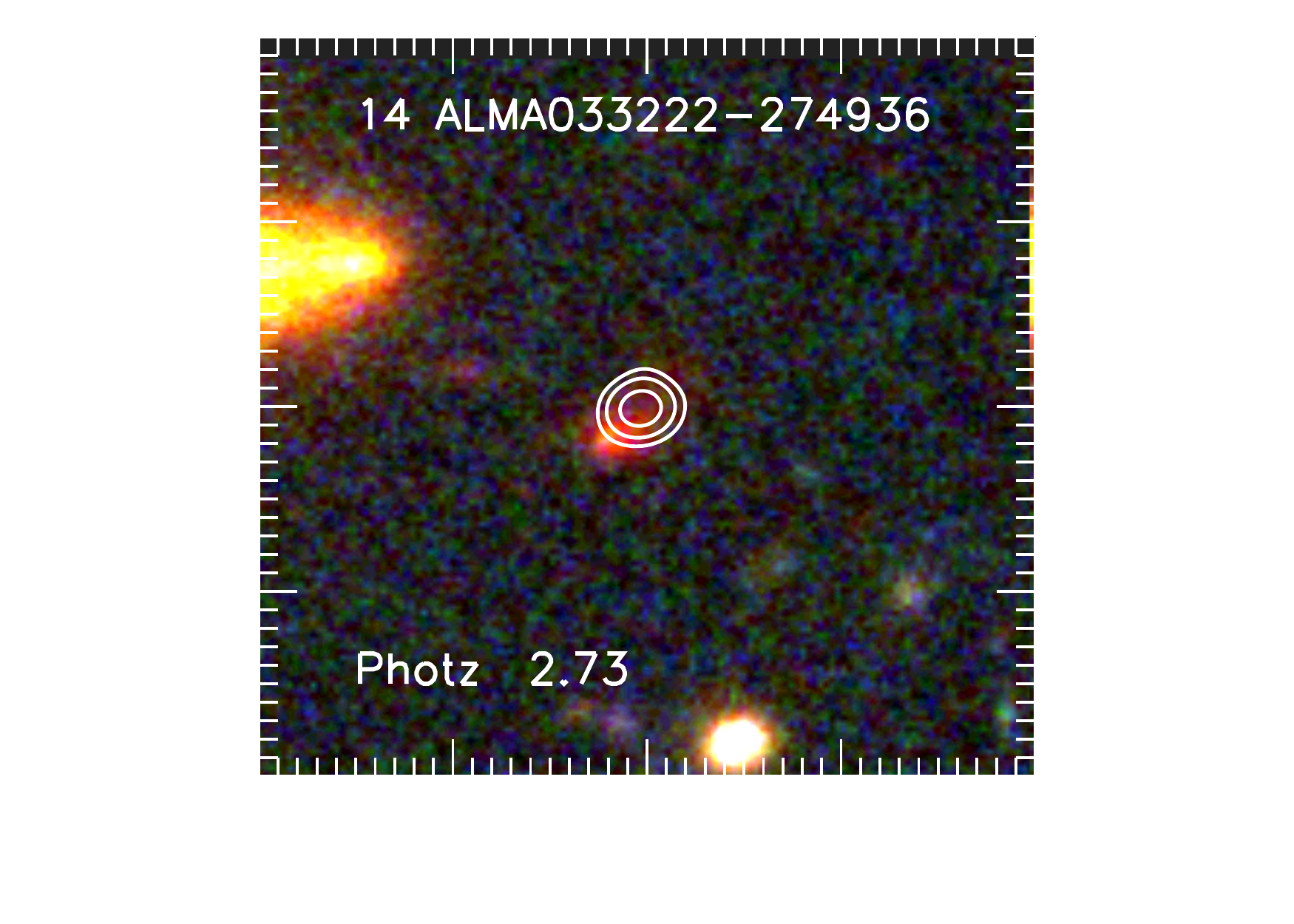}
\hspace{-3.3cm}\includegraphics[width=2.5in,angle=0]{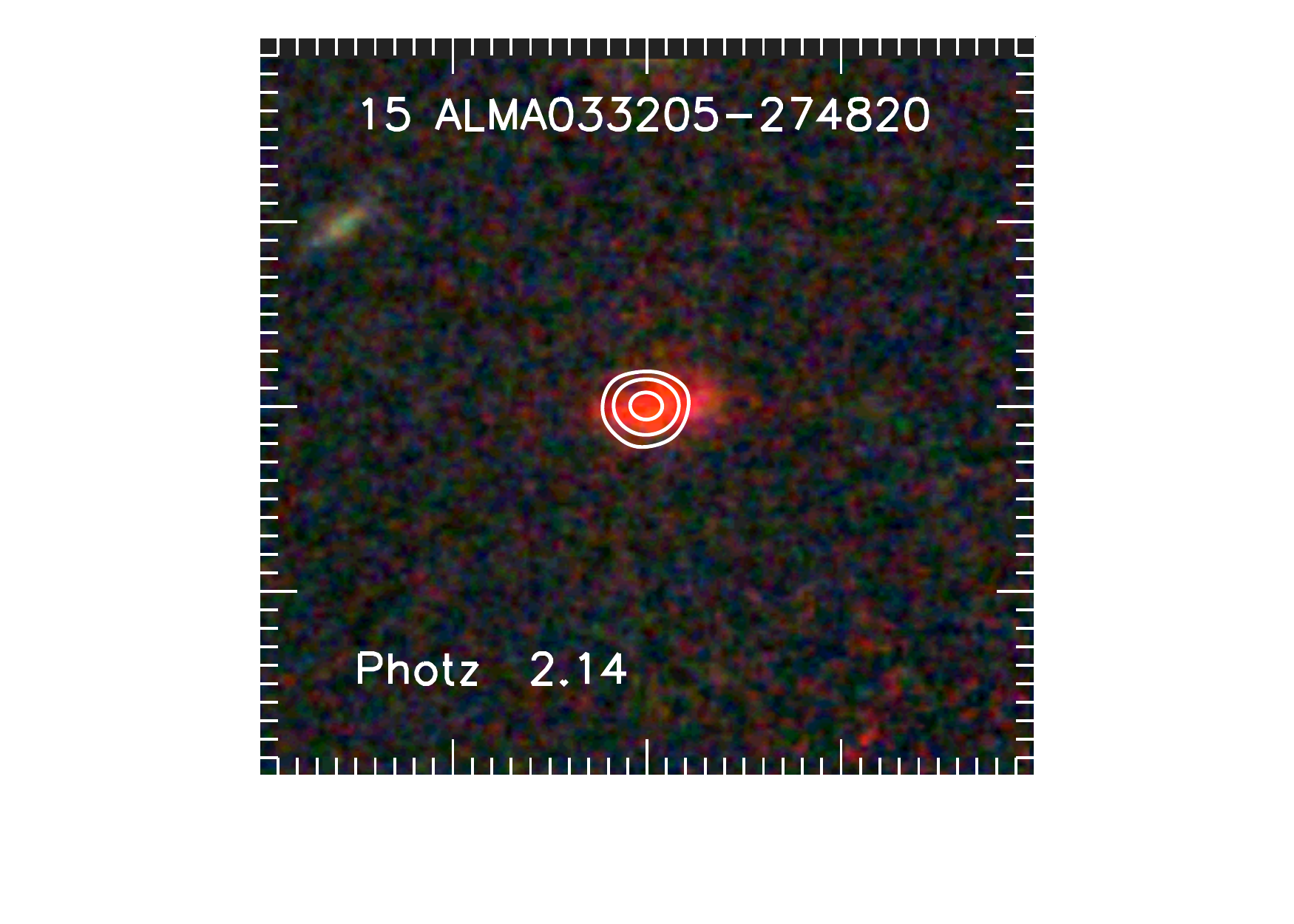}
\hspace{-3.3cm}\includegraphics[width=2.5in,angle=0]{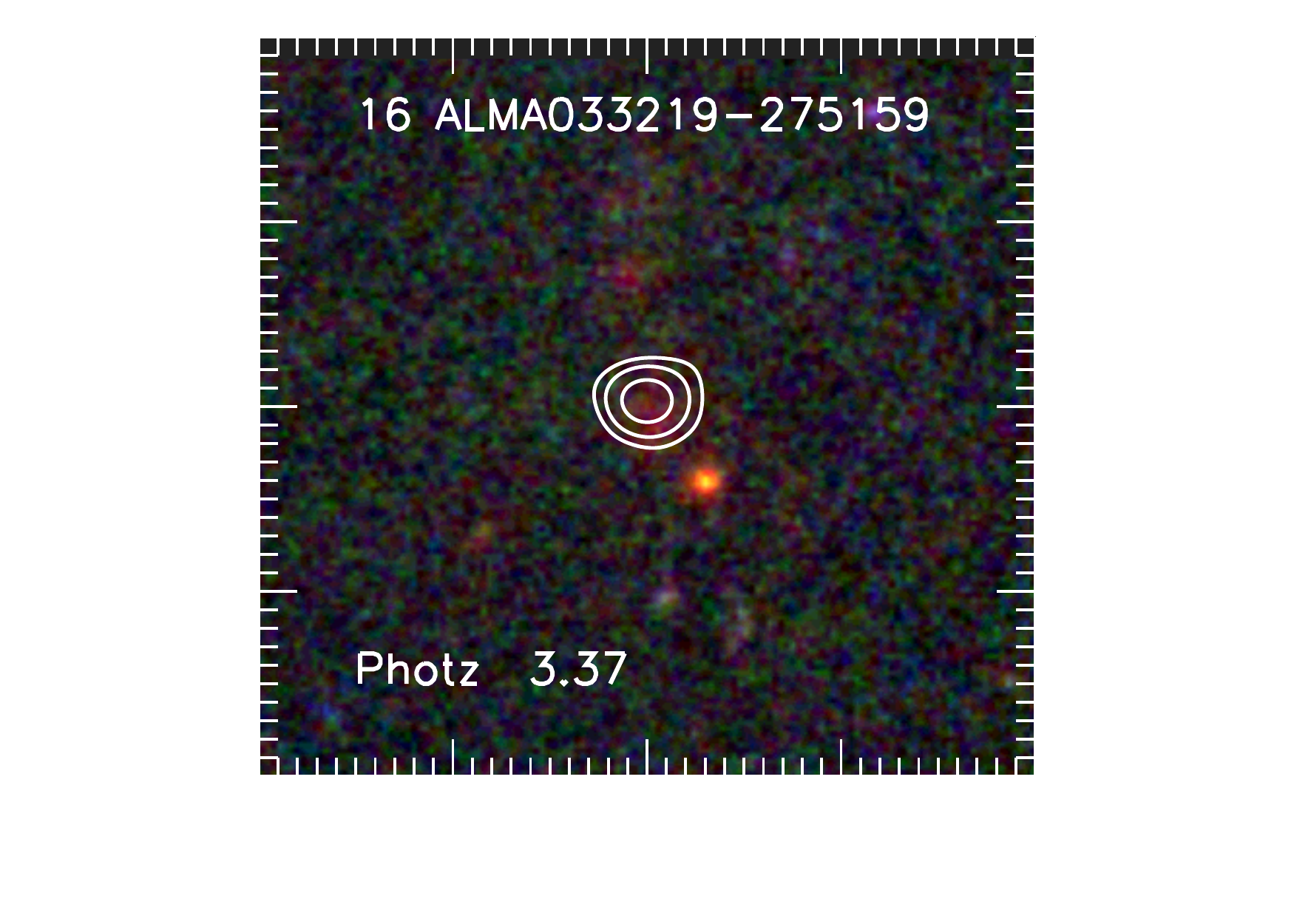}
\includegraphics[width=2.5in,angle=0]{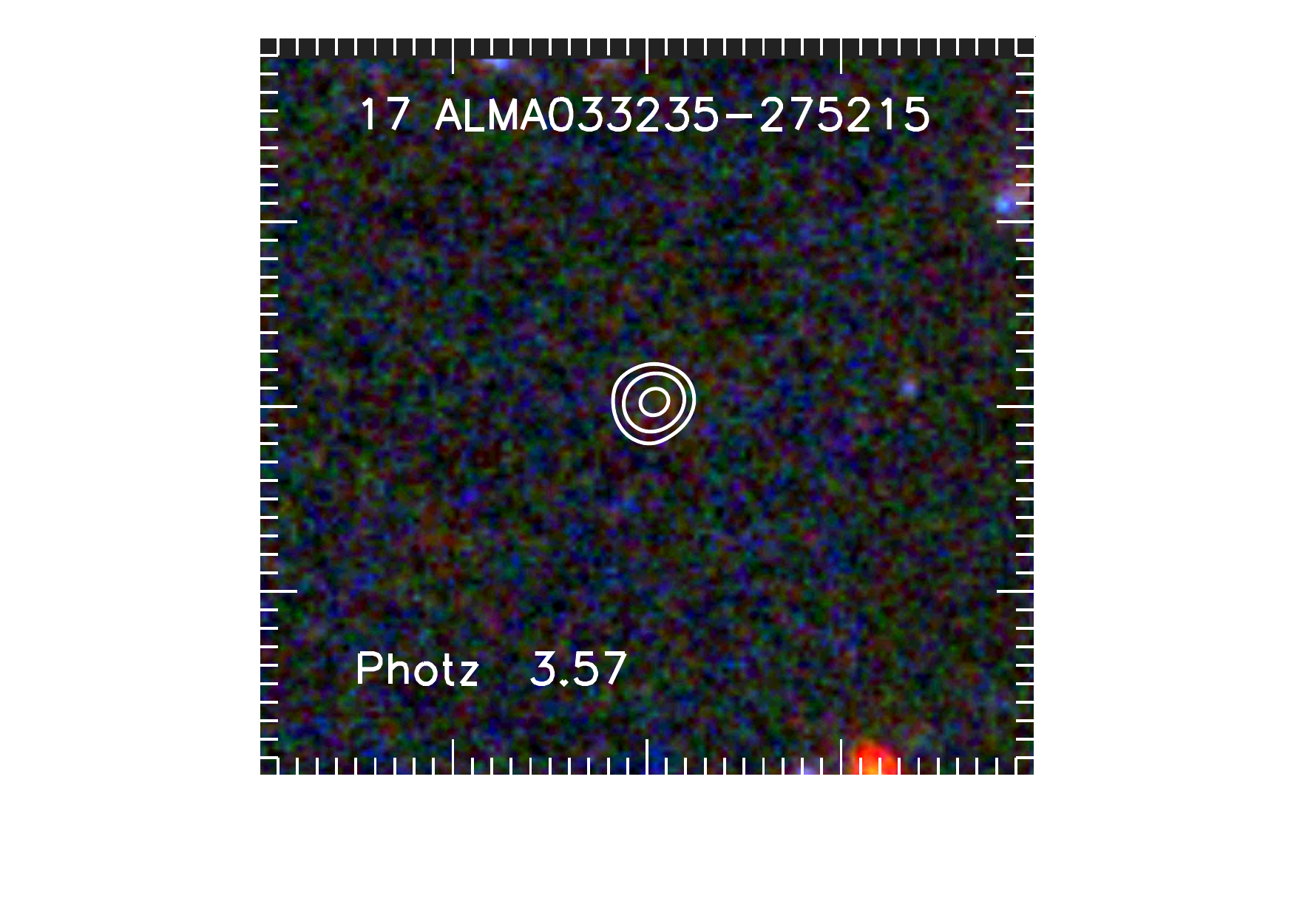}
\hspace{-2.55cm}\includegraphics[width=2.5in,angle=0]{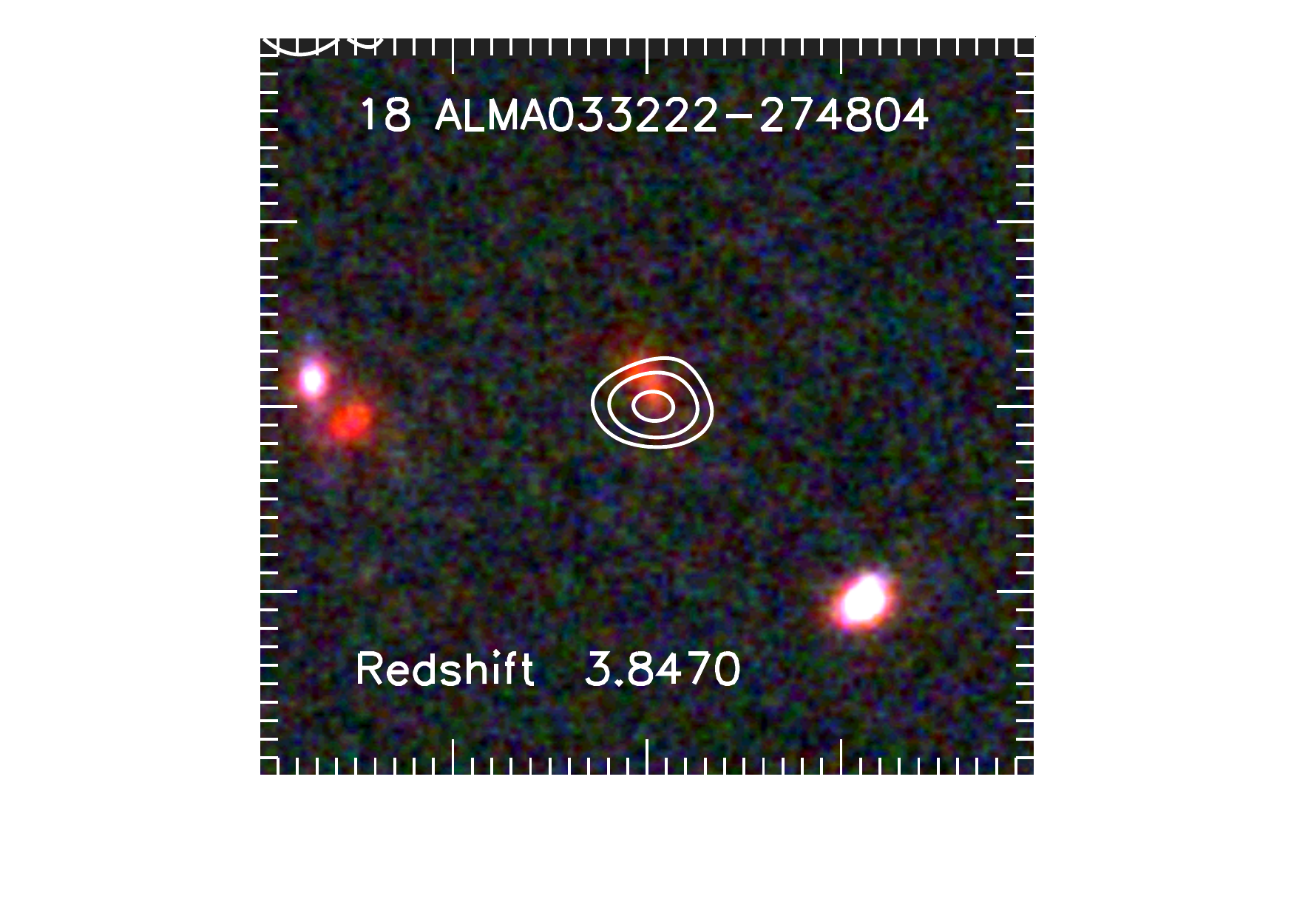}
\hspace{-2.55cm}\includegraphics[width=2.5in,angle=0]{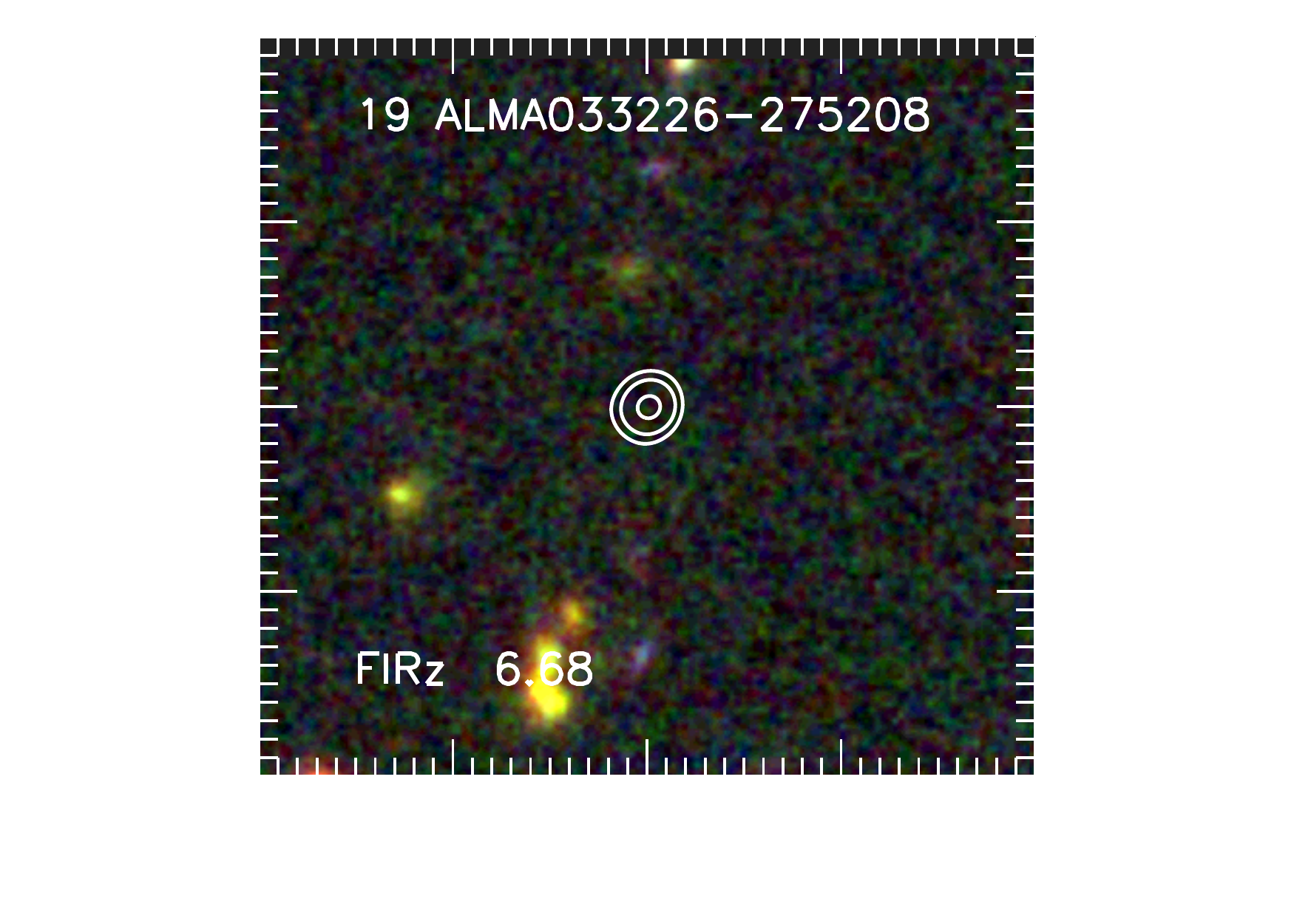}
\hspace{-2.55cm}\includegraphics[width=2.5in,angle=0]{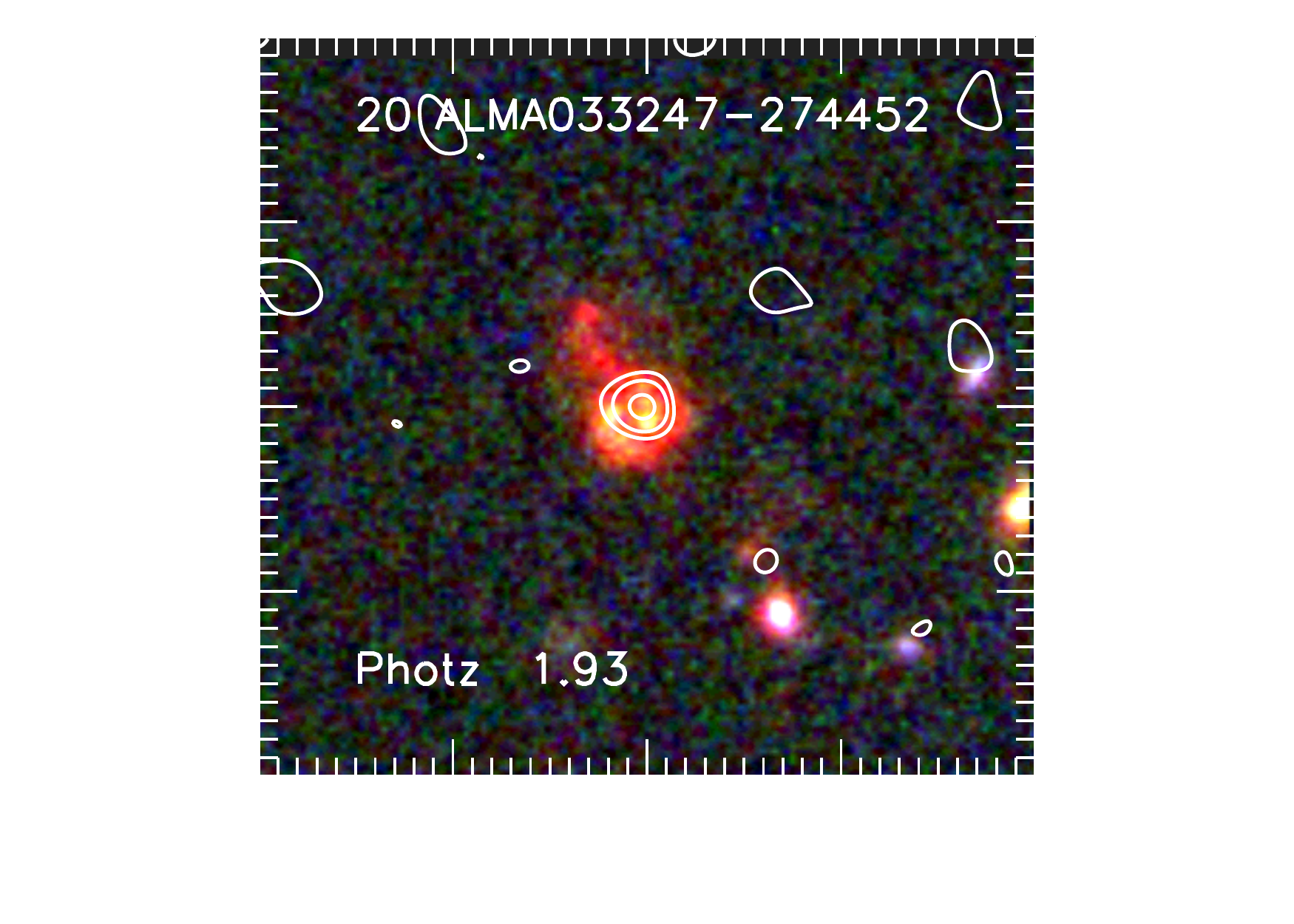}
\vskip -0.75cm
\caption{Three color {\em HST\/} images (blue = F450W, green = F814W, 
and red = F160W) for the ALMA sources in Table~4.
The thumbnails are $8''$ on a side, or roughly 70~kpc at $z=2$.
The ALMA continuum emission is  shown with white contours using the
$0\farcs5$ FWHM tapered images contoured at 0.6, 1.2, 2.4 and 4.8~mJy/beam.
The numbers refer to Table~4, and the redshifts are marked as spectroscopic
(Redshift), standard photometric (Photz), or FIR photometric (FIRz). Sources with
none of these are marked with ``Photz ?''. The redshift given was chosen
in the order of first specz, then $Q<3$ photz from S16,
and finally FIRz when there was no available specz or photz.
\label{basic_alma_images}}
\end{figure*}
\begin{figure*}
\setcounter{figure}{9}
\includegraphics[width=2.5in,angle=0]{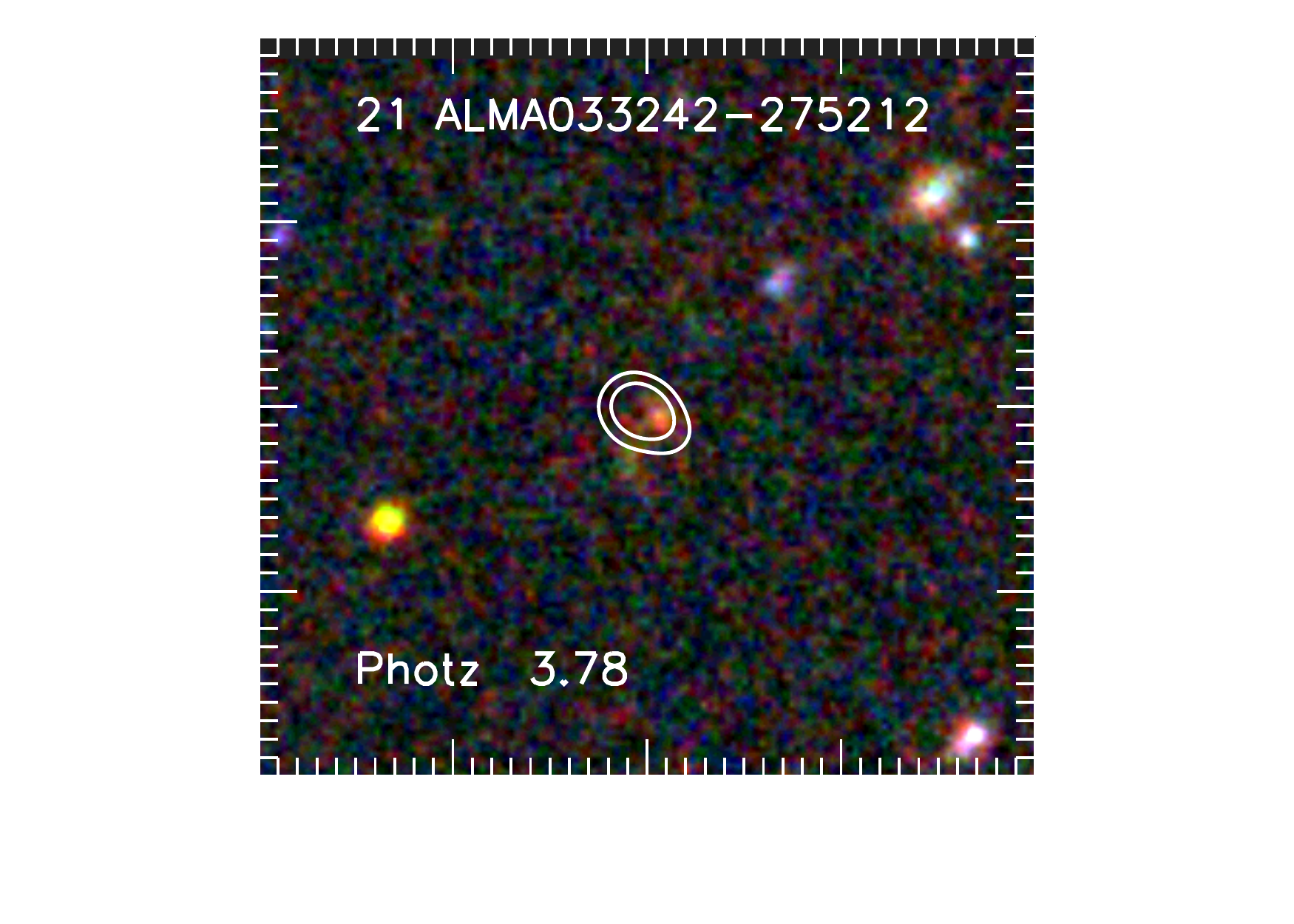}
\hspace{-3.3cm}\includegraphics[width=2.5in,angle=0]{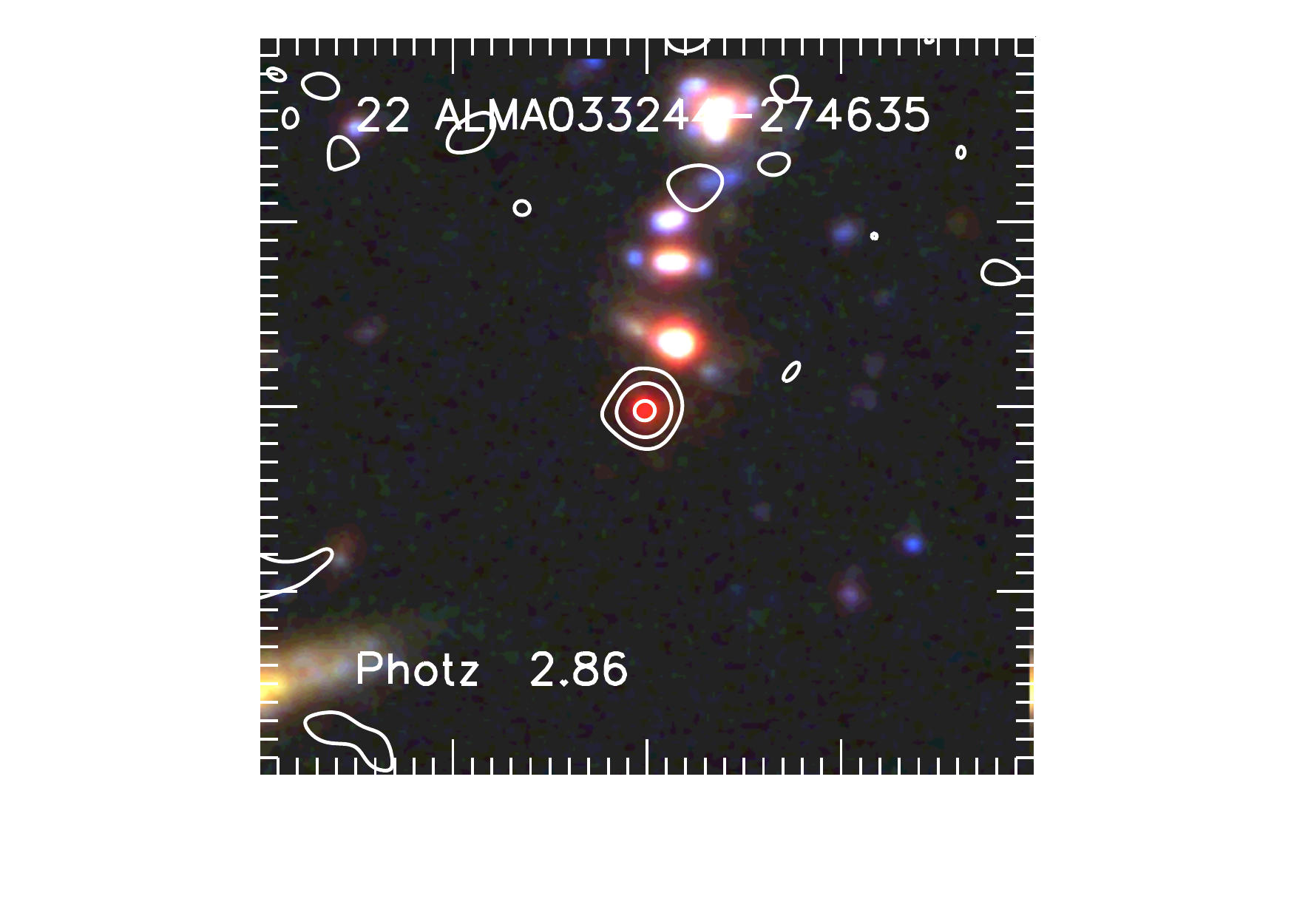}
\hspace{-3.3cm}\includegraphics[width=2.5in,angle=0]{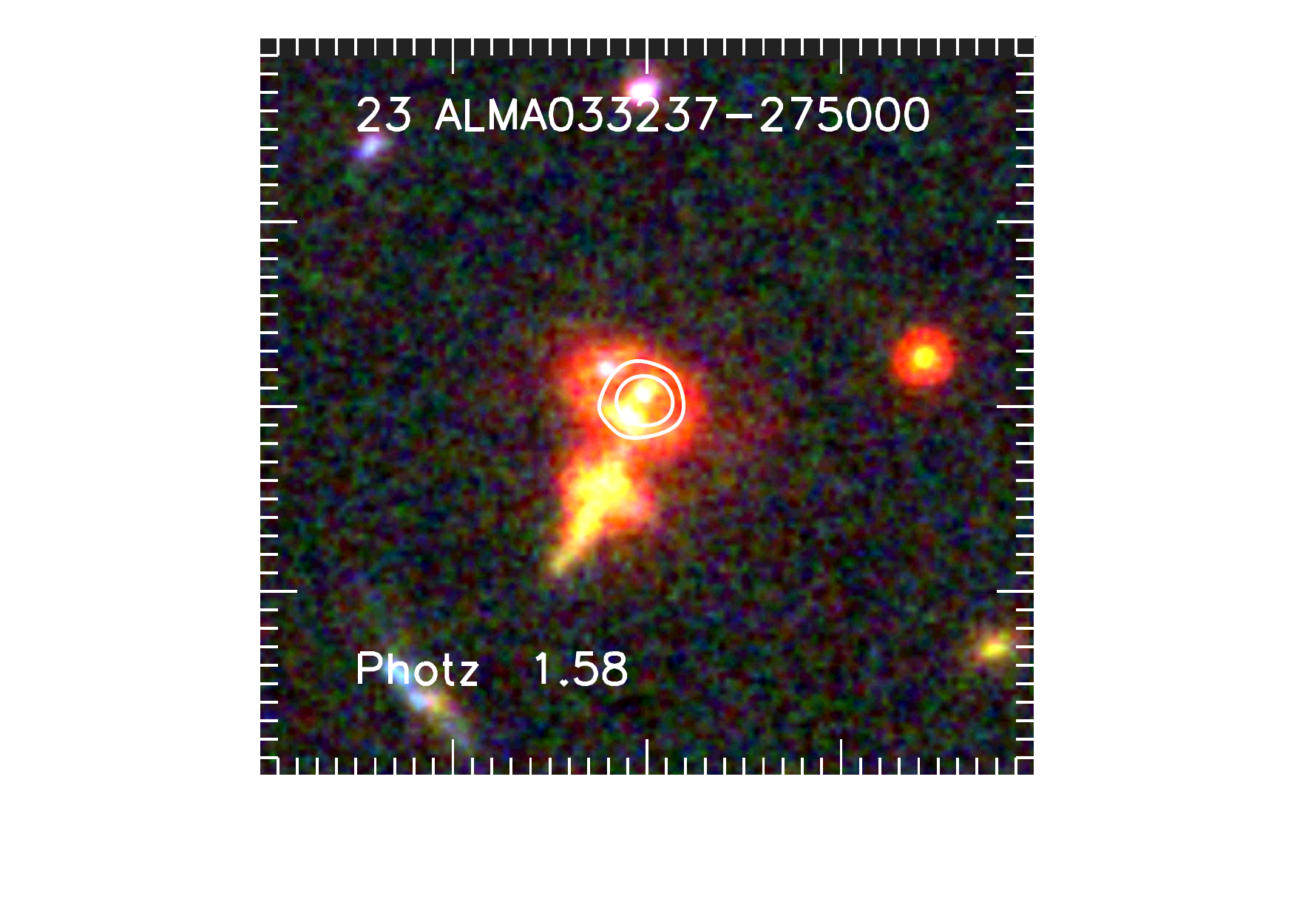}
\hspace{-3.3cm}\includegraphics[width=2.5in,angle=0]{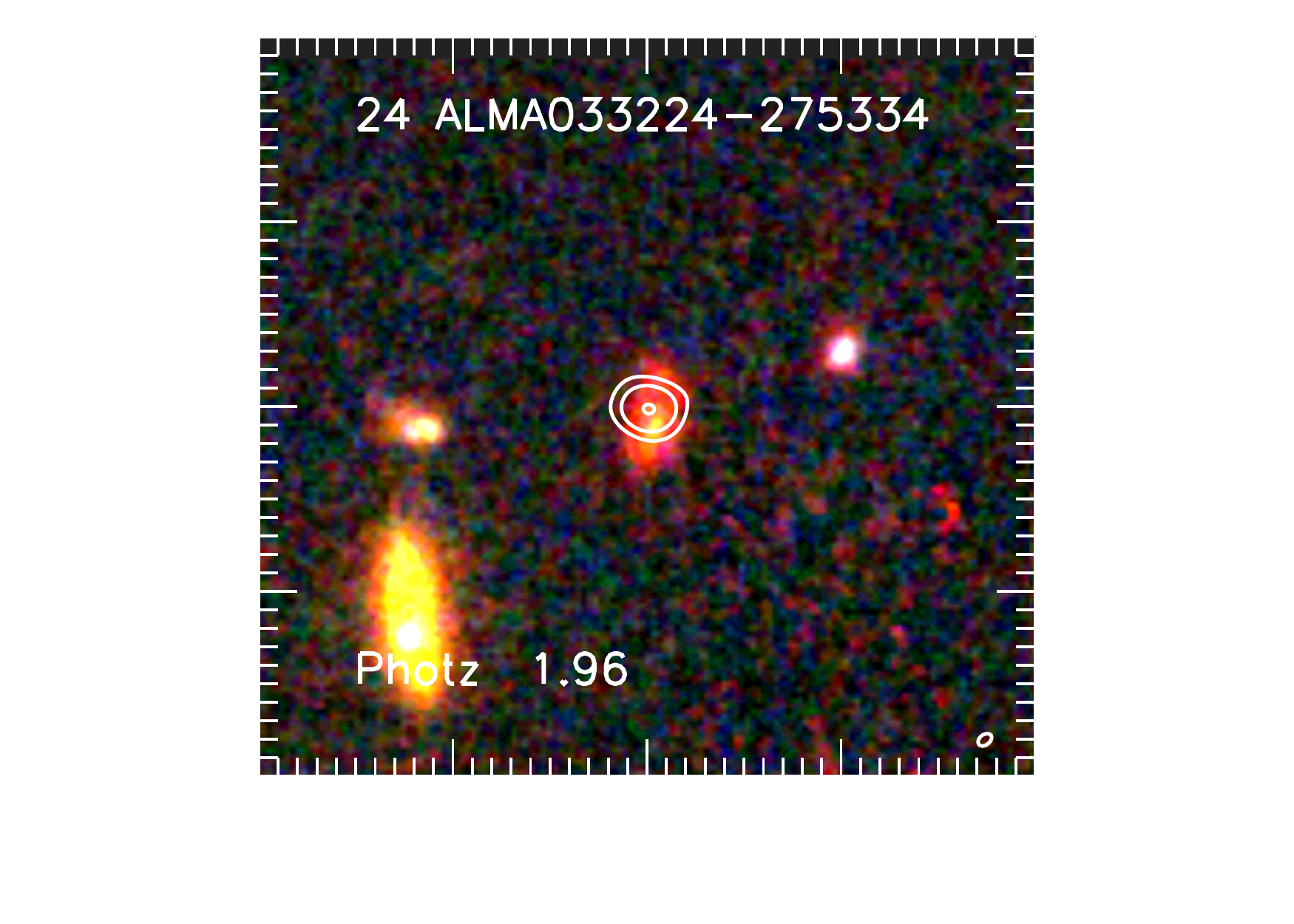}
\includegraphics[width=2.5in,angle=0]{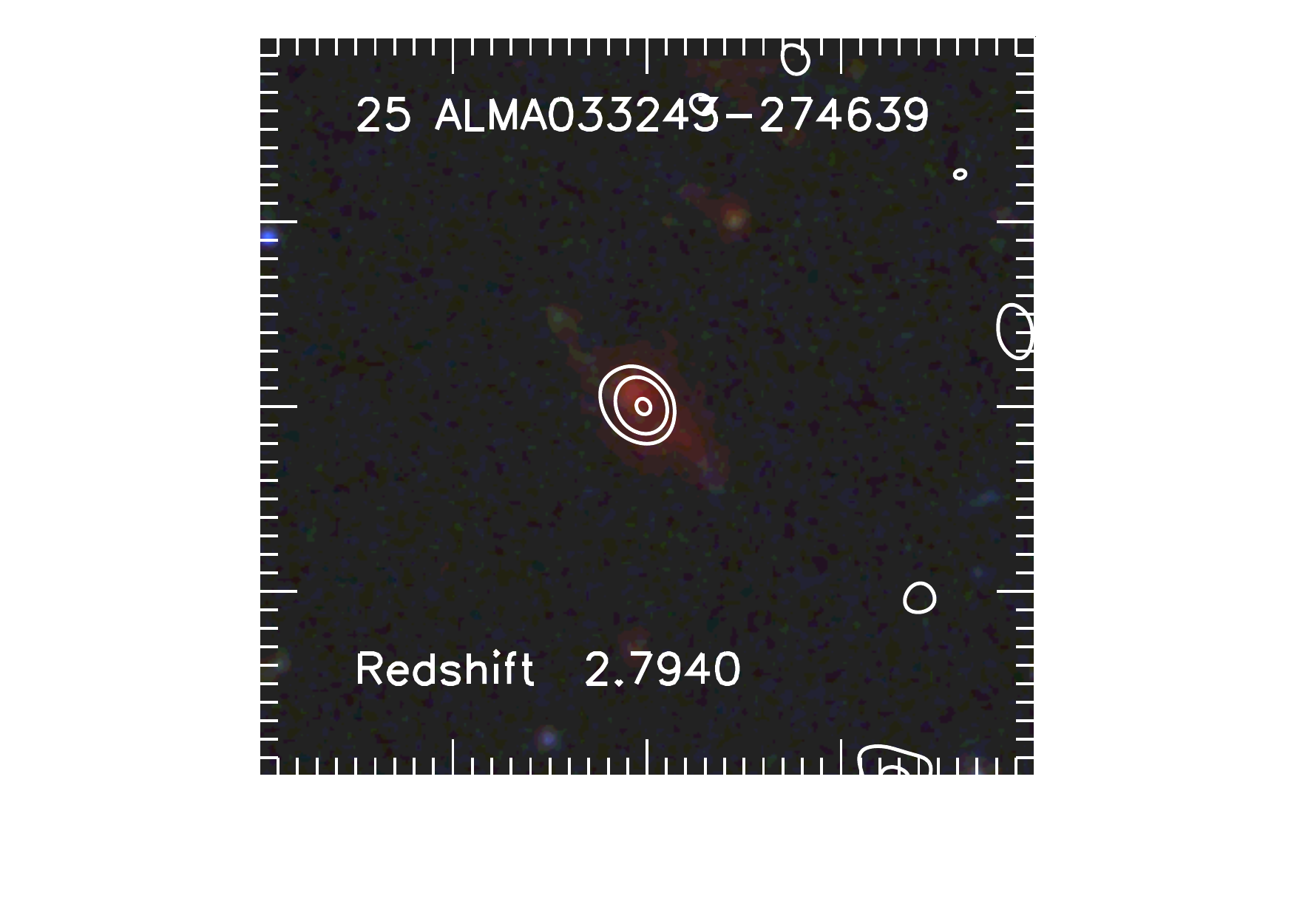}
\hspace{-3.3cm}\includegraphics[width=2.5in,angle=0]{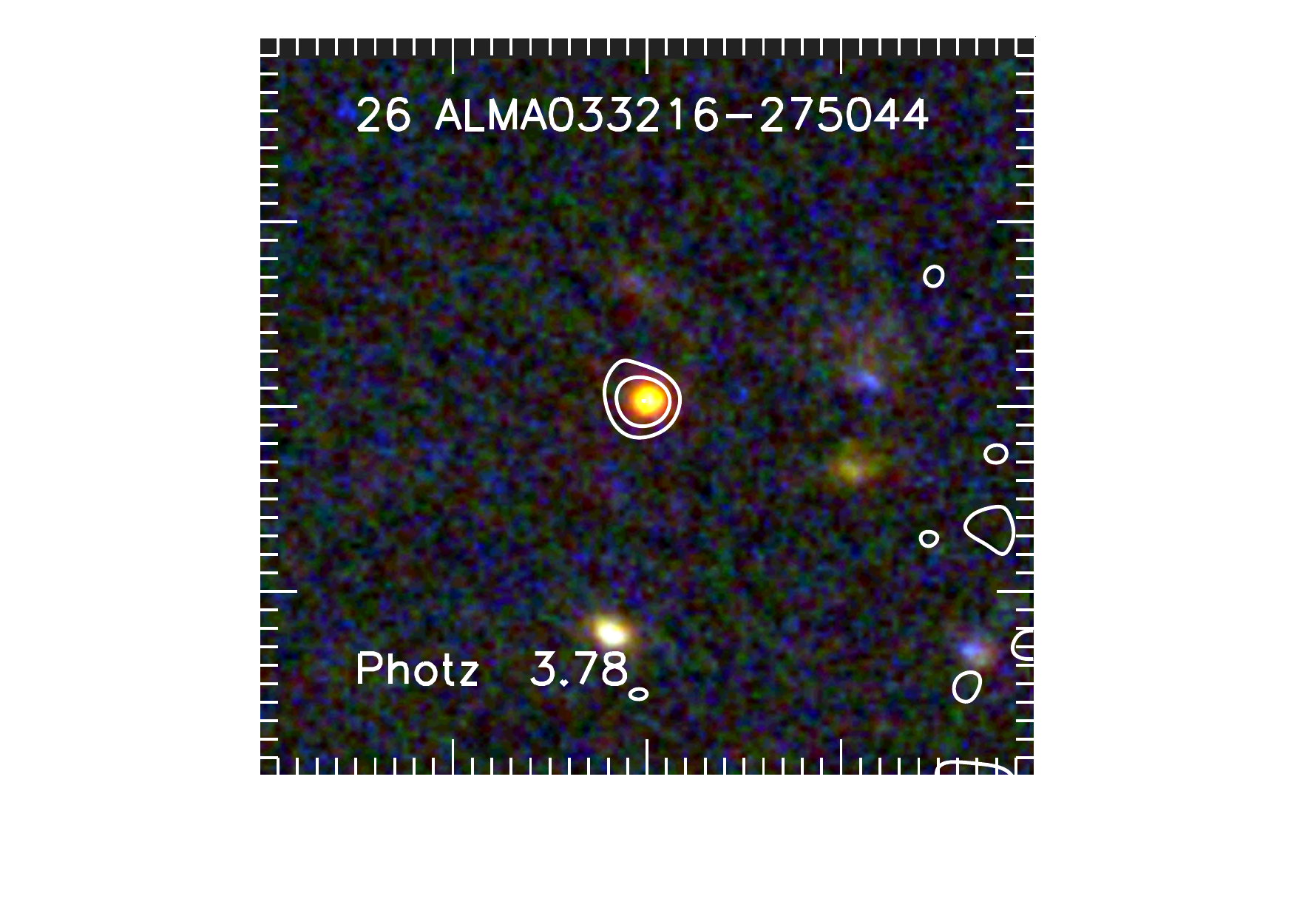}
\hspace{-3.3cm}\includegraphics[width=2.5in,angle=0]{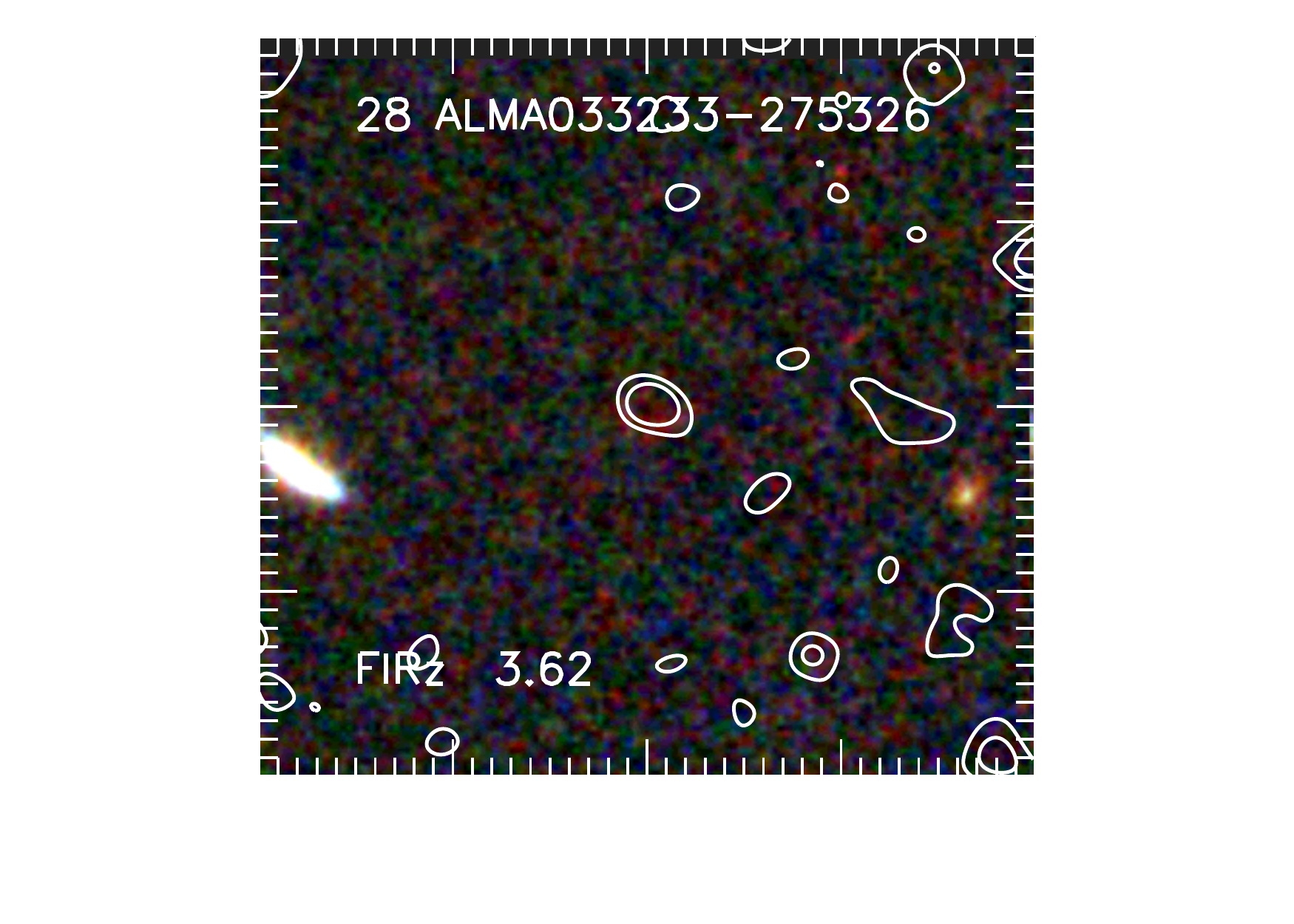}
\hspace{-3.3cm}\includegraphics[width=2.5in,angle=0]{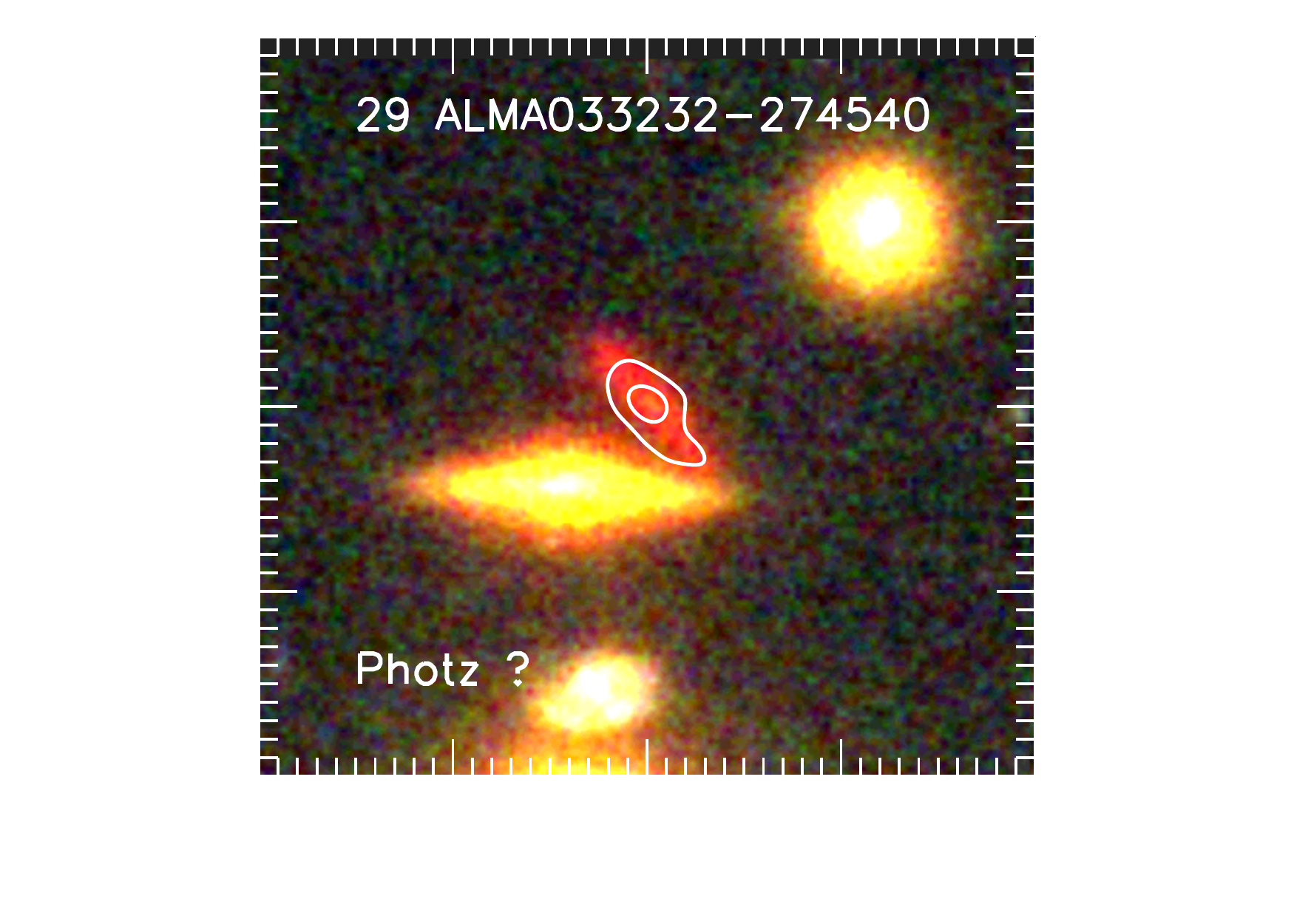}
\includegraphics[width=2.5in,angle=0]{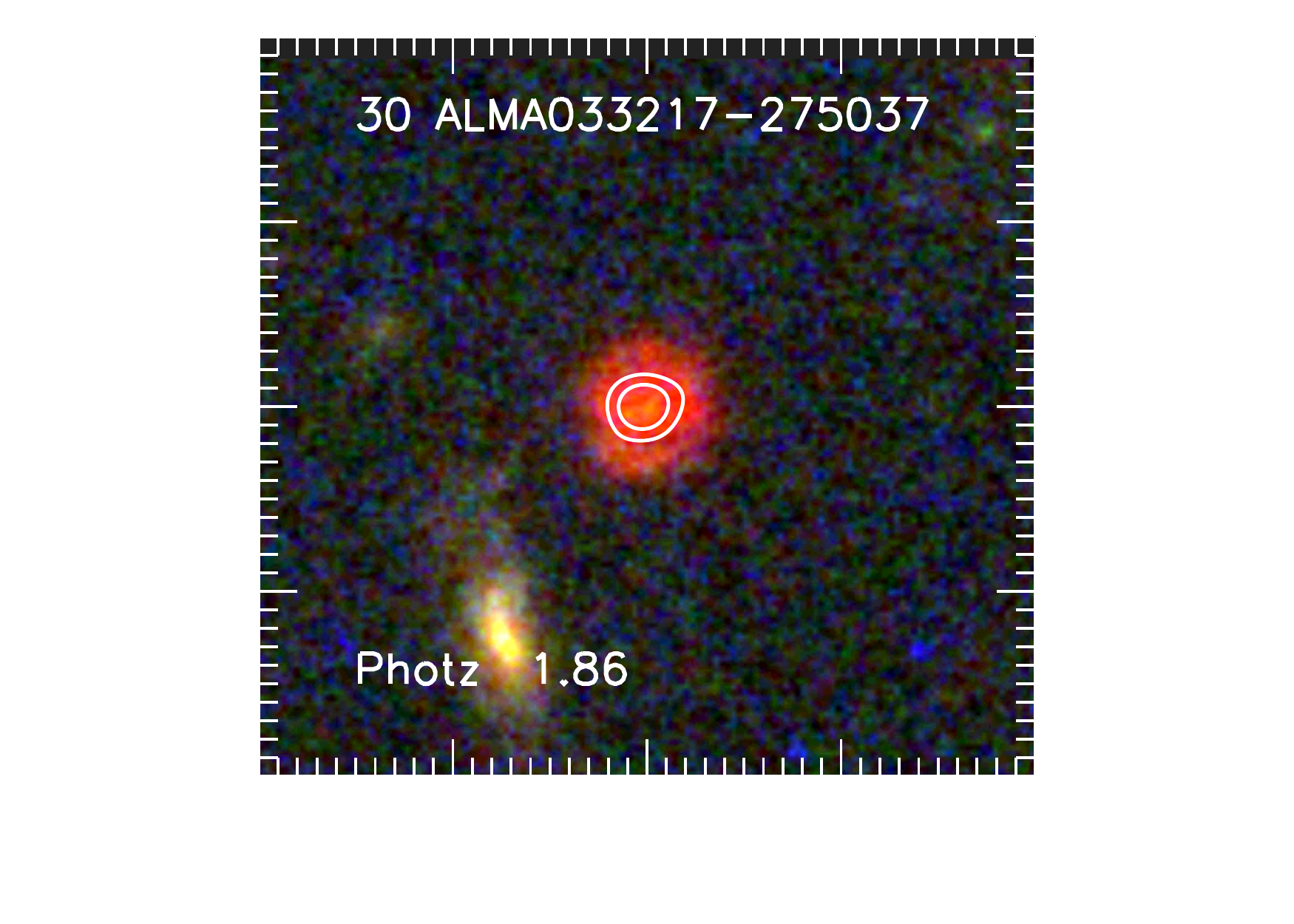}
\hspace{-3.3cm}\includegraphics[width=2.5in,angle=0]{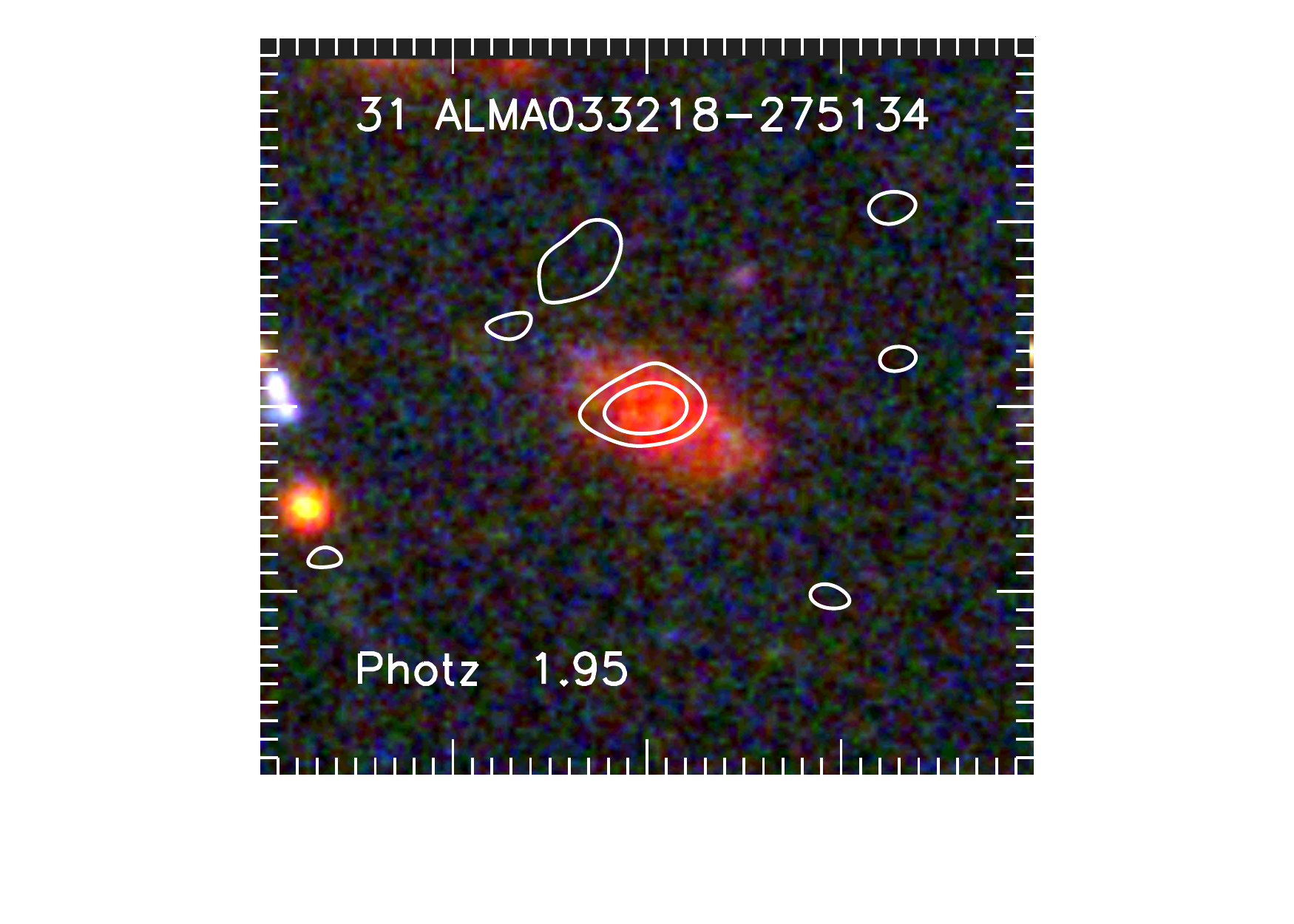}
\hspace{-3.3cm}\includegraphics[width=2.5in,angle=0]{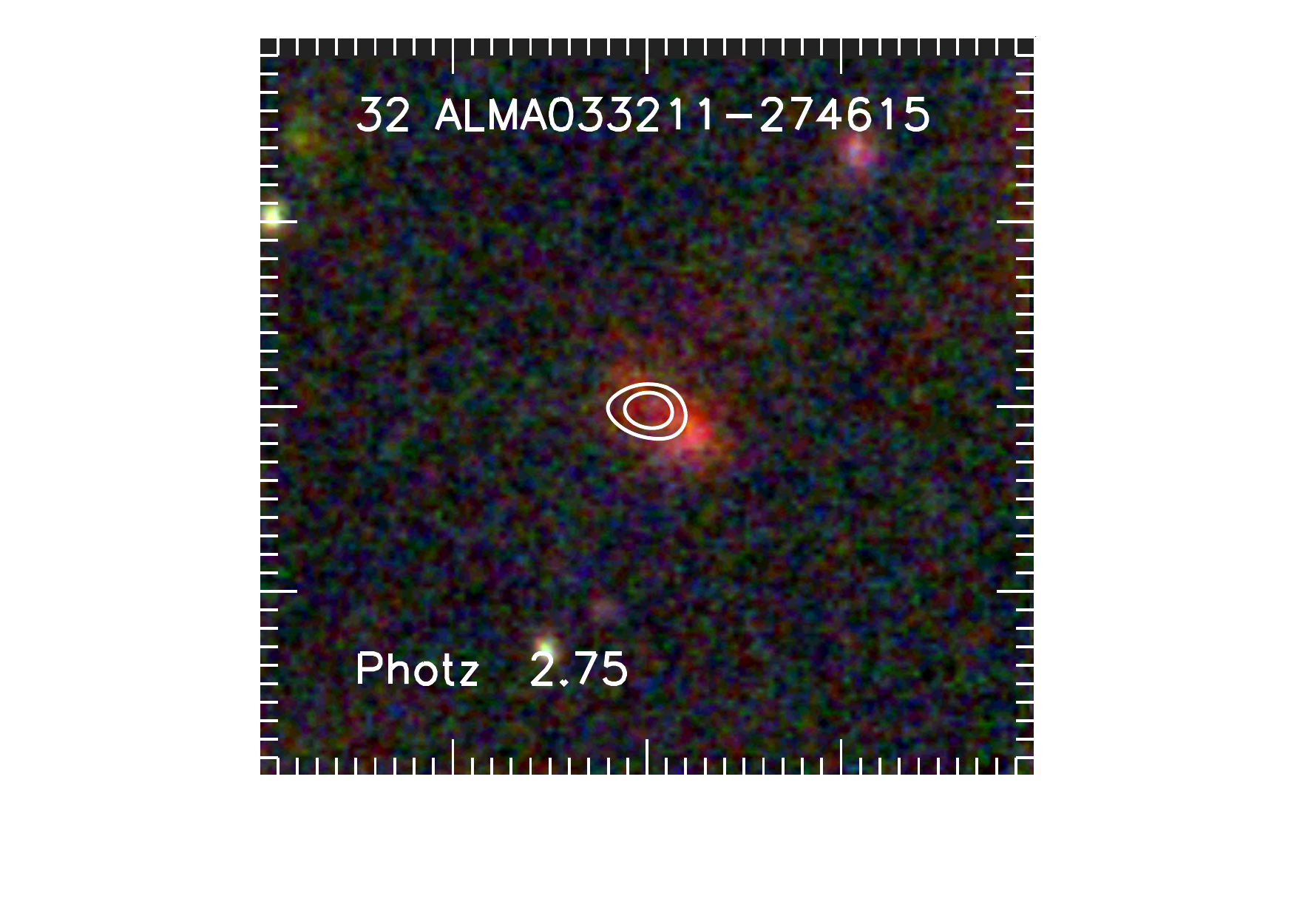}
\hspace{-3.3cm}\includegraphics[width=2.5in,angle=0]{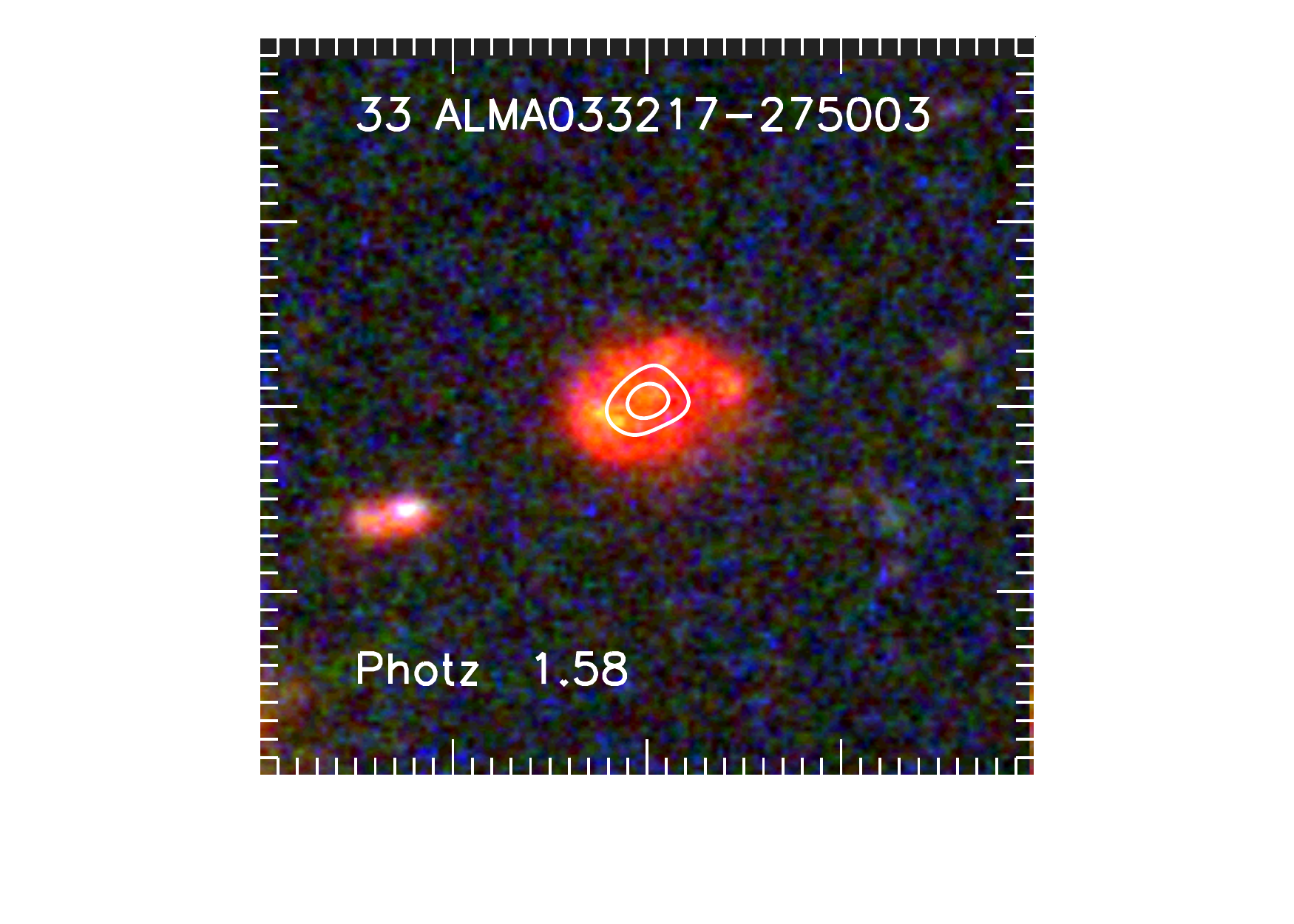}
\includegraphics[width=2.5in,angle=0]{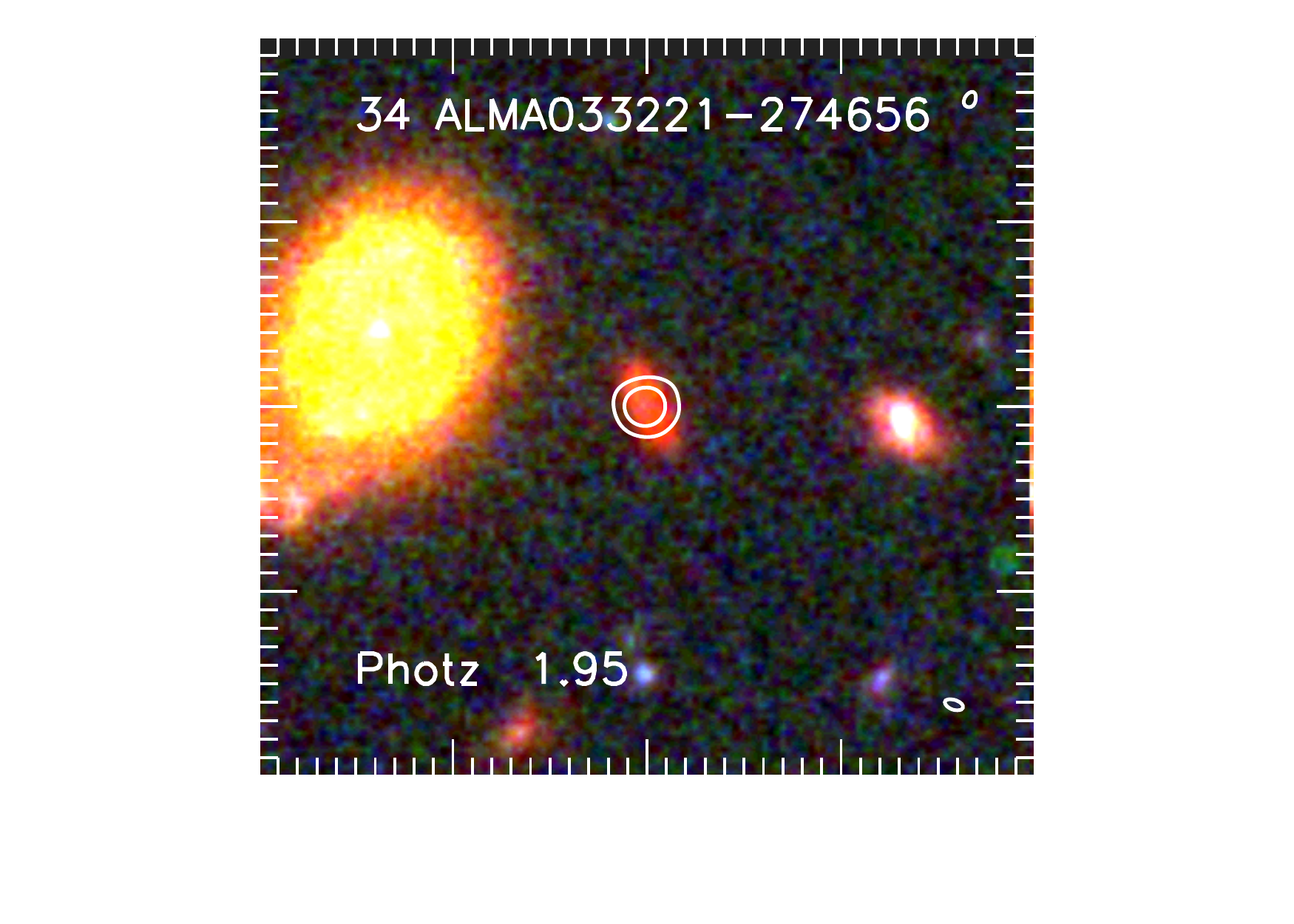}
\hspace{-3.3cm}\includegraphics[width=2.5in,angle=0]{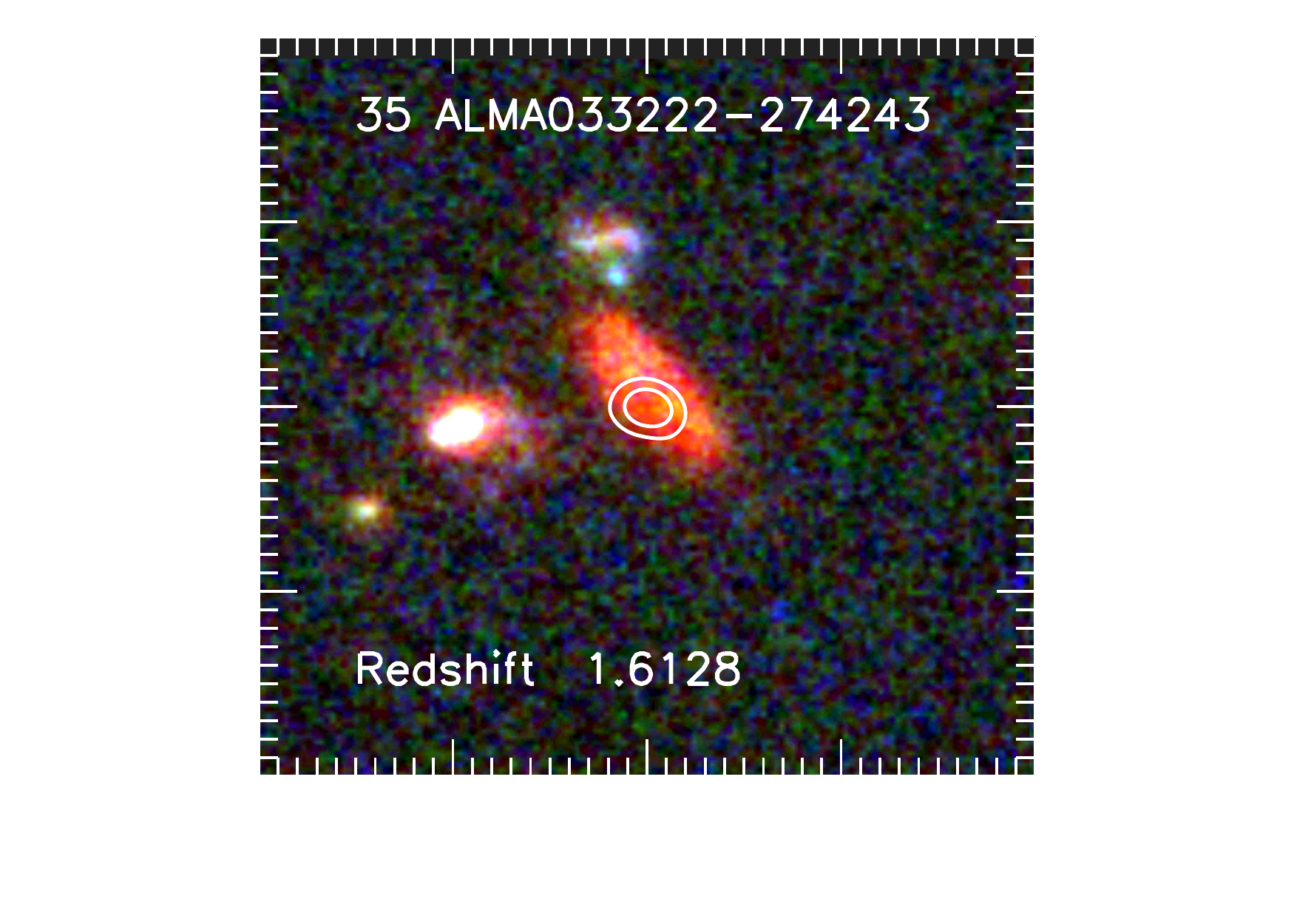}
\hspace{-3.3cm}\includegraphics[width=2.5in,angle=0]{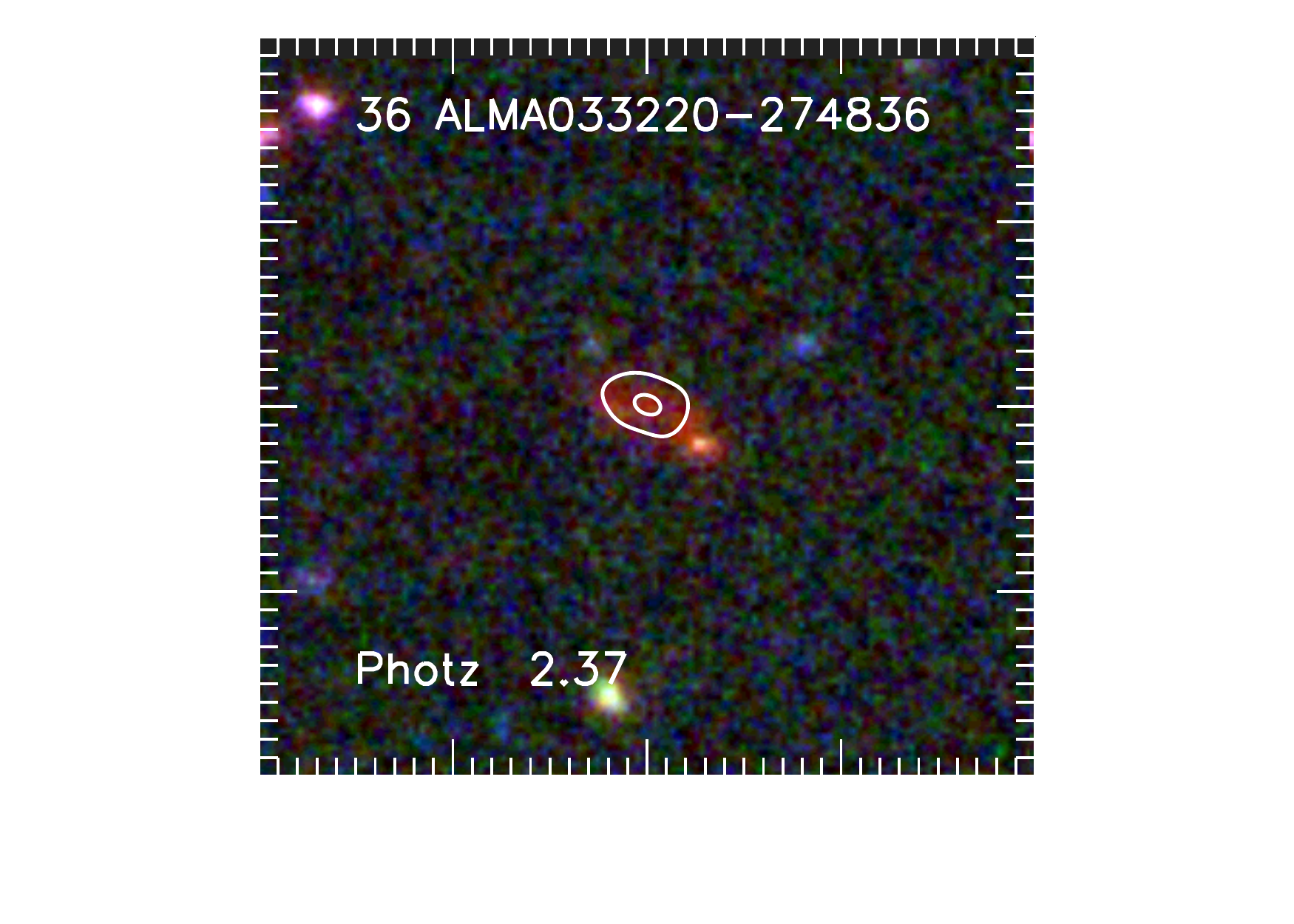}
\hspace{-3.3cm}\includegraphics[width=2.5in,angle=0]{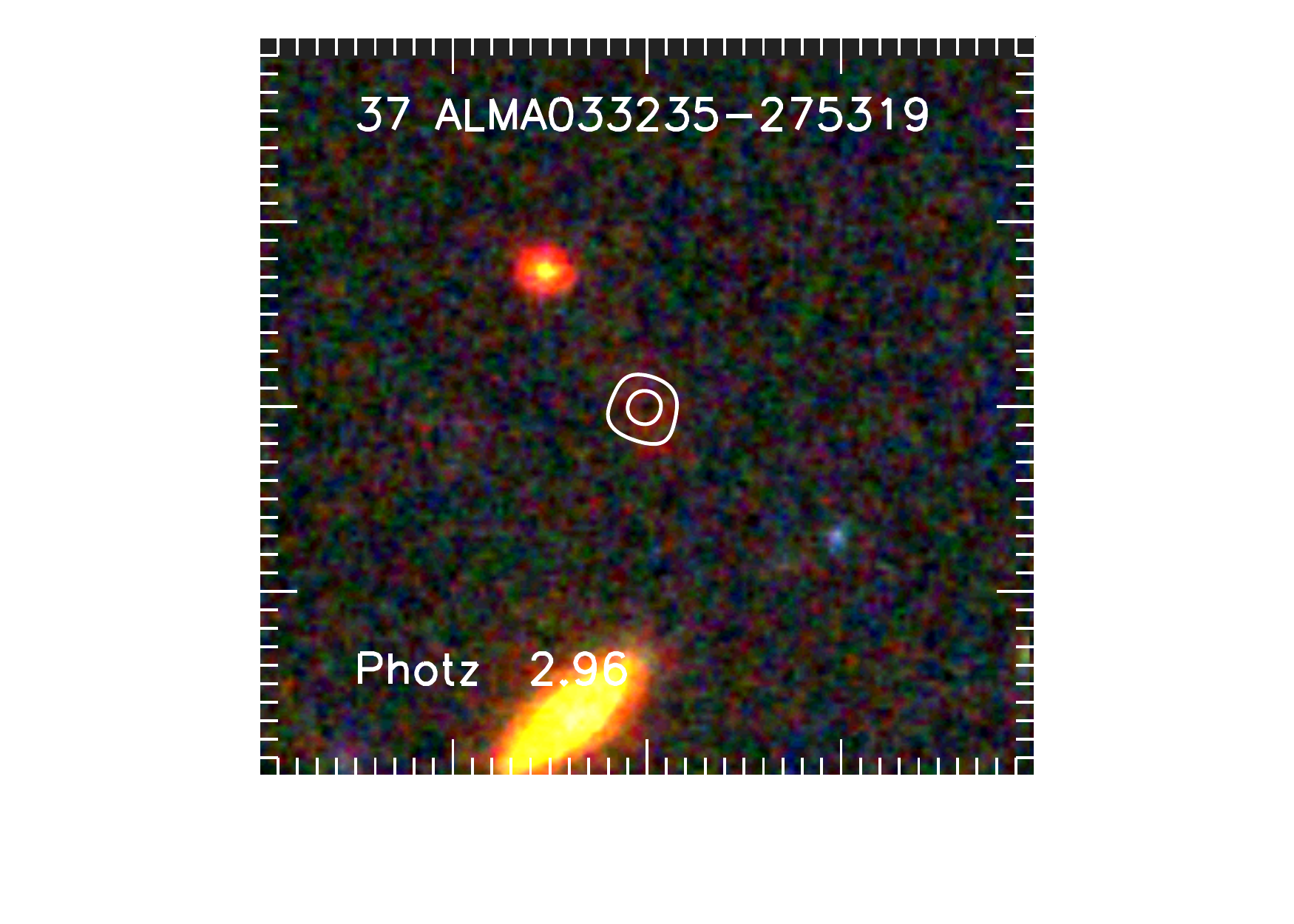}
\includegraphics[width=2.5in,angle=0]{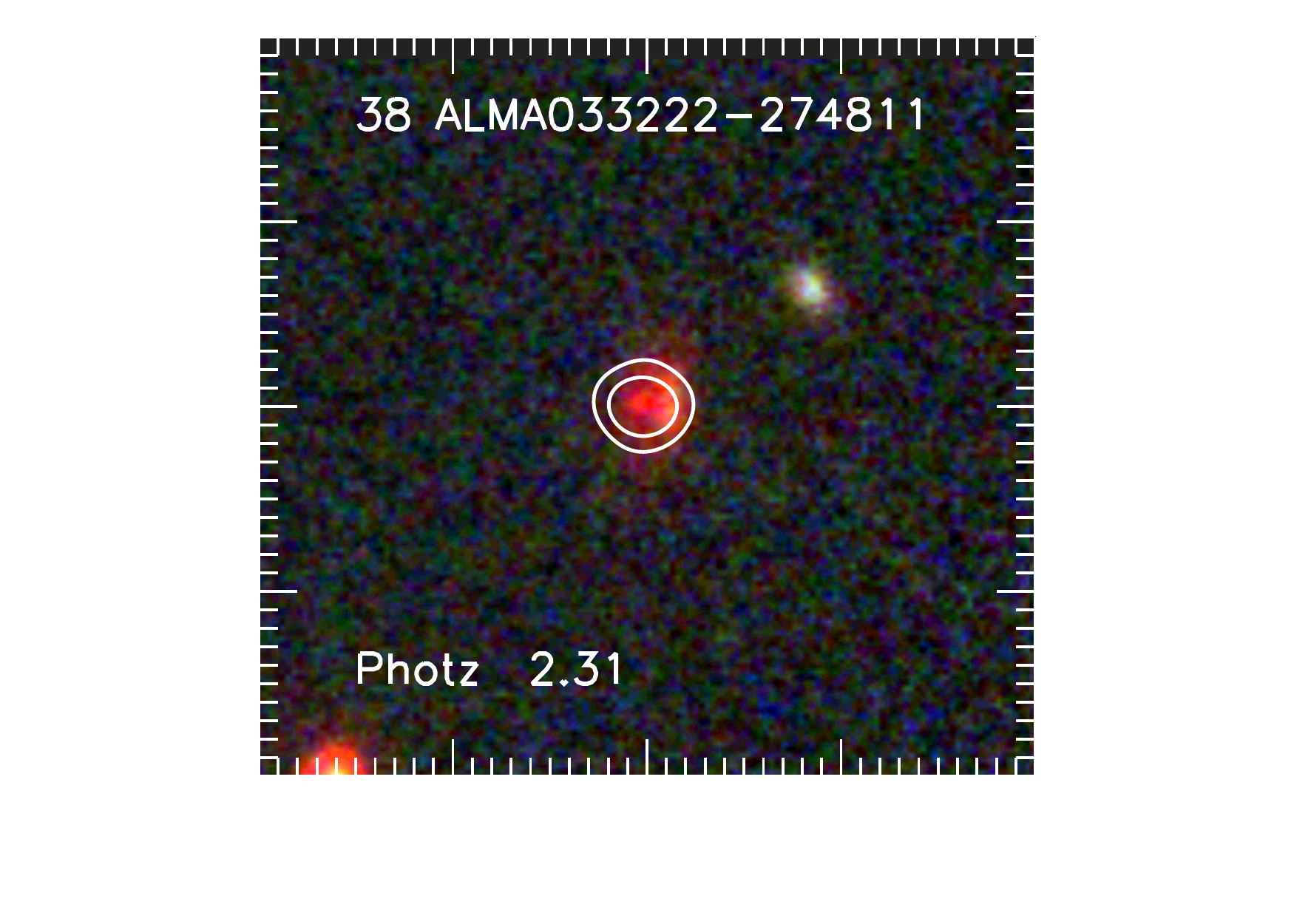}
\hspace{-2.55cm}\includegraphics[width=2.5in,angle=0]{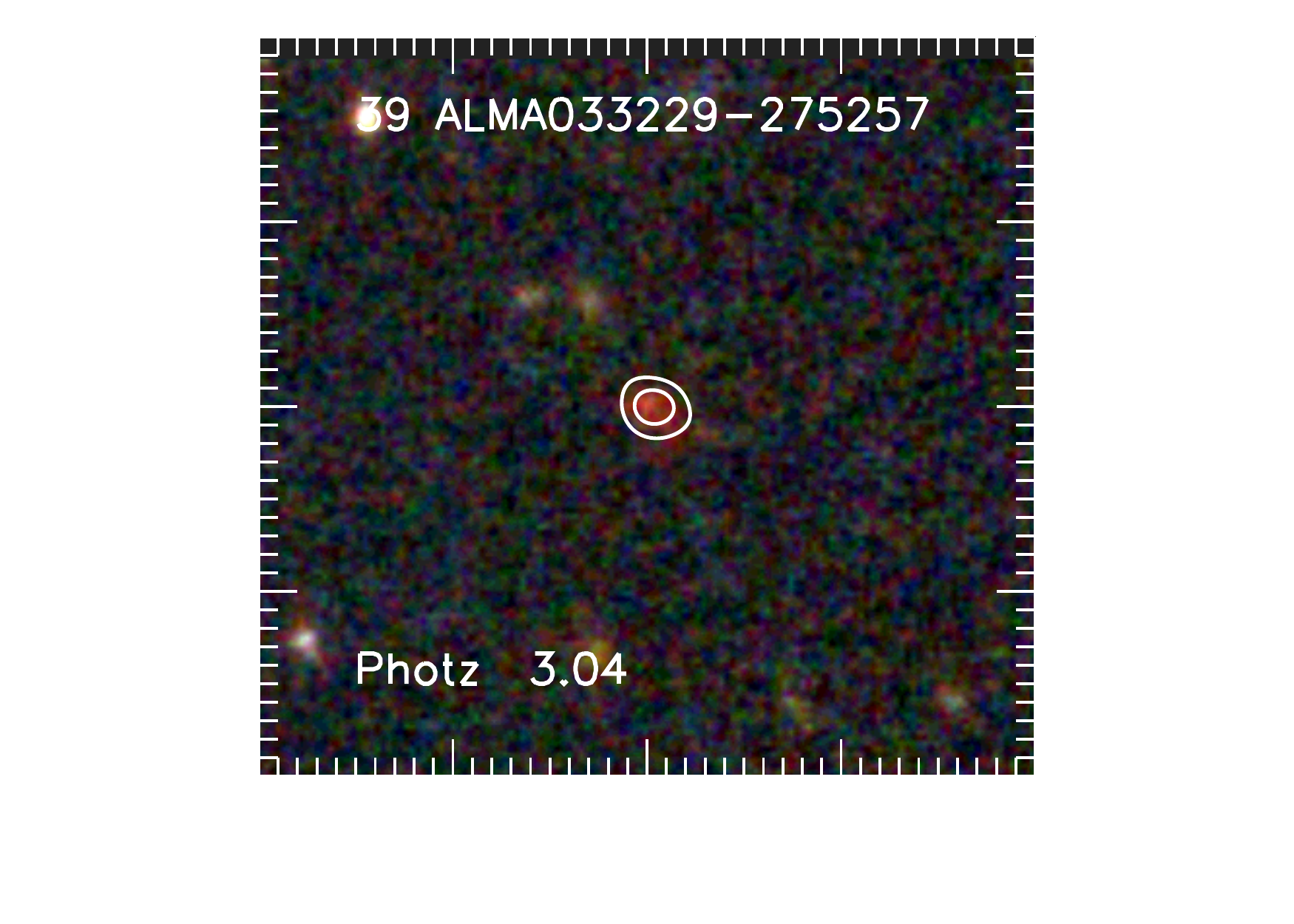}
\hspace{-2.55cm}\includegraphics[width=2.5in,angle=0]{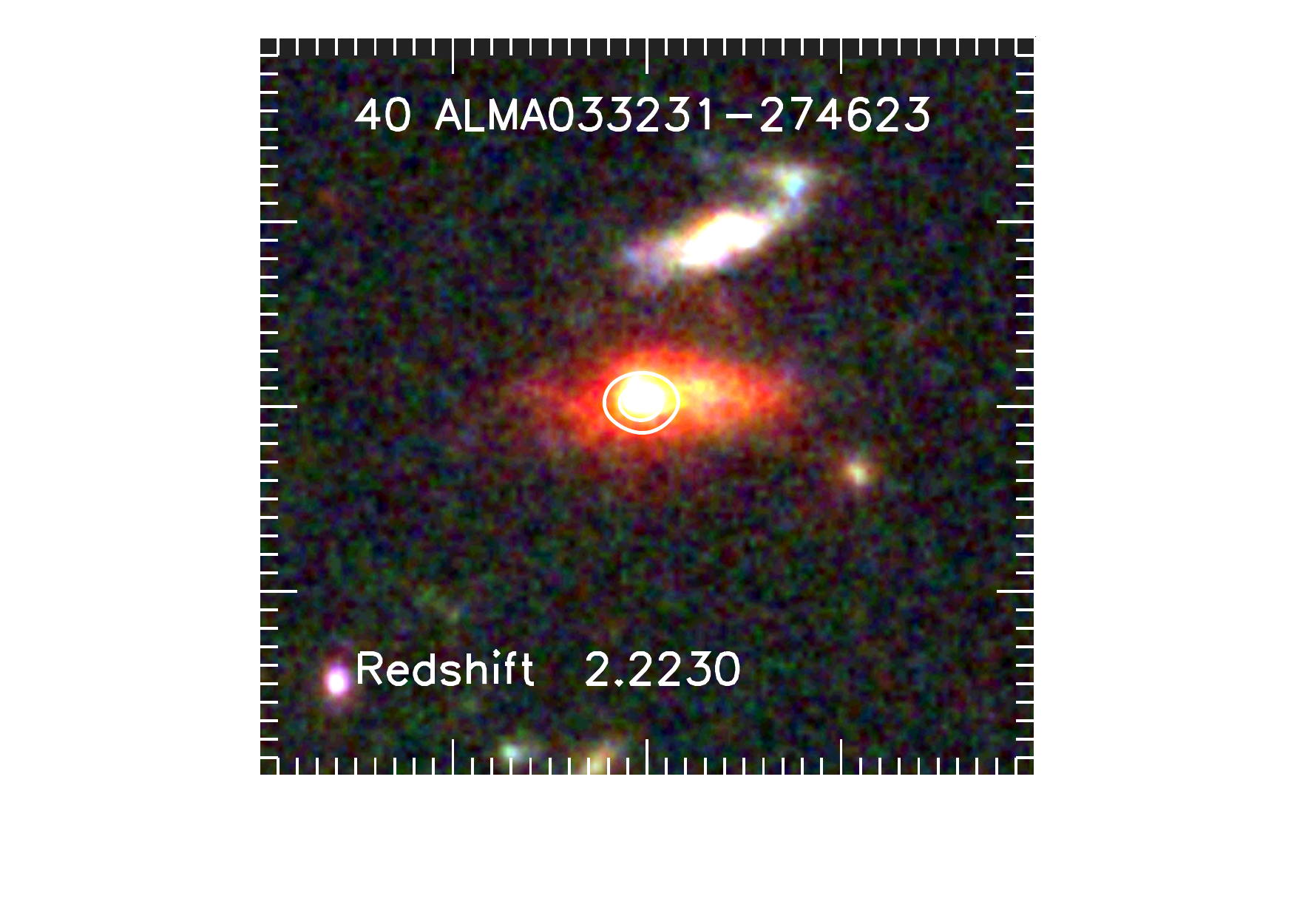}
\hspace{-2.55cm}\includegraphics[width=2.5in,angle=0]{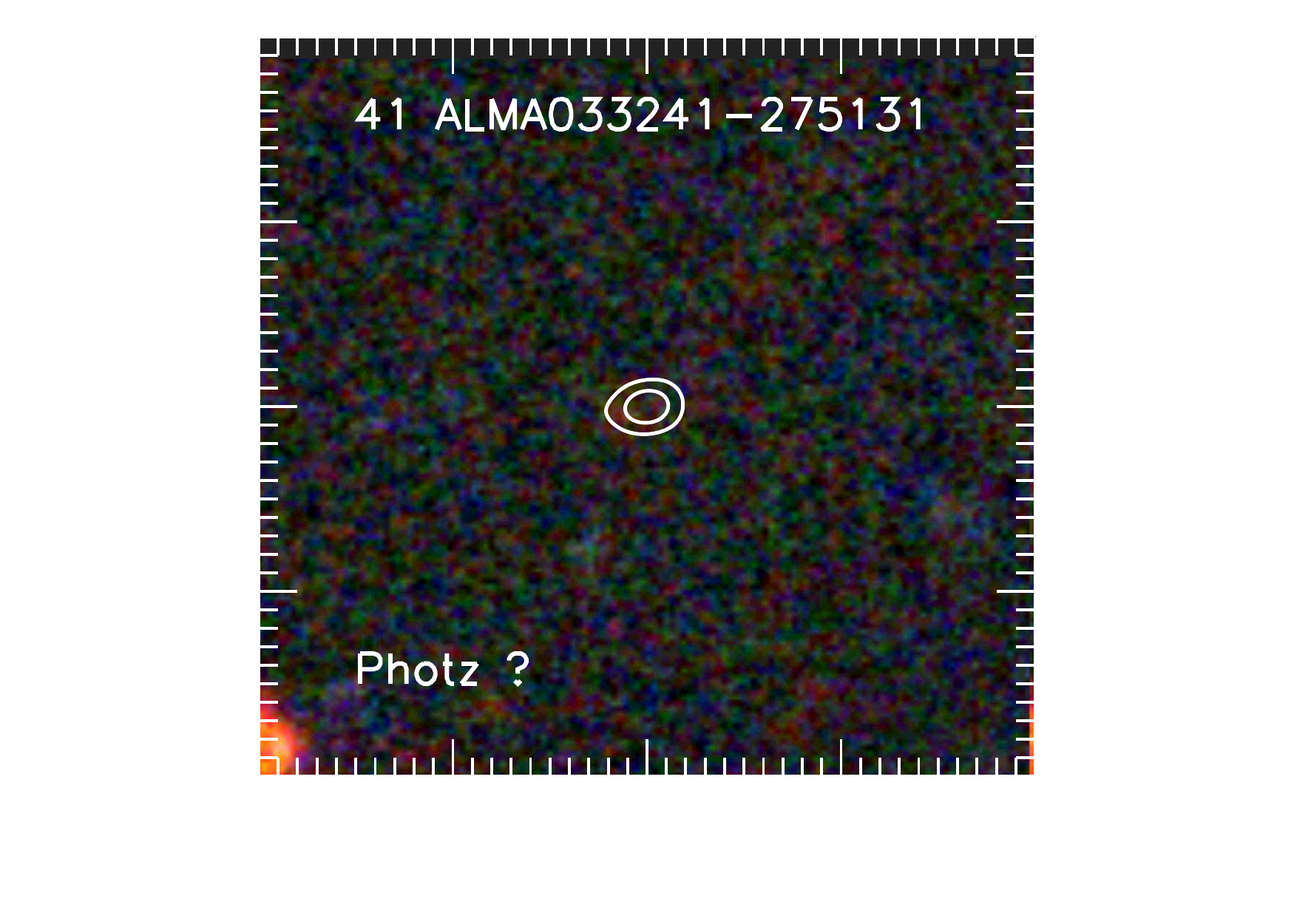}
\vskip -0.75cm
\caption{Continued
\label{basic_alma_images_2}}
\end{figure*}
\begin{figure*}
\setcounter{figure}{9}
\includegraphics[width=2.5in,angle=0]{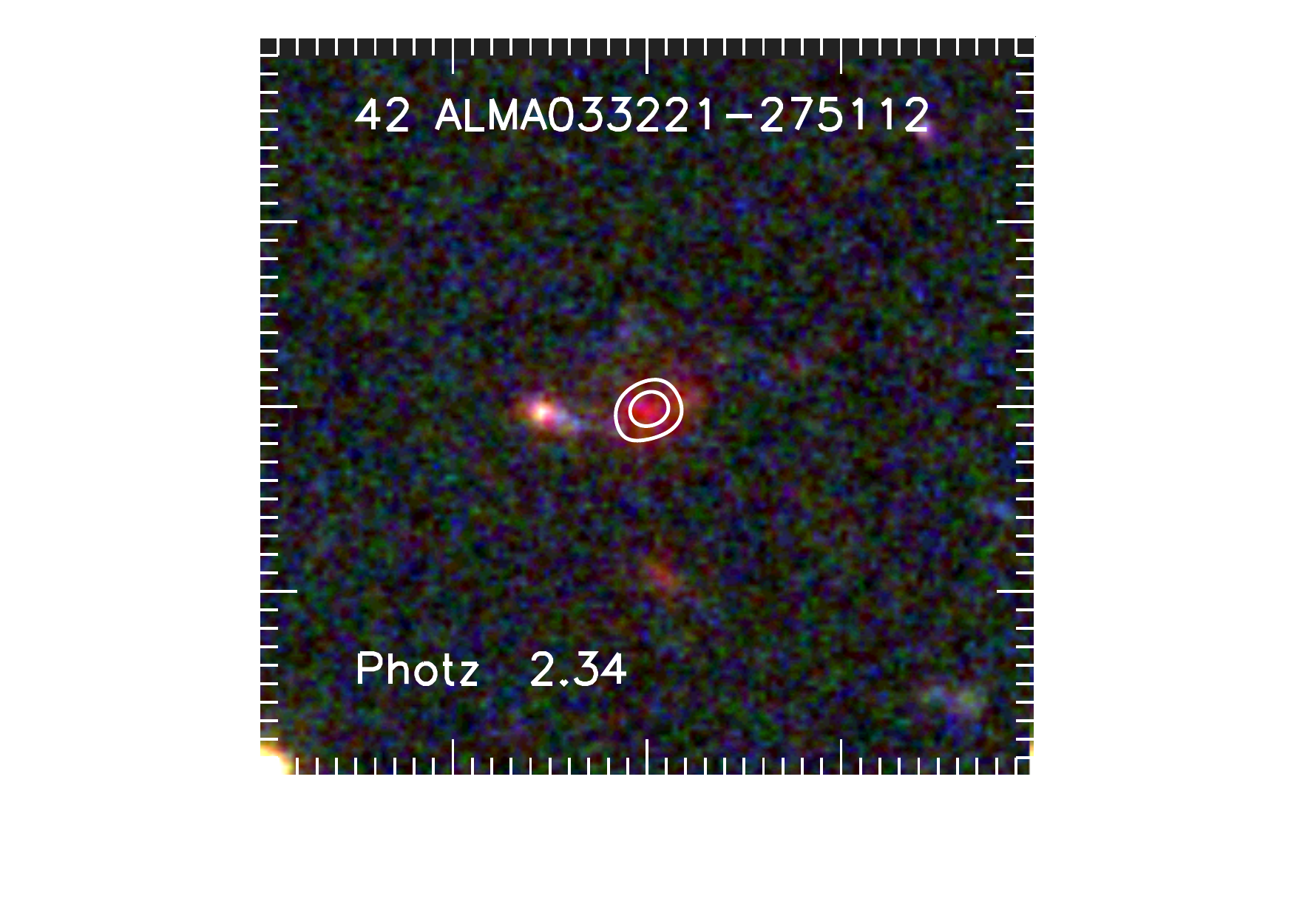}
\hspace{-3.3cm}\includegraphics[width=2.5in,angle=0]{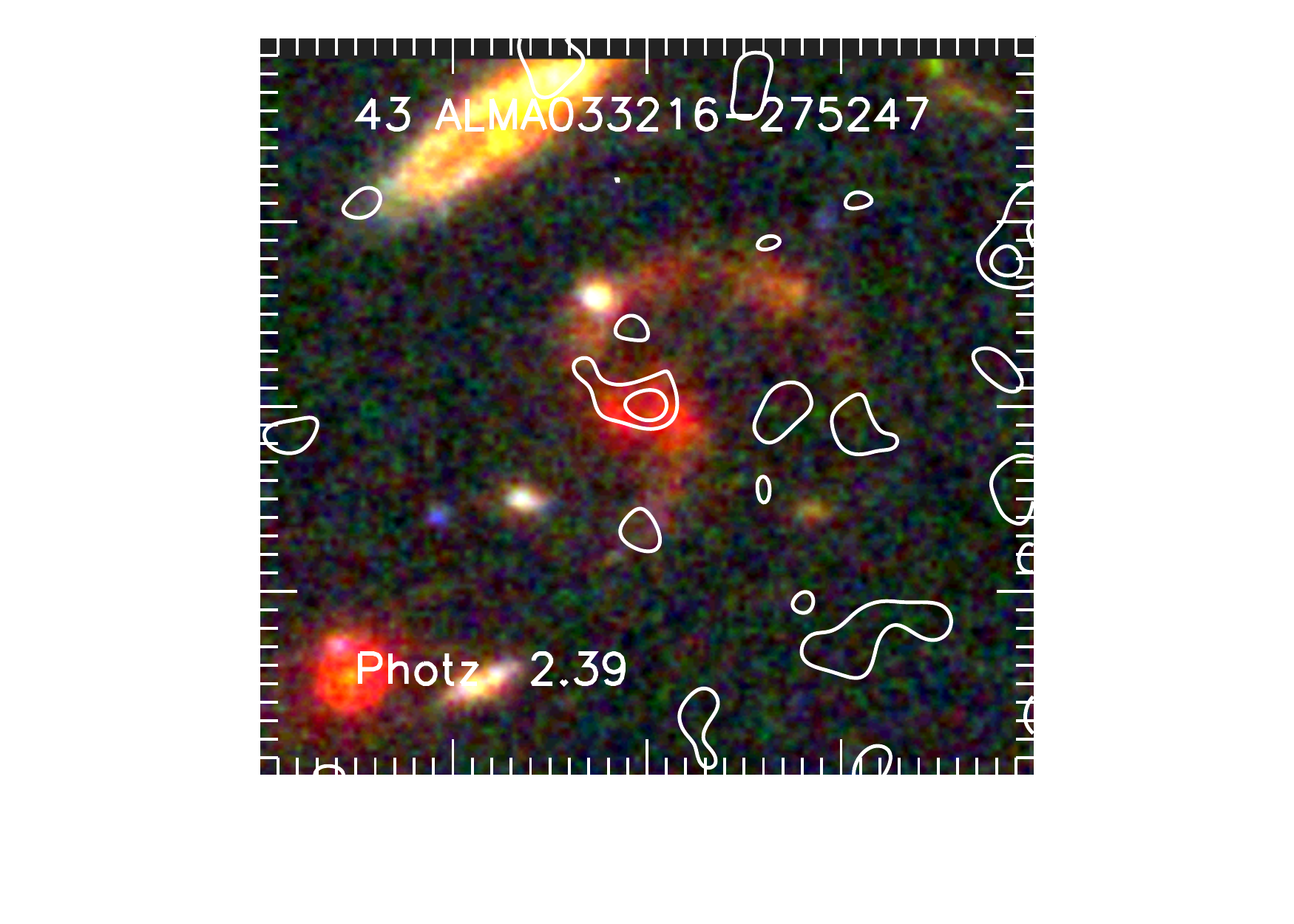}
\hspace{-3.3cm}\includegraphics[width=2.5in,angle=0]{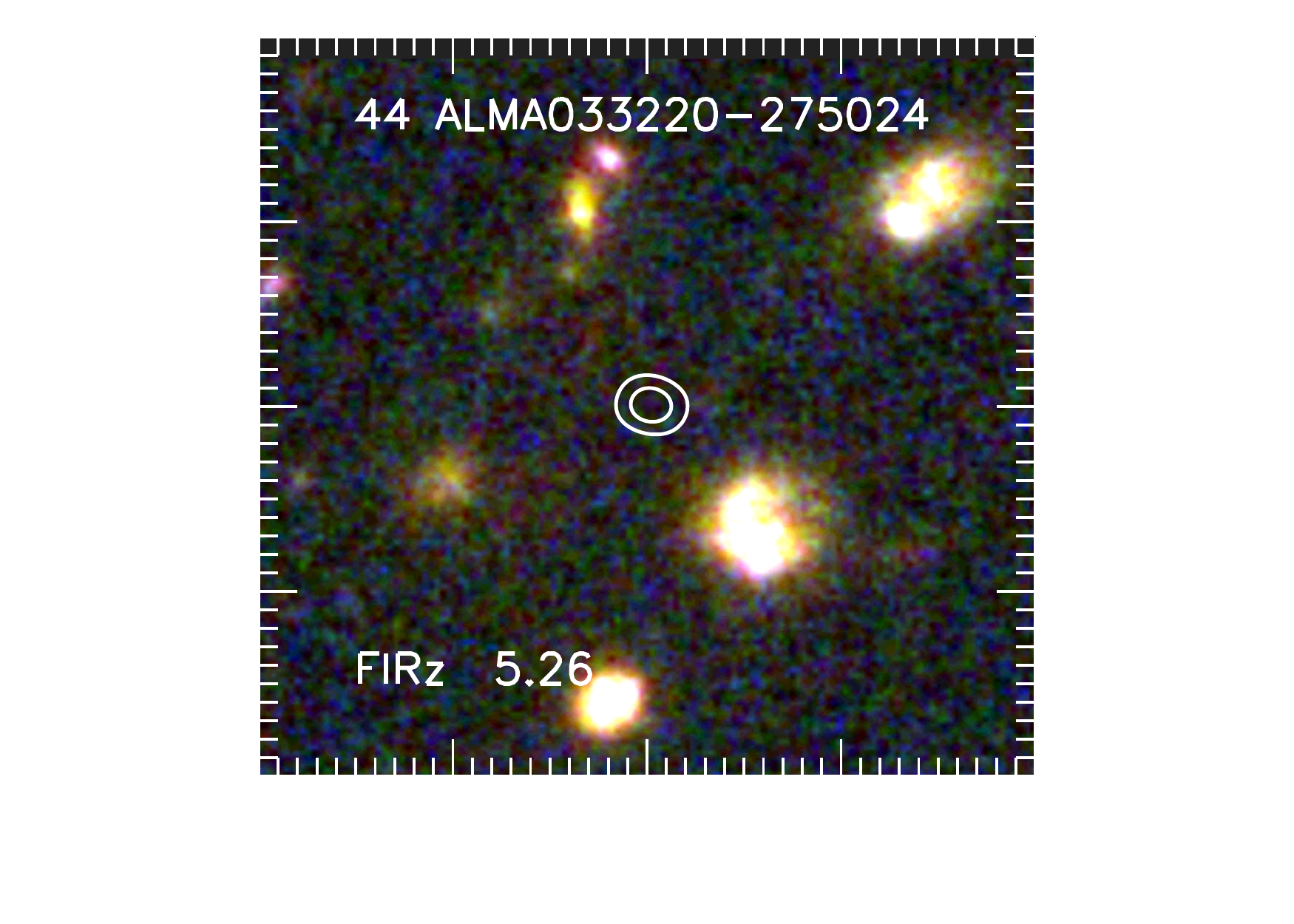}
\hspace{-3.3cm}\includegraphics[width=2.5in,angle=0]{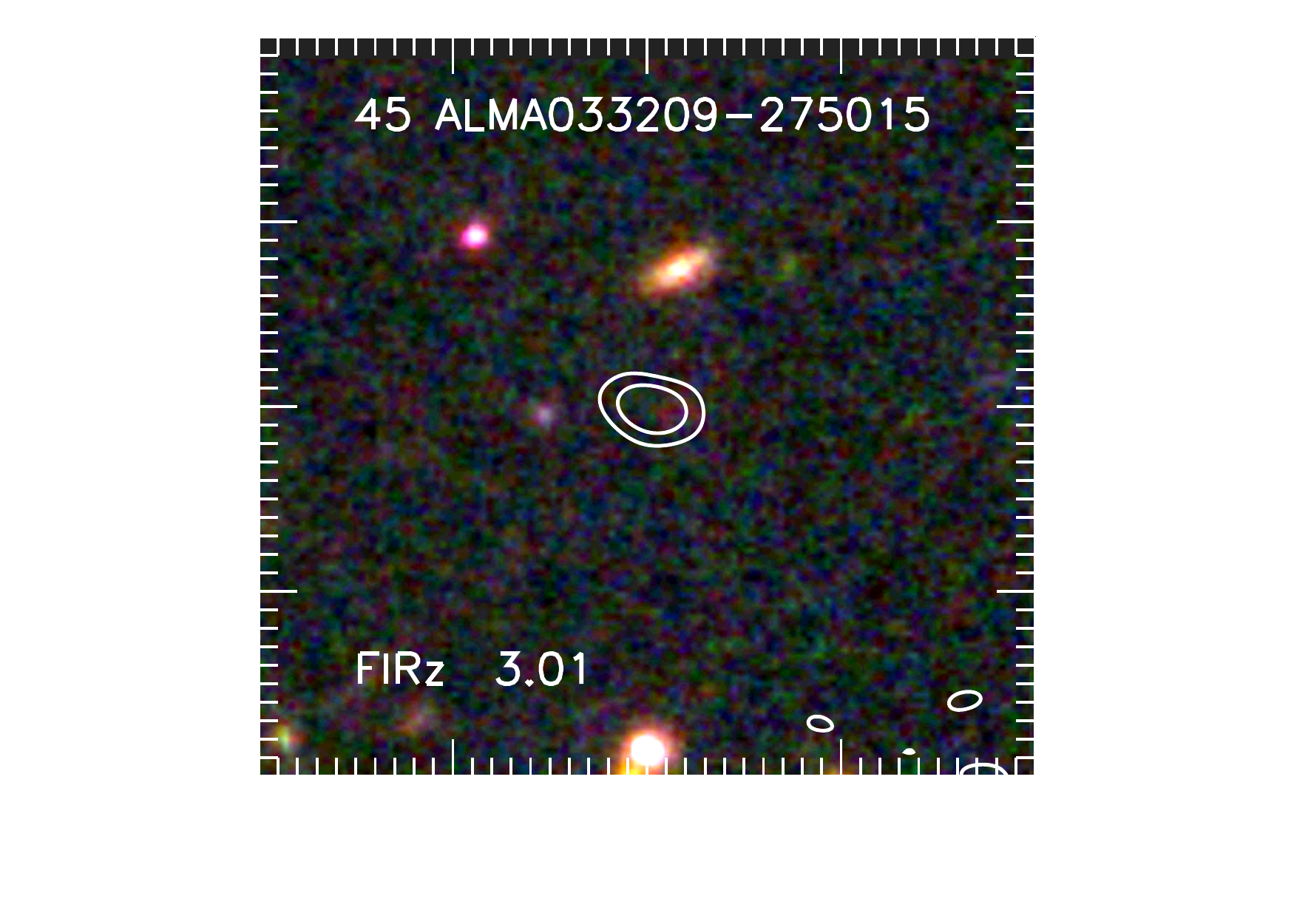}
\includegraphics[width=2.5in,angle=0]{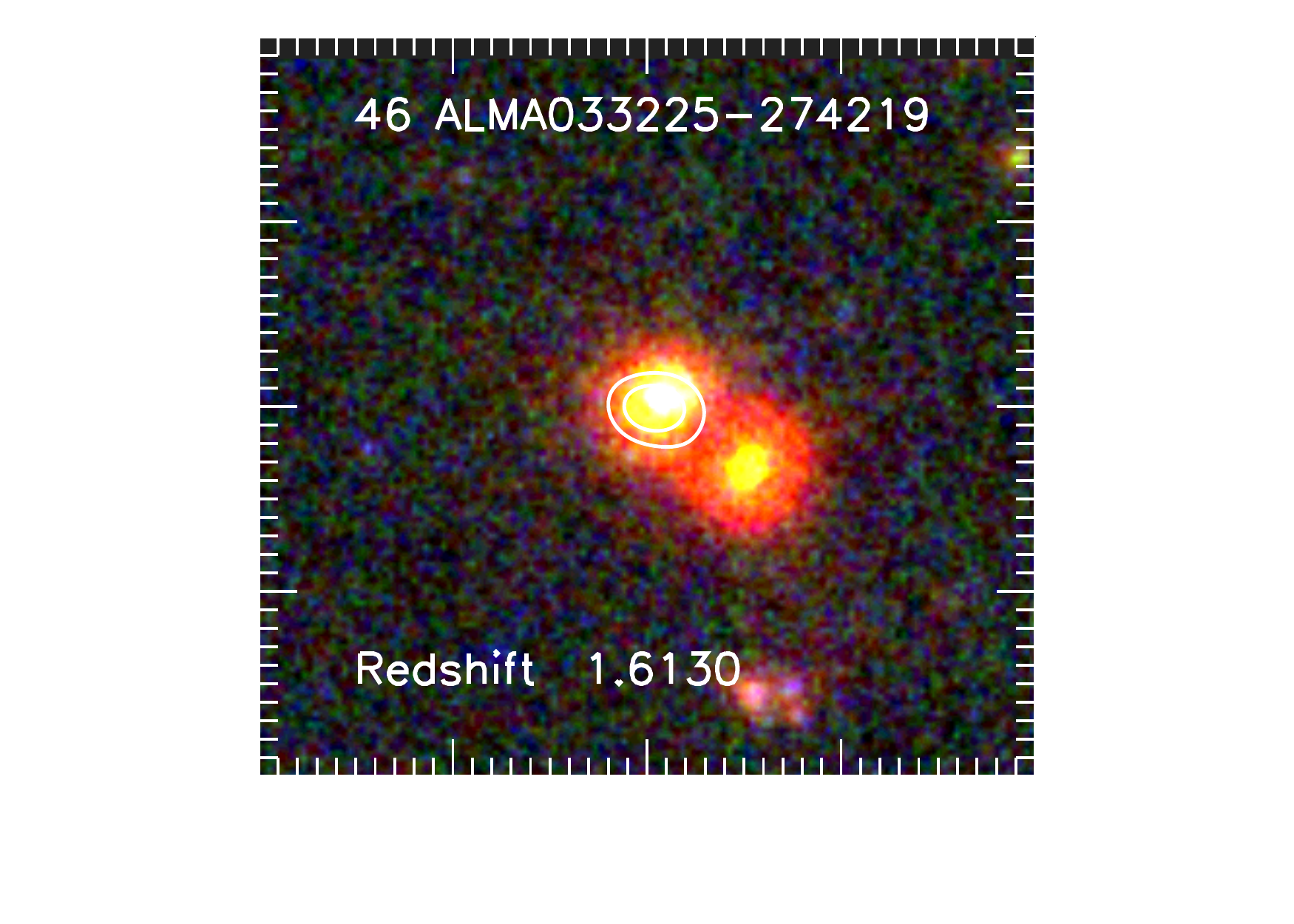}
\hspace{-3.3cm}\includegraphics[width=2.5in,angle=0]{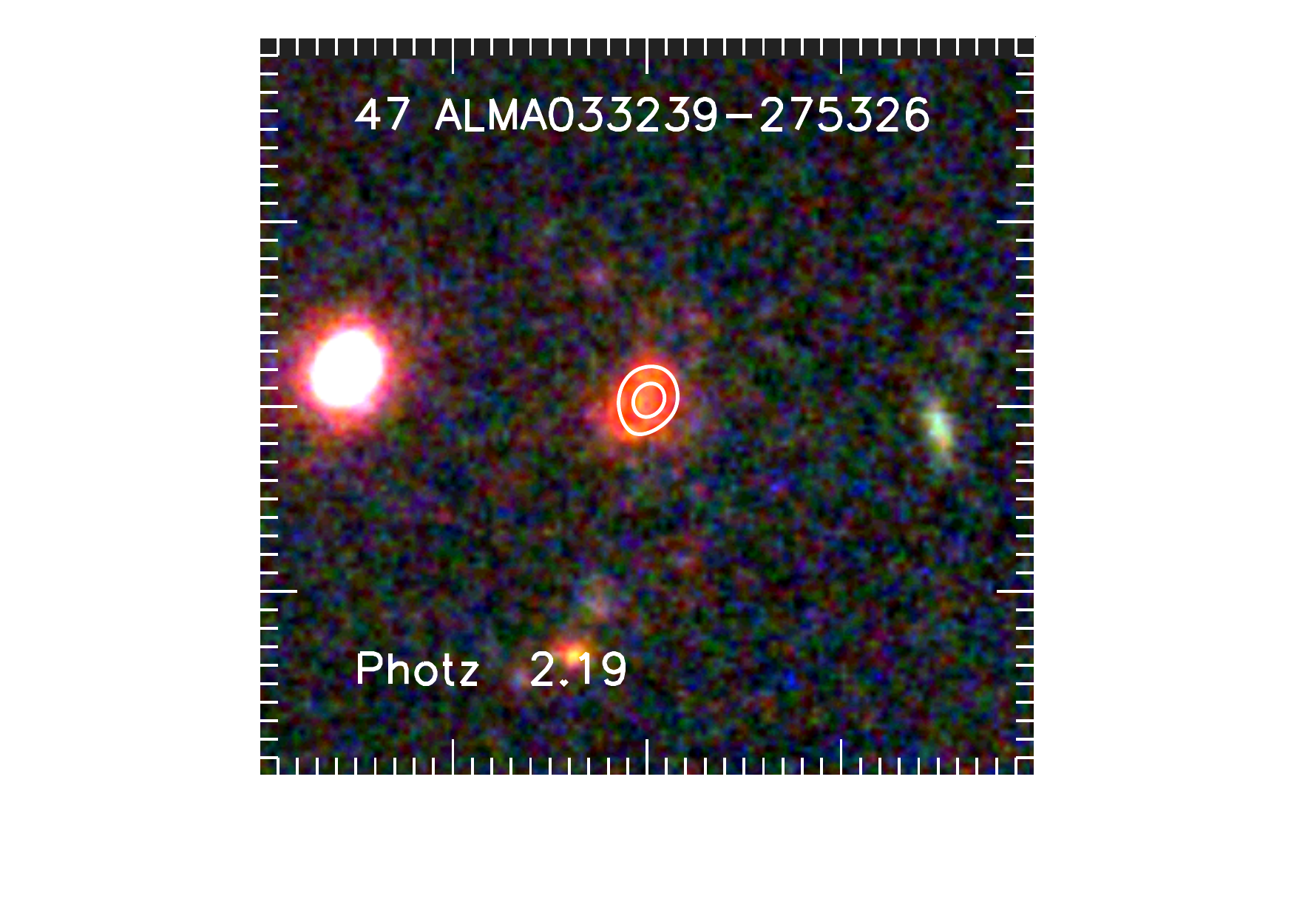}
\hspace{-3.3cm}\includegraphics[width=2.5in,angle=0]{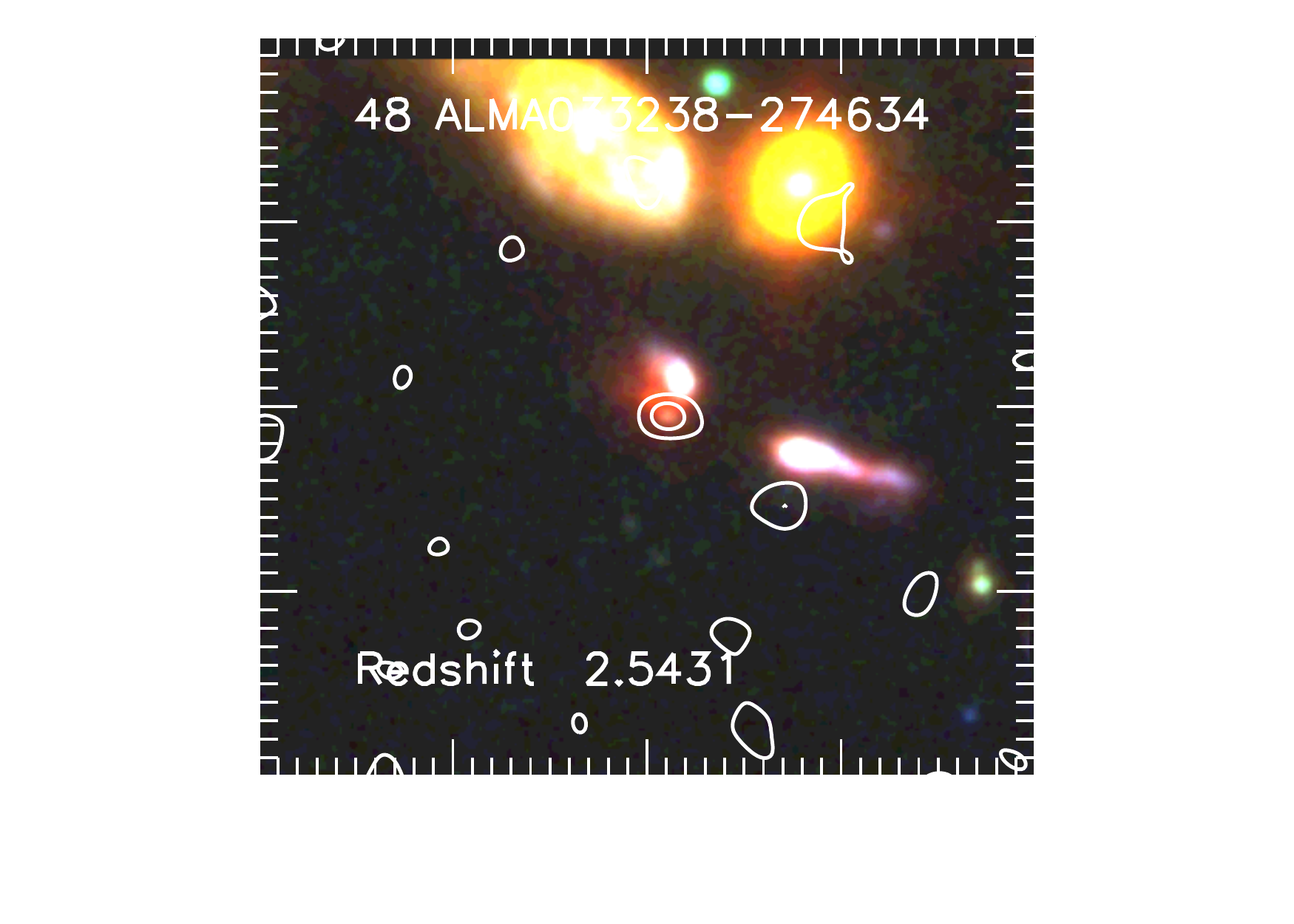}
\hspace{-3.3cm}\includegraphics[width=2.5in,angle=0]{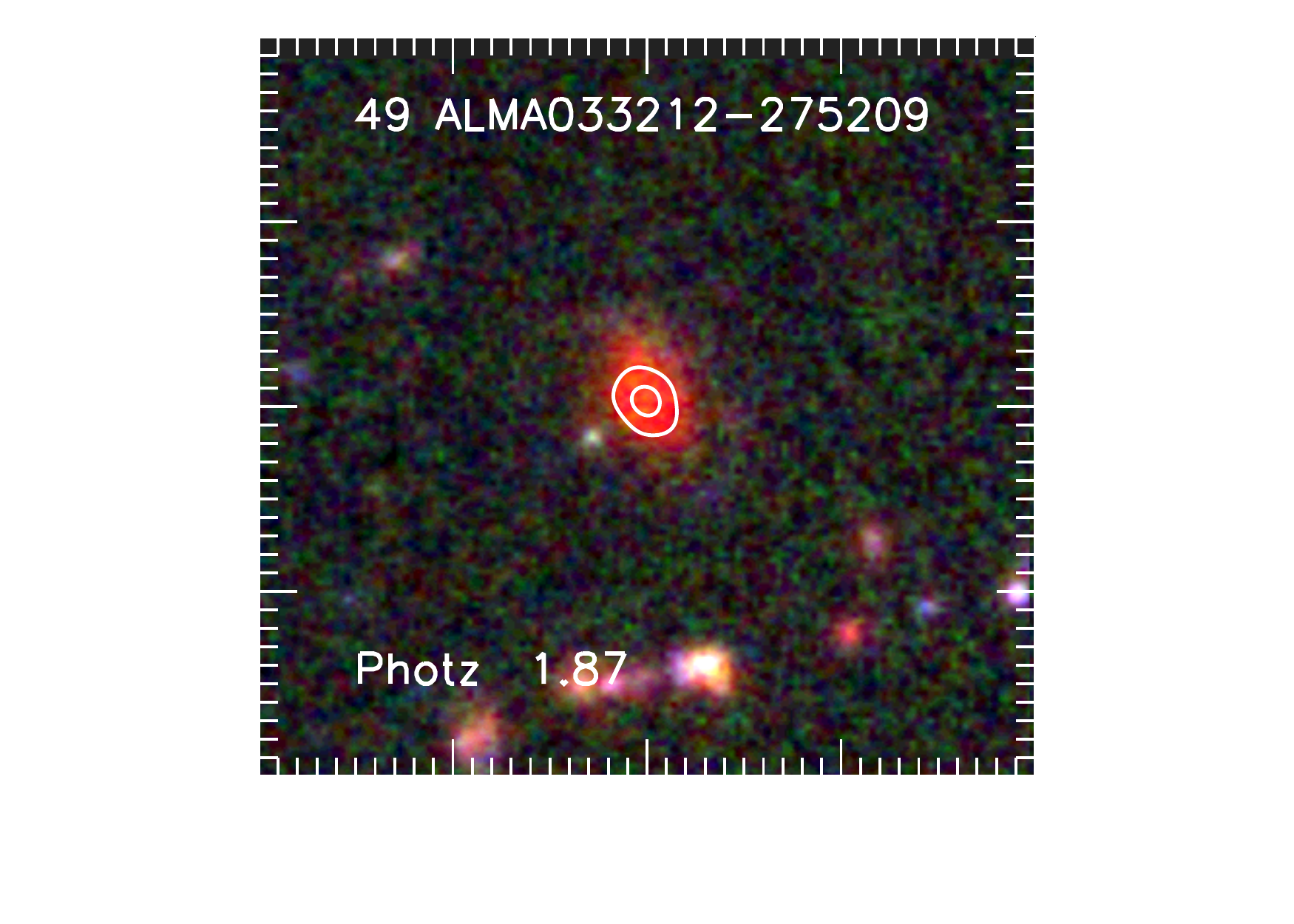}
\hspace{-3.3cm}\includegraphics[width=2.5in,angle=0]{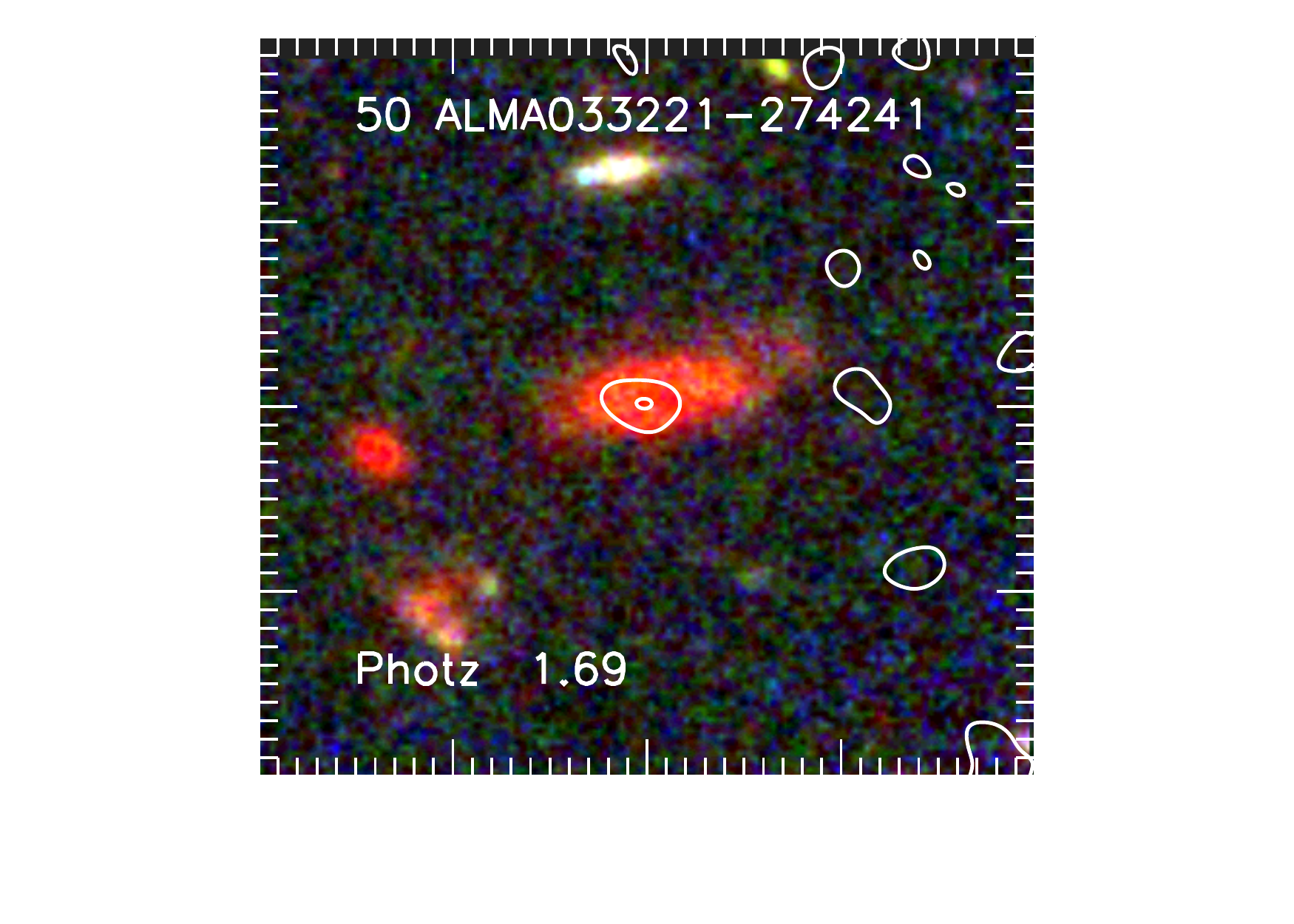}
\hspace{-3.3cm}\includegraphics[width=2.5in,angle=0]{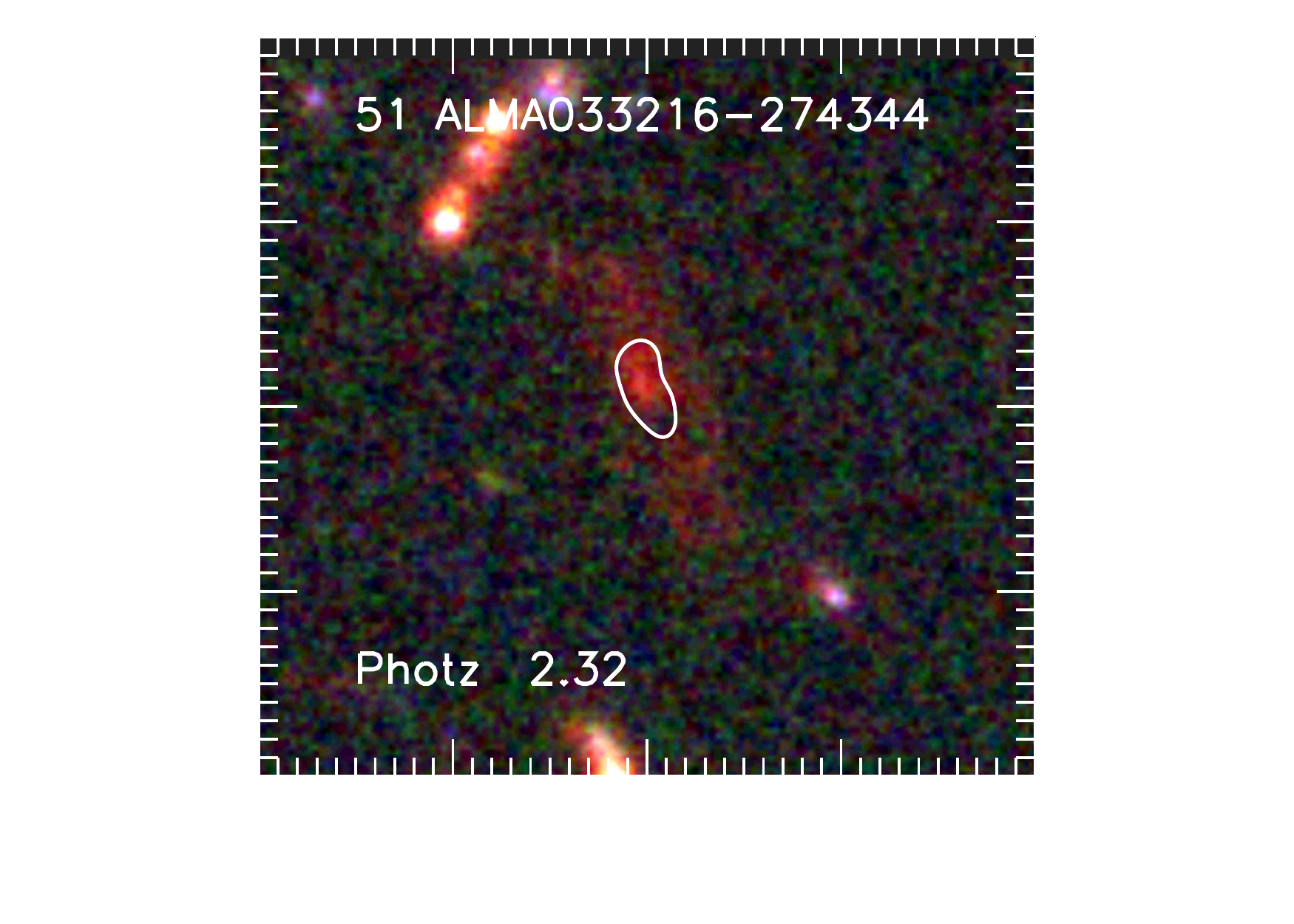}
\hspace{-3.3cm}\includegraphics[width=2.5in,angle=0]{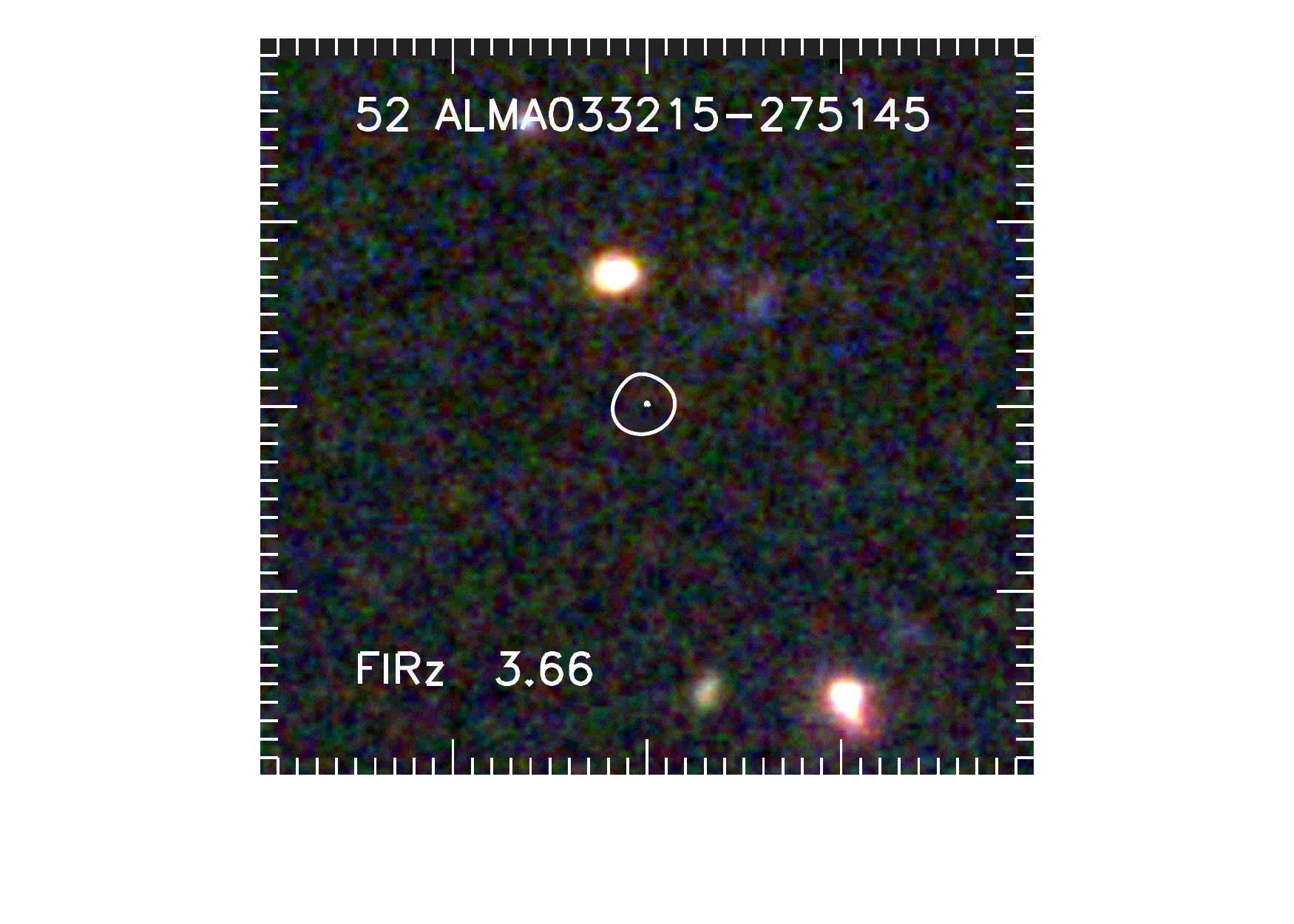}
\hspace{-3.3cm}\includegraphics[width=2.5in,angle=0]{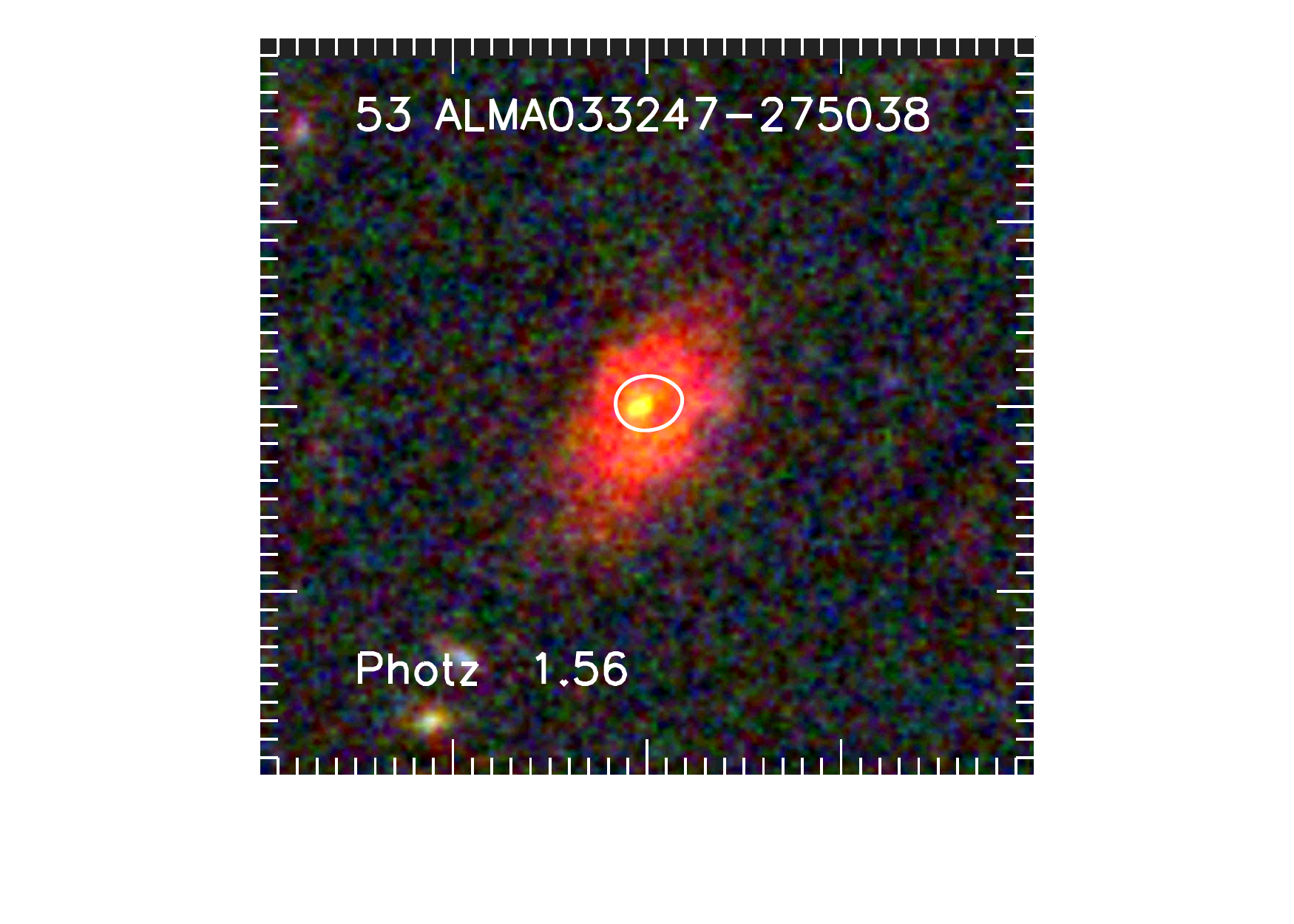}
\hspace{-3.3cm}\includegraphics[width=2.5in,angle=0]{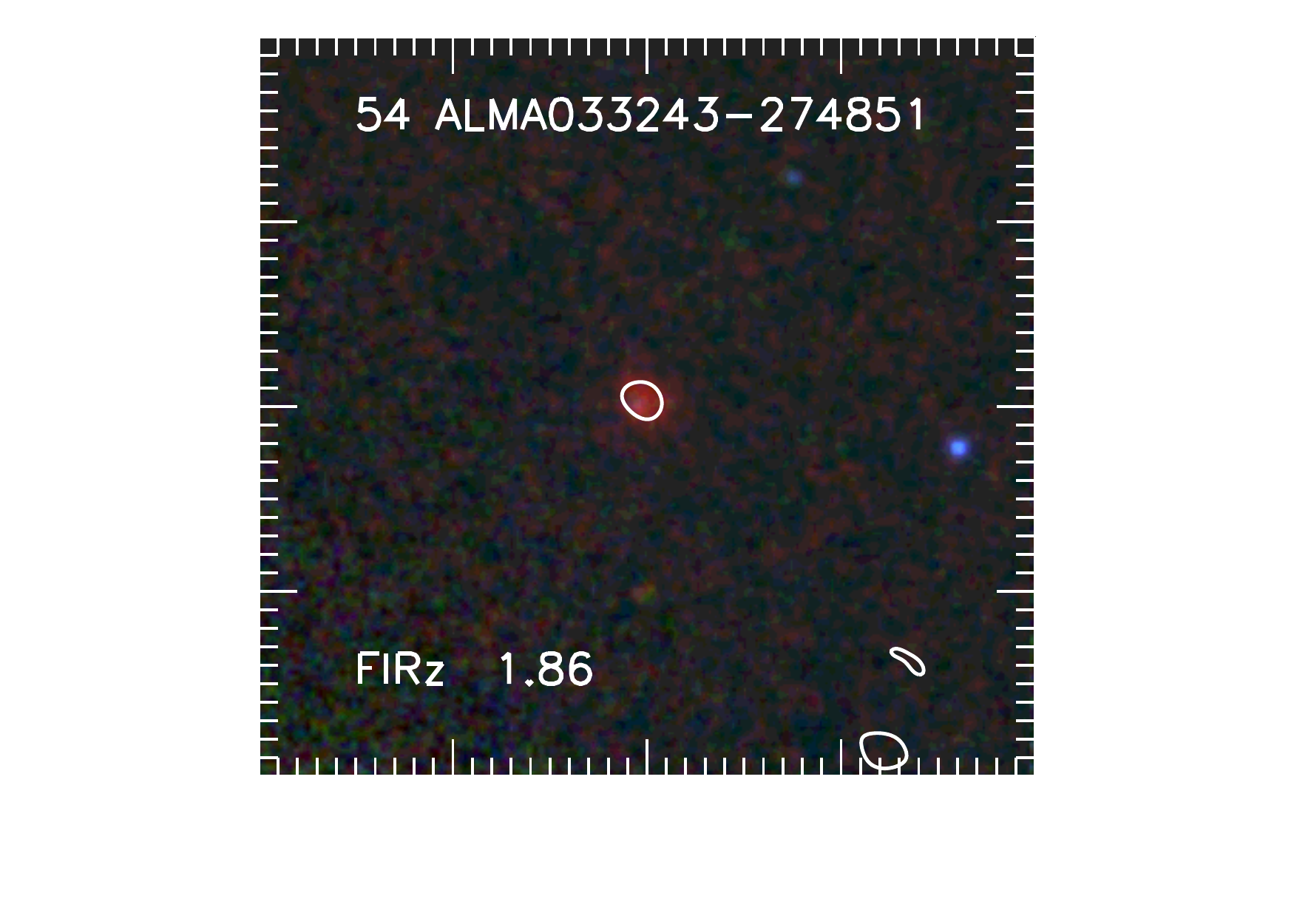}
\hspace{-3.3cm}\includegraphics[width=2.5in,angle=0]{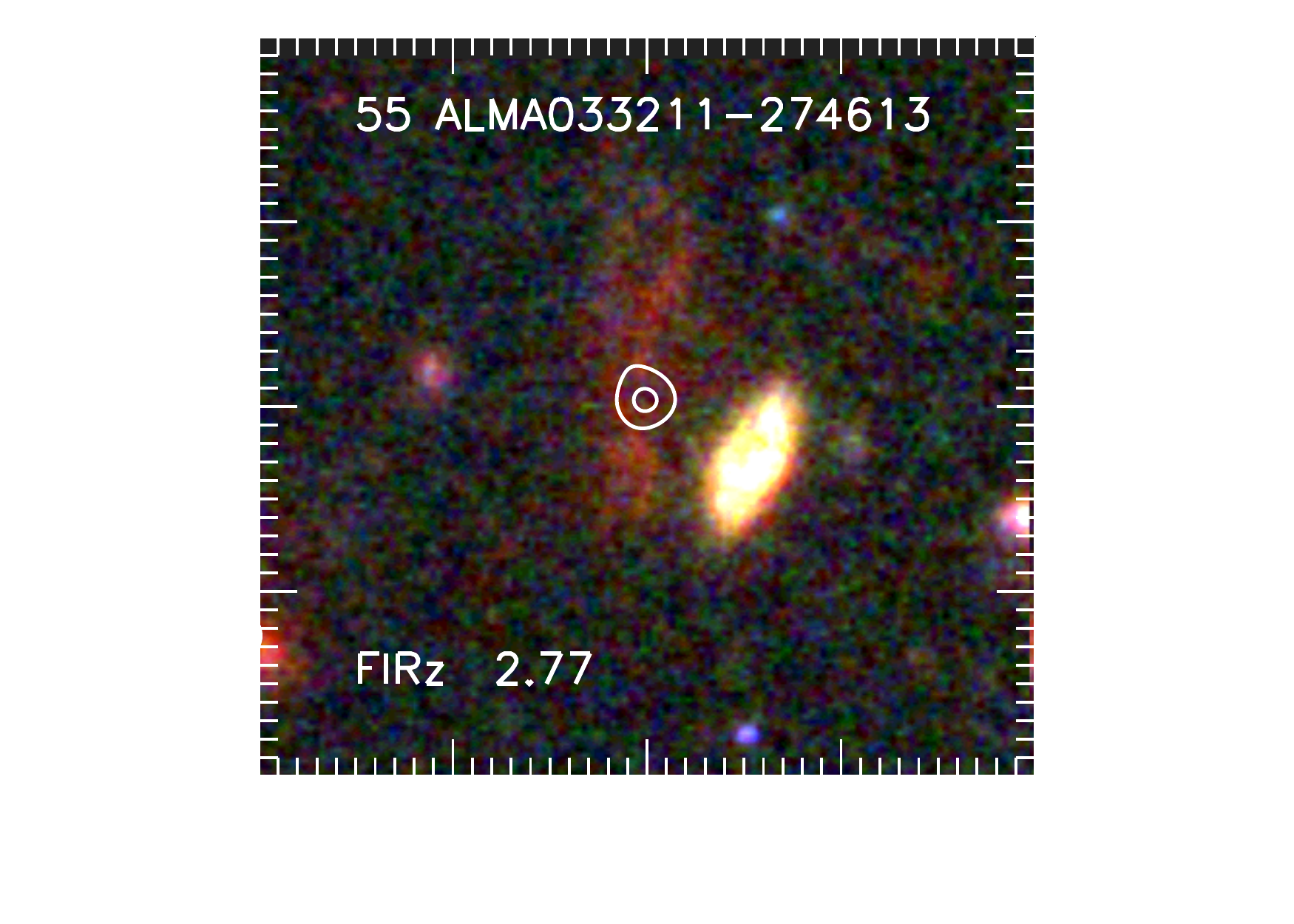}
\hspace{-3.3cm}\includegraphics[width=2.5in,angle=0]{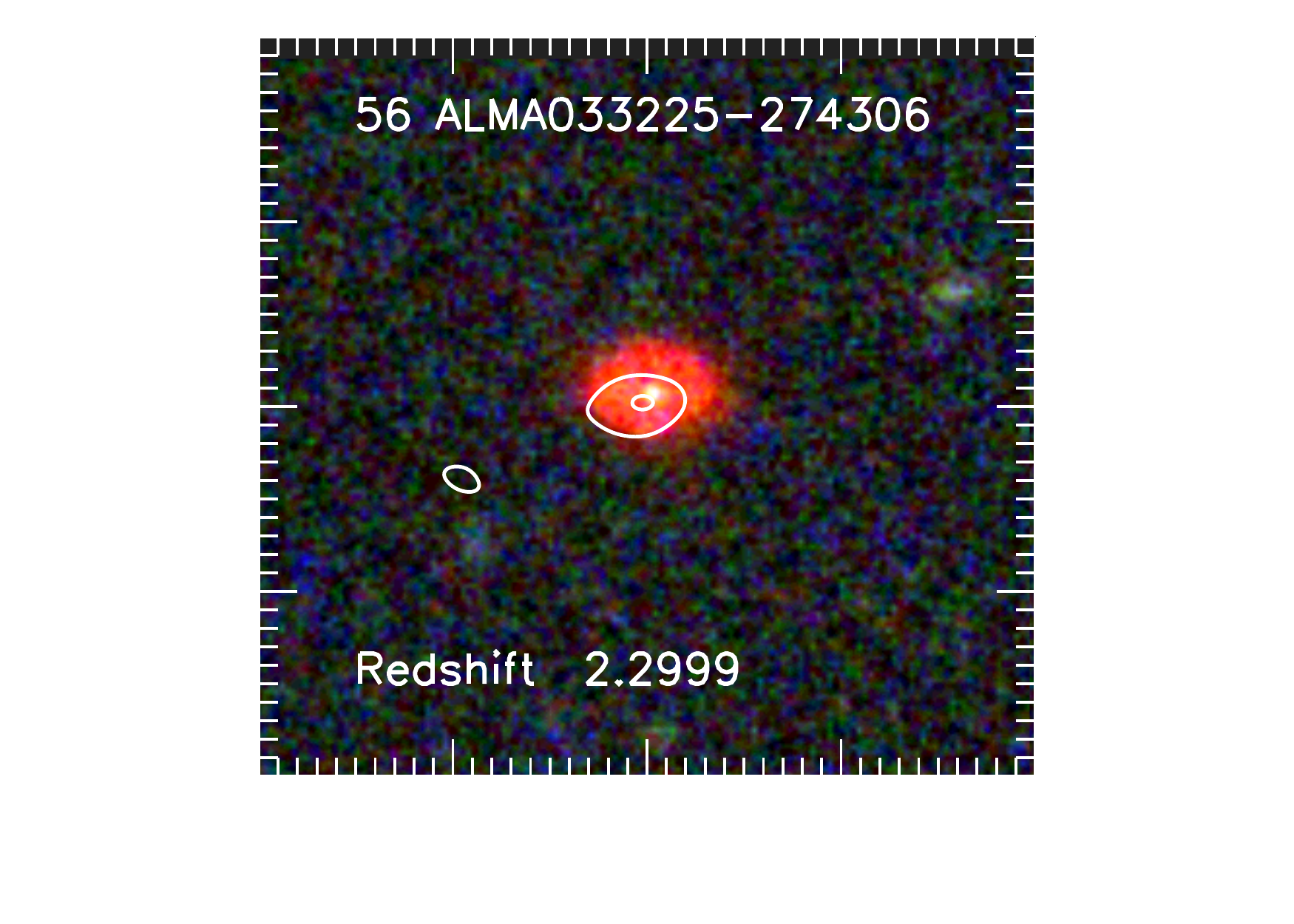}
\hspace{-3.3cm}\includegraphics[width=2.5in,angle=0]{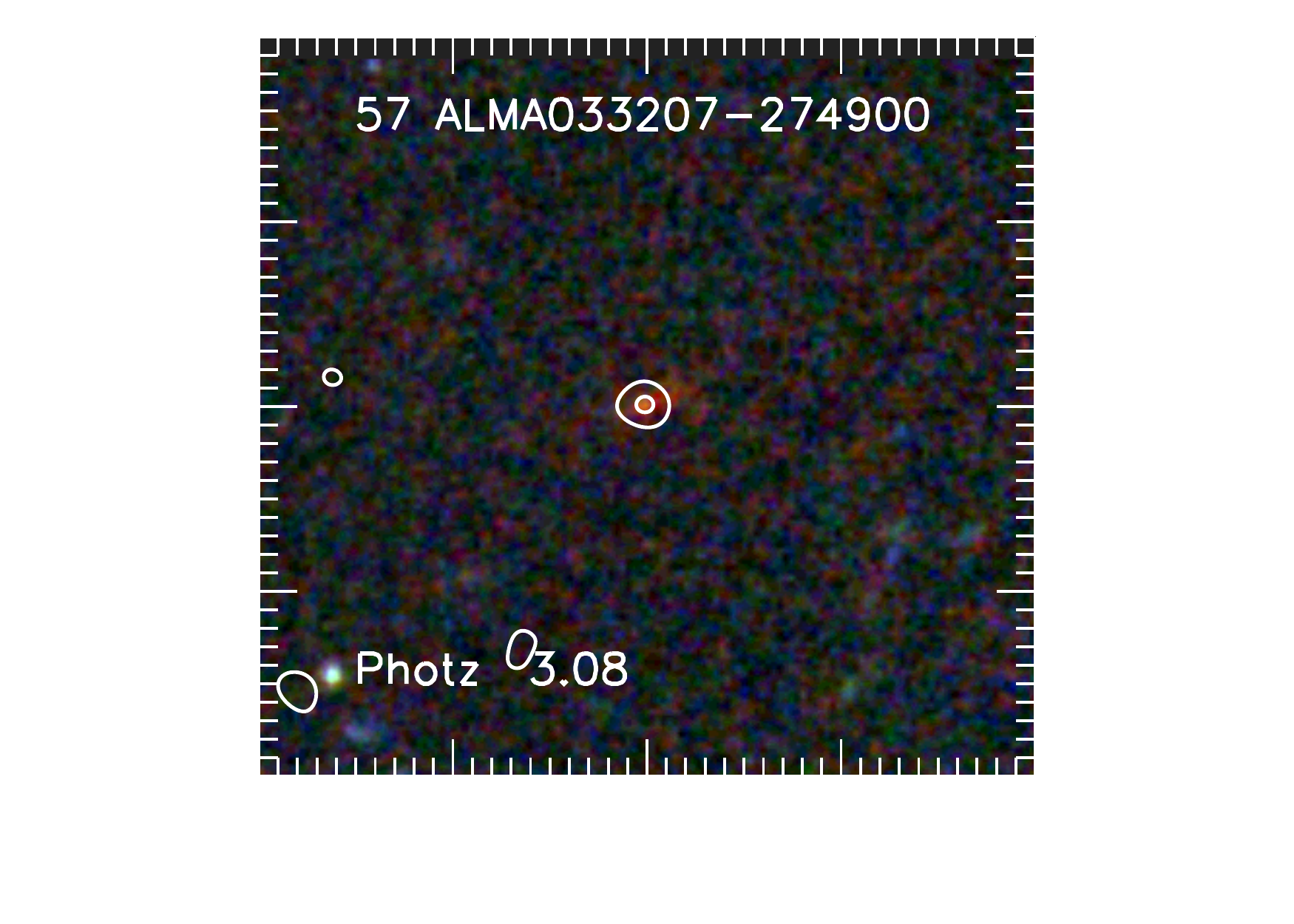}
\hspace{-2.55cm}\includegraphics[width=2.5in,angle=0]{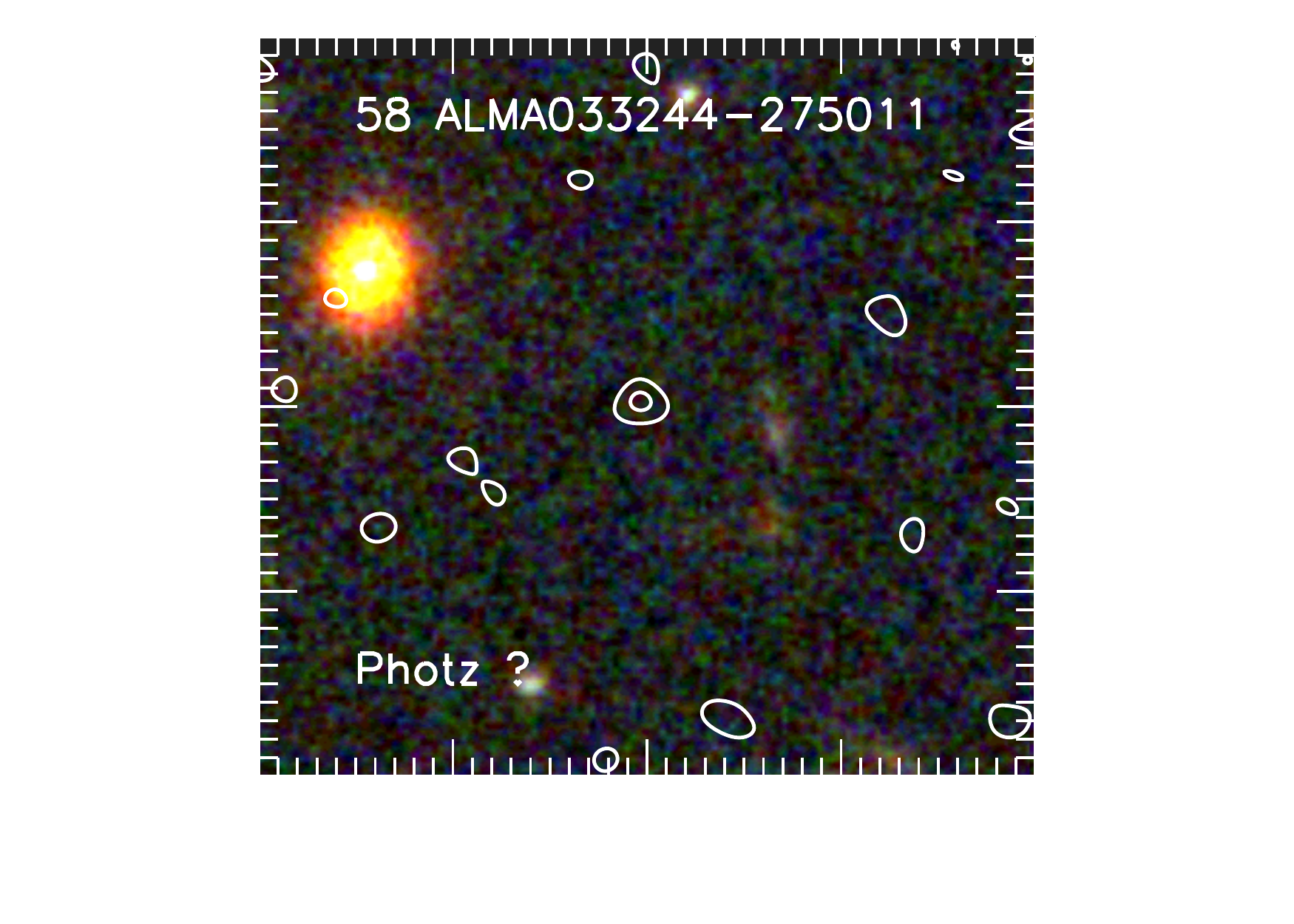}
\hspace{-2.55cm}\includegraphics[width=2.5in,angle=0]{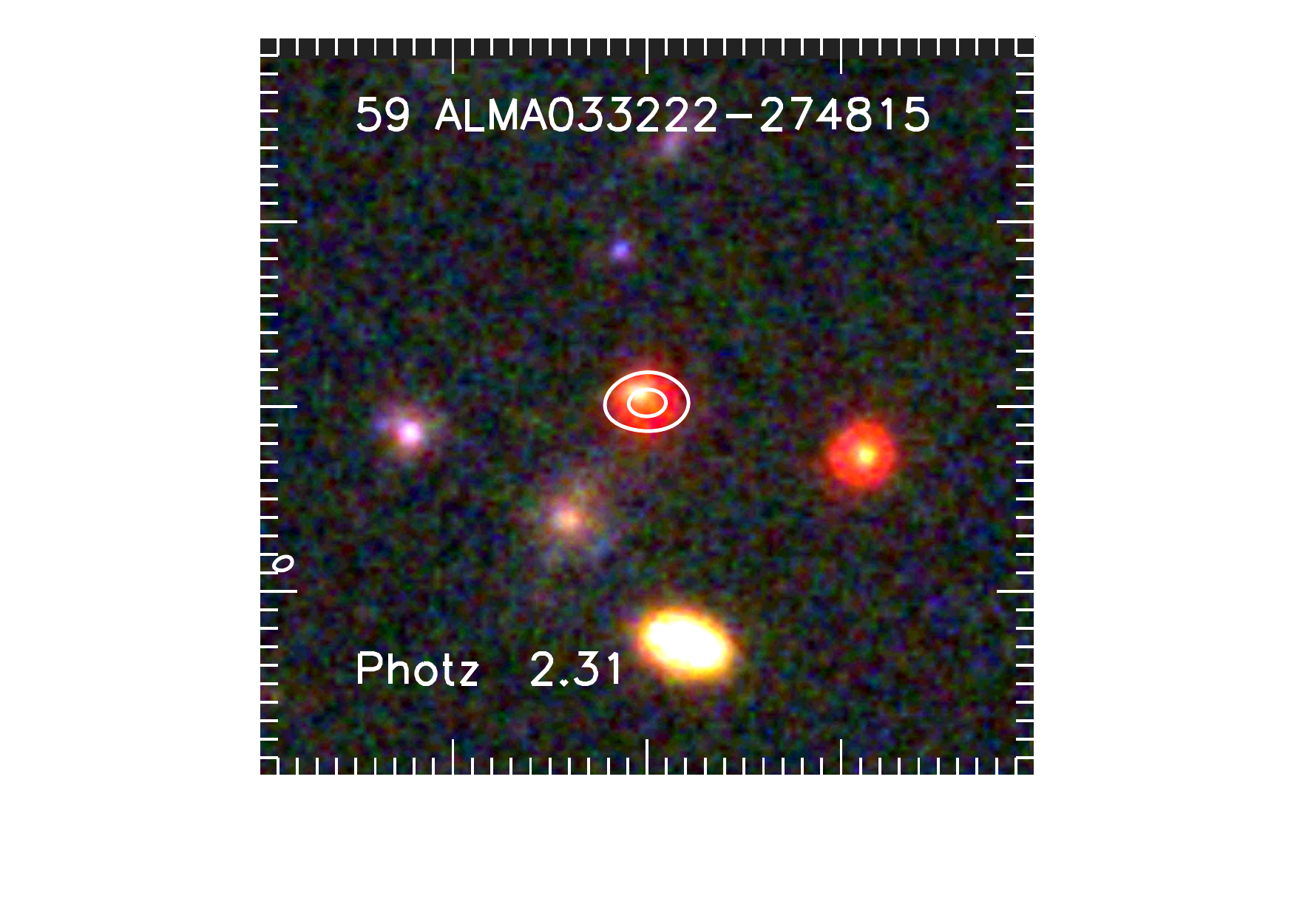}
\hspace{-2.55cm}\includegraphics[width=2.5in,angle=0]{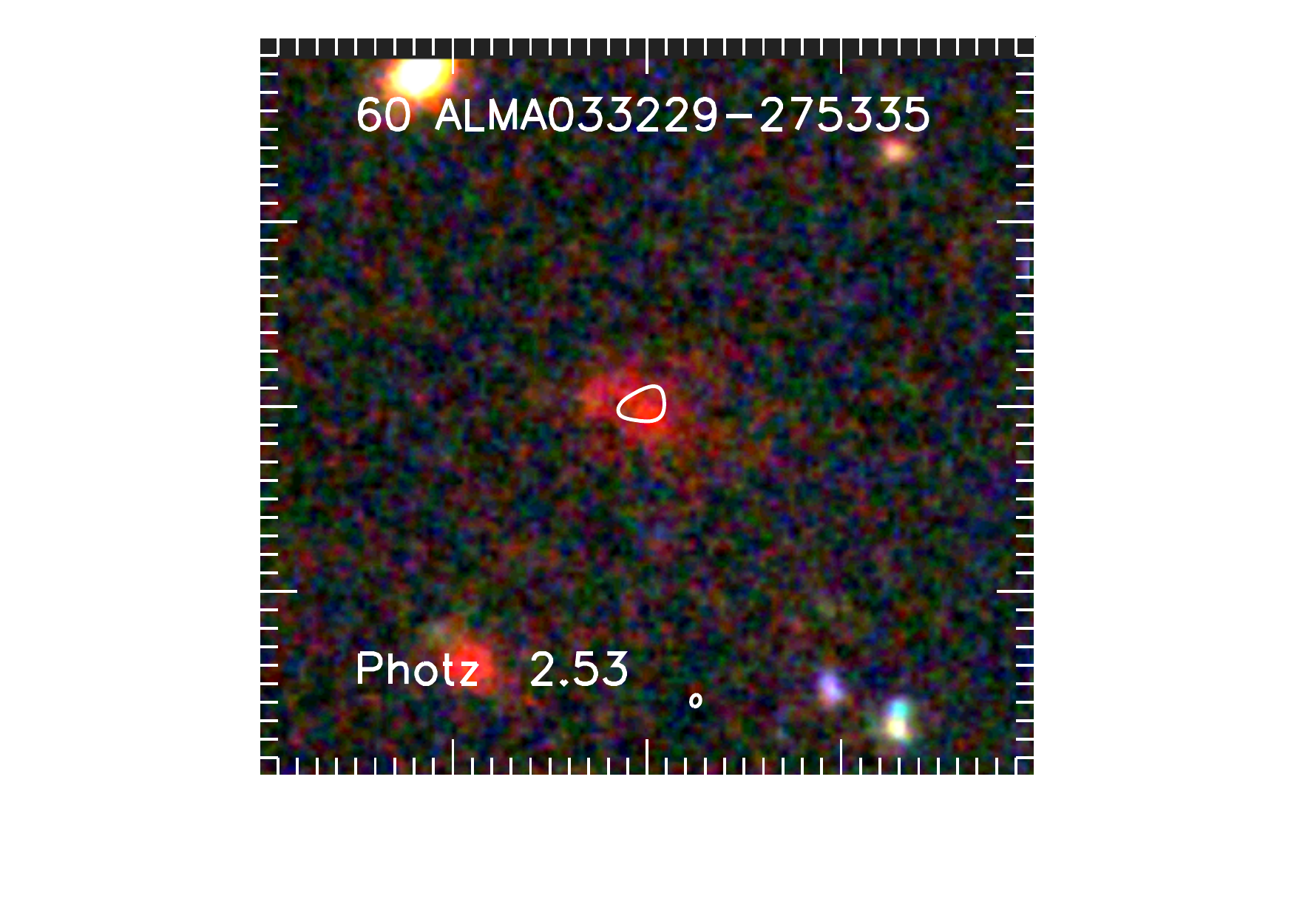}
\hspace{-2.55cm}\includegraphics[width=2.5in,angle=0]{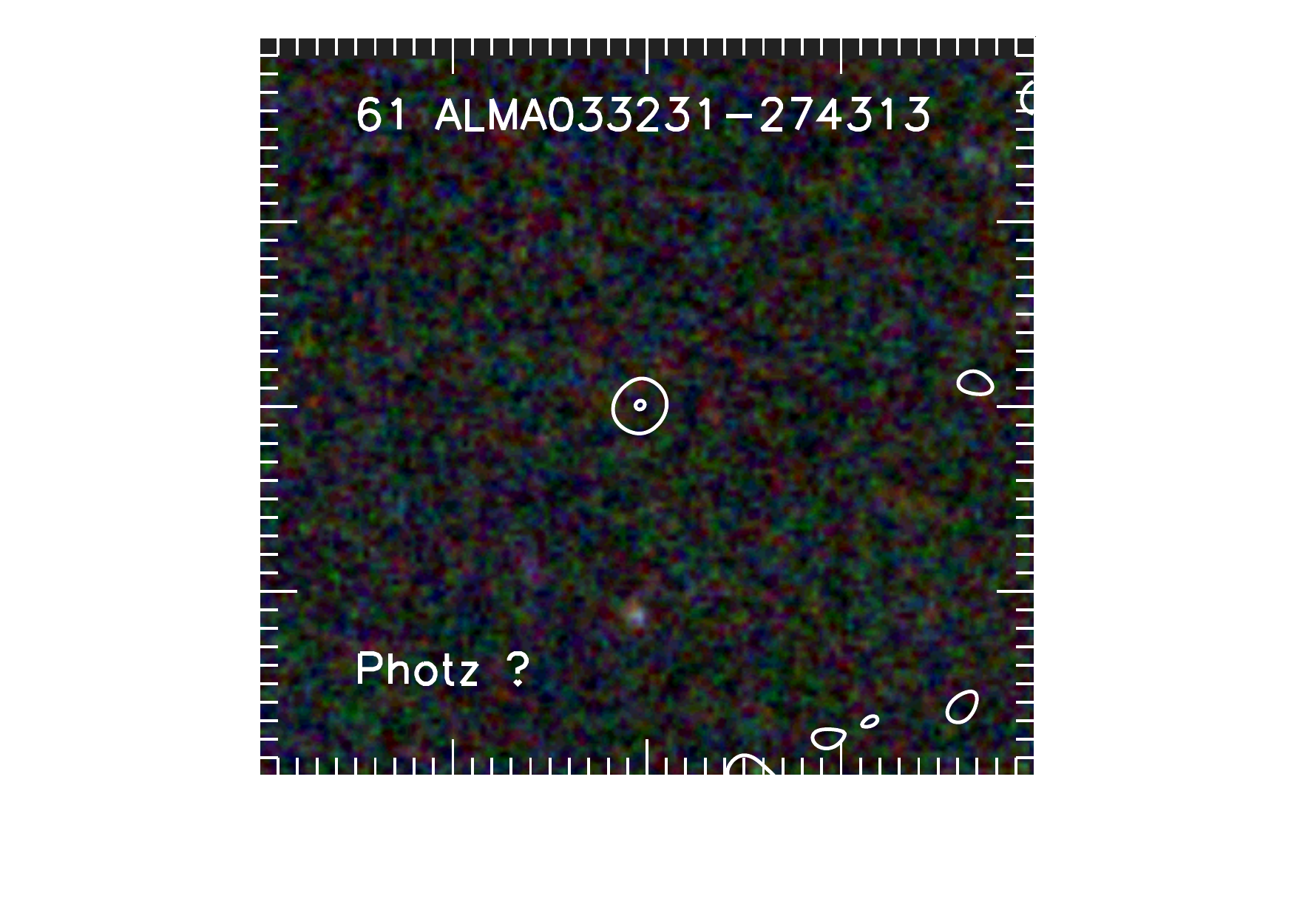}
\vskip -0.75cm
\caption{Continued
\label{basic_alma_images_3}}
\end{figure*}
\begin{figure*}
\setcounter{figure}{9}
\includegraphics[width=2.5in,angle=0]{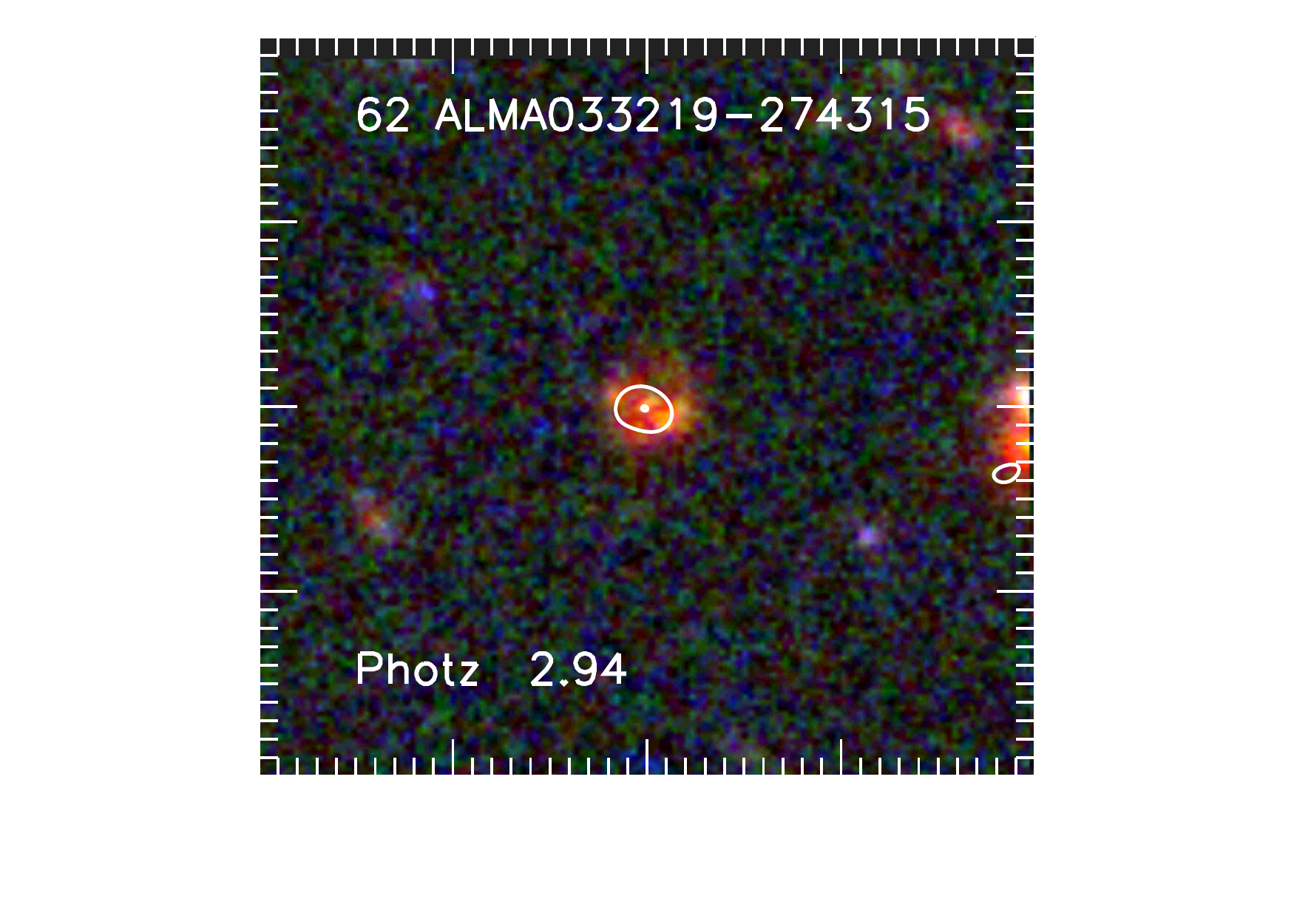}
\hspace{-3.3cm}\includegraphics[width=2.5in,angle=0]{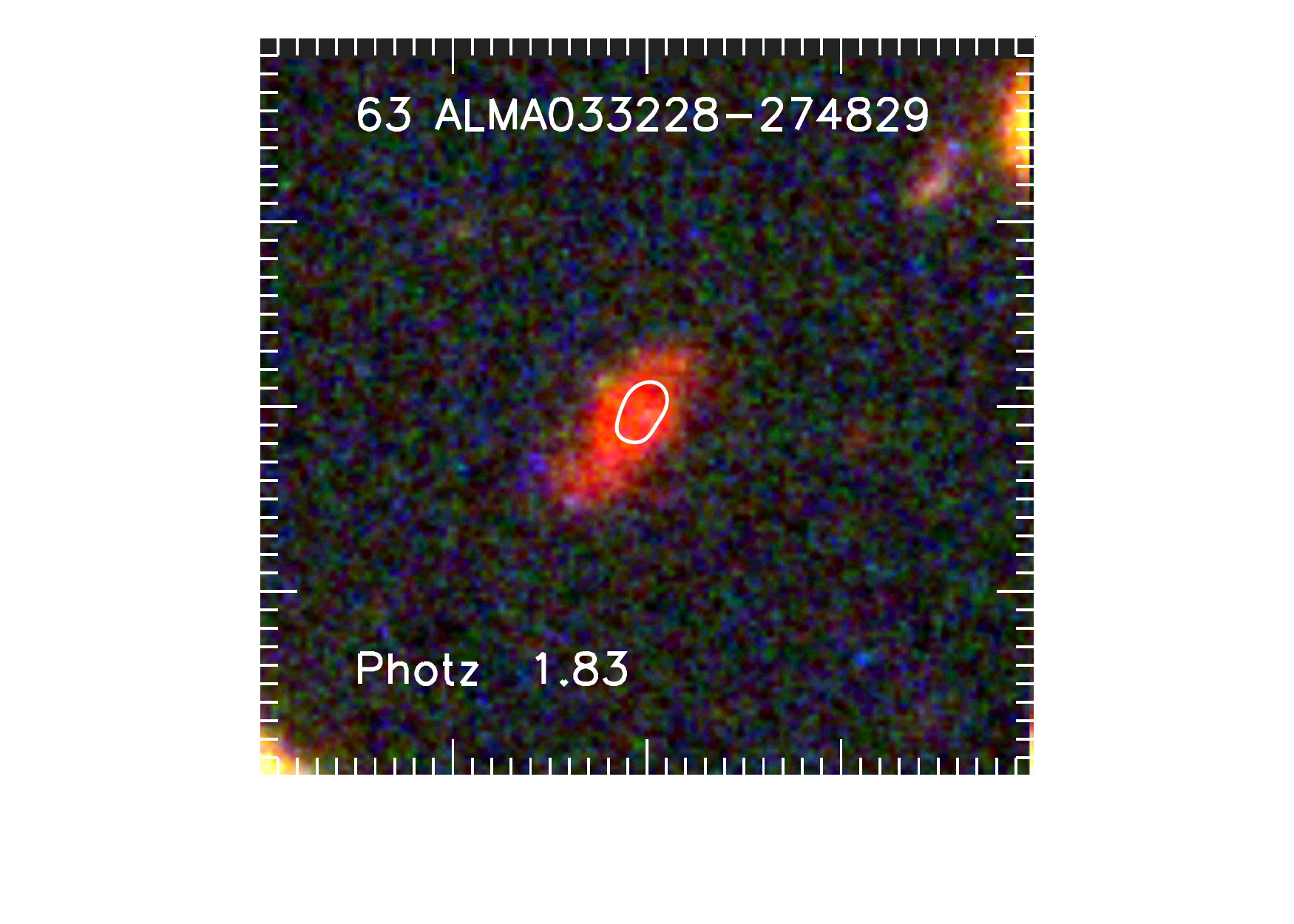}
\hspace{-3.3cm}\includegraphics[width=2.5in,angle=0]{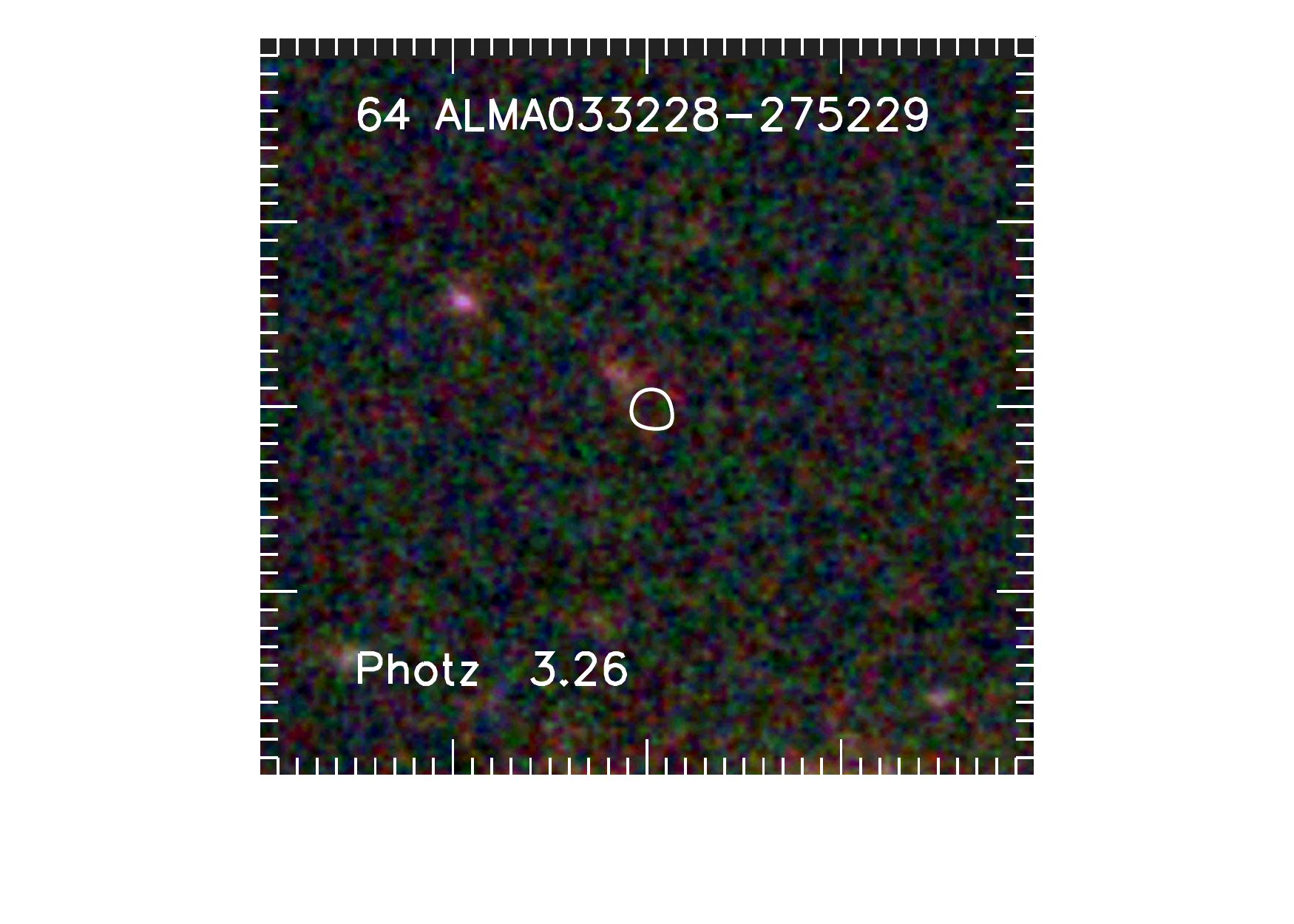}
\hspace{-3.3cm}\includegraphics[width=2.5in,angle=0]{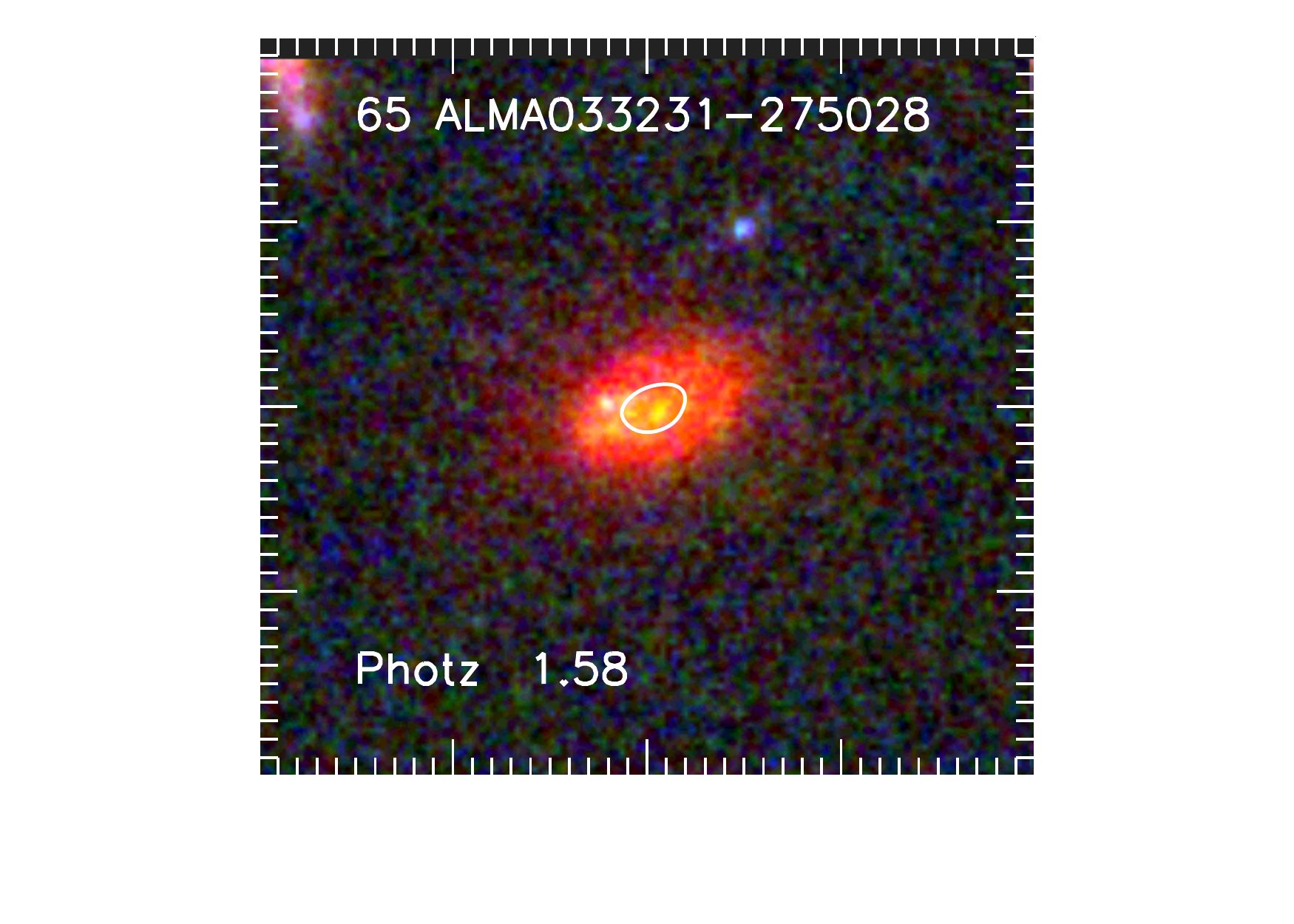}
\hspace{-3.3cm}\includegraphics[width=2.5in,angle=0]{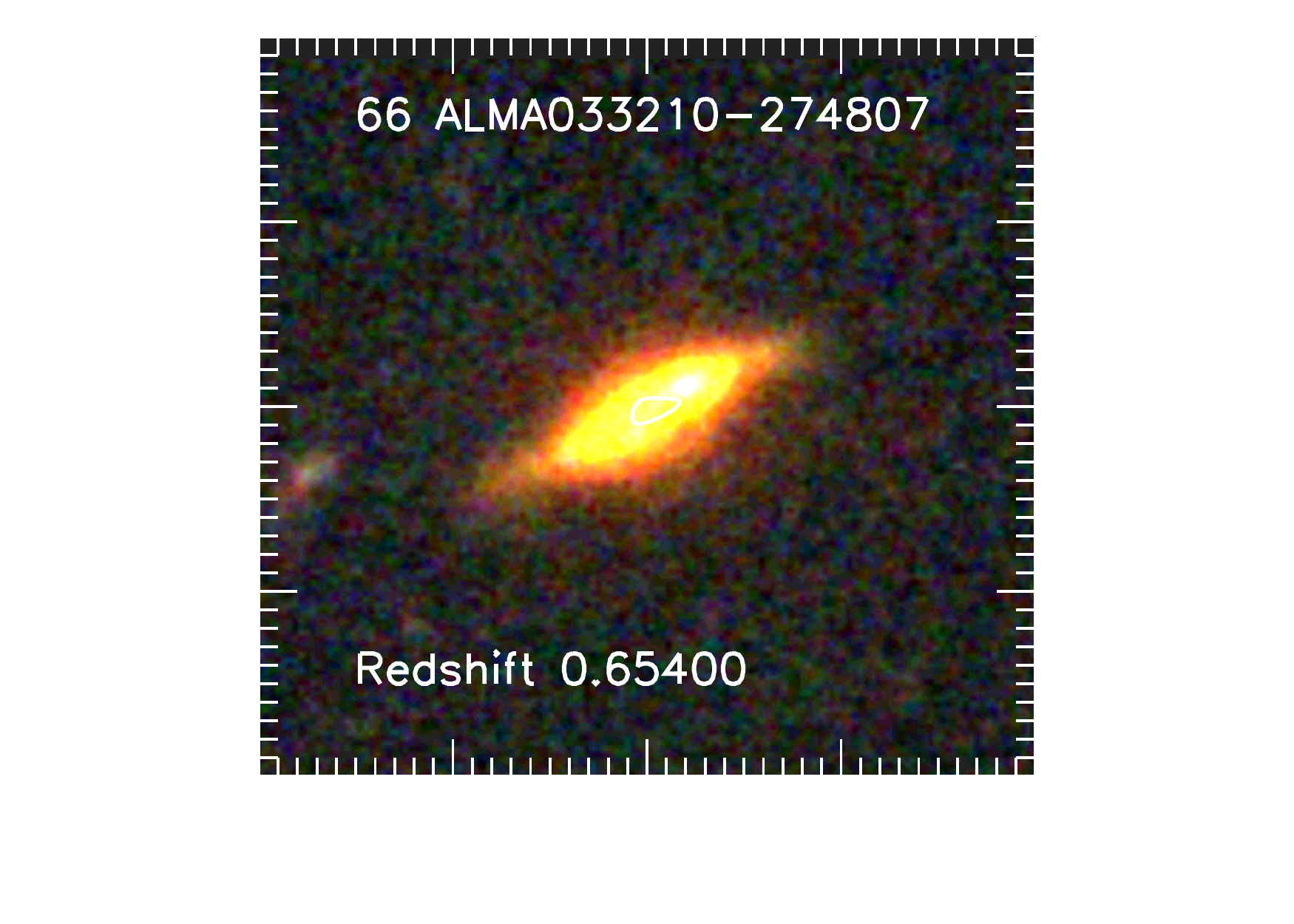}
\hspace{-3.3cm}\includegraphics[width=2.5in,angle=0]{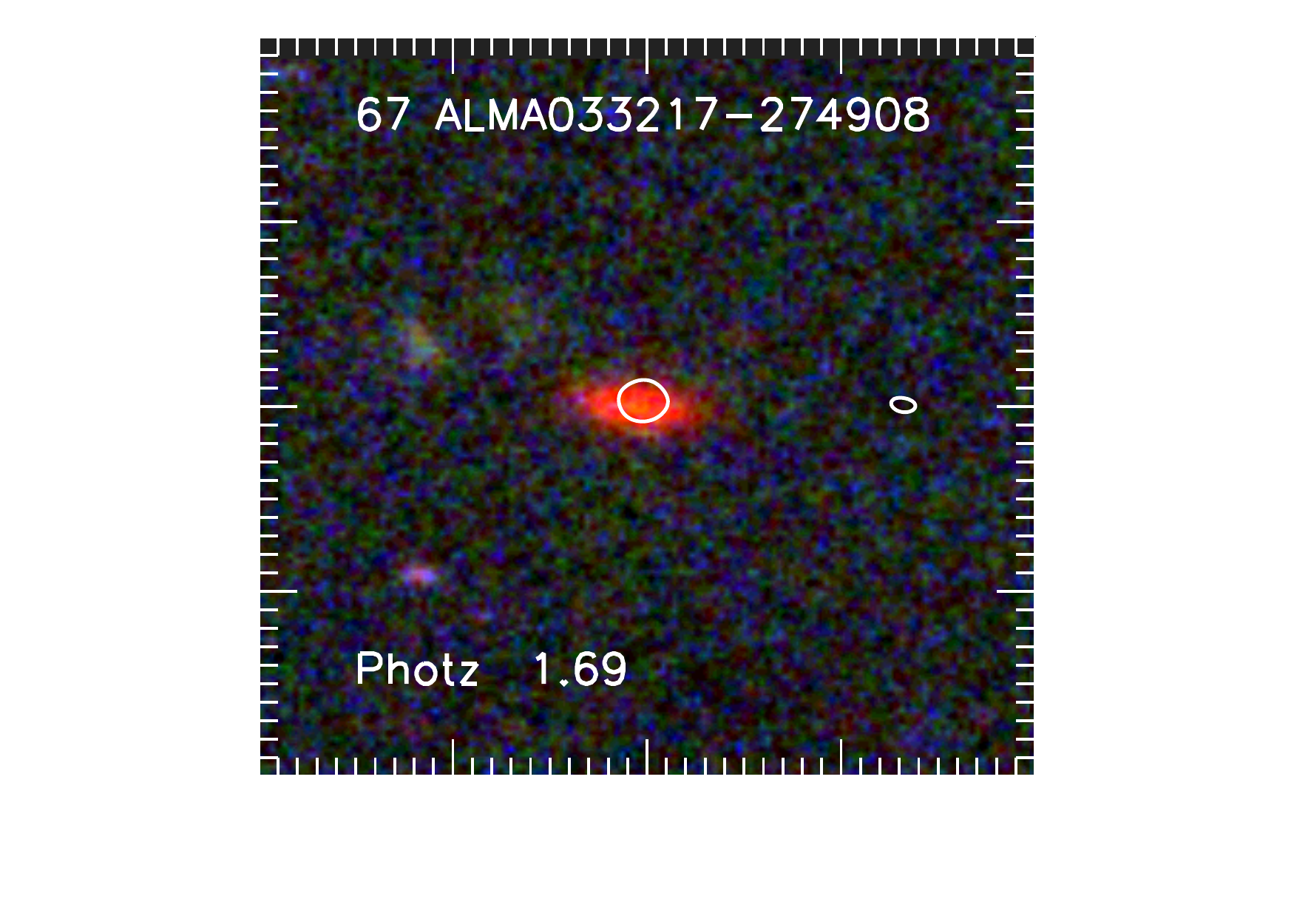}
\hspace{-3.3cm}\includegraphics[width=2.5in,angle=0]{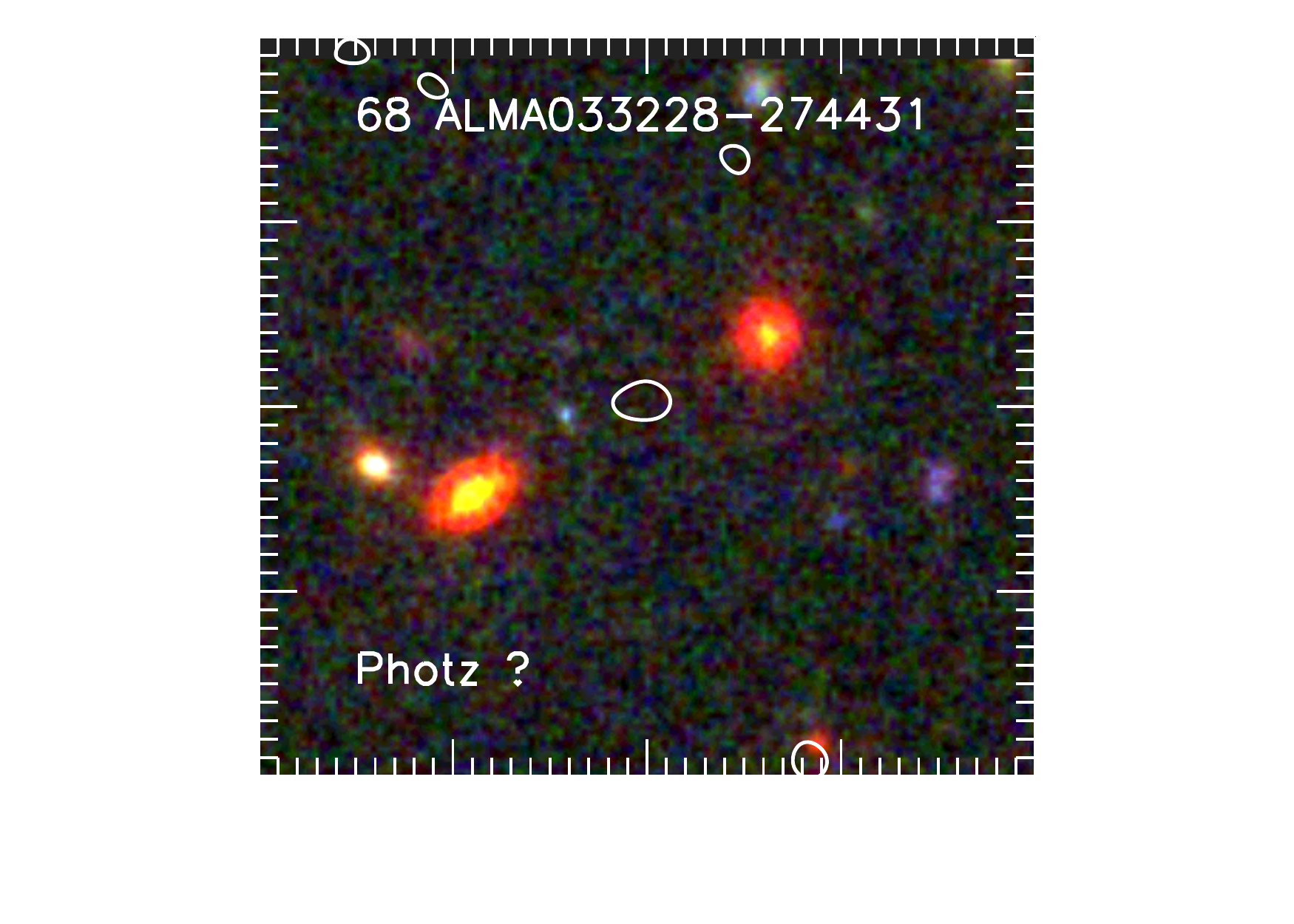}
\hspace{-3.3cm}\includegraphics[width=2.5in,angle=0]{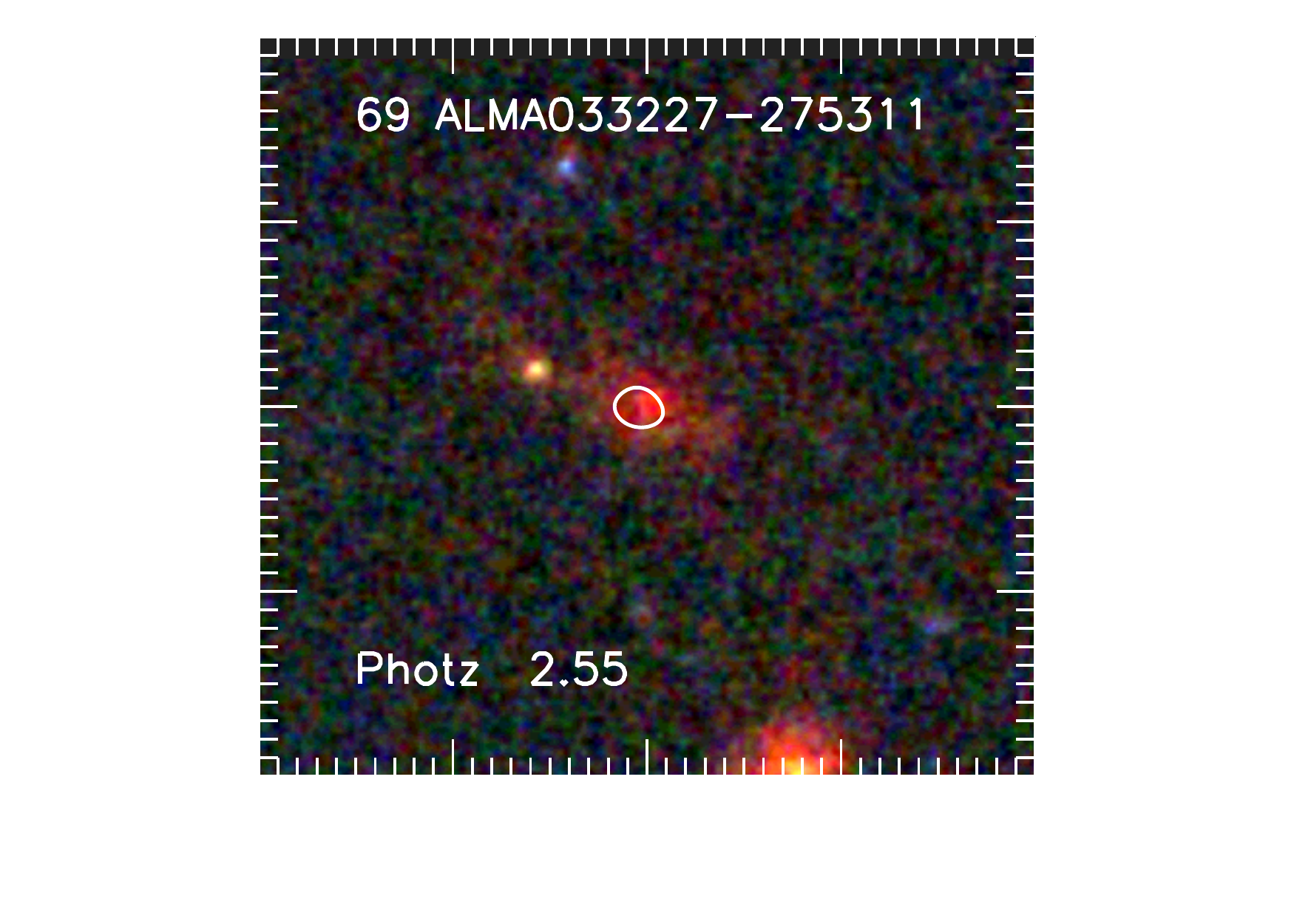}
\hspace{-2.55cm}\includegraphics[width=2.5in,angle=0]{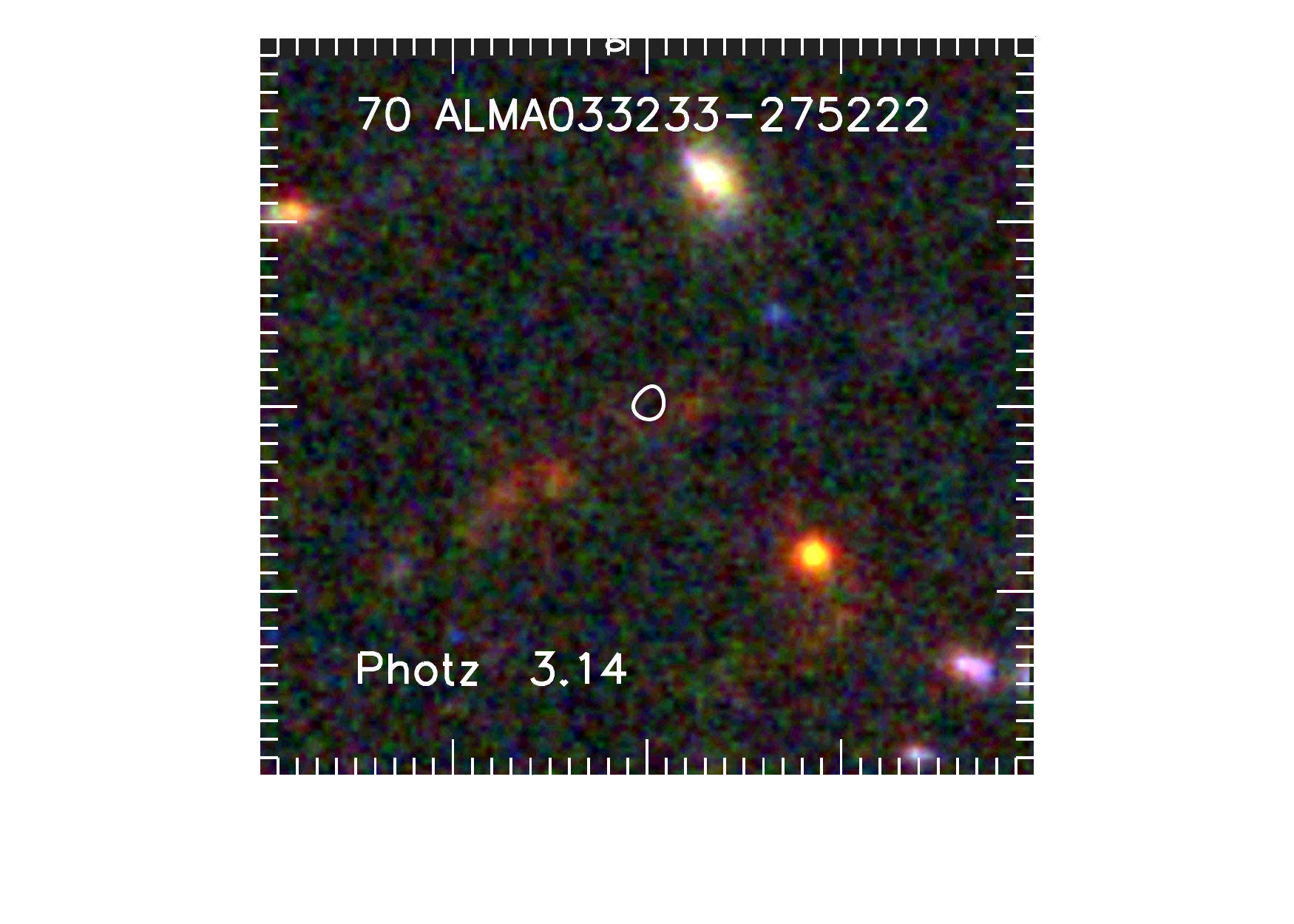}
\hspace{-2.55cm}\includegraphics[width=2.5in,angle=0]{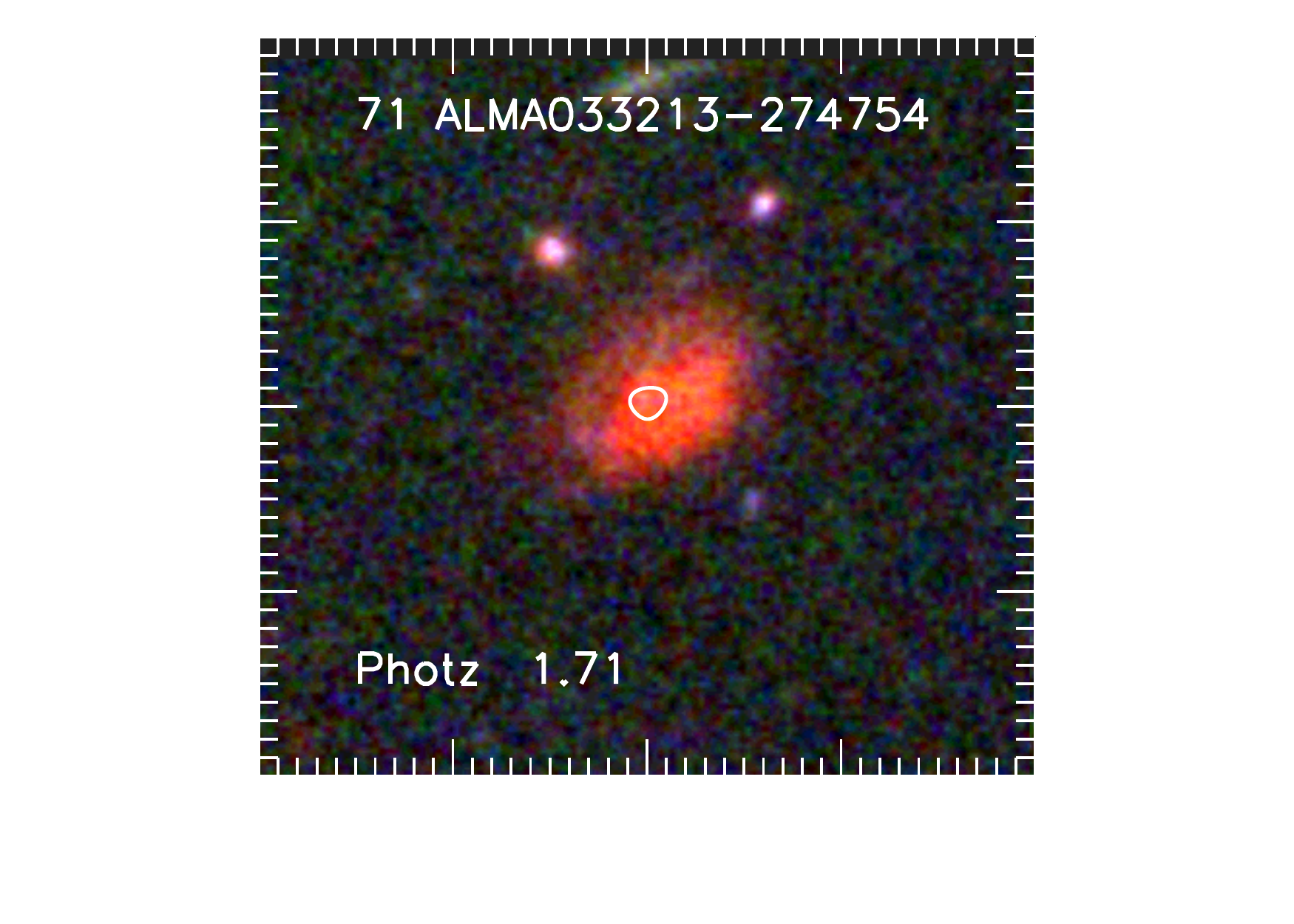}
\hspace{-2.55cm}\includegraphics[width=2.5in,angle=0]{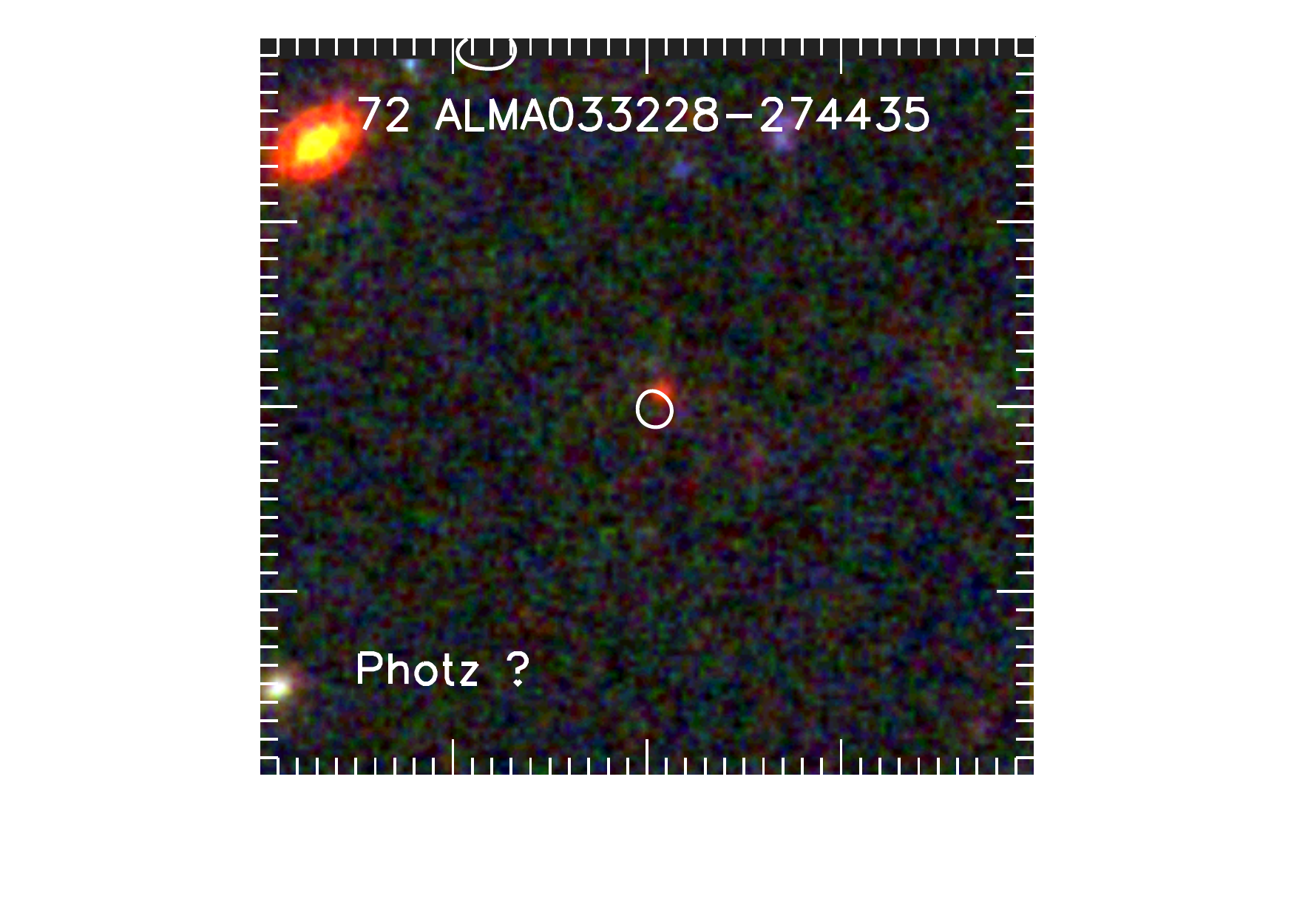}
\hspace{-2.55cm}\includegraphics[width=2.5in,angle=0]{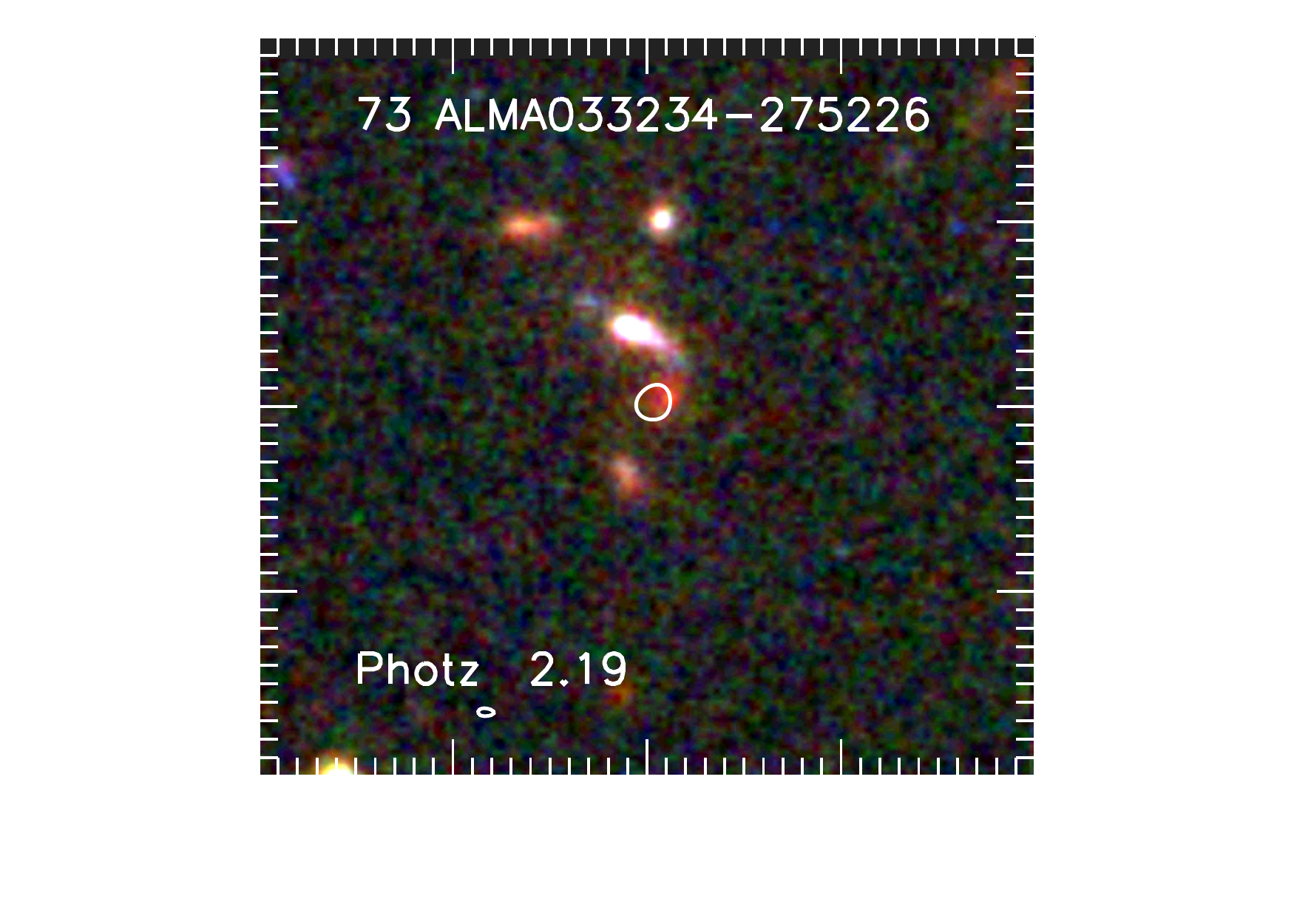}
\hspace{-2.55cm}\includegraphics[width=2.5in,angle=0]{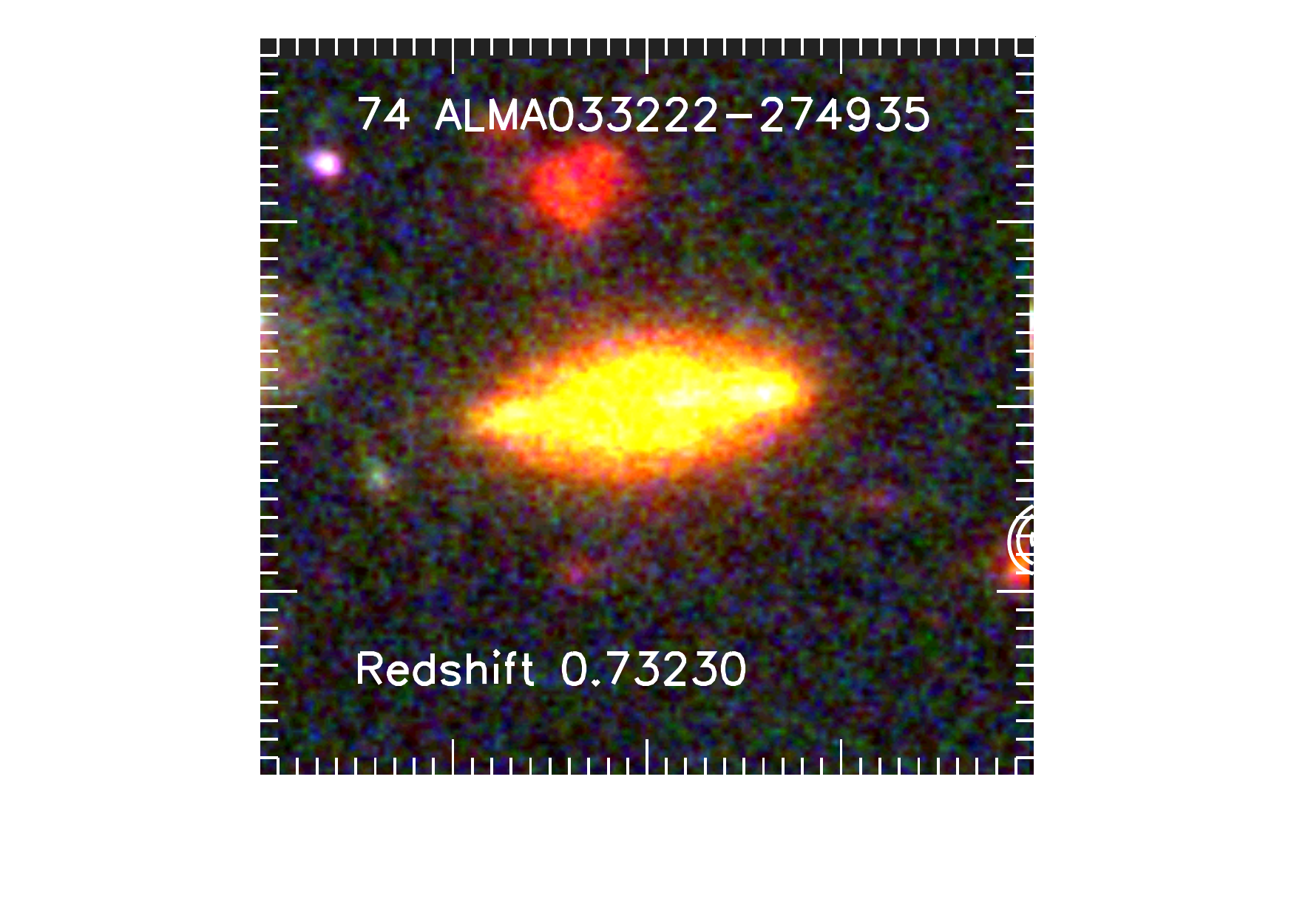}
\hspace{-2.55cm}\includegraphics[width=2.5in,angle=0]{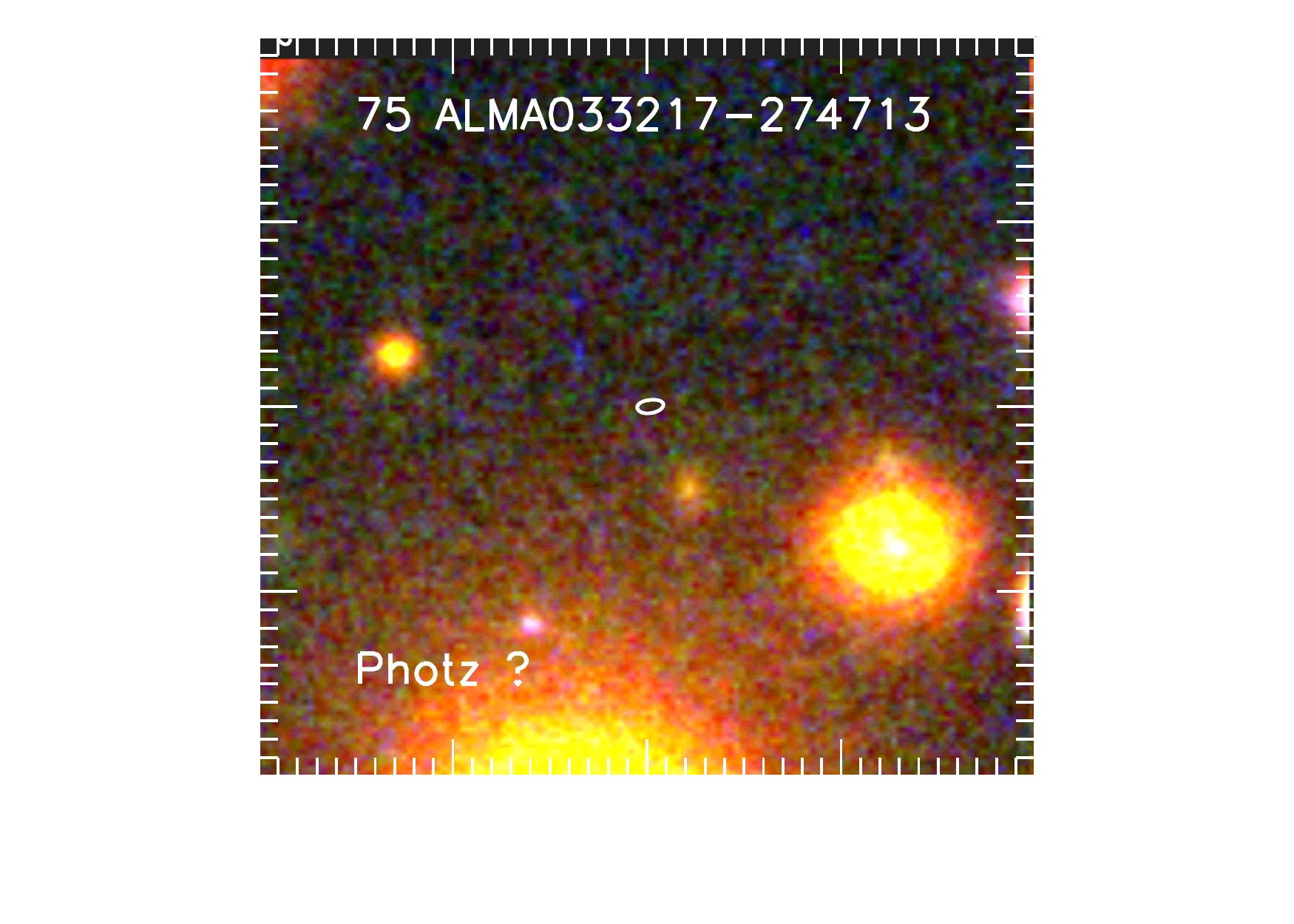}
\vskip -0.75cm
\caption{Continued
\label{basic_alma_images_4}}
\end{figure*}

\clearpage

\startlongtable
\begin{deluxetable}{ccccrrcccc}
\setcounter{table}{1}
\renewcommand\baselinestretch{1.0}
\tablewidth{0pt}
\tablecaption{Central CDF-S 850~$\mu$m Sample (rms $<0.56$~mJy, flux $>2.25$~mJy)}
\scriptsize
\tablehead{No. & Name & R.A. & Decl. &  Flux  & Error & R.A. (ALMA) & Decl. & offset & Flux (ALMA)\\ & & J2000.0 & J2000.0 & (mJy) & (mJy) & J2000.0 & J2000.0 & (arcsec) & (mJy)  \\ (1) & (2) & (3) & (4) & (5) & (6) & (7) & (8) & (9) & (10)}
\startdata
 1& SMM033211-275210 &        3       32 11.47 &      -27       52 10.8 & 10.10 & 0.45  &        3       32 11.33 &      -27       52 12.0 &  2.2 &   8.8 \cr
   2  &  SMM033207-275120 &        3       32 7.490 &      -27       51 20.8 & 9.25 & 0.47  &        3       32 7.289 &      -27       51 20.8 &  2.6 &   8.9 \cr
   3  &  SMM033205-274648 &        3       32 5.019 &      -27       46 48.8 & 7.51 & 0.47  &        3       32 4.889 &      -27       46 47.7 &  2.0 &   6.4 \cr
 4& SMM033246-275119 &        3       32 46.85 &      -27       51 19.9 & 6.76 & 0.53  &        3       32 46.83 &      -27       51 20.8 &  1.0 &   5.9 \cr
 5& SMM033222-274808 &        3       32 22.41 &      -27       48 8.99 & 6.04 & 0.31  &        3       32 22.16 &      -27       48 11.6 &  4.3 &   2.2 \cr
 5& \nodata& \nodata& \nodata & 6.04 & 0.31  &        3       32 22.28 &      -27       48 4.79 &  4.5 &   3.6\cr
 6& SMM033219-275214 &        3       32 19.24 &      -27       52 14.0 & 5.93 & 0.41  &        3       32 19.05 &      -27       52 14.9 &  2.7 &   4.7 \cr
 7& SMM033215-275038 &        3       32 15.33 &      -27       50 38.0 & 5.58 & 0.38  &        3       32 15.33 &      -27       50 37.6 & 0.4 &   6.6 \cr
 8& SMM033222-274936 &        3       32 22.41 &      -27       49 36.9 & 5.41 & 0.33  &        3       32 22.47 &      -27       49 35.2 &  1.9 &  0.93 \cr
 8& \nodata& \nodata& \nodata & 5.41 & 0.33  &        3       32 22.16 &      -27       49 36.6 &  3.4 &   4.6\cr
 9& SMM033225-275231 &        3       32 25.36 &      -27       52 31.0 & 5.18 & 0.42  &        3       32 25.25 &      -27       52 30.7 &  1.3 &   5.1 \cr
10& SMM033243-274639 &        3       32 43.82 &      -27       46 39.9 & 5.11 & 0.46  &        3       32 43.53 &      -27       46 39.2 &  3.9 &   3.1 \cr
10& \nodata& \nodata& \nodata & 5.11 & 0.46  &        3       32 44.03 &      -27       46 35.8 &  4.8 &   3.3\cr
11& SMM033234-274941 &        3       32 34.48 &      -27       49 41.0 & 4.86 & 0.38  &        3       32 34.27 &      -27       49 40.3 &  2.8 &   4.7 \cr
12& SMM033235-274914 &        3       32 35.83 &      -27       49 14.9 & 4.84 & 0.38  &        3       32 35.73 &      -27       49 16.2 &  1.8 &   5.0 \cr
13& SMM033243-275212 &        3       32 43.00 &      -27       52 12.8 & 4.69 & 0.51  &        3       32 42.80 &      -27       52 12.7 &  2.7 &   3.5 \cr
14& SMM033228-274658 &        3       32 28.60 &      -27       46 58.0 & 4.50 & 0.34  &        3       32 28.50 &      -27       46 58.3 &  1.2 &   6.3 \cr
15& SMM033216-274344 &        3       32 16.39 &      -27       43 44.0 & 4.17 & 0.48  &        3       32 16.27 &      -27       43 44.0 &  1.5 &   1.9 \cr
16& SMM033211-274614 &        3       32 11.80 &      -27       46 14.8 & 4.10 & 0.42  &        3       32 11.94 &      -27       46 15.4 &  1.9 &   2.5 \cr
16& \nodata& \nodata& \nodata & 4.10 & 0.42  &        3       32 11.61 &      -27       46 13.1 &  3.0 &   1.7\cr
17& SMM033217-275039 &        3       32 17.22 &      -27       50 39.9 & 3.85 & 0.38  &        3       32 17.21 &      -27       50 37.2 &  2.7 &   2.7 \cr
18& SMM033226-275206 &        3       32 26.27 &      -27       52 6.99 & 3.84 & 0.42  &        3       32 26.11 &      -27       52 8.50 &  2.5 &   3.6 \cr
19& SMM033218-275228 &        3       32 18.12 &      -27       52 28.9 & 3.68 & 0.42  &        3       32 17.96 &      -27       52 33.2 &  4.7 &   4.6 \cr
20& SMM033238-274401 &        3       32 38.08 &      -27       44 1.00 & 3.65 & 0.48  &        3       32 38.00 &      -27       44 1.00 &  1.0 &   5.6 \cr
21& SMM033219-274604 &        3       32 19.79 &      -27       46 4.00 & 3.62 & 0.39  &        3       32 19.70 &      -27       46 2.20 &  2.1 &   4.9 \cr
22& SMM033237-275000 &        3       32 37.87 &      -27       50 0.99 & 3.49 & 0.40  &        3       32 37.72 &      -27       50 0.59 &  2.0 &   3.3 \cr
23& SMM033247-275038 &        3       32 47.90 &      -27       50 38.9 & 3.48 & 0.52  &        3       32 47.73 &      -27       50 38.2 &  2.3 &   1.8 \cr
24& SMM033230-275334 &        3       32 30.11 &      -27       53 34.9 & 3.42 & 0.47  &        3       32 29.90 &      -27       53 35.8 &  2.9 &   1.6 \cr
25& SMM033217-275002 &        3       32 17.60 &      -27       50 2.00 & 3.42 & 0.37  &        3       32 17.45 &      -27       50 3.40 &  2.4 &   2.4 \cr
26& SMM033221-275024 &        3       32 21.13 &      -27       50 24.0 & 3.38 & 0.37  &        3       32 20.91 &      -27       50 24.7 &  2.9 &   2.2 \cr
27& SMM033224-275334 &        3       32 24.91 &      -27       53 34.9 & 3.31 & 0.46  &        3       32 24.66 &      -27       53 34.2 &  3.2 &   3.2 \cr
28& SMM033216-275247 &        3       32 16.69 &      -27       52 47.0 & 3.30 & 0.44  &        3       32 16.52 &      -27       52 47.0 &  2.1 &   2.2 \cr
29& SMM033227-275312 &        3       32 27.24 &      -27       53 12.0 & 3.29 & 0.44  &        3       32 27.15 &      -27       53 11.8 &  1.3 &   1.2 \cr
  30  &  SMM033205-274819 &        3       32 5.990 &      -27       48 19.8 & 3.23 & 0.45  &        3       32 5.830 &      -27       48 20.5 &  2.2 &   3.9 \cr
31& SMM033234-275225 &        3       32 34.11 &      -27       52 25.0 & 3.19 & 0.43  &        3       32 34.28 &      -27       52 26.7 &  2.9 &   1.0 \cr
31& \nodata& \nodata& \nodata & 3.19 & 0.43  &        3       32 33.90 &      -27       52 22.2 &  3.8 &   1.1\cr
32& SMM033212-275209 &        3       32 12.91 &      -27       52 9.89 & 3.17 & 0.44  &        3       32 12.88 &      -27       52 9.40 & 0.71 &   1.9 \cr
33& SMM033219-275201 &        3       32 19.92 &      -27       52 1.00 & 3.08 & 0.41  &        3       32 19.86 &      -27       51 59.7 &  1.5 &   3.9 \cr
  34  &  SMM033203-275039 &        3       32 3.349 &      -27       50 39.8 & 3.02 & 0.52  &        3       32 3.500 &      -27       50 39.8 &  1.9 &   3.0 \cr
35& SMM033219-274314 &        3       32 19.33 &      -27       43 14.9 & 2.98 & 0.49  &        3       32 19.36 &      -27       43 15.1 & 0.29 &   1.5 \cr
36& SMM033233-274544 &        3       32 33.04 &      -27       45 44.0 & 2.80 & 0.42  &        3       32 32.91 &      -27       45 40.9 &  3.5 &   2.8 \cr
37& SMM033231-274624 &        3       32 31.32 &      -27       46 24.9 & 2.78 & 0.40  &        3       32 31.46 &      -27       46 23.5 &  2.4 &   2.2 \cr
38& SMM033229-274433 &        3       32 29.04 &      -27       44 33.0 & 2.76 & 0.47  &        3       32 28.91 &      -27       44 31.4 &  2.3 &   1.3 \cr
38& \nodata& \nodata& \nodata & 2.76 & 0.47  &        3       32 28.78 &      -27       44 35.2 &  4.1 &   1.1\cr
39& SMM033235-275318 &        3       32 35.24 &      -27       53 18.9 & 2.71 & 0.49  &        3       32 35.13 &      -27       53 19.7 &  1.6 &   2.3 \cr
40& SMM033229-275258 &        3       32 29.96 &      -27       52 58.0 & 2.68 & 0.44  &        3       32 29.83 &      -27       52 57.7 &  1.6 &   2.2 \cr
41& SMM033221-274244 &        3       32 21.90 &      -27       42 44.0 & 2.68 & 0.52  &        3       32 22.01 &      -27       42 43.7 &  1.5 &   2.4 \cr
42& SMM033235-275217 &        3       32 35.32 &      -27       52 17.0 & 2.64 & 0.43  &        3       32 35.19 &      -27       52 15.7 &  2.1 &   3.8 \cr
43& SMM033241-275133 &        3       32 41.71 &      -27       51 33.0 & 2.57 & 0.45  &        3       32 41.47 &      -27       51 31.8 &  3.3 &   2.2 \cr
44& SMM033215-274629 &        3       32 15.27 &      -27       46 29.0 & 2.55 & 0.40  & \nodata & \nodata &\nodata & [ 0.12] \cr
45& SMM033231-275027 &        3       32 31.69 &      -27       50 27.0 & 2.54 & 0.39  &        3       32 31.55 &      -27       50 28.9 &  2.7 &   1.4 \cr
46& SMM033229-274301 &        3       32 29.57 &      -27       43 1.99 & 2.49 & 0.49  & \nodata & \nodata &\nodata & [ 0.22] \cr
47& SMM033233-275326 &        3       32 33.58 &      -27       53 26.9 & 2.45 & 0.49  &        3       32 33.42 &      -27       53 26.6 &  2.0 &   2.8 \cr
48& SMM033225-274306 &        3       32 25.88 &      -27       43 6.00 & 2.38 & 0.48  &        3       32 25.69 &      -27       43 6.00 &  2.6 &   1.7 \cr
  49  &  SMM033204-275007 &        3       32 4.629 &      -27       50 7.89 & 2.32 & 0.47  & \nodata & \nodata &\nodata & [ 0.10] \cr
50& SMM033215-274514 &        3       32 15.80 &      -27       45 14.0 & 2.32 & 0.41  & \nodata & \nodata &\nodata & [ 0.12] \cr
51& SMM033238-274634 &        3       32 38.61 &      -27       46 34.9 & 2.32 & 0.42  &        3       32 38.55 &      -27       46 34.5 & 0.94 &   2.0 \cr
52& SMM033218-275131 &        3       32 18.72 &      -27       51 31.9 & 2.29 & 0.40  &        3       32 18.57 &      -27       51 34.6 &  3.3 &   2.7 \cr
  53  &  SMM033208-274646 &        3       32 8.789 &      -27       46 46.8 & 2.25 & 0.45  & \nodata & \nodata &\nodata & [ 0.37] \cr
\enddata
\tablecomments{The columns are (1) the SCUBA-2 source number, (2) the SCUBA-2 source name, (3 and 4)
the SCUBA-2 R.A. and Decl., (5 and 6) the flux and error measured from the SCUBA-2 
matched-filter image, (7 and 8) the R.A. and Decl. of the 
corresponding ALMA source(s), (9) the offset between the SCUBA-2 and ALMA source positions, 
and (10) the ALMA 850~$\mu$m flux. In Column~(10), for the 5 cases where the ALMA observations
yielded no ALMA detections, we list the central $1\sigma$ noise of the ALMA image (enclosed in 
square brackets) instead of an ALMA flux detection.
}
\label{tab2}
\end{deluxetable}

\clearpage

\startlongtable
\pagestyle{empty}
\begin{deluxetable*}{ccccccclcccccc}
\setcounter{table}{3}
\renewcommand\baselinestretch{1.0}
\tablewidth{0pt}
\tablecaption{CDF-S Band 7 ALMA $>4.5\sigma$}
\scriptsize
\tablehead{No. & Name & R.A. & Decl. &  Pk Flux  & Error & S/N & Flux & Error & 1.13~mm  & SCUBA-2 & offset \\ & & J2000.0 & J2000.0 & (mJy) & (mJy) &  & (mJy) & (mJy) & (mJy)  & No. & (arcsec) & \\ (1) & (2) & (3) & (4) & (5) & (6) & (7) & (8) & (9) & (10) & (11) & (12)  }
\startdata
   1 &  ALMA033207-275120 &        3       32 7.289 &      -27       51 20.8 &  5.97 & 0.14 &  42.1 &  8.93 H & 0.21 & \nodata &      2 &   2.6\cr
   2 &  ALMA033211-275212 &        3       32 11.33 &      -27       52 12.0 &  6.00 & 0.12 &  47.2 &  8.83 H & 0.26 & \nodata &      1 &   2.2\cr
   3 &  ALMA033215-275037 &        3       32 15.33 &      -27       50 37.6 &  5.02 & 0.12 &  41.1 &  6.61 & 0.16 &  2.1 &      7 &  0.39\cr
   4 &  ALMA033204-274647 &        3       32 4.889 &      -27       46 47.7 &  4.39 & 0.15 &  28.1 &  6.45 B & 0.41 &  1.9 &      3 &   2.0\cr
   5 &  ALMA033228-274658 &        3       32 28.50 &      -27       46 58.3 &  4.86 & 0.12 &  38.2 &  6.39 & 0.16 &  1.9 &     14 &   1.2\cr
   6 &  ALMA033246-275120 &        3       32 46.83 &      -27       51 20.8 &  4.10 & .084 &  48.7 &  5.90 S & 0.18 & \nodata &      4 &   1.0\cr
   7 &  ALMA033238-274401 &        3       32 38.00 &      -27       44 1.00 &  4.25 & 0.10 &  39.7 &  5.60 & 0.14 &  1.4 &     20 &   1.0\cr
   8 &  ALMA033225-275230 &        3       32 25.25 &      -27       52 30.7 &  3.45 & 0.10 &  32.5 &  5.18 H & 0.22 & \nodata &      9 &   1.3\cr
   9 &  ALMA033235-274916 &        3       32 35.73 &      -27       49 16.2 &  3.86 & .096 &  39.9 &  5.09 & 0.12 &  2.1 &     12 &   1.8\cr
  10 &  ALMA033219-274602 &        3       32 19.70 &      -27       46 2.20 &  3.15 & 0.14 &  21.8 &  4.90 & 0.29 &  1.0 &     21 &   2.1\cr
  11 &  ALMA033219-275214 &        3       32 19.05 &      -27       52 14.9 &  2.79 & 0.14 &  19.3 &  4.76 H & 0.29 &  1.3 &      6 &   2.7\cr
  12 &  ALMA033234-274940 &        3       32 34.27 &      -27       49 40.3 &  3.59 & 0.12 &  28.0 &  4.73 & 0.16 &  1.6 &     11 &   2.8\cr
  13 &  ALMA033217-275233 &        3       32 17.96 &      -27       52 33.2 &  3.56 & 0.30 &  11.7 &  4.69 & 0.81 & 0.64 &     19 &   4.7\cr
  14 &  ALMA033222-274936 &        3       32 22.16 &      -27       49 36.6 &  3.52 & 0.13 &  25.9 &  4.64 & 0.17 &  1.1 &      8 &   3.4\cr
  15 &  ALMA033205-274820 &        3       32 5.830 &      -27       48 20.5 &  2.98 & 0.11 &  25.8 &  3.93 & 0.15 & \nodata &     30 &   2.1\cr
  16 &  ALMA033219-275159 &        3       32 19.86 &      -27       51 59.7 &  3.52 & 0.11 &  31.7 &  3.91 M & 0.13 &  1.2 &     33 &   1.5\cr
  17 &  ALMA033235-275215 &        3       32 35.19 &      -27       52 15.7 &  2.89 & 0.14 &  20.1 &  3.80 & 0.18 & \nodata &     41 &   2.1\cr
  18 &  ALMA033222-274804 &        3       32 22.28 &      -27       48 4.79 &  2.87 & 0.12 &  22.8 &  3.64 M & 0.13 &  1.3 &      5 &   4.5\cr
  19 &  ALMA033226-275208 &        3       32 26.11 &      -27       52 8.50 &  2.75 & 0.12 &  21.2 &  3.62 & 0.17 &  1.5 &     18 &   2.5\cr
  [20] &  ALMA033247-274452 &        3       32 47.59 &      -27       44 52.3 &  2.88 & 0.24 &  11.7 &  3.61 & 0.30 & \nodata & \nodata & \nodata\cr
  21 &  ALMA033242-275212 &        3       32 42.80 &      -27       52 12.7 &  2.36 & .098 &  24.0 &  3.55 & 0.20 & \nodata &     13 &   2.7\cr
  22 &  ALMA033244-274635 &        3       32 44.03 &      -27       46 35.8 &  2.69 & 0.26 &  10.2 &  3.38 & 0.32 &  1.2 &     10 &   4.8\cr
  23 &  ALMA033237-275000 &        3       32 37.72 &      -27       50 0.59 &  2.33 & 0.14 &  16.4 &  3.32 & 0.29 & \nodata &     22 &   2.0\cr
  24 &  ALMA033224-275334 &        3       32 24.66 &      -27       53 34.2 &  2.47 & 0.10 &  22.7 &  3.25 & 0.14 & \nodata &     27 &   3.2\cr
  25 &  ALMA033243-274639 &        3       32 43.53 &      -27       46 39.2 &  2.54 & 0.18 &  13.3 &  3.18 & 0.23 & 0.79 &     10 &   3.8\cr
  26 &  ALMA033216-275044 &        3       32 16.86 &      -27       50 44.2 &  2.40 & 0.19 &  12.3 &  3.15 & 0.25 & 0.88 & \nodata & \nodata\cr
  27 &  ALMA033203-275039 &        3       32 3.500 &      -27       50 39.8 &  2.32 & 0.14 &  15.5 &  3.05 & 0.19 & \nodata &     34 &   1.9\cr
  28 &  ALMA033233-275326 &        3       32 33.42 &      -27       53 26.6 &  2.30 & 0.29 &  7.7 &  2.89 & 0.37 & \nodata &     46 &   2.0\cr
  29 &  ALMA033232-274540 &        3       32 32.91 &      -27       45 40.9 &  1.56 & 0.13 &  11.4 &  2.82 & 0.28 & \nodata &     36 &   3.4\cr
  30 &  ALMA033217-275037 &        3       32 17.21 &      -27       50 37.2 &  2.11 & 0.11 &  18.1 &  2.78 & 0.15 & \nodata &     17 &   2.6\cr
  31 &  ALMA033218-275134 &        3       32 18.57 &      -27       51 34.6 &  1.96 & 0.19 &  10.0 &  2.72 M & 0.51 & \nodata &     51 &   3.3\cr
  32 &  ALMA033211-274615 &        3       32 11.94 &      -27       46 15.4 &  1.95 & 0.12 &  15.9 &  2.56 & 0.16 & \nodata &     16 &   1.9\cr
  33 &  ALMA033217-275003 &        3       32 17.45 &      -27       50 3.40 &  1.63 & 0.11 &  14.4 &  2.49 & 0.23 & \nodata &     25 &   2.4\cr
  34 &  ALMA033221-274656 &        3       32 21.78 &      -27       46 56.9 &  1.88 & 0.16 &  11.5 &  2.47 & 0.21 & \nodata & \nodata & \nodata\cr
  35 &  ALMA033222-274243 &        3       32 22.01 &      -27       42 43.7 &  1.88 & 0.10 &  17.9 &  2.47 & 0.13 & \nodata & \nodata & \nodata\cr
  36 &  ALMA033220-274836 &        3       32 20.78 &      -27       48 36.8 &  1.30 & 0.12 &  10.5 &  2.41 & 0.25 &  1.0 & \nodata & \nodata\cr
  37 &  ALMA033235-275319 &        3       32 35.13 &      -27       53 19.7 &  1.52 & 0.13 &  11.0 &  2.35 & 0.28 & \nodata &     39 &   1.6\cr
  38 &  ALMA033222-274811 &        3       32 22.16 &      -27       48 11.6 &  2.16 & .091 &  23.5 &  2.27 M & .090 & \nodata &      5 &   4.2\cr
  39 &  ALMA033229-275257 &        3       32 29.83 &      -27       52 57.7 &  1.72 & 0.14 &  12.1 &  2.26 & 0.18 & \nodata &     40 &   1.6\cr
  40 &  ALMA033231-274623 &        3       32 31.46 &      -27       46 23.5 &  1.72 & 0.13 &  12.8 &  2.26 & 0.17 & 0.47 &     37 &   2.4\cr
  41 &  ALMA033241-275131 &        3       32 41.47 &      -27       51 31.8 &  1.71 & 0.14 &  11.8 &  2.25 & 0.18 & \nodata &     42 &   3.3\cr
  42 &  ALMA033221-275112 &        3       32 21.99 &      -27       51 12.2 &  1.71 & 0.13 &  12.4 &  2.25 & 0.18 & \nodata &     54 &   3.1\cr
  43 &  ALMA033216-275247 &        3       32 16.52 &      -27       52 47.0 &  1.78 & 0.32 &  5.4 &  2.23 & 0.41 & \nodata &     28 &   2.1\cr
  44 &  ALMA033220-275024 &        3       32 20.91 &      -27       50 24.7 &  1.68 & .096 &  17.3 &  2.21 & 0.12 & \nodata &     26 &   2.9\cr
  45 &  ALMA033209-275015 &        3       32 9.860 &      -27       50 15.7 &  2.05 & 0.19 &  10.5 &  2.20 M & 0.23 & \nodata & \nodata & \nodata\cr
  [46] &  ALMA033225-274219 &        3       32 25.17 &      -27       42 19.0 &  1.88 & .069 &  26.9 &  2.08 M & .085 & \nodata & \nodata & \nodata\cr
  [47] &  ALMA033239-275326 &        3       32 39.24 &      -27       53 26.0 &  1.56 & 0.11 &  13.0 &  2.05 & 0.15 & \nodata & \nodata & \nodata\cr
  48 &  ALMA033238-274634 &        3       32 38.55 &      -27       46 34.5 &  1.63 & 0.28 &  5.6 &  2.04 & 0.36 &  1.1 &     50 &  0.93\cr
  49 &  ALMA033212-275209 &        3       32 12.88 &      -27       52 9.40 &  1.41 & 0.11 &  12.5 &  1.98 & 0.23 & \nodata &     32 &  0.71\cr
  50 &  ALMA033221-274241 &        3       32 21.49 &      -27       42 41.9 &  1.23 & 0.22 &  5.4 &  1.97 & 0.45 & \nodata & \nodata & \nodata\cr
  51 &  ALMA033216-274344 &        3       32 16.27 &      -27       43 44.0 & 0.85 & 0.10 &  7.9 &  1.94 & 0.22 & \nodata &     15 &   1.5\cr
  52 &  ALMA033215-275145 &        3       32 15.55 &      -27       51 45.4 &  1.20 & 0.11 &  10.2 &  1.88 & 0.24 & \nodata & \nodata & \nodata\cr
  53 &  ALMA033247-275038 &        3       32 47.73 &      -27       50 38.2 &  1.09 & 0.15 &  6.9 &  1.86 & 0.32 & \nodata &     23 &   2.3\cr
  54 &  ALMA033243-274851 &        3       32 43.67 &      -27       48 51.1 &  1.45 & 0.24 &  5.9 &  1.82 & 0.30 & \nodata & \nodata & \nodata\cr
  55 &  ALMA033211-274613 &        3       32 11.61 &      -27       46 13.1 &  1.36 & 0.11 &  11.8 &  1.79 & 0.15 & \nodata &     16 &   3.0\cr
  56 &  ALMA033225-274306 &        3       32 25.69 &      -27       43 6.00 &  1.24 & 0.16 &  7.6 &  1.78 M & 0.43 & \nodata &     47 &   2.6\cr
  57 &  ALMA033207-274900 &        3       32 7.950 &      -27       49 0.40 &  1.30 & 0.20 &  6.4 &  1.72 & 0.26 & \nodata & \nodata & \nodata\cr
  58 &  ALMA033244-275011 &        3       32 44.07 &      -27       50 11.4 &  1.37 & 0.24 &  5.5 &  1.72 & 0.31 & \nodata & \nodata & \nodata\cr
  59 &  ALMA033222-274815 &        3       32 22.57 &      -27       48 15.1 &  1.43 & 0.13 &  10.2 &  1.66 M & 0.13 & \nodata & \nodata & \nodata\cr
  60 &  ALMA033229-275335 &        3       32 29.90 &      -27       53 35.8 & 0.73 & 0.12 &  5.9 &  1.61 & 0.25 & \nodata &     24 &   2.9\cr
  [61] &  ALMA033231-274313 &        3       32 31.86 &      -27       43 13.0 &  1.22 & 0.19 &  6.4 &  1.61 & 0.25 & \nodata & \nodata & \nodata\cr
  62 &  ALMA033219-274315 &        3       32 19.36 &      -27       43 15.1 &  1.20 & 0.13 &  8.8 &  1.59 & 0.17 & \nodata &     35 &  0.29\cr
  63 &  ALMA033228-274829 &        3       32 28.80 &      -27       48 29.7 & 0.76 & 0.12 &  5.9 &  1.57 & 0.26 & \nodata & \nodata & \nodata\cr
  [64] &  ALMA033228-275229 &        3       32 28.10 &      -27       52 29.7 & 0.82 & 0.15 &  5.3 &  1.53 & 0.31 & \nodata & \nodata & \nodata\cr
  65 &  ALMA033231-275028 &        3       32 31.55 &      -27       50 28.9 &  1.11 & 0.11 &  9.7 &  1.46 & 0.14 & \nodata &     44 &   2.7\cr
  66 &  ALMA033210-274807 &        3       32 10.73 &      -27       48 7.29 & 0.67 & 0.13 &  5.1 &  1.44 & 0.26 & \nodata & \nodata & \nodata\cr
  [67] &  ALMA033217-274908 &        3       32 17.28 &      -27       49 8.40 &  1.03 & 0.14 &  7.0 &  1.36 & 0.19 & \nodata & \nodata & \nodata\cr
  68 &  ALMA033228-274431 &        3       32 28.91 &      -27       44 31.4 &  1.02 & 0.18 &  5.5 &  1.35 & 0.24 & \nodata &     38 &   2.3\cr
  69 &  ALMA033227-275311 &        3       32 27.15 &      -27       53 11.8 & 0.78 & 0.13 &  5.7 &  1.25 & 0.27 & \nodata &     29 &   1.3\cr
  70 &  ALMA033233-275222 &        3       32 33.90 &      -27       52 22.2 & 0.73 & 0.12 &  5.7 &  1.18 & 0.25 & \nodata &     31 &   3.8\cr
  71 &  ALMA033213-274754 &        3       32 13.64 &      -27       47 54.2 & 0.77 & 0.14 &  5.2 &  1.16 & 0.30 & \nodata & \nodata & \nodata\cr
  72 &  ALMA033228-274435 &        3       32 28.78 &      -27       44 35.2 & 0.75 & 0.14 &  5.2 &  1.11 & 0.29 & \nodata &     38 &   4.1\cr
  73 &  ALMA033234-275226 &        3       32 34.28 &      -27       52 26.7 & 0.81 & 0.13 &  6.1 &  1.07 & 0.17 & \nodata &     31 &   2.8\cr
  74 &  ALMA033222-274935 &        3       32 22.47 &      -27       49 35.2 & 0.55 & 0.11 &  4.8 & 0.93 & 0.23 & \nodata &      8 &   1.9\cr
  75 &  ALMA033217-274713 &        3       32 17.96 &      -27       47 13.6 & 0.64 & 0.11 &  5.6 & 0.84 & 0.14 & \nodata & \nodata & \nodata\cr
\enddata
\tablecomments{
The columns are (1) ALMA source number, (2) ALMA source name, (3 and 4) ALMA R.A. and Decl., 
(5, 6, and 7) peak flux, error, and S/N measured from the ALMA images,
(8 and 9) best ALMA flux estimates and errors (see text for details),
(10) ALMA 1.13~mm flux from F18 based on their GALFIT values and without applying any
deboosting, (11) SCUBA-2 counterpart source number from 
Table~2 (for offsets less than $5''$), and (12) offset between the
nearest SCUBA-2 source position and the ALMA position for these sources.
In Column~1, six ALMA source numbers (20, 46, 47, 61, 64, 67) are in square brackets, 
since they do not come from the full central SCUBA-2 sample (see Table~3 for description).
In Column~8, the four archival sources from Hodge et al.\ (2013)
are marked with an `H', the eight Mullaney et al.\ (2015) sources with an `M', the Barro
source with a `B', and the Schreiber et al.\ (2017) source with an `S'.
}
\label{tab4}
\end{deluxetable*}

\clearpage

\startlongtable
\begin{deluxetable*}{ccccccccccc}
\renewcommand\baselinestretch{1.0}
\tablewidth{0pt}
\tablecaption{ALMA Redshifts}
\scriptsize
\tablehead{No. & Name & $m_{850lp}$ & $m_{160w}$ &  $K_s$ & 4.5~$\mu$m &  8~$\mu$m  & CANDELS & Specz & Photz & FIRz\\ & & & & & & & offset (arcsec)  & & & \\ (1) & (2) & (3) & (4) & (5) & (6) & (7) & (8) & (9) & (10) & (11)}
\startdata
   1 &  ALMA033207275120 & \nodata &  21.3 &  20.8 &  19.8 &   20.0 & 0.67  & \nodata & \nodata &      3.26\cr
   2 &  ALMA033211275212 & \nodata & \nodata &  23.5 & \nodata &  \nodata & 0.33  & \nodata &   4.45 &      3.75\cr
   3 &  ALMA033215275037 & \nodata &  22.3 &  21.3 &  21.1 &   20.2 & 0.17  & \nodata &   3.12 &      2.80\cr
   4 &  ALMA033204274647 &  25.1 &  21.9 &  21.3 &  19.7 &   19.7 & 0.14  &   2.252 4 &   1.95 &      2.84\cr
   5 &  ALMA033228274658 &  25.7 &  22.9 &  22.2 &  20.9 &   20.6 & 0.08  &   2.309 7 &   2.29 &      2.16\cr
   6 &  ALMA033246275120 &  23.5 &  22.7 &  22.0 &  20.8 &   20.2 & 0.15  & \nodata & \nodata &      3.11\cr
   7 &  ALMA033238274401 & \nodata &  24.4 &  23.3 &  21.7 &   20.8 & 0.15  & \nodata &   3.48 &      3.36\cr
   8 &  ALMA033225275230 &  27.3 &  23.7 &  22.7 &  20.8 &   20.3 & 0.10  & \nodata &   2.69 &      3.26\cr
   9 &  ALMA033235274916 &  24.4 &  22.2 &  21.7 &  20.6 &   19.6 & 0.11  &   2.576 5 &   2.55 &      2.65\cr
  10 &  ALMA033219274602 & \nodata &  23.6 &  22.6 &  20.7 &   20.4 & 0.39  & \nodata &   2.41 &      2.78\cr
  11 &  ALMA033219275214 &  26.0 &  24.5 &  23.1 &  21.5 &   21.0 & 0.20  & \nodata & \nodata &      3.32\cr
  12 &  ALMA033234274940 &  25.4 &  24.3 &  23.5 &  21.9 &   21.1 & 0.05  & \nodata &   3.76 &      3.32\cr
  13 &  ALMA033217275233 &  27.1 &  26.3 &  26.3 &  23.2 &   22.6 & 0.47  & \nodata & [  2.73] &      3.69\cr
  14 &  ALMA033222274936 &  26.7 &  23.9 &  23.3 &  21.7 &   21.2 & 0.45  & \nodata &   2.73 &      2.11\cr
  15 &  ALMA033205274820 & \nodata &  23.4 &  22.5 &  20.8 &   20.7 & 0.09  & \nodata &   2.14 &      2.96\cr
  16 &  ALMA033219275159 &  26.9 &  24.2 &  22.5 &  21.2 &   20.9 & 0.19  & \nodata &   3.37 &      4.34\cr
  17 &  ALMA033235275215 & \nodata &  26.3 &  24.5 &  22.4 &   21.5 & 0.13  & \nodata &   3.57 &      7.99\cr
  18 &  ALMA033222274804 &  26.8 &  24.2 &  23.7 &  22.3 &   21.6 & 0.31  &   3.847 8 &   2.96 &      3.51\cr
  19 &  ALMA033226275208 & \nodata & \nodata & -25.7 &  23.5 &  \nodata & \nodata  & \nodata & [  4.47] &      6.68\cr
  20 &  ALMA033247274452 &  24.9 &  22.4 &  21.8 &  20.7 &   20.8 & 0.07  & \nodata &   1.93 &      2.02\cr
  21 &  ALMA033242275212 &  26.6 &  24.7 &  24.0 &  22.3 &   21.3 & 0.18  & \nodata &   3.78 &      4.17\cr
  22 &  ALMA033244274635 &  24.3 &  24.8 &  22.8 &  21.3 &   20.7 & 0.08  & \nodata & \nodata &      2.60\cr
  23 &  ALMA033237275000 &  23.3 &  21.8 &  20.9 &  19.9 &   20.2 & 0.14  & \nodata &   1.58 &      2.03\cr
  24 &  ALMA033224275334 &  25.4 &  23.1 &  22.6 &  21.1 &   21.2 & 0.25  & \nodata &   1.96 &      2.11\cr
  25 &  ALMA033243274639 & \nodata &  24.6 &  23.6 &  21.7 &   21.0 & 0.13  &   2.794 8 &   2.92 &      3.56\cr
  26 &  ALMA033216275044 &  25.2 &  24.3 &  23.6 &  22.5 &   21.7 & 0.02  & \nodata &   3.78 & \nodata\cr
  27 &  ALMA033203275039 & \nodata & \nodata &  23.5 &  22.1 &  \nodata & \nodata  & \nodata & [  1.78] &      4.60\cr
  28 &  ALMA033233275326 & \nodata &  25.8 &  24.5 &  22.1 &   21.4 & 0.21  & \nodata & [  3.33] &      3.62\cr
  29 &  ALMA033232274540 &  21.6 &  20.3 &  20.0 &  20.0 &   20.2 & 0.06  & \nodata & \nodata & \nodata\cr
  30 &  ALMA033217275037 &  25.2 &  22.5 &  21.7 &  20.5 &   20.8 & 0.01  & \nodata &   1.86 &      2.74\cr
  31 &  ALMA033218275134 &  26.9 &  22.3 &  21.6 &  20.3 &   20.4 & 0.19  & \nodata &   1.95 &      2.85\cr
  32 &  ALMA033211274615 &  25.8 &  23.6 &  23.4 &  22.2 &   21.8 & 0.41  & \nodata &   2.75 &      2.52\cr
  33 &  ALMA033217275003 &  24.1 &  21.8 &  21.0 &  19.9 &   20.1 & 0.21  & \nodata &   1.58 &      2.09\cr
  34 &  ALMA033221274656 &  26.1 &  23.5 &  22.8 &  21.0 &   21.3 & 0.12  & \nodata &   1.95 &      2.23\cr
  35 &  ALMA033222274243 &  24.0 &  22.3 &  21.7 &  20.5 &   20.4 & 0.18  &   1.612 7 &   1.71 &      2.22\cr
  36 &  ALMA033220274836 &  26.9 &  24.1 &  22.9 &  21.2 &   20.9 & 0.24  & \nodata &   2.37 &      3.30\cr
  37 &  ALMA033235275319 & \nodata &  24.7 &  23.8 &  20.9 &   20.9 & 0.12  & \nodata &   2.96 & \nodata\cr
  38 &  ALMA033222274811 &  26.3 &  23.6 &  22.6 &  20.9 &   20.6 & 0.05  & \nodata &   2.31 &      3.07\cr
  39 &  ALMA033229275257 & \nodata &  25.0 &  24.2 &  22.3 &   21.6 & 0.03  & \nodata &   3.04 &      2.20\cr
  40 &  ALMA033231274623 &  22.8 &  21.3 &  20.8 &  20.0 &   19.1 & 0.06  &   2.223 5 &   2.22 &      2.53\cr
  41 &  ALMA033241275131 & \nodata &  26.5 &  25.4 &  23.3 &   22.3 & 0.34  & \nodata & [  4.13] & \nodata\cr
  42 &  ALMA033221275112 &  25.6 &  24.1 &  23.0 &  20.8 &   20.8 & 0.12  & \nodata &   2.34 &      2.61\cr
  43 &  ALMA033216275247 &  25.8 &  22.8 &  22.1 &  20.8 &   20.8 & 0.11  & \nodata &   2.39 &      3.01\cr
  44 &  ALMA033220275024 & \nodata & \nodata &  23.0 &  22.6 &  \nodata & \nodata  & \nodata & \nodata &      5.26\cr
  45 &  ALMA033209275015 & \nodata &  26.3 &  25.3 &  22.6 &   21.0 & 0.22  & \nodata & [  7.62] &      3.01\cr
  46 &  ALMA033225274219 &  22.1 &  21.4 &  20.2 &  19.2 &   18.8 & 0.11  &   1.613 1 &   1.69 &      1.73\cr
  47 &  ALMA033239275326 &  25.4 &  22.8 &  22.0 &  20.9 &   19.5 & 0.08  & \nodata &   2.19 &      2.06\cr
  48 &  ALMA033238274634 &  24.4 &  23.3 &  22.9 &  21.8 &   21.3 & 0.32  &   2.543 6 &   2.58 &      1.85\cr
  49 &  ALMA033212275209 &  25.9 &  22.9 &  21.9 &  20.4 &   20.6 & 0.10  & \nodata &   1.87 &      1.91\cr
  50 &  ALMA033221274241 &  25.2 &  22.1 &  21.6 &  20.1 &   20.2 & 0.14  & \nodata &   1.69 &      2.59\cr
  51 &  ALMA033216274344 & \nodata &  23.1 &  22.7 &  20.7 &   20.4 & 0.28  & \nodata &   2.32 &      3.30\cr
  52 &  ALMA033215275145 & \nodata &  26.7 &  24.6 &  22.7 &   22.0 & 0.23  & \nodata & [  4.78] &      3.66\cr
  53 &  ALMA033247275038 &  23.7 &  21.5 &  20.9 &  19.7 &   19.9 & 0.07  & \nodata &   1.56 &      2.08\cr
  54 &  ALMA033243274851 & \nodata &  25.3 &  24.1 &  22.6 &   20.8 & 0.05  & \nodata & [  9.42] &      1.86\cr
  55 &  ALMA033211274613 &  22.7 &  22.9 &  21.8 &  21.2 &   21.3 & 0.26  & \nodata & \nodata &      2.77\cr
  56 &  ALMA033225274305 &  24.8 &  22.3 &  21.7 &  20.2 &   18.3 & 0.12  &   2.299 3 & [  1.55] &      2.92\cr
  57 &  ALMA033207274900 & \nodata &  25.0 &  24.8 &  23.1 &   22.2 & 0.05  & \nodata &   3.08 & \nodata\cr
  58 &  ALMA033244275011 & \nodata &  26.1 &  26.9 &  22.9 &   99.0 & \nodata  & \nodata & [  4.73] & \nodata\cr
  59 &  ALMA033222274815 &  25.8 &  23.2 &  22.1 &  20.9 &   20.8 & 0.06  & \nodata &   2.31 & \nodata\cr
  60 &  ALMA033229275335 & \nodata &  23.4 &  22.7 &  21.0 &   20.8 & 0.05  & \nodata &   2.53 & \nodata\cr
  61 &  ALMA033231274313 & \nodata & \nodata &  24.7 &  23.5 &  \nodata & \nodata  & \nodata & [  4.67] & \nodata\cr
  62 &  ALMA033219274315 &  25.4 &  23.7 &  23.4 &  21.9 &   21.4 & 0.17  & \nodata &   2.94 & \nodata\cr
  63 &  ALMA033228274829 &  26.3 &  22.7 &  21.8 &  20.3 &   20.4 & 0.20  & \nodata &   1.83 & \nodata\cr
  64 &  ALMA033228275229 & \nodata &  25.5 &  24.8 &  23.3 &   22.6 & 0.37  & \nodata &   3.26 & \nodata\cr
  65 &  ALMA033231275028 &  23.9 &  21.3 &  20.8 &  19.7 &   19.9 & 0.10  & \nodata &   1.58 & \nodata\cr
  66 &  ALMA033210274807 &  21.9 &  20.6 &  19.9 &  19.8 &   19.4 & 0.05  &  0.6540 5 &  0.680 & \nodata\cr
  67 &  ALMA033217274908 &  27.1 &  23.5 &  22.6 &  21.2 &   21.3 & 0.19  & \nodata &   1.69 & \nodata\cr
  68 &  ALMA033228274431 &  27.3 & \nodata &  23.0 & \nodata &  \nodata & \nodata  & \nodata & \nodata & \nodata\cr
  69 &  ALMA033227275311 &  26.3 &  23.1 &  22.7 &  21.5 &   21.2 & 0.03  & \nodata &   2.55 & \nodata\cr
  70 &  ALMA033233275222 &  27.9 &  25.1 &  23.5 &  22.3 &   21.7 & 0.38  & \nodata &   3.14 & \nodata\cr
  71 &  ALMA033213274754 &  23.7 &  22.0 &  21.6 &  20.5 &   20.7 & 0.10  & \nodata &   1.71 & \nodata\cr
  72 &  ALMA033228274435 & \nodata &  25.3 &  23.7 &  21.8 &   20.8 & 0.16  & \nodata & [  3.76] & \nodata\cr
  73 &  ALMA033234275226 &  26.1 &  25.6 &  22.7 &  21.8 &   22.5 & 0.19  & \nodata &   2.19 & \nodata\cr
  74 &  ALMA033222274935 &  21.5 &  20.2 &  19.6 &  19.5 &   19.4 & 0.10  &  0.7323 1 &  0.766 & \nodata\cr
  75 &  ALMA033217274713 &  25.0 & \nodata &  22.5 & \nodata &  \nodata & \nodata  & \nodata & \nodata & \nodata\cr
\enddata
\tablecomments{
The columns are (1) ALMA source number, (2) ALMA source name, (3 and 4) {\em HST\/} ACS
F850LP and WFC3 F160W magnitudes from G13, (5) ground-based $K_s$ magnitude
measured from the image of Hsieh et al.\ (2012),
(6 and 7) {\em Spitzer\/} 4.5 and 8~$\mu$m magnitudes from Ashby et al.\ (2015), 
(8) offset between the ALMA and G13 source positions, 
(9) specz and reference number (see details below),
(10) photz from S16 (poor quality flag $Q>3$ estimates are in 
square brackets to distinguish them from the more reliable $Q<3$ estimates; 
see Section~\ref{secredshifts}), and (11) FIRz for the
59 sources with ALMA fluxes above 1.65~mJy. There is no entry in Column~(11) for
the seven excluded sources (26, 29, 37, 41, 57, 58, and 59) with probability $<5$\% 
for the $\chi^2$ fits, as discussed in Section~\ref{secFIRz}.
In Column~9, `1' indicates the redshift is from our own Keck DEIMOS observations, 
`2' from K20 (Mignoli et al.\ 2005), `3' from MOSDEF (Kriek et al.\ 2015), 
`4' from Casey et al.\ (2012), `5' from Szokoly et al.\ (2004), `6' from Inami et al.\ (2017), 
`7' from Kurk et al.\ (2013), and `8' from F18 (private communications from Mobasher and Brammer).
}
\label{tab5}
\end{deluxetable*}

\begin{deluxetable*}{cccccccccc}
\renewcommand\baselinestretch{1.0}
\tablewidth{0pt}
\tablecaption{Candidate High-Redshift ALMA Sources Above 1.65~mJy}
\scriptsize
\tablehead{No. & Name & $f_{850\,\mu {\rm m}}$  & $f_{4.5}/f_{850}$  &  $f_{24}/f_{850}$  & $f_{100}/f_{850}$  & $f_{250}/f_{850}$ & $\log f_{2-7~{\rm keV}}$ & Photz & FIRz \\ & & (mJy) & & & & &(erg~cm$^{-2}$~s$^{-1}$) & & \\ (1) & (2) & (3) & (4) & (5) & (6) & (7) & (8) & (9) & (10) }
\startdata
   2 &  ALMA033211275212 &   8.83 & 0.00069 & 0.011 &  0.113 &  1.81 & \nodata &   4.45 (4.02 -  4.74) &      3.75 (3.30 -  4.71)\cr
  16 &  ALMA033219275159 &   3.91 & 0.0030 & 0.0039 &  0.0309 &  1.50 & -15.8 &   3.37 (3.08 -  3.61) &      4.34 (4.23 -  4.50)\cr
  17 &  ALMA033235275215 &   3.80 & 0.00097 & 0.0069 &  0.0243 & 0.179 & -15.4 &   3.57 (3.29 -  4.09) &      7.99 (5.79 -  7.99)\cr
  19 &  ALMA033226275208 &   3.62 & 0.00038 & 0.0024 &  0.0847 & 0.287 & -16.0 &   4.47 (3.60 -  6.63) &      6.68 (5.07 -  7.99)\cr
  21 &  ALMA033242275212 &   3.55 & 0.0011 & 0.0047 &  0.140 &  1.79 & \nodata &   3.78 (3.48 -  4.07) &      4.17 (3.05 -  5.50)\cr
  27 &  ALMA033203275039 &   3.05 & 0.0016 & 0.011 &  0.163 &  1.12 & \nodata &   1.78 (1.50 -  4.21) &      4.60 (3.05 -  5.79)\cr
  44 &  ALMA033220275024 &   2.21 & 0.0013 & 0.0057 &  0.198 &  1.08 & \nodata & \nodata &      5.26 (3.23 -  7.26)\cr
\enddata
\tablecomments{
The columns are (1) ALMA source number, (2) ALMA source name, 
(3) ALMA 850~$\mu$m flux, (4-7) 4.5, 24, 100, and 250~$\mu$m to 850~$\mu$m flux ratios,
(8) logarithm of the observed $2-7$~keV flux, (9) photz and 95\% confidence range,
and (10) FIRz and 95\% confidence range.
}
\label{tab6}
\end{deluxetable*}

\begin{deluxetable*}{ccccccc}
\renewcommand\baselinestretch{1.0}
\tablewidth{0pt}
\tablecaption{Hard Band Detected X-ray Subsample of 21 Sources}
\scriptsize
\tablehead{No. & Name & Luo+17 & $f_{850\,\mu {\rm m}}$ & $\log L_{\rm 8-28~keV}$ & $\Gamma$ & Spectral \\  
& &  & (mJy) & (erg~s$^{-1}$) & & Class \\ 
(1) & (2) & (3) & (4) & (5) & (6) & (7)}
\startdata
   1 &  ALMA033207-275120 &  161 &  8.93 &   43.1 & 0.47  & \nodata \cr
   9 &  ALMA033235-274916 &  665 &  5.09 &   43.2 & 0.51  & Type~2 \cr
  16 &  ALMA033219-275159 &  355 &  3.91 &   43.2 & 0.95  & \nodata \cr
  17 &  ALMA033235-275215 &  656 &  3.80 &   44.3 & 0.10 & \nodata \cr
  19 &  ALMA033226-275208 &  471 &  3.62 &   43.4 & 0.11  & \nodata \cr
  22 &  ALMA033244-274635 &  804 &  3.38 &   43.4 &  1.9  & \nodata \cr
  26 &  ALMA033216-275044 &  298 &  3.15 &   42.8 & 0.85  & \nodata\cr
  34 &  ALMA033221-274656 &  385 &  2.47 &   43.0 & 0.83  & \nodata \cr
  38 &  ALMA033222-274811 &  392 &  2.27 &   43.6 & 0.49   & \nodata \cr
  40 &  ALMA033231-274623 &  586 &  2.26 &   42.9 & 0.83  & SF \cr
  42 &  ALMA033221-275112 &  386 &  2.25 &   43.3 & 0.22  & \nodata \cr
  45 &  ALMA033209-275015 &  194 &  2.20 &   43.4 & 0.85 & \nodata \cr
  46 &  ALMA033225-274219 &  448 &  2.08 &   43.7 &  1.8  & Type~2 \cr
  47 &  ALMA033239-275326 &  738 &  2.05 &   43.2 & 0.11 & \nodata \cr
  53 &  ALMA033247-275038 &  853 &  1.86 &   43.5 & 0.53 & \nodata \cr
  54 &  ALMA033243-274851 &  801 &  1.82 &   42.1 & 0.19 & \nodata \cr
  56 &  ALMA033225-274305 &  457 &  1.78 &   44.2 &  1.0 & \nodata \cr
  59 &  ALMA033222-274815 &  400 &  1.66 &   42.7 & 0.48 & SF \cr
  65 &  ALMA033231-275028 &  587 &  1.46 &   42.6 & .017 & \nodata \cr
  66 &  ALMA033210-274807 &  202 &  1.44 &   41.1 & 0.88 & SF \cr
  67 &  ALMA033217-274908 &  309 &  1.36 &   42.2 & 0.92 & \nodata \cr
\enddata
\tablecomments{
The columns are (1) ALMA source number, (2) ALMA source name, 
(3) corresponding Luo et al.\ (2017) X-ray catalog source number,
(4) ALMA 850~$\mu$m flux, (5) logarithm of the $8-28$~keV luminosity, (6) photon index $\Gamma$,
and (7) classification from spectroscopy.
}
\label{tab7}
\end{deluxetable*}

\end{document}